\documentclass[11pt,a4paper]{article}
\pdfoutput=1

\usepackage{jheppub}

\usepackage{makecell} 

\usepackage[T1]{fontenc}

\usepackage{amsmath,braket}
\usepackage{amssymb}
\usepackage{amsthm}
\usepackage{amsfonts}
\usepackage{amscd}
\usepackage{bbm}
\usepackage{array}
\usepackage{blindtext}
\usepackage{booktabs}
\usepackage{enumerate}
\usepackage[shortlabels]{enumitem}
\usepackage{fancyhdr} 
\usepackage{float}
\usepackage{mdframed}
\usepackage{graphicx}
\usepackage{latexsym}
\usepackage{lmodern}
\usepackage{mathrsfs}
\usepackage{makeidx}
\usepackage{dsfont}
\usepackage{multirow}
\usepackage{tensor}
\usepackage{slashed}
\usepackage{url}
\usepackage{xspace}
\usepackage{tikz-cd}
\usepackage{mathtools}
\usepackage{subcaption}
\usetikzlibrary{arrows.meta, positioning,decorations.pathmorphing}

\DeclareMathOperator{\Tr}{Tr}

\renewcommand{\O}{{\mathcal O}}

  \newcommand{\volS}{\Omega_{d-2}}
 \newcommand{\be}{\begin{equation}}
 \newcommand{\ee}{\end{equation}}
 \newcommand{\bea}{\begin{eqnarray}}
 \newcommand{\eea}{\end{eqnarray}}

 \newcommand{\bm}[1]{\mathbf{#1}}

    \newcommand{\ZZ}{\mathbb{Z}}

\newtheorem*{theorem*}{Theorem}
\newcommand{\PPP}{\mathcal{P}}

\usepackage{simplewick}
\usepackage{stmaryrd}

\title{Positivity in energy correlators and \\ the event distribution formula}

\author[1,2]{Alexandre Belin,}
\author[3]{Nathan Borak,}
\author[4]{Johan Henriksson,}
\author[5]{Romain Piron,}
\author[4]{Alexander Zhiboedov}

\affiliation[1]{Dipartimento di Fisica, Universit\`a di Milano - Bicocca \\
I-20126 Milano, Italy}
\affiliation[2]{INFN, sezione di Milano-Bicocca, I-20126 Milano, Italy}
\affiliation[3]{Department of Physics, Yale University, New Haven, Connecticut, USA}
\affiliation[4]{Theoretical Physics Department, CERN, 1211, Geneva, Switzerland}
\affiliation[5]{National Institute of Informatics, Chiyoda-ku, Tokyo, Japan}

\emailAdd{alexandre.belin@unimib.it}
\emailAdd{nathan.borak@yale.edu}
\emailAdd{johan.henriksson@cern.ch}
\emailAdd{piron@nii.ac.jp}
\emailAdd{alexander.zhiboedov@cern.ch}

\numberwithin{equation}{section}

\abstract{Energy correlators are universal observables, well defined across a wide range of theories and spacetime dimensions, from gauge theory and conformal field theory to string theory. Energy correlators are constrained by three fundamental properties: pointwise positivity of the energy flux, global positivity originating from their interpretation as state norms in a unitary theory, and energy conservation  organizing multi-point energy correlators into an infinite consistent hierarchy. We argue that the most general solution to the infinite hierarchy positivity problem of energy correlators is given by the event distribution formula, which expresses energy correlators as moments of the measure on the space of probability measures on the celestial sphere. We work out in detail implications of positivity for two- and three-point energy correlators and show that it implies nontrivial two-sided bounds on their multipole expansion coefficients. The bounds obtained by requiring consistency of the infinite
hierarchy of energy correlators are strictly stronger than those obtained by imposing positivity of the two- and three-point correlators alone. For the low-spin multipole coefficients studied in the paper, the derived bounds are optimal in the sense that their extrema are realized by finite mixtures of finite-particle events. We further demonstrate consequences of positivity in energy correlators in collider physics and conformal field theories. 
}

\begin{document}

\maketitle

\section{Introduction}

A collider experiment is usually pictured in terms of particles and their momenta. This definition is natural in theories with an $S$-matrix, but it is not fundamental. Interacting conformal field theories do not have particle asymptotic states, and four-dimensional theories with long-range forces need not admit a simple Fock-space description. The notion of a collider experiment and a collider event, however, still applies to these theories. Energy correlators provide a natural characterization of a collider experiment even in situations when the simple particle picture does not apply \cite{Hofman:2008ar}. They are defined as matrix element of the energy flux operator ${\cal E}(\vec n)$
\be
\langle {\cal E}(\vec n_1) \cdots  {\cal E}(\vec n_k) \rangle \equiv {\langle \psi| {\cal E}(\vec n_1) \cdots  {\cal E}(\vec n_k) | \psi \rangle \over \langle \psi| \psi \rangle}, ~~~ \vec n_i \in S^{d-2} .
\ee
The energy flux detector ${\cal E}(\vec n)$ measures how much energy reaches a given angular direction $\vec n$ at infinity.\footnote{In particle physics, a detector that measures the energy of particles is called a calorimeter. In this paper, we instead talk about the energy flux detectors to model idealized calorimeters with infinite angular and energy resolutions.} In what follows we will assume that the state $| \psi \rangle $ carries a definite momentum $q^\mu = (E_{\text{tot}}, \vec 0 )$ and is rotationally invariant.

In this paper we explore positivity properties of the energy correlators. They are positive in two ways:
\begin{itemize}
    \item \emph{Locally}, as matrix elements of the positive operator ${\cal E} (\vec n) \geq 0$, they are pointwise positive\footnote{Strictly speaking, \eqref{eq:multipos} also requires strong commutativity of energy flux operators which we assume in this paper and discuss in more detail below.}
\be
\label{eq:multiposintro}
\langle \mathcal E(\vec n_1)\mathcal E(\vec n_2)\cdots \mathcal E(\vec n_k)\rangle \geq 0 \ .
\ee
\item \emph{Globally}, when interpreted as norms in a unitary theory: for any $f(\vec n_1, ... , \vec n_{\ell})$, we define $| \psi' \rangle = \int \left(\prod_{i=1}^{\ell}d \Omega_{\vec n_i} {\cal E}(\vec n_i) \right) f(\vec n_1, \ldots  , \vec n_{\ell}) | \psi \rangle $, then
\begin{equation}
\begin{aligned}
\label{eq:unitaritycondintro}
\langle \psi' | \psi' \rangle &\geq0, \qquad \langle \psi' | {\cal E}(\vec n) | \psi' \rangle \geq 0 . 
\end{aligned}
\end{equation}
\end{itemize}
The combination of \eqref{eq:multiposintro} and \eqref{eq:unitaritycondintro} implies nontrivial constraints on the possible shapes of the energy correlators. 

In addition, in a physical theory each $k$-point energy correlator should be part of a consistent family of energy correlators:
\begin{itemize}
    \item \emph{Hierarchy consistency} expresses the fact that the integral of the energy flux operator over the celestial sphere is the time-translation symmetry generator
    \be
    \label{eq:hierconsist}
    \langle \mathcal E(\vec n_1)\mathcal E(\vec n_2)\cdots \mathcal E(\vec n_k)\rangle = {1 \over E_{\text{tot}}} \int d \Omega_{\vec n_{k+1}} \langle \mathcal E(\vec n_1)\mathcal E(\vec n_2)\cdots \mathcal E(\vec n_{k+1})\rangle .
    \ee
    \end{itemize}

\noindent 
To understand what is the most general solution to \eqref{eq:multiposintro}, \eqref{eq:unitaritycondintro}, \emph{and} \eqref{eq:hierconsist}, it is useful to recall how these conditions are satisfied in theories with particles. In this case, one expresses energy correlators through the manifestly positive probability of producing $n$-particle state, which are eigenstates of the energy flux operators 
\be
{\cal E}(\vec n) | \vec p_1 , ... , \vec p_n \rangle = \Big( \sum_{i=1}^n E_{\vec p_i} \ \delta^{(d-2)}\big(\vec n - {\vec p_i \over | \vec p_i|}\big) \Big) | \vec p_1 , ... , \vec p_n  \rangle \ .
\ee
This immediately suggests defining \emph{a collider event} more broadly as an eigenstate of the infinite family of commuting positive energy flux operators ${\cal E}(\vec n)$.

With this perspective in mind, we can immediately write the following generalization of the textbook cross section formula for energy correlators
\begin{equation}
\label{eq:representationEnCol}
\langle {\cal E}(f_1) \cdots  {\cal E}(f_k) \rangle  = \int \sigma(d\varepsilon) \prod_{i=1}^k \int_{S^{d-2}}  \varepsilon(d \Omega_{\vec n_i}) f_i(\vec n_i) , ~~~ \sigma(d\varepsilon) \geq 0, ~~~\varepsilon(d \Omega_{\vec n_i}) \geq 0, 
\end{equation}
where we have defined a smeared energy-flux operator ${\cal E}(f) \equiv \int_{S^{d-2}} d\Omega_{\vec n} f(\vec n) {\cal E}(\vec n)$. In the formula above $\varepsilon(d \Omega_{\vec n_i}) \in {\cal P}(S^{d-2})$ represents the energy distribution on the celestial sphere in a given collider event, and $\sigma(d\varepsilon)\in {\cal P}({\cal P}(S^{d-2}))$ is a positive measure on the space of energy distributions on the celestial sphere. In particular, energy conservation implies that $\int_{S^{d-2}} \varepsilon(d \Omega_{\vec n})= E_{\text{tot}}$, where $E_{\text{tot}}$ is the energy of the state $| \psi \rangle$.

As we will argue below, \emph{the event distribution formula} \eqref{eq:representationEnCol} does not rely on the existence of particles and follows from consistency of the \emph{infinite} hierarchy of energy correlators (with the precise assumptions specified in the main text). It presents the most general solution to the positivity constraints described above.  Our primary goal in this paper is to explore the constraints that this ``energy correlators as positive moments'' perspective places on the space of energy correlators and on the space of theories.

The event distribution viewpoint on energy correlators \eqref{eq:representationEnCol}  has been discussed in the past. Most directly, Appendix B of
Hofman and Maldacena \cite{Hofman:2008ar} introduced a formal probability functional for energy
distributions and used it to discuss energy correlators at strong coupling. Earlier, with the standard cross section representation of the energy correlators in mind, Sveshnikov and Tkachov \cite{Sveshnikov:1995vi,Tkachov:1995kk} formulated a
calorimetric $C$-algebra in which multi-point $C$-correlators, expressible through energy correlators, encode infrared-safe information about hadronic final states. More recently,
Komiske, Metodiev, and Thaler \cite{Komiske:2017aww} showed that $C$-correlators span the space of infrared- and collinear-safe observables and introduced energy-flow polynomials as a linear basis on this space. However, to the best of our knowledge, the constraints that follow from the positivity properties of energy correlators have not yet been systematically explored, see  \cite{Fox:1978vu,Fox:1978vw,Mecaj:2025ecl,Dempsey:2025yiv,Mecaj:2026kji} for work in this direction. We also note that the positive moments perspective on observables has been recently intensively explored in the context of scattering amplitudes starting from \cite{Adams:2006sv,Bellazzini:2020cot,Tolley:2020gtv,Caron-Huot:2020cmc,Arkani-Hamed:2020blm}. 

Given the event distribution representation \eqref{eq:representationEnCol}, the study of consistency of energy correlators is reduced to an interesting moment problem on the space of measures on the celestial sphere.\footnote{As we review in detail, the class of distributions defined by energy correlators is not exactly one of the standard classes studied in the mathematical literature, for instance those discussed in \cite{Gneiting2011}. We thank T. Gneiting for useful correspondence. Nevertheless, we will benefit from mathematical results obtained in the closely related context of the study of positive functions on spheres (a notion that should be appropriately extended in the context of energy correlators).}  A particularly convenient set of moments for energy correlators that we will focus on is given by their expansion in spherical harmonics.\footnote{These are linear combinations of the energy flow polynomials studied in \cite{Komiske:2017aww}.} By the Richter--Tchakaloff theorem, see e.g. Theorem 1.24 in 
\cite{Schmudgen2017}, 
every event distribution measure \eqref{eq:representationEnCol} can be replaced, while preserving any prescribed finite collection of moments, by a positive atomic measure with at most as many atoms as there are independent moments being kept.\footnote{An atomic measure is a finite positive-weighted sum of Dirac-delta measures.} By applying this theorem twice, first to $\sigma(d\varepsilon)\rightarrow\sum_{a=1}^{N_{\rm ev}}
w_a\,\delta_{\varepsilon_a}(d\varepsilon)$ (which replaces the measure by a finite number of events) and then $\varepsilon_a(d\Omega_{\vec n})\rightarrow\sum_{i=1}^{N_a}E_{ai}\,\delta_{\vec n_{ai}}(d\Omega_{\vec n})$ (which replaces each event by an energy flow generated by a finite number of particles or atoms), allows one to effectively explore the space of moments. As a result, whenever a bound depending on finitely many moments is
attained, there exists an atomic optimizer: a finite positive combination of events, each of which is a finite-particle event. In the examples below we construct such atomic optimizers explicitly. 

\subsection{Main results}

\begin{figure}
    \centering 
    \includegraphics[width=0.96\textwidth]{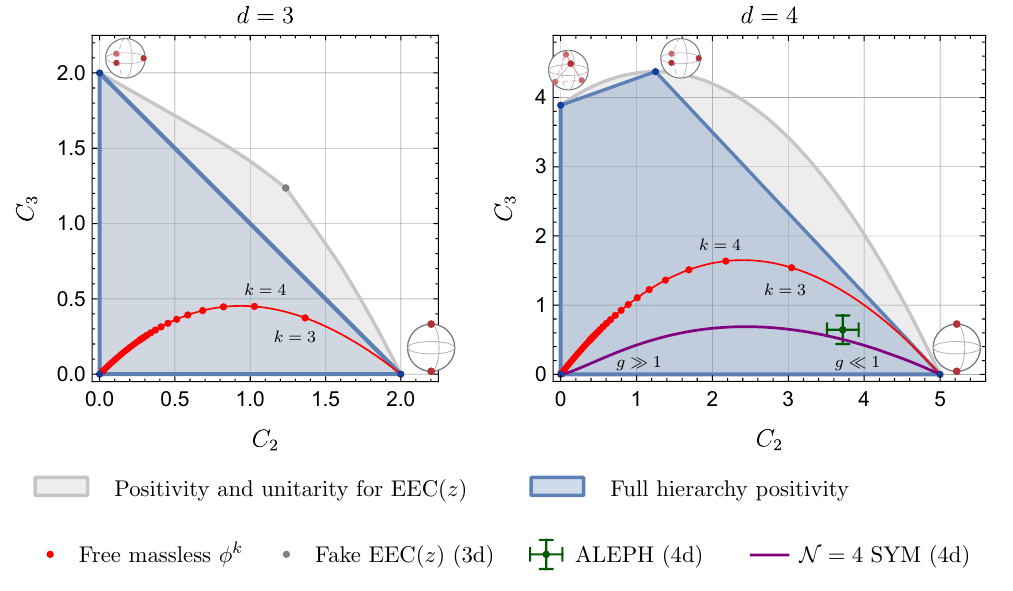}
    \caption{Two-sided bounds on the energy multipoles $(C_2,C_3)$ in $d=3$ and $d=4$, assuming the masslessness condition $C_1=0$ (for ALEPH, $C_1$ is non-zero but small). The blue region assumes full hierarchy and improves on previous bounds using positivity for two-point $\text{EEC}(z)$ only (gray region) in $d=3$ \cite{Mecaj:2026kji} and $d=4$ \cite{Dempsey:2025yiv} respectively. For these plots, the allowed regions are spanned by atomic models, and uniform distribution (at the origin). 
    Free theory: \eqref{eq:C2in3d}--\eqref{eq:C3in3d} (3d), \eqref{eq:C2anyk}--\eqref{eq:C3anyk} (4d). Fake $\text{EEC}(z)$ \eqref{eq:fakeAtom}, ALEPH (normalized to $C_0=1$) \eqref{eq:finalCj}, $\mathcal N=4$ SYM Pad\'{e} from \cite{Dempsey:2025yiv}.
    }
\label{fig:twosidedc2c3}
\end{figure}

\begin{figure}
    \centering 
    \includegraphics[width=0.88\textwidth]{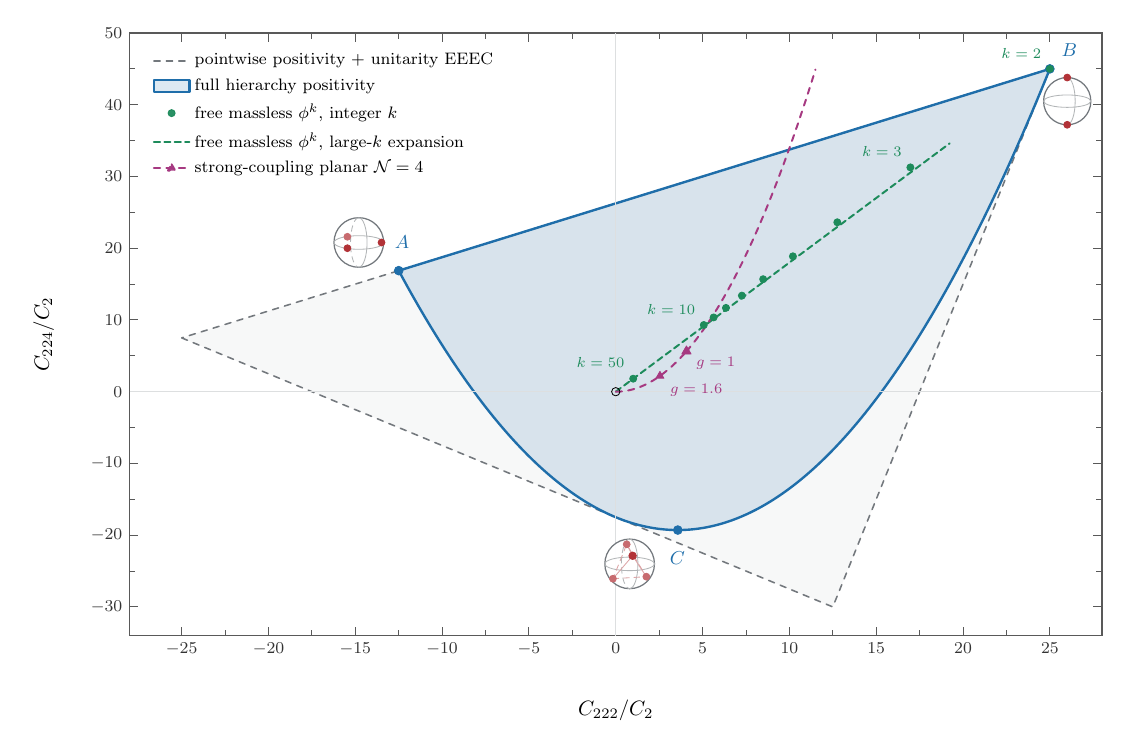}
    \caption{Universal two-sided bounds on the energy multipoles of the two- and three-point energy correlator (assuming the masslessness condition $C_1=0$). The gray triangle comes from imposing \emph{local} and \emph{global positivity} of the three-point energy correlator. The blue region is a stronger constraint that arises from imposing in addition \emph{the infinite hierarchy positivity}, or, equivalently, from the existence of the event distribution formula. We also plot the result in the free scalar theory for the state $\phi^k$ (green), where the dashed line is the large-$k$ prediction, and the leading stringy correction in ${\cal N}=4$ SYM at strong coupling $\lambda \to \infty$. These models are discussed in more detail in Section \ref{sec:physmodels}.}
\label{fig:twosidedbispectrumbound}
\end{figure}

\begin{enumerate}
  \item Assuming that energy correlators form a positive hierarchy that is permutation-symmetric in the detector insertions and consistent under integration of any detector over the celestial sphere, we show that this hierarchy is generated by a unique probability law on the space of probability measures on the celestial sphere, as expressed by the event distribution formula~\eqref{eq:representationEnCol}.  
  \item We consider the two-point energy multipoles $C_J$ and the three-point energy multipoles  $C_{J_1J_2J_3}$. It is known that the two-point energy correlator and its multipole expansion coefficients satisfy
  \begin{equation}
    \label{eq:conditions-2pt-intro}
        \mathrm{EEC}(z)\geq  0 \quad (\text{pointwise}), \qquad C_J\geq 0\quad (\text{unitarity}).
    \end{equation}
    We show how these two positivity conditions (local and global) constrain the possible shapes of the two-point energy correlator, as illustrated in Table~\ref{tab:admissible} in the main text. The two-sided bounds on the energy multipoles $C_J$ have been discussed in \cite{Fox:1978vu,Fox:1978vw,Dempsey:2025yiv,Mecaj:2025ecl,Mecaj:2026kji}, see Figure \ref{fig:twosidedc2c3}. Here, we show that the three-point multipoles $C_{J_1 J_2 J_3}$ also obey two-sided bounds in terms of the $C_J$. 
    
    The two-sided bounds following from local and global positivity correspond to the \emph{gray regions} in Figure \ref{fig:twosidedc2c3} (two-point multipoles) and Figure \ref{fig:twosidedbispectrumbound} (three-point multipoles). 
  
  \item We study the more stringent constraints coming from consistency of the full hierarchy to place stronger bounds on the multipoles: the \emph{blue regions} in Figure \ref{fig:twosidedc2c3} and Figure \ref{fig:twosidedbispectrumbound}. These are strictly smaller than the gray regions and constitute our main conceptual improvement over the two-sided bounds previously studied  in \cite{Fox:1978vu,Fox:1978vw,Dempsey:2025yiv,Mecaj:2025ecl,Mecaj:2026kji}. The positivity bounds on the energy multipoles can be saturated by finite mixtures of atomic events, as illustrated in these figures.

  \item We discuss different models of the energy correlators that approach the flat distribution: free gas model, stringy corrections, and matter loops in AdS. We show that at the event level they can be interpreted as soft and hard (rare) events respectively, see Table \ref{tab:modelsofhierarchies}.

  \begin{table}[ht]
    \centering
\begin{center}
\begin{tabular}{ | m{5cm} | m{1.5cm}| m{1.5cm} | m{1.5cm} | m{2.0cm} |} 
  \hline
  Model & $C_J$ & $C_{J_1 J_2 J_3}$ & $\hat C_{J_1 J_2 J_3}$ & Event \\ 
  \hline\hline
  Free gas $\phi^k$ $(k \to \infty)$ & $O(k^{-1})$ & $O(k^{-2})$ & $O(k^{-{1/2}})$ & Soft  \\ 
  \hline
  Stringy corrections $(\lambda \to \infty)$ & $O(\lambda^{-1})$ & $O(\lambda^{-3/2})$ &  $O(1)$ & Soft  \\ 
  \hline
  Loops in AdS $(c_T \to \infty)$ & $O(c_T^{-1})$ & $O(c_T^{-1})$ & $O(c_T^{1/2})$ & Hard (rare) \\ 
  \hline
\end{tabular}
\end{center}
    \caption{Event interpretation of different physical models discussed in the present section. The normalized non-Gaussianity is defined as $\hat C_{J_1 J_2 J_3} = {C_{J_1 J_2 J_3} \over \sqrt{C_{J_1} C_{J_2} C_{J_3}}}$. In the models considered it either goes to $0$, stays $O(1)$, or approaches $\infty$ and thus discriminates between the underlying physics.}
        \label{tab:modelsofhierarchies}
\end{table}

  \item We apply the two-point constraints to experimental $e^+e^-$ data from charged-track measurements by the ALEPH collaboration. Using the positivity properties of the two-point energy correlator, we derive a lower bound on the size of the contact term at zero angles.

  \item In a CFT, we consider polarization-averaged stress-tensor state, and show that detector positivity
        gives positive quadratic sum rules for $TT{\cal O}$ couplings, including
        generic spin-2 and spin-4 primaries, see Figure \ref{fig:two-panels-spin2}. We discuss implications of the energy-correlator positivity for large-$c_T$ CFTs and its natural relation to the species bound.
          \end{enumerate}

\noindent 
The plan of the paper is as follows. In Section~\ref{sec:posHierarchy} we review the basic properties of the energy correlators and reconstruct the event distribution formula from the full positive hierarchy of energy correlators. In Section~\ref{sec:twoPoint} and Section~\ref{sec:threePoint} we derive constraints on the energy multipole coefficients of the two- and three-point energy correlator respectively. In Section~\ref{sec:physmodels} we present three different physical models of energy-correlator hierarchies, which exhibit different relative scalings of energy-correlator non-Gaussianity, and we present an underlying event interpretation of these results. In Section \ref{sec:experimental} and Section \ref{sec:Ads} we apply the constraints to collider data and CFTs. We collect many of the technical derivations in the dedicated Appendices.

\section{Energy correlators as positive moments}
\label{sec:posHierarchy}

In this section, we introduce energy correlators and review their basic properties. We then argue that consistency of the infinite hierarchy of energy correlators leads to the event distribution representation which manifestly solves all the positivity constraints.

\subsection{Energy correlators: basic properties}
In quantum field theory, we can define the energy flux operator using the conserved stress-energy tensor, first introduced and studied in \cite{Sveshnikov:1995vi,Korchemsky:1997sy,Korchemsky:1999kt,Hofman:2008ar},
\begin{equation}
    \mathcal{E}(\vec{n}) = \lim_{r \to \infty} r^{d-2} \int^\infty_{0} dt \,T_{0i}n^i(t,r \vec{n}) \,,
\end{equation}
where $\vec n \in S^{d-2}$, i.e. $\vec n^2 =1$, is a point on the celestial sphere. The integral $\int_0^\infty dt$ stands for the working time of the detector, and $r$ can be thought of as a distance from the center of the experiment. In  gravitational theories, the energy flux operator can be expressed directly in terms of the asymptotic fields at infinity \cite{Gonzo:2020xza,Herrmann:2024yai}.

The most fundamental property of the energy flux operator is that it is nonnegative
\begin{equation}
\label{eq:pos}
{\cal E}(\vec n) \geq 0 . 
\end{equation}
This is an obvious statement when applied to theories with particles, and highly nontrivial in interacting conformal field theories (CFTs) \cite{Faulkner:2016mzt,Hartman:2016lgu}, see also \cite{Hofman:2016awc,Komargodski:2016gci,Kravchuk:2018htv}.\footnote{In CFTs, \emph{the average null energy condition} (ANEC) proven in \cite{Faulkner:2016mzt,Hartman:2016lgu} is equivalent to \eqref{eq:pos}, see e.g. \cite{Li:2025knf} for a detailed discussion of this point. In general QFT, e.g. in QCD, the collider experiment positivity \eqref{eq:pos} and the ANEC \emph{are not} equivalent. For example, the latter has been recently used to rederive the well-known monotonicity properties of the RG flows \cite{Hartman:2023qdn,Hartman:2023ccw}.} The energy flux operator represents the density of the energy-momentum in the following sense 
\begin{equation}
\label{eq:WI}
P^0 = \int_{S^{d-2}} d \Omega_{\vec n} {\cal E}(\vec n),
\end{equation}
where we have defined $n^\mu = (1, \vec n)$ and $P^\mu$ is the symmetry generator of translations. 
The positivity property \eqref{eq:pos} together with \eqref{eq:WI} imply that ${\cal E}(\vec n)$ annihilates the vacuum $|\Omega \rangle$, i.e. the translation-invariant state $P^\mu |\Omega \rangle = 0$ of the theory.

Another important property of the energy flux operators that we will assume in this paper is that they commute
\begin{equation}
\label{eq:comm}
[{\cal E}(\vec n_1) , {\cal E}(\vec n_2)]=0 .
\end{equation}
For $\vec n_1 \neq \vec n_2$, intuitively this is the statement about independence of measurements at spacelike separated points, but because the energy flux operators are integrated over infinite time, showing \eqref{eq:comm} requires extra care \cite{Kologlu:2019bco}. Similarly, contact terms $\vec n_1 = \vec n_2$ can be argued to vanish \cite{Cordova:2018ygx} (under plausible assumptions).

A related condition that holds is commutativity of the flux operators with the generators of translations\footnote{For $P^0$, it follows from \eqref{eq:comm} and \eqref{eq:WI}. For $P^i$ in CFTs it follows from the fact that $P^i = \int_{S^{d-2}} d \Omega_{\vec n} n^i {\cal E}(\vec n)$. For theories which admit the S-matrix description, it can also be easily checked.}
\begin{equation}
\label{eq:momentumcons}
[P^\mu, {\cal E}(\vec n)] = 0 .
\end{equation}
As a result, to study energy correlators it is convenient to consider momentum eigenstates
\begin{equation}
P^\mu | q \rangle = q^\mu | q \rangle ,
\end{equation}
which we could create, for example, by (the Fourier transform of) a local operator acting on the vacuum $\mathcal O(q) |\Omega \rangle$. 

A natural way to probe a quantum state, such as one produced by a local measurement, is through the pattern of energy flux it generates at infinity. Correlations of the energy flux measured in different directions define energy correlators. In high-energy physics, energy correlators play important roles in settings
ranging from precision QCD and collider event shapes to conformal collider
physics, holography, and quantum gravity; see \cite{Moult:2025nhu} for a recent
review. Thanks to \eqref{eq:momentumcons}, we then have
\begin{equation}
\langle q' | {\cal E}(\vec n_1) \cdots  {\cal E}(\vec n_k) | q \rangle =\delta^{(d)}(q'-q) \langle {\cal E}(\vec n_1) \cdots  {\cal E}(\vec n_k) \rangle \ , 
\end{equation}
where we normalize the states as $\langle q' | q \rangle = \delta^{(d)}(q'-q)$.\footnote{These are (as usual) the so-called improper states, and to construct a normalized state we need to consider a wave packet $\int d^d q \psi(q)| q \rangle$, with $\int d^d q |\psi(q)|^2 = 1$.} In this paper, we will consider $q^\mu = (E_{\text{tot}}, \vec 0)$, and focus on rotation-invariant states only.\footnote{In an actual collider experiment, the state is not rotationally invariant, however a situation equivalent to rotationally invariant states occurs if we average over all configurations with respect to the collider beam, keeping the angles between energy operators (detectors) fixed.} The one-point function in a rotation-invariant state is fixed by symmetries to be 
\begin{equation}
\label{eq:onepoint}
\langle {\cal E}(\vec n)\rangle = {E_{\text{tot}} \over \Omega_{d-2}} \ , 
\end{equation}
where $\Omega_{d-2} \equiv {\rm Vol}(S^{d-2}) = {2 \pi^{(d-1)/2} \over \Gamma({d-1 \over 2})}$, i.e. $\Omega_2=4\pi$ in $d=4$. 
More general rotation non-invariant states have been analyzed, for example, in \cite{Hofman:2008ar,Cordova:2017zej,Riembau:2025wjc,Riembau:2025isw}.

We will assume that the multi-point energy correlators are \emph{pointwise positive}
\be
\label{eq:multipos}
\langle \mathcal E(\vec n_1)\mathcal E(\vec n_2)\cdots \mathcal E(\vec n_k)\rangle \geq 0 \, .
\ee
Strictly speaking, pointwise positivity of multi-point energy correlators does not follow from \eqref{eq:comm} alone. A sufficient condition is, for example, \emph{strong commutativity}, $[e^{i t {\cal E}(f)},e^{i s {\cal E}(g)}]=0$, where ${\cal E}(f) \equiv \int d\Omega_{\vec n} f(\vec n) {\cal E}(\vec n)$, for suitable smearing functions $f,g$ and all $s,t\in \mathbb{R}$, see e.g. \cite{Schmudgen2012}. This condition is physically equivalent to the shock-wave commutativity
discussed in \cite{Kologlu:2019bco}. We will \emph{assume} in this paper that the multi-point energy correlators are pointwise nonnegative.

\subsection{Positive hierarchy of energy correlators}

Let us consider a rotation-invariant state with zero spatial momentum and energy $E_{\text{tot}} = 1$. Energy conservation and rotation symmetry fix 
\begin{equation} \label{1ptfctsec2}
\langle {\cal E}(\vec n) \rangle = {1 \over \Omega_{d-2}} \ ,
\end{equation}
where $\vec n_1^2 =1$ is a unit vector labeling the location of the energy detector $\vec n \in S^{d-2}$.
Let us define a smeared energy detector
\begin{equation}
{\cal E}(A) \equiv \int_{A} d \Omega_{\vec n} {\cal E}(\vec n) .
\end{equation}
Starting from the two-point function and higher, energy correlators are not fixed by symmetries. They however obey a series of inequalities
\begin{equation} \label{mostgenineq}
1 \geq \langle {\cal E}(A_1) \rangle \geq \langle {\cal E}(A_1) {\cal E}(A_2) \rangle \geq \ldots  \geq \langle {\cal E}(A_1) {\cal E}(A_2) \cdots  {\cal E}(A_k) \rangle \geq 0 \ . 
\end{equation}
This inequality follows from positivity of the multi-point energy correlators \eqref{eq:multipos}, and conservation of energy which implies that
\begin{equation}
\int d \Omega_{\vec n_k}\langle {\cal E}(\vec n_1) {\cal E}(\vec n_2) \cdots  {\cal E}(\vec n_k) \rangle = \langle {\cal E}(\vec n_1) {\cal E}(\vec n_2) \cdots  {\cal E}(\vec n_{k-1}) \rangle \ , 
\end{equation}
where the integration is over the full sphere.

A second set of inequalities comes from \emph{unitarity} of the underlying theory. For concreteness, let us start with the two- and three-point correlators. Then for any $f(\vec n)$ we have
\begin{equation} \label{eq:unitarity}
\begin{split}
\int d \Omega_{\vec n_1} d \Omega_{\vec n_2} f^*(\vec n_1) \langle {\cal E}(\vec n_1) {\cal E}(\vec n_2) \rangle f(\vec n_2) &\geq 0 \, , \\
\int d \Omega_{\vec n_1} d \Omega_{\vec n_3} f^*(\vec n_1) \langle {\cal E}(\vec n_1) {\cal E}(\vec n_2) {\cal E}(\vec n_3) \rangle f(\vec n_3) &\geq 0 \, \ .    
\end{split}
\end{equation}
The first inequality follows from the fact that it computes a norm $\langle \psi'| \psi' \rangle$ of the state $| \psi' \rangle = \int d \Omega_{\vec n_2}  f(\vec n_2) {\cal E}(\vec n_2) | \psi \rangle $. The second inequality follows from the positivity of the energy flux operator $\langle \psi' | {\cal E}(\vec n) | \psi' \rangle \geq 0$. The analogs of \eqref{eq:unitarity} exist for the even- and odd- $k$-point energy correlators respectively. We can consider an arbitrary function $f(\vec n_1, \ldots  , \vec n_{\ell})$ to get unitarity constraints (analogous to \eqref{eq:unitarity}) on the $2\ell$- and $(2\ell+1)$-point energy correlators: 
\begin{equation}
\begin{aligned}
\label{eq:unitaritycond}
\langle \psi' | \psi' \rangle &\geq0, \qquad \langle \psi' | {\cal E}(\vec n) | \psi' \rangle \geq 0 , \\
| \psi' \rangle &= \int \left(\prod_{i=1}^{\ell}d \Omega_{\vec n_i} {\cal E}(\vec n_i) \right) f(\vec n_1, \ldots  , \vec n_{\ell}) | \psi \rangle \ . 
\end{aligned}
\end{equation}
Finally, in certain cases there is an additional \emph{masslessness} condition
\begin{equation}
\label{eq:spatialmomentum}
\int d \Omega_{\vec n_1} n_1^i \langle {\cal E}(\vec n_1) {\cal E}(\vec n_2) \cdots  {\cal E}(\vec n_k) \rangle = 0  \qquad \text{(massless)} \,, 
\end{equation}
which is realized in physical theories with massless degrees of freedom \emph{only}, such as for example CFTs. This condition emerges as follows: in a theory with only massless degrees of freedom we have $P^i = \int d \Omega_{\vec n} n^i {\cal E}(\vec n)$, therefore if we consider states with zero spatial momentum \eqref{eq:spatialmomentum} follows.

We will be interested in exploring the constraints above in situations, where they lead to nontrivial predictions. However, let us first review how these positive hierarchies can be realized microscopically.

\subsection{Example: gapped theories}

The simplest way to realize all the conditions above is to consider a gapped theory that admits a description in terms of the S-matrix of massive particles, such as QCD. In this case we assume that behind the energy correlators, there is a positive cross section which can be used to calculate them. More precisely, we can write 
\begin{equation}
\label{eq:crosss}
\langle {\cal E}(\vec n_1) {\cal E}(\vec n_2) \cdots  {\cal E}(\vec n_k) \rangle = \sum_{n=2}^\infty \int d\sigma_{\psi \to n} \delta^{(d)}( q_{\text{tot}} - \sum_{j=1}^n q_i) \prod_{i=1}^k \left(\sum_{j=1}^n E_j \delta({\vec q_j \over |\vec q_j|} - \vec n_i) \right), ~~~ d\sigma_{\psi \to n} \geq 0 ,
\end{equation}
where $n$ is the number of particles in the final state and the integral goes over the relativistic $n$-particle phase space, and $q_{\text{tot}}$ is the energy-momentum vector of the state the correlator is calculated in. For any choice of nonnegative $d\sigma_{\psi \to n}$ the formula above defines a positive hierarchy described in the previous section. Let us check this explicitly.

First of all, we consider normalized states, such that $\langle 1 \rangle = 1$. We then get for the one-point function
\begin{equation}
\langle {\cal E}(\vec n_1) \rangle = \sum_{n=2}^\infty \int d\sigma_{\psi \to n} \delta^{(d)}(  q_{\text{tot}} - \sum_{j=1}^n q_i) \left(\sum_{j=1}^n E_j \delta({\vec q_j \over |\vec q_j|} - \vec n_1) \right).
\end{equation}
This expression is manifestly rotation-invariant and therefore is simply a constant. To fix it, let us notice that
\begin{equation}
\int d \Omega_{\vec n_1} \langle {\cal E}(\vec n_1) \rangle  =  \sum_{n=2}^\infty \int d\sigma_{\psi \to n} \delta^{(d)}( q_{\text{tot}} - \sum_{j=1}^n q_i) \left(\sum_{j=1}^n E_j \right) = E_{{\rm tot}} \langle 1 \rangle = 1 ,
\end{equation}
where we used that under the sum $E_{{\rm tot}} =\sum_i E_i$ due to energy conservation. 
Therefore, as expected, we recover $\langle {\cal E}(\vec n_1) \rangle  = {1 \over \Omega_{d-2}}$. 

Second, let us check the inequalities \eqref{mostgenineq}. They simply follow from the fact that in the sense of distributions we have
\begin{equation}
E_{{\rm tot}} =\sum_{j=1}^n E_j  \geq \int_{A} d \Omega_{\vec n} \sum_{j=1}^n E_j \delta\left({\vec q_j \over |\vec q_j|} - \vec n\right) \ .
\end{equation}
Finally, let us also check that it satisfies unitarity \eqref{eq:unitaritycond}. We choose $k$ to be even and we can perform the spherical integrals to get for the integrand
\begin{align} \label{eq:result of 2k integrals}
    &\sum_{j_1, \ldots , j_{k/2}=1}^n E_{j_1} \dots E_{j_{k/2}} f^* \left({\vec q_{j_1} \over |\vec q_{j_1}|}, \ldots  , {\vec q_{j_{k/2}} \over |\vec q_{j_{k/2}}|} \right) \nonumber \\
    \times &\sum_{j_{k/2+1}, \ldots , j_k=1}^n E_{j_{k/2 + 1}} \dots E_{j_{k}} f \left({\vec q_{j_{k/2+1}} \over |\vec q_{j_{k/2+1}}|}, \ldots  , {\vec q_{j_{k}} \over |\vec q_{j_{k}}|} \right)  \geq 0 , 
\end{align}
where we used $\delta({\vec q_j \over |\vec q_j|} - \vec n)$ to perform all the integrals over the sphere in \eqref{eq:unitaritycond}. When $k$ is odd, we have an additional nonnegative factor $\sum_{j=1}^n E_j \delta({\vec q_j \over |\vec q_j|} - \vec n_k)$, which does not change the conclusion.
This shows that nonnegative cross sections automatically satisfy all the constraints. 

A distinctive feature of the realization \eqref{eq:crosss} is that such energy correlators contain $\delta(\vec n_i - \vec n_j)$ contact terms at coincident points. These contact terms are important for ensuring the positivity properties of the energy correlators.

\subsection{Event distribution representation}
\label{sec:mathematical-char}

While physical theories with particles present concrete realizations of the solutions to the consistency conditions above, it is also interesting to see how the event distribution representation emerges abstractly from the basic properties of the energy correlators. 

It turns out that the existence of such a representation follows from classical theorems in probability theory. The most famous one, the so-called de Finetti theorem, states that if an infinite sequence of $\{ 0 , 1 \}$-valued random variables $X$ is \emph{exchangeable} (meaning that joint probability distributions are permutation-invariant), then it is representable as a mixture (or weighted average with some measure $\mu(p)$) of the probability distributions of independent and identically distributed sequences with $\text{P}(X=0)=p$ and $\text{P}(X=1)=1-p$. In this theorem, the positive measure $\mu(p)$ plays the role of the cross section, and independent identically distributed sequences play the role of collider events.

The key point is that energy correlators naturally define exchangeable joint probability distributions with the auxiliary random variable taking values on the celestial sphere $X = S^{d-2}$. The existence of the event distribution representation is then the statement of the Hewitt--Savage theorem \cite{Hewitt1955}, which is a continuous version of the de Finetti theorem. We review the theorem, its relationship to energy correlators, and its proof in Appendix \ref{app:Hewitt--Savage}.

When applied to the energy correlators, this theorem states that given the basic properties of multi-point positivity, commutativity and energy conservation,\footnote{In mathematical terms, we assume that the energy correlators define symmetric positive Borel measures on $(S^{d-2})^k$, consistent under marginalization.} the energy correlators admit the following representation
\begin{equation}
\label{eq:representation}
\langle {\cal E}(A_1) \cdots  {\cal E}(A_k) \rangle = \int \sigma(d\varepsilon) \prod_{i=1}^k \varepsilon(A_i) ,\qquad \sigma(d\varepsilon) \geq0,
\end{equation}
where $\varepsilon(A) \equiv \int_A \varepsilon(d \Omega_{\vec n})$.

The integral in \eqref{eq:representation} is over the space of energy distributions on the celestial sphere $\sigma(d\varepsilon)$, and $\varepsilon(A_i)$ measures the energy flux through region $A_i$ in a given energy distribution. In the example of the previous section, $\sigma(d\varepsilon)$ is the differential cross section and $\varepsilon(A_i)$ is the energy flux of a sum of free particles through a given region on the celestial sphere. The probability distribution on the space of events $\sigma(d\varepsilon)$ depends both on the theory and the state. Equivalently, we can also write a formula for the energy correlator smeared against arbitrary functions
\be
\label{eq:representationsmeared}
\langle {\cal E}(f_1) \cdots  {\cal E}(f_k) \rangle = \int \sigma(d\varepsilon) \prod_{i=1}^k \left( \int_{S^{d-2}} \varepsilon(d \Omega_{\vec n_i})  f_i (\vec n_i)  \right) ,
\ee
where we have used the notation
\be
\mathcal{E}(f) \equiv \int d \Omega_{\vec n} f(\vec n) {\cal E}(\vec n) \ .
\ee
We call \eqref{eq:representationsmeared}  \textbf{the event distribution formula}.

The representation \eqref{eq:representation} for energy correlators was discussed in Appendix B of \cite{Hofman:2008ar}, where it was assumed and then used to formally express $\sigma(d\varepsilon)$ through the path integral over the celestial sphere of the generating functional for energy correlators
\be
\label{eq:pathintegralcs}
\sigma(d\varepsilon) = \int \mathcal{D}\lambda  e^{- i\int d \Omega_{\vec n} \lambda(\vec n) \varepsilon(\vec n)} \langle e^{i \int d \Omega_{\vec n} \lambda(\vec n) \mathcal{E}(\vec n) } \rangle \ d\varepsilon .
\ee

Here we argued that the event distribution representation of the energy correlators follows naturally from their fundamental properties without referring to the formal path integral expression \eqref{eq:pathintegralcs}. In particular, such a representation should exist in CFTs and other interacting theories in the IR, such as 4d QED or gravity. In all of these cases, simple Fock-space formulas analogous to \eqref{eq:crosss} are not available,\footnote{In practice, when performing perturbative calculations one still formally expresses them in this way at the intermediate stages of the calculation using an IR regulator.} however we expect the event distribution formula \eqref{eq:representation} to hold. 
The advantage of \eqref{eq:representation} is that it makes the study of consistency of energy correlators a moment problem in the space of measures on the celestial sphere, and we can benefit from the available results in the mathematical literature, as well as from the collider experiment intuition. For the two-point energy correlator, we will use the results of Gneiting \cite{Gneiting2011}. For the three-point energy correlator, we will make use of the analysis of Buhmann and J\"ager \cite{BuhmannJager2022}.

Let us also quickly demonstrate that the unitarity conditions \eqref{eq:unitaritycond} immediately \emph{follow} from the positive measure representation of the energy correlators \eqref{eq:representationsmeared}. By plugging \eqref{eq:representationsmeared} into \eqref{eq:unitaritycond}, we get
\be
\langle \psi' | \psi' \rangle = \int\sigma(d \varepsilon) \left|\prod_{i=1}^{\ell} \int_{S^{d-2}} \varepsilon(d\Omega_{\vec n_i}) f(\vec n_1,\cdots, \vec n_{\ell}) \right|^2 \geq 0 , 
\ee
and 
\be
\langle \psi' | {\cal E}(\vec n) | \psi' \rangle = \int \sigma(d \varepsilon) \varepsilon(\vec n) \left|\prod_{i=1}^{\ell} \int_{S^{d-2}} \varepsilon(d\Omega_{\vec n_i}) f(\vec n_1,\cdots, \vec n_{\ell}) \right|^2 \geq 0 .
\ee

In this section we focused on the properties of the energy correlators in a given state. More generally, and assuming the existence of an underlying Hilbert space of the theory of interest, we expect that the energy correlators admit the event distribution representation \emph{at the operator level} via the application of the joint spectral theorem to the product of energy flux operators. In this case, instead of a positive measure $\sigma(d \varepsilon)$ we get projectors on energy-flux eigenstates on the celestial sphere.

\section{Constraints on the two-point energy correlator}
\label{sec:twoPoint}

In this section we study the constraints on the two-point
energy correlator in a rotation-invariant state, which depends on a single variable $z=\frac{1-\cos\theta}2$. 
The consistency, or bootstrap, problem for general
energy-energy correlators is therefore reduced to analyzing
functions, and more generally distributions, on the celestial
sphere that admit the \textbf{multipole expansion}
\eqref{eq:eec}, are locally \textbf{nonnegative}
\eqref{eq:positivityeec}, obey
\textbf{unitarity} \eqref{eq:unitarity2pt}, and \textbf{can be embedded} into a consistent hierarchy. As we review
next, these requirements lead to nontrivial constraints.

\subsection{Basics and definitions}

We consider the two-point energy correlator in a rotation-invariant state. In the previous sections, $\langle {\cal E}(\vec n_1) {\cal E}(\vec n_2) \rangle$ denoted the corresponding energy-flux density with respect to the standard measure on $S^{d-2}$. Rotation invariance implies that it depends only on the relative angle $\theta$ between the two detectors. We therefore introduce the $z$ variable defined as
\begin{equation}
\cos \theta \equiv \vec n_1 \cdot \vec n_2 \, , \qquad
z \equiv \frac{1-\cos\theta}{2} \in [0,1] \ .
\end{equation}
Here, $z=0$ corresponds to coincident detectors, while $z=1$ corresponds to back-to-back detectors. Thus, instead of treating $\langle {\cal E}(\vec n_1) {\cal E}(\vec n_2) \rangle$ as a density over $S^{d-2} \times S^{d-2}$, one can define a density over $[0,1]$, referred to as the normalized energy-energy correlator and denoted by $\text{EEC}(z)$. The two distributions are related by \begin{equation}
\label{eq:firstDefn-2pt}
\langle {\cal E}(\vec n_1) {\cal E}(\vec n_2) \rangle
=
\frac{E_{\rm tot}^2}{(\Omega_{d-2})^2}\,\text{EEC}(z) \ ,
\end{equation}
so that $\text{EEC}(z)=1$ for a homogeneous flux distribution.
Recall that throughout this paper we set $E_{\rm tot}=1$ unless explicitly stated otherwise. Energy conservation, see \eqref{eq:WI}, then implies the normalization condition
\begin{equation}
\int_0^1 d\mu(z)\,\text{EEC}(z) = 1 \ ,
\end{equation}
where the measure on the celestial sphere takes the form\begin{equation}
\label{eq:measure}
d \mu(z) =
\frac{2^{d-3} \Gamma \left(\frac{d-1}{2}\right)}
{\sqrt{\pi } \Gamma \left(\frac{d}{2}-1\right)}
\bigl(z (1-z)\bigr)^{{d-4 \over 2}} d z \, .
\end{equation}
In $d=4$, this reduces to $d\mu(z)=dz$.

We can next introduce the multipole expansion of the energy-energy correlator:
\begin{equation}
\label{eq:eec}
    \text{\bfseries Multipole expansion:}\qquad
    \text{EEC}(z)=1+ \sum_{J=1}^\infty C_J
    P_J^{(d)}(1-2z), \qquad 0\leq z\leq1 \ .
\end{equation}
where $P_J^{(d)}(\cos \theta)$ are orthogonal polynomials (proportional to the Gegenbauer polynomials $C_J^{({d-3 \over 2})}(\cos \theta)$):
\begin{equation}
\label{eq:PJdef}
P^{(d)}_{J}(x)= {}_{2}F_{1}\!\left(-J,\, J+d-3,\, \frac{d-2}{2},\, \frac{1-x}{2}\right)  .
\end{equation}

The $d$-dimensional partial waves $P_J^{(d)}(x)$ obey the orthonormality condition
\begin{equation}
\label{eq:PJorthogonality}
\int_0^1 d \mu(z) P_J^{(d)}(1-2z) P_{J'}^{(d)}(1-2z)
=
\frac{\delta_{J,J'}}{N_J^{(d)}} \, , \qquad
N_J^{(d)}=\frac{(d+2J-3) \Gamma(d+J-3)}{\Gamma(d-2)\Gamma(J+1)} \, .
\end{equation}
The constant $N_J^{(d)}$ computes the number of different spherical harmonics of spin $J$ on $S^{d-2}$. Therefore, the multipole coefficients can be recovered from the two-point energy correlator by the inversion formula
\begin{equation}
C_J= N_J^{(d)}\int_0^1 d\mu(z)\, P_J^{(d)}(1-2z)\,\text{EEC}(z) \, .
\label{eq:cj_inversion_pj}
\end{equation}

Now, recall that the EEC is locally positive
\begin{equation}
\label{eq:positivityeec}
\text{\bfseries Positivity:}\qquad \text{EEC}(z) \geq 0 \, .
\end{equation}
In addition due to the unitarity condition \eqref{eq:unitarity}, we have
\begin{equation}
\label{eq:unitarity2pt}
\text{\bfseries Unitarity:}\qquad C_J \geq 0 \, . 
\end{equation}
The unitarity statement can be immediately seen starting from
\eqref{eq:unitarity} and choosing $f(\vec n) = Y_{J, \textbf{m}}^{(d)}(\vec n)$,
which furnish a complete basis of functions on the sphere. Due to the orthogonality of the spherical harmonics, \eqref{eq:unitarity2pt} follows. 

In CFTs, and more generally in theories where the only stable asymptotic states are massless, we have an additional constraint
\begin{equation}
\label{eq:CFT-massless}
C_1 = 0   \qquad \text{(massless)}\,,
\end{equation}
which follows from the fact that we have chosen the state to carry zero spatial momentum. 

For some purposes, we will also need the inverse definition of \eqref{eq:firstDefn-2pt}, 
\begin{equation}
    \label{eq:precise2ptinz}
    \text{EEC}(z)= \frac{1}{E_{\mathrm{tot}}^2 \frac{d\mu}{dz}} \, \int d\Omega_{\vec n_1} d\Omega_{\vec n_2} \delta\left(z - \frac{1 - \vec n_1 \cdot \vec n_2}{2} \right) \langle \mathcal E(\vec n_1)\mathcal E (\vec n_2)\rangle \,,
\end{equation}
which is the $d$-dimensional version of the definition of the normalized energy correlator used \emph{e.g.} in \cite{Jaarsma:2025tck}. One can check that inserting \eqref{eq:firstDefn-2pt} into \eqref{eq:precise2ptinz} indeed gives back $\text{EEC}(z)$. Using \eqref{eq:precise2ptinz}, we can also write the following useful formula for $C_J$
 \be
 \label{eq:cJfromni}
 C_J=N^{(d)}_J\int \sigma(d\varepsilon) \int_{S^{d-2}}\varepsilon(d\Omega_{\vec n_1})\int_{S^{d-2}}\varepsilon(d\Omega_{\vec n_2})P_J(\vec n_1\cdot \vec n_2)
, \ee 
where $\sigma(d\varepsilon)$ is the event distribution measure introduced in Section~\ref{sec:posHierarchy}.

\paragraph{Relation to positive-definite functions.} 
A mathematical class of functions related to the two-point energy correlator is the class of positive-definite zonal kernels on the sphere,
originally studied by Schoenberg \cite{Schoenberg1942} and recently reviewed by Gneiting
\cite{Gneiting2011}. A kernel \(K\) is positive definite if
\begin{equation}
    \sum_{i,j=1}^{N}
    c_i c_j\,
    K(\vec n_i,\vec n_j)
    \geq 0
    \qquad
    \text{for all }
    N,\ \vec n_i\in S^{d-2},\ c_i\in\mathbb{R}.
    \label{eq:positive-definite-kernel}
\end{equation}
For a rotationally invariant kernel, the Schoenberg theorem states that
this condition is equivalent to the existence of a Gegenbauer expansion
with non-negative coefficients,
\begin{equation}
    K(\hat n_1,\hat n_2)
    =
    \sum_{J=0}^{\infty}
    C_J
    P_J^{(d)}(\vec n_1\cdot\vec n_2),
    \qquad
    C_J\geq 0.
    \label{eq:schoenberg-expansion}
\end{equation}
The condition \eqref{eq:positive-definite-kernel} is equivalent to unitarity \eqref{eq:unitarity} with $f(\vec n)=\sum_{i=1}^n c_i \delta(\vec n - \vec n_i)$.
Positive definiteness, however, does not imply pointwise positivity. For example,
\begin{equation}
    K(\vec n_1,\vec n_2)
    =
    \vec n_1\cdot\vec n_2
    \label{eq:gram-kernel-example}
\end{equation}
is positive definite, but it is negative when
\(\vec n_1\cdot\vec n_2<0\).

Energy correlators obey the additional local-positivity condition \eqref{eq:positivityeec}, which is normally not imposed in the mathematical literature on positive-definite functions. In addition, local positivity and unitarity are
not sufficient to characterize energy correlators. The two-point
correlator must belong to a symmetric, positive, and consistent hierarchy
generated by a single event distribution.
Thus the class of functions (or, more precisely, distributions) relevant for energy correlators is more constrained than
the intersection of pointwise-positive and positive-definite kernels: it
must also satisfy full-hierarchy extendibility. Nevertheless, membership in the class of positive-definite functions imposes some constraints on the global shape of energy correlators, which we shall discuss next.

\subsection{Unitarity of the two-point energy correlator: intuition and general results}
\label{sec:global-two-point-constraints}

In this section, we will make various statements about the general shape of two-point energy correlators, which follow from \textbf{unitarity}. Indeed, pointwise positivity is trivial to check, whereas the constraints coming from unitarity are less transparent.  

The general intuition behind these statements is the following: 
\begin{equation}
    \begin{matrix}\text{If $\text{EEC}(z)$ has a certain ``feature,'' e.g. a bump or non-analyticity, at $z=z_0\neq0$,}\\\text{it must admit the same or more prominent feature at $z=0$}\end{matrix}
    \label{eq:principles}
\end{equation}
This statement mimics a similar statement in \cite{Gneiting2011} for positive-definite functions, and leads to the following constraints.
\begin{enumerate}
    \item The EEC reaches its maximum at $z=0$. This follows from the trivial fact that    
    \begin{equation}
    \label{eq:maxorigin}
    \text{EEC}(z)=\sum_{J=0}^\infty C_J P_J^{(d)}(1-2z)\leq \sum_J C_J = \text{EEC}(0) \ .
\end{equation}
However, such an argument should be applied with care, since nonperturbative energy correlators are never finite at $z=0$.\footnote{The standard example of a flat energy correlator produced by the strong-coupling computation in \cite{Hofman:2008ar} is an artifact of the large $N_c$ perturbation theory, see \cite{Chen:2024iuv}. At finite $N_c$, the energy-energy correlator is singular at $z=0$. This follows from the light-ray OPE, and convexity of the leading Regge trajectory proven in \cite{Costa:2017twz}, see also \cite{Franken:2025gwr}.} A smeared version of \eqref{eq:maxorigin} takes the form
\begin{equation}
\label{eq:maxoriginsmeared}
\langle \left(\int_{\delta \Omega_{\vec n_1}} d \Omega_{n} {\cal E}(\vec n)- \int_{\delta \Omega_{\vec n_2}} d \Omega_{n} {\cal E}(\vec n) \right)^2 \rangle \geq 0 \ ,
\end{equation}
where $\delta \Omega_{n_i}$ are small regions centered around $\vec n_i$. For physical energy-energy correlators, the condition \eqref{eq:maxoriginsmeared} implies that a smooth energy-energy correlator that does not attain a maximum at $z=0$ must be accompanied by a contact term $\delta(z)$ that restores \eqref{eq:maxoriginsmeared}. We will encounter a refined version of this phenomenon in Section~\ref{sec:experimental} when discussing the ALEPH data.

\item If $\text{EEC}(z)$ is $k$-differentiable at the origin, i.e. $\text{EEC}^{(k)}(0)<\infty$, it is $k$-differentiable everywhere, namely for $0 \leq z \leq 1$. This follows from the same argument as above, when applied to the $k$th derivative of the EEC and using the fact that
\be
\label{eq:legder}
(-1)^{k} \partial_z^k P_J^{(d)}(1-2z) \leq (-1)^{k} \partial_z^k P_J^{(d)}(1-2z) \Big|_{z=0} \ . 
\ee
This property is analogous to Theorem~8a of \cite{Gneiting2011}.\footnote{The theorem states that 
    \emph{if $f:[0,\pi]\to\mathbb R$ is the restriction of an even continuous function $\phi:\mathbb R\to\mathbb R$ for which the derivative $\phi^{(2k)}(0)$ exists, but which fails to be $2k$ times differentiable on $(0,\pi)$, then $f(\theta)$ is not a positive-definite function.}}
  As explained above this condition can be either applied to perturbative energy correlators, or one needs to use its smeared version. The property \eqref{eq:legder} implies that the EEC cannot have sharp features at $z=z_0>0$ without having equally sharp features at $z=0$. In Section~\ref{sec:deltafunctionmodels}, we explore this quantitatively by showing that if $\text{EEC}(z)$ has distributional terms $\delta(z-z_0)$---the simplest model of a sharp feature---for some $z_0\neq0$, it must also have a term $\delta(z)$ with a  coefficient bounded from below by the size of the feature at $z_0$.

\item If $\text{EEC}(z)$ is non-analytic at $z=1$, it must be non-analytic at $z=0$ as well. This property is based on the inversion formula for the energy multipoles $C_J$ discussed in Section 5 of \cite{Dempsey:2025yiv}. This formula reads (in $d=4$)
\begin{equation}
    C_J=\frac{2(2J+1)}\pi\int\limits_{-\infty}^0dz\,Q_J(1-2z)\left(\mathrm{Disc}_{z=0}\text{EEC}(z)+(-1)^J\mathrm{Disc}_{z=0}\text{EEC}(1-z)\right)\,,
\end{equation}
where the precise definitions of $Q_J$ and the discontinuity are given in \cite{Dempsey:2025yiv}. 
For $\text{EEC}(z)$ satisfying \eqref{statementGneiting8d}, the first term vanishes, leaving only the second term which probes the singularity at $z=1$, and which is of alternating sign for even and odd $J$. These alternations would make some $C_J$ negative, unless the term multiplying $(-1)^J$ itself has a compensating $(-1)^J$ behavior. Using the asymptotics of $Q_J$, we can exclude such behavior. 

There is a version of this statement that does not rely on the inversion formula. It is a consequence of theorem 8d from \cite{Gneiting2011}, originally proven in \cite{Devinatz1959}:
\begin{equation}
    \begin{matrix}\text{\emph{
If $f:[0,\pi]\to\mathbb R$ is the restriction of an analytic $\phi:\mathbb R\to\mathbb R$ of}}\\\text{\emph{period not $2\pi$, then $f(\theta)$ is not a positive-definite function. 
     }}\end{matrix}
    \label{statementGneiting8d}
\end{equation}

\end{enumerate}
These statements are illustrated by listing admissible and inadmissible shapes of the EEC in Table~\ref{tab:admissible}.

\begin{table}[]
    \centering
    \begin{tabular}{|p{0.45\textwidth}|p{0.45\textwidth}|}
    \hline
       \textbf{Admissible} & \textbf{Inadmissible}
       \\
       \hline
    \includegraphics[height=28mm]{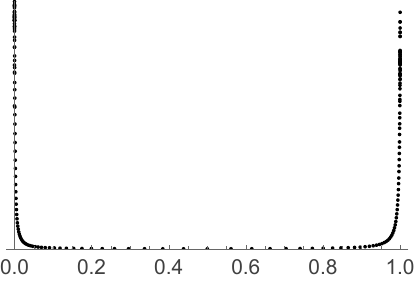}& \includegraphics[height=28mm]{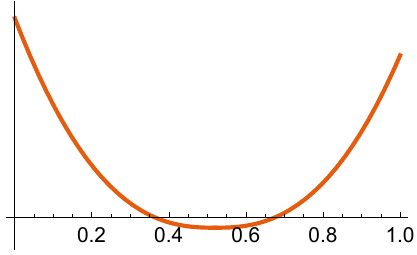}    \\  Experimentally measured (binned data)&  Negative at some $z_0$
       \\\hline
    \includegraphics[height=28mm]{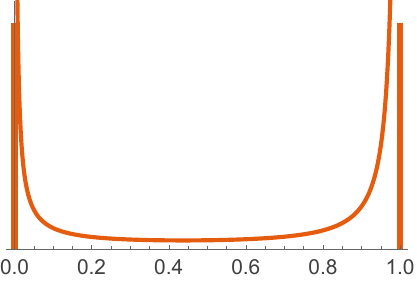} & \includegraphics[height=28mm]{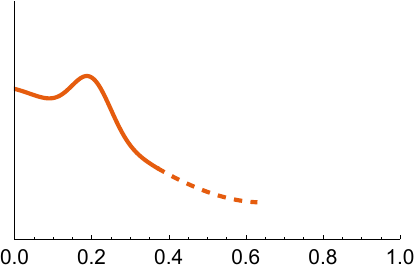}  \\ One-loop QCD & Not attaining maximal value at $z=0$
             \\\hline
    \includegraphics[height=28mm]{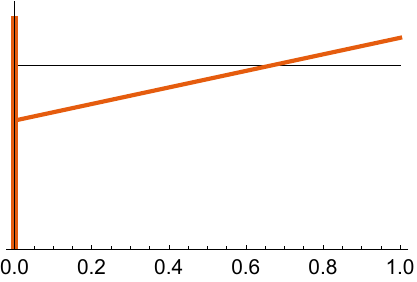} & \includegraphics[height=28mm]{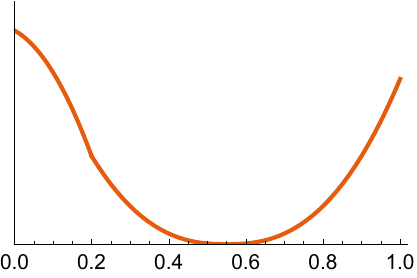}  \\ Free particles at large multiplicity & $\mathcal C^k$ discontinuity at $z_0\neq0$ despite being regular at $z=0$  
         \\\hline
    \includegraphics[height=28mm]{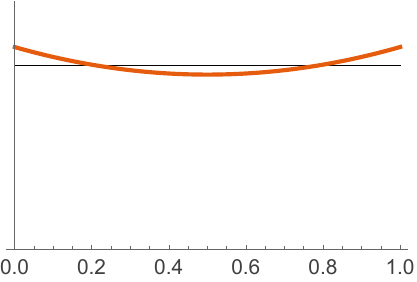} &    \includegraphics[height=28mm]{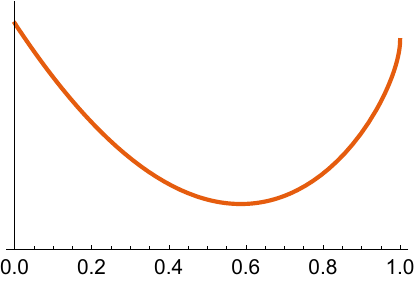}  \\ Strings in AdS (planar $\mathcal N=4$) & Discontinuity at $z=1$ and analyticity at $z=0$
         \\\hline
    \includegraphics[height=28mm]{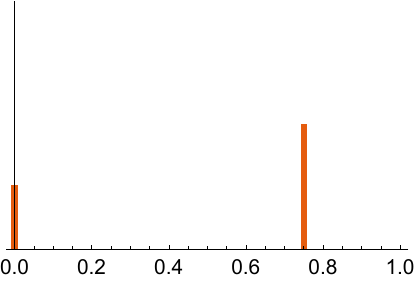} 
   &     \includegraphics[height=28mm]{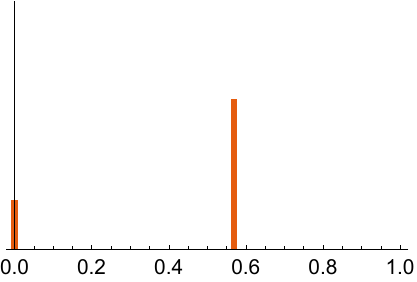}  \\ Extremal solution maximizing $C_3$ (triangular atomic event) & Coefficient of $\delta(z-z_0)$ violating bound \eqref{eq:boundOnDeltaModel} compared to $\delta(z)$
         \\\hline
    \end{tabular}
\caption{Some admissible and inadmissible two-point energy correlators. $C_3$ refers to a coefficient in the partial wave decomposition. Thick vertical lines represent Dirac delta functions. The conditions ruling out the shapes in the right column are discussed in Section~\ref{sec:global-two-point-constraints}.}
\label{tab:admissible}
\end{table}

Here we have considered the constraints from $C_J\geq0$ on the complete (resummed) two-point energy correlator. We can also consider the case where the EEC admits a perturbative expansion in an infinitesimal parameter $g$ (it could be a coupling constant, or a large quantum number):
\begin{equation}
\label{eq:boundInPertTheory}
    \text{EEC}(z)=\text{EEC}^{(0)}(z)+g\,\text{EEC}^{(1)}(z)+g^2\,\text{EEC}^{(2)}(z)+\ldots
\end{equation}
If $\text{EEC}^{(0)}(z)$ is generic with all $C_J^{(0)}\geq c_0>0$, then there are no constraints on the higher-order functions $\text{EEC}^{(1)}(z)$, etc. However, if for instance $C_J^{(i)}=0$ for some $i=0,1,2,\ldots, l-1$, then the constraint $C_J\geq0$ would imply $C_J^{(l)}\geq 0$, implying constraints on the shape of the order-$l$ correction $\text{EEC}^{(l)}(z)$.
This is what happens for energy correlators computed in low-energy effective theories in AdS \cite{Hofman:2008ar}, and energy correlators in heavy states \cite{Chicherin:2023gxt,Firat:2023lbp,Cuomo:2025pjp}, see also models discussed in Section~\ref{sec:physmodels}. 
Moreover, if $\text{EEC}^{(0)}$ is equal to the back-to-back model $\sim \delta(z)+\delta(z-1)$, which saturates the upper bounds $C_J^{(0)}=N_J^{(d)}$ for even $J$ (see Section~\ref{sec:deltafunctionmodels} immediately below), we get alternating sign constraints on the leading correction for all spins:
\begin{equation}
    C_J^{(1)}\leq 0 \quad \text{($J$ even)}, \qquad\qquad  C_J^{(1)}\geq 0 \quad \text{($J$ odd)}. 
\end{equation}
It can be checked that for instance the $C_J$ extracted from the one-loop QCD satisfies this infinite sequence of alternating conditions. 

\subsection{Upper bounds on $C_J$ and atomic measures}
\label{sec:deltafunctionmodels}

An important consequence of the pointwise positivity of the energy correlators is the existence of upper bounds on the energy multipoles $C_J$. The simplest upper bound follows from the inversion formula \eqref{eq:cj_inversion_pj}, where the property $P_J^{(d)}(1-2z) \leq 1$ immediately leads to
\begin{align}
    C_J = N_J^{(d)} \int_0^1 d\mu(z)\, P_J^{(d)}(1-2z)\,\operatorname{EEC}(z) \leq N_J^{(d)},
    \label{eq:general_bound_cj_nj}
\end{align}
where we also used total normalization $\int_0^1d\mu(z)\text{EEC}(z)=1$. 
This generalizes the Fox and Wolfram result in $d=4$~\cite{Fox:1978vu} to any $d$. In the absence of the masslessness constraint $C_1=0$, the bound \eqref{eq:general_bound_cj_nj} is optimal and it is saturated by 
\begin{equation}
\label{eq:collinear}
\text{EEC}(z) ={1 \over {d \mu(z) \over d z}} \delta(z).
\end{equation}
In the presence of the masslessness constraint $C_1 =0$, for even $J$, the bound \eqref{eq:general_bound_cj_nj} is still optimal; it is saturated by
\begin{equation}
\label{eq:backtobackGend}
\text{EEC}(z) =\frac12{1 \over {d \mu(z) \over d z}} ( \delta(z) + \delta(1-z) ) .
\end{equation}
The physical interpretation of this model is a process where all events consist of two particles in the back-to-back configuration. For odd $J$, the bound can be improved, see \cite{Dempsey:2025yiv}. For example, for $J=3$, we have found the upper bound
\begin{equation}
\label{eq:c3max}
    C^{(d)}_3\leq N^{(d)}_3\frac{d+1}{4(d-2)}=\frac{(d+3)(d+1)(d-1)}{24} < N^{(d)}_3 \ , 
\end{equation}
with the extremal form
\begin{equation}
    \mathrm{EEC}(z)={1 \over {d \mu(z) \over d z}}\left(\frac13\delta(z)+\frac23\delta\left(z-\frac34\right)\right)
\end{equation}
for any $d$. In Appendix~\ref{app:multipoles}, we derive the bound on $J=5$, and an asymptotic bound at large $J$, both for arbitrary spacetime dimension $d$. 

We have found that the bounds on simple $C_J$ are saturated by simple atomic measures, i.e. sums over Dirac delta distributions. This is not a coincidence. 
To understand it, it is very instructive to think about the problem of $C_J$ extremization from the point of view of the event distribution representation of the energy-energy correlator.

Let us imagine that we are interested in deriving bounds on a linear combination of the two-point energy multipoles
\begin{equation}
{\cal X} = \sum_a \lambda_a C_{J_a} \ .
\end{equation}
Using \eqref{eq:representationEnCol}, we can write 
\begin{equation}
{\cal X} =\int  \sigma(d \varepsilon) {\cal X}[\varepsilon],
\end{equation}
where the value of the observable of interest on a given energy-flow measure is
\begin{equation}
\label{eq:definitionXvareps}
{\cal X}[\varepsilon] = \iint_{S^{d-2}\times S^{d-2}}   \varepsilon(d \Omega_{\vec n_1})\varepsilon(d \Omega_{\vec n_2}) \left( \sum_a \lambda_a N_{J_a}^{(d)} P_{J_a}^{(d)}(\vec n_1\!\cdot\!\vec n_2)\ \right)  \ .
\end{equation}
It is then clear from this representation that \begin{equation}
\sup_{{\cal P}} {\cal X}=\sup_{\varepsilon} {\cal X}[\varepsilon] \ ,
\label{eq:singleEventReduction}
\end{equation}
where on the RHS we are extremizing over \emph{single-event energy flows}.
Indeed, a mixture of events with nonnegative coefficients can never exceed $\sup_{\varepsilon} {\cal X}[\varepsilon]$. Therefore we need to take $\sigma(d\varepsilon)$ to be the Dirac delta-function localized on a single collider event that maximizes the quantity of interest.\footnote{By our assumption the state is rotation-invariant, whereas the single event-shape $\varepsilon_*$ that maximizes ${\cal X}$ need not be. We can trivially fix it by averaging over the rotation orbit of a given event energy flow $\varepsilon_*$. From \eqref{eq:definitionXvareps}, it is immediately clear that for any rotation $R$ we have $
{\cal X}[\varepsilon^{R}] = {\cal X}[\varepsilon] ,$
where we have defined
$\varepsilon^{R}(\vec n) \equiv \varepsilon(R^{-1} \vec n)
$. 
For this reason averaging over the rotation group is trivial
$
\int_{\mathrm{SO}(d-1)} dR {\cal X}[\varepsilon^{R}]={\cal X}[\varepsilon] 
$, 
where we used that $\int_{\mathrm{SO}(d-1)} dR = 1$.
}

We would like next to argue that $\sup_{\varepsilon} {\cal X}[\varepsilon]$ is \emph{atomic}, i.e. it is a finite-particle event. It is convenient to organize this discussion by considering separately several classes of observables.

\paragraph{Case I:} $\max {\cal X} = \sum_a \lambda_a C_{J_a}, ~~~ \lambda_a \geq 0.$
Imagine we would like to maximize a linear combination of $C_J$ with nonnegative coefficients (generalizing the analysis of the previous section).
In this case the problem is convex in the following sense
\begin{equation}
\label{eq:inequalityconvex}
{\cal X}[t\varepsilon_1+(1-t)\varepsilon_2]\le t\,{\cal X}[\varepsilon_1]+(1-t)\,{\cal X}[\varepsilon_2] \ .
\end{equation}
This follows from the fact that\footnote{This normalization above corresponds to 
$\int d\Omega_{\vec n}\,
\overline{Y^{(d)}_{J\textbf{m}}(\vec n)}\,
Y^{(d)}_{J'\textbf{m}'}(\vec n)
=
\Omega_{d-2}\,
\delta_{J J'}\,
\delta_{\textbf{m} \textbf{m}'}
$.
} 
\begin{equation}
P_J^{(d)}(\vec n_1\!\cdot\!\vec n_2)
=
\frac{1}{N_J^{(d)}}\sum_{\textbf{m}}
Y_{J \textbf{m}}^{(d)}(\vec n_1)\,
\overline{Y_{J\textbf{m}}^{(d)}(\vec n_2)} \ ,
\end{equation}
and the elementary identity
\begin{equation}
\left|t\,a+(1-t)\,b\right|^2
=
t|a|^2+(1-t)|b|^2-t(1-t)|a-b|^2
\le
t|a|^2+(1-t)|b|^2 \ .
\end{equation}

\paragraph{Without the constraint $C_1=0$.}
It is then clear that the maximum $\max {\cal X}$ is attained at the extreme points of single-event energy flows, in other words energy flows that cannot be written as a positive linear combination of several single-event energy flows.
If we do not have any further constraints, such extremal flows are concentrated at a point (or, equivalently, are one-atomic) 
\begin{equation}
\label{eq:singleatom}
\varepsilon_* = \delta_{\vec n_*} \,,
\end{equation}
which corresponds to the EEC
\begin{equation}
\text{EEC}(z) = \delta(z) \ .
\end{equation}
Therefore in the unconstrained case $\max {\cal X}$ is attained on this trivial energy flux.

Notice that \eqref{eq:singleatom} does not mean that this single-atom measure is unique. For example, if we maximize any linear combination of $C_{J \in \text{even}}$, the same maximum is attained by any $\text{EEC}(z) = a\delta(z) +(1-a) \delta(1-z)$.

\paragraph{With the constraint $C_1=0$.}

Consider next a more nontrivial case, where we impose the masslessness constraint $C_1=0$. In terms of the underlying events we can think of this as a set of $d-1$ linear constraints that each underlying event must satisfy
\begin{equation}
\int_{S^{d-2}} d\Omega_{\vec n} n^i\,\varepsilon(\vec n)=0 , ~~~ 1 \le i \le d-1 \ .
\end{equation}
Theorem 2.1 by Winkler \cite{Winkler1988} states that with this extra constraint added, the extremal measures are \emph{at most} $d$-atomic, meaning that they are in the class
\begin{equation}
\label{eq:winklerprediction}
\varepsilon_*=\sum_{a=1}^{d} w_a\,\delta_{\vec n_a} \ .
\end{equation}
The simple example of the previous section, where we were maximizing a single $C_J$, indeed produced two-atomic measures of this type.\footnote{The basic idea behind this result is the following. If a measure were supported on
more than $d$ points, the vectors $(1,\vec n_a)\in\mathbb{R}^d$ would be
linearly dependent. There would therefore exist a nonzero deformation of
the weights satisfying
\be
\sum_a \delta w_a=0,
\qquad
\sum_a \delta w_a\,\vec n_a=0.
\ee
For sufficiently small $\eta$, both $w_a+\eta\delta w_a$ and
$w_a-\eta\delta w_a$ remain nonnegative and obey the same normalization and
momentum constraints. The original measure is then the average of two
distinct admissible measures and is therefore not extremal.} 

We can directly give an example where \eqref{eq:winklerprediction} is saturated, namely the maximization of $C_3$ in \eqref{eq:c3max}. The measure that produces this result is $3$-atomic and is supported on an equilateral triangle
\begin{equation}
\label{eq:triangle1}
\varepsilon_{\triangle}=
\frac13\left(\delta_{\phi_0}+\delta_{\phi_0+2\pi/3}+\delta_{\phi_0+4\pi/3} \right) 
\end{equation}
embedded for instance on the equator in $S^{d-2}$. 
Extracting the $\text{EEC}(z)$ from this event indeed gives back \eqref{eq:c3max}.

\paragraph{Case II:} $\max/\min {\cal X} = \sum_a \lambda_a C_{J_a}$.
In this most general case the convexity inequality \eqref{eq:inequalityconvex} need not hold. Consequently, the preceding argument does not guarantee that an optimizer is an extreme point of the space of event measures.

To illustrate this point, let us consider the problem of minimization of $C_2$. By unitarity it is minimized by $C_2=0$, which can be attained for example by considering a homogeneous energy flux
\begin{equation}
\varepsilon_{\mathrm{hom}}(\vec n)=\frac{1}{\Omega_{d-2}} \ .
\end{equation}
However, we can also consider an atomic measure that achieves $C_2=0$. Indeed, the lower bound on $C_2$ is saturated by the regular simplex (hypertetrahedral) measure \begin{equation}
\varepsilon_{\mathrm{simp}}=\frac1d\sum_{a=1}^{d}\delta_{\vec v_a} \ ,
\qquad
\sum_{a=1}^{d}\vec v_a=0 \ ,
\qquad
\frac1d\sum_{a=1}^{d}\vec v_a \vec v_a^{\,T}=\frac1{d-1} I_{d-1} \ .
\end{equation}

The fact that the extremum can be achieved on an atomic measure is in fact very general and can be simply understood again using the same Winkler theorem as we used above. Let us consider the problem of ${\cal X} = \sum_a \lambda_a C_{J_a}$ extremization and let us assume that the extremum is attained at $\mu_*$. If the sum over $a$ is \emph{finite}, we can consider a class of measures with a set of additional linear constraints
\begin{equation}
\int_{S^{d-2}} d \Omega_{\vec n} Y_{J_a\textbf{m}}^{(d)}(\vec n)\,\varepsilon(\vec n) = \int_{S^{d-2}} d \Omega_{\vec n} Y_{J_a\textbf{m}}^{(d)}(\vec n)\,\varepsilon_*(\vec n) \ .
\end{equation}
On the one hand, this class of measures by construction attains the same value of the target observable ${\cal X}$. On the other hand, we can apply Winkler's theorem to this set to argue that its extreme points are atomic measures 
\begin{equation}
\varepsilon = \sum_{i=1}^N w_i \delta_{\vec n_i},~~~ N \leq \sum_{a} N_{J_a}^{(d)}+1 \ . 
\end{equation}
In case we want to include the masslessness condition $C_1=0$, the bound instead becomes
\begin{equation}
N \leq \sum_{a} N_{J_a}^{(d)}+d \ .
\end{equation}
This reduction to the atomic measure is also closely related to Tchakaloff's theorem \cite{Tchakaloff1957}. In the context of the present discussion, it 
states that any measure can be replaced by a finitely supported one with exactly the same relevant harmonic moments, hence the same value of any \emph{finite} combination $\sum_a \lambda_a C_{J_a}$ can be attained.

\subsection{Strengthening with the hierarchy}
\label{sec:strength-hierarchy}

In the section above, we explored upper bounds on the energy multipoles of $C_J$ which follow from the pointwise positivity. We also argued that these upper bounds are saturated by atomic measures. Next we would like to give a simple example in which the existence of \emph{the full consistent hierarchy} leads to the stronger results. 

\subsubsection{Two-delta model}

\begin{figure}
    \centering
   \includegraphics[width=0.68\textwidth]{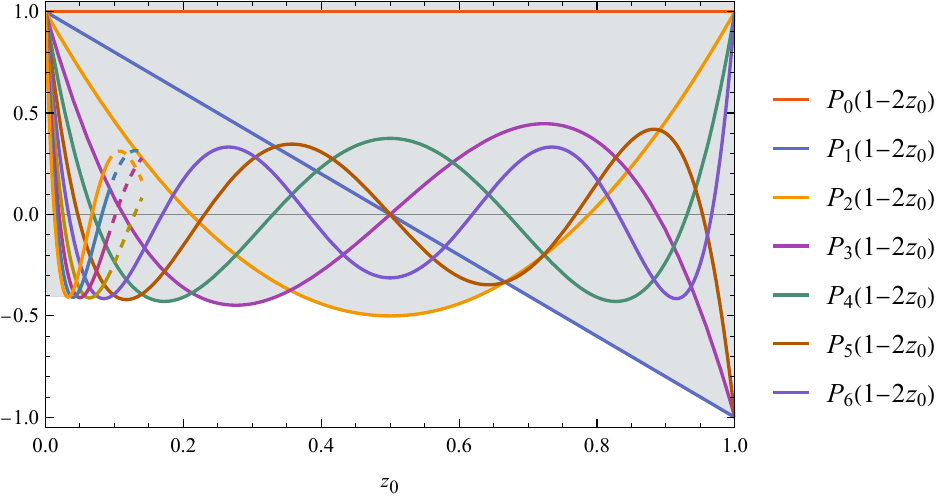}
    \caption{The maximum allowed value of $B$ in \eqref{eq:TwoDeltas} is directly related to the minimum value of $P^{(d)}_J(1-2z_0)$ for all $J$. Here we plot the Legendre polynomials $P_J(1-2z)$ in $d=4$.}
    \label{fig:plottingLegendres}
\end{figure}

\begin{figure}
    \centering
   \includegraphics[width=0.92\textwidth]{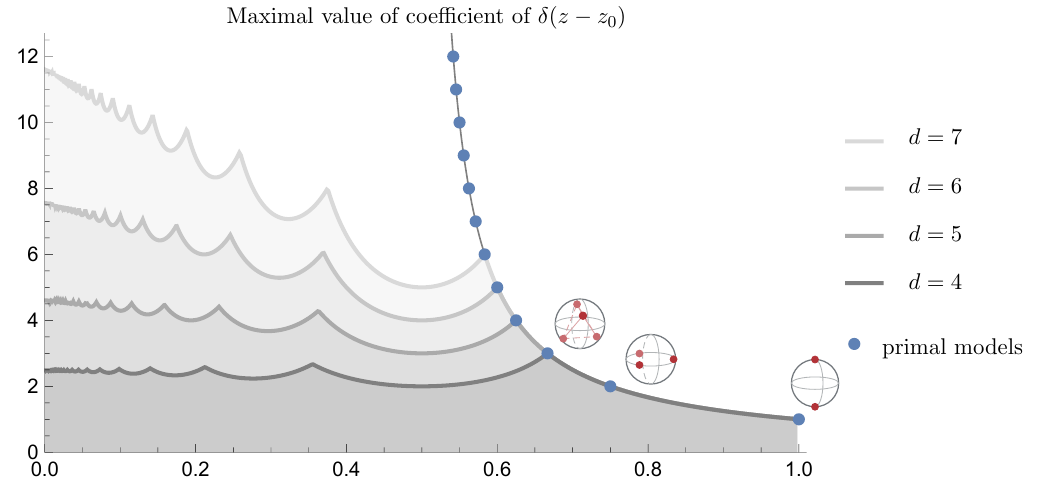}
    \caption{Bounds on the coefficient $B$ in the non-normalized model $\text{EEC}(z)=\delta(z)+B\delta(z-z_0)$ \eqref{eq:TwoDeltas} for different $d$. The rightmost peak in the $d$-dimensional bound is the atomic model corresponding to the $(d-1)$ simplex (tetrahedron for $d=4$, $5$-cell for $d=5$, etc.)}
    \label{fig:tetrahedra}
\end{figure}

First we will consider in detail the simplest two-delta model (we will see below that it naturally arises when extremizing multipoles of the EEC). We are interested in deriving a bound on the coefficient $B$ in the shape
\begin{equation}
\label{eq:TwoDeltas}
\text{EEC}(z)=  {1 \over {d \mu(z) \over d z}} \big(  \delta(z)+B\delta(z-z_0)\big)
\end{equation}
which is not yet subject to normalization $C_0=1$. From the principles stated in \eqref{eq:principles}, it is clear that there must exist an upper bound on $B$ as a function of $z_0$.\footnote{One way to derive an upper bound would be to start from the manifestly positive expression \eqref{eq:maxoriginsmeared}. It turns out that the appropriate regions to choose in this expression are not infinitesimally small circles separated by an angle with $\cos\theta=1-2z$, but instead large tangential discs. We discuss this in detail in Appendix~\ref{app:twodeltamodel}. } 
The exact solution to this problem can be found by considering the partial-wave decomposition, where $C_J$ take the form
\begin{equation}
    C_J=N^{(d)}_J+N^{(d)}_JBP_J^{(d)}(1-2z_0)\,.
\end{equation}
The positivity constraint $C_J\geq0$ then yields the bound
\begin{equation}
\label{eq:boundOnDeltaModel}
    B\leq B_{\mathrm{max}}(z_0), \qquad B_{\mathrm{max}}(z_0)=-\left(\min_J P_J^{(d)}(1-2z_0)\right)^{-1}.
\end{equation}
This gives a non-trivial function $B_{\mathrm{max}}(z_0)$ that is saturated by different $J$ at different angles. In Figure~\ref{fig:plottingLegendres}, we show the $P_J(1-2z_0)$ in $d=4$. 
Inverting the gray region, one gets the $d=4$ bound in Figure~\ref{fig:tetrahedra}. Repeating this using $d$-dimensional Legendre polynomials carves out similar regions for $d=5$, $d=6$, etc.

Having derived an exact bound on the two-delta model \eqref{eq:TwoDeltas}, we would now like to consider this problem from a primal perspective. Here we consider not only the expression \eqref{eq:TwoDeltas} itself, but also how it can arise from an underlying event distribution. Since $\text{EEC}(z)$ is given by a sum of $\delta$-functions, we conclude that the underlying event distribution corresponds to events with particles.
The term $\delta(z)$ corresponds to measuring the same particle twice. The fact that there is only one other $\delta(z-z_0)$ means that all particles in the event must have equal relative angles $\vec n_i\cdot\vec n_j=1-2z_0$, $i\neq j$. Together with momentum conservation, we conclude that in $d=4$ the underlying events can be any of the following:
\begin{itemize}
    \item Two-particle events back-to-back: $\text{EEC}(z)=\frac12(\delta(z)+\delta(z-1))$
    \item Three-particle events on an equilateral triangle: $\text{EEC}(z)=\frac13(\delta(z)+2\delta(z-3/4))$
    \item Four-particle events on a regular tetrahedron: $\text{EEC}(z)=\frac14(\delta(z)+3\delta(z-2/3))$
\end{itemize}
In higher $d$, one can have $d$-particle events placed in the shape of a $(d-1)$-dimensional hypertetrahedron (simplex). All these primal models are shown as blue dots in Figure~\ref{fig:tetrahedra}.

The first peak in the $d=4$ bound is saturated by the tetrahedral model. It is exactly in the transition between the vanishing of $C_1$ and $C_2$, and in fact also, accidentally, $C_5=0$ for this particular model. The vanishing of $C_1$, $C_2$ and $C_5$ at $z_0=\frac23$ is the primal picture of the fact that $P_1(1-2z_0)$, $P_2(1-2z_0)$ and $P_5(1-2z_0)$ all are equal at that point. We also note that all primal models above have $C_1=0$, so they can be realized by massless particles.

 \subsubsection{Bounds on multipoles}

Let us study how the consistency of the full hierarchy imposes stronger constraints on the multipole coefficients $C_J$. We illustrated this in Figure~\ref{fig:twosidedc2c3} in the Introduction, which we will now explain in full detail. We discuss here the case $d=4$, leaving $d=3$ to Appendix~\ref{app:d3bounds}. 

We consider bounds on $(C_2,C_3)$ under the masslessness condition $C_1=0$. Using positivity $\text{EEC}(z)\geq0$ and unitarity $C_J\geq0$, a dual optimization produces the bound
\begin{equation}
    C_3\leq \frac7{45}(2C_2+5)(5-C_2),
\end{equation}
as derived in \cite{Dempsey:2025yiv}. This bound is shown in gray in Figure~\ref{fig:twosidedc2c3}, and is satisfied by the distribution
\begin{equation}
    \text{EEC}(z)=\frac\alpha{1+\alpha}\delta(z)+\frac1{1+\alpha}\delta\big(z-(1+\alpha)/2\big). 
\end{equation}
As discussed above, this interpolates between the back-to-back, triangular, and tetrahedral model for $\alpha=1$, $\frac12$, and $\frac13$ respectively, but does not derive from a consistent hierarchy for other values of $\alpha$. 

In order to derive a bound using the full hierarchy, we first make the following definitions:
\begin{align}
M^{i j}[\varepsilon] &= \int_{S^2} \varepsilon(d \Omega_{\vec n}) n^i n^j , \quad  Q = M - {1 \over 3} 1 \, , \quad T^{ijk}= \int_{S^2} \varepsilon(d \Omega_{\vec n}) n^i n^jn^k, 
\end{align}
and
\begin{align}
    s = \Tr Q^2,\qquad t=\Tr Q^3,\qquad u = T^{ijk}T^{ijk}.
\end{align}
Using \eqref{eq:cJfromni}, one can check that
\begin{equation}
    C_2=\frac{15}2\langle s\rangle, \qquad C_3=\frac{35}2\langle u\rangle, 
\end{equation}
whereas $t$ will be used in intermediate steps. 

We will now derive two linear inequalities of on $C_3$, which will determine the shape of the blue allowed region. Let $\lambda_i\geq0$ be the eigenvalues of the matrix $M$, on a single event, and let $X=\sum_{i<j}\lambda_i\lambda_j$, $Y=\lambda_1\lambda_2\lambda_3$. Then
$
    s=\frac23-2X$, $t=\frac29-X+3Y
$. 
We then apply the following inequality $1-4X+9Y\geq0$ to get
\begin{equation}
\label{eq:tgtr}
    t\geq-\frac s6\,. 
\end{equation}
Moreover, one can show that
\begin{equation}
\label{eq:uless}
u\leq \Tr M^2-\Tr M^3=\frac29-t\,,
\end{equation}
 where we used that $\Tr M^2=s+\frac13$ and $\Tr M^3=s+t+\frac19$. We prove \eqref{eq:tgtr} and \eqref{eq:uless} in Appendix \ref{app:moments1}.

 Combining \eqref{eq:uless} with \eqref{eq:tgtr}, we get 
\begin{equation}
\label{eq:firstLine}
    u\leq \frac29+\frac s6 \quad \leftrightarrow \quad C_3\leq \frac{35}9+\frac7{18}C_2 \,.
\end{equation}
Using instead positivity of $M$ we find
\begin{equation}
\label{eq:posM}
    3\det M=t-\frac s2+\frac19\geq0\,,
\end{equation}
which combined with \eqref{eq:uless} gives
\begin{equation}
\label{eq:secondLine}
    u\leq\frac13-\frac s2\quad \leftrightarrow\quad C_3\leq \frac76(5-C_2).
\end{equation}
Together \eqref{eq:firstLine} and \eqref{eq:secondLine} carve out the blue allowed region in Figure~\ref{fig:twosidedc2c3}.

The bound is also optimal, since it can be spanned by the following events:
\begin{align}
&    \text{back-to-back}:& (C_2,C_3)&=(5,0)\,,
\\
&    \text{triangle}:& (C_2,C_3)&=\left(\frac54,\frac{35}8\right),
\\
&    \text{tetrahedron}:& (C_2,C_3)&=\left(0,\frac{35}9\right),
\\
&    \text{uniform}:& (C_2,C_3)&=(0,0)\,,
\end{align}
where the uniform event could for instance be replaced with a regular octahedral atomic event.
Thus the full allowed region can be characterized by a finite number of atomic events, just as predicted by Winkler's theorem as discussed above.

\section{Constraints on the three-point energy correlator}
\label{sec:threePoint}

In this section we discuss positivity constraints on the three-point energy correlator. We show that the multipole expansion coefficients of the three-point energy correlator are two-sided bounded by the two-point energy correlator multipoles. Similarly, the bounds are saturated by atomic measures.

In a rotation-invariant state, the three-point energy correlator is a function of three relative angles
\begin{equation}
z_{i j} = {1 - \vec n_i \cdot \vec n_j \over 2} \ , ~~~ i<j \ . 
\end{equation}
As before, the three-point energy correlator is pointwise nonnegative. In addition, however, there are nontrivial global constraints, which we characterize below. 

It is convenient to introduce the following notation
\begin{equation}
\langle {\cal E}(\vec n_1) {\cal E}(\vec n_2) {\cal E}(\vec n_3) \rangle
=
\frac{E_{\rm tot}^3}{(\Omega_{d-2})^3}\,\text{EEEC}(z_{12},z_{13},z_{23}) \,.
\end{equation}
In $d=3$, the three-point energy correlator depends on two variables only due to the fact that $\phi_{12}+\phi_{23}+\phi_{31}=0$, where $\vec n_i = (\cos \phi_i, \sin \phi_i)$, which implies the following identity
\be
z_{12}^{2}+z_{13}^{2}+z_{23}^{2}-2 z_{12} z_{13}-2 z_{12} z_{23}-2 z_{13} z_{23}+4 z_{12} z_{13} z_{23}=0 \ .
\ee

\subsection{Multipole expansion}
As a function of three insertions $\vec n_i$, the three-point energy correlator admits the multipole expansion\footnote{In the case of $d=3$ and $d=4$, this expression is only valid for three-point energy correlators that are invariant under parity $\vec n_i\to-\vec n_i$. For $d=4$, we can also consider a parity-odd structure $\epsilon_{i j k} n_1^i n_2^j n_3^k (\vec n_1 \cdot \vec n_2 - \vec n_1 \cdot \vec n_3)  (\vec n_1 \cdot \vec n_2 - \vec n_2 \cdot \vec n_3) (\vec n_1 \cdot \vec n_3 - \vec n_2 \cdot \vec n_3) \Phi_{\text{sym}} \left( \vec n_1 \cdot \vec n_2 , \vec n_1 \cdot \vec n_3, \vec n_2 \cdot \vec n_3 \right)$, where $\Phi_{\text{sym}} \left( \vec n_1 \cdot \vec n_2 , \vec n_1 \cdot \vec n_3, \vec n_2 \cdot \vec n_3 \right)$ is a symmetric function of its arguments. Similarly, in $d=3$, we could have $\epsilon_{i j} n_1^i n_2^j \epsilon_{kl} n_1^k n_3^l \epsilon_{mn} n_2^m n_3^n (\vec n_1 \cdot \vec n_2 - \vec n_1 \cdot \vec n_3)  (\vec n_1 \cdot \vec n_2 - \vec n_2 \cdot \vec n_3) (\vec n_1 \cdot \vec n_3 - \vec n_2 \cdot \vec n_3) \Phi_{\text{sym}}$. For simplicity, we assume that these terms are absent, but it would be interesting to generalize our analysis to include them.} 
\begin{align}
\text{EEEC}(z_{12},z_{13},z_{23}) &= 1 + \sum_{J=1}^\infty C_J \left( P_J^{(d)}(1-2z_{12}) + P_J^{(d)}(1-2z_{13}) +P_J^{(d)}(1-2z_{23})  \right) \nonumber\\
&+ \sum_{J_1,J_2,J_3 = 1}^\infty C_{J_1 J_2 J_3} \Phi^{(d)}_{J_1 J_2 J_3}(z_{12},z_{13},z_{23}) \,,
\label{eq:threePointParam}
\end{align}
discussed already in \cite{Fox:1978vw}. 
Other parametrizations for the three-point energy correlator include \cite{Chen:2019bpb,Yan:2022cye,Chen:2022swd,Gong:2025jqi,Budhraja:2026pyi}. 
In this expansion we have separated the two-point energy-correlator contribution in the first line, and the rotation-invariant polynomials $\Phi_{J_1,J_2,J_3}(z_{12},z_{13},z_{23})$ are defined as follows:
\begin{equation}
\Phi^{(d)}_{J_1J_2J_3}(z_{12},z_{13},z_{23}) ={1+(-1)^{J_1+J_2+J_3} \over 2} {1 \over \volS}\int d \Omega_{\vec n} P_{J_1}^{(d)}( \vec n_1 \vec n) P_{J_2}^{(d)}( \vec n_2 \vec n)  P_{J_3}^{(d)}( \vec n_3 \vec n) \ . 
\end{equation}
They are nonzero as long as $J_i$ satisfy the triangle inequality 
\begin{equation}
\label{eq:3jbounds}
|J_1 - J_2 | \leq J_3 \leq J_1+J_2 ,
\end{equation}
and $J_1+J_2+J_3$ is even.\footnote{In $d=3$, the condition is stronger $J_3 = |J_1 \pm J_2|$.} Thus, permutation symmetry of the detectors implies
\be
C_{J_1 J_2 J_3} = C_{J_2 J_1 J_3} = C_{J_1 J_3 J_2} \ .
\ee
The expansion can be inverted by the following projection formula \begin{equation}
\label{eq:threePointProjectionRaw}
C_{J_1 J_2 J_3} ={N_{J_1}^{(d)} N_{J_2}^{(d)} N_{J_3}^{(d)} \over \Phi^{(d)}_{J_1J_2J_3}(0,0,0)} {1 \over \volS^3} \int \prod_{i=1}^3 d\Omega_{\vec n_i}\,
\text{EEEC}(z_{12},z_{13},z_{23})\,
\prod_{i=1}^3 P_{J_i}^{(d)}(\vec n_i \vec n)
 \ ,
\end{equation}
where $\vec n$ is a reference vector and 
\begin{equation}
\label{eq:NJ}
N_J^{(d)}=(2J+d-3)\frac{\Gamma(J+d-3)}{\Gamma(d-2)\Gamma(J+1)} 
\end{equation}
is the expression appearing in the orthogonality relation \eqref{eq:PJorthogonality} for the Legendre polynomials. 
For $d=4$, we have $N_J^{(4)}=2J+1$.

In CFTs (or theories where all radiation is carried by massless particles), the spatial momentum Ward identity implies in addition that 
\be
\label{eq:masslessEEEC}
C_{1 J J+1}=0 \ . 
\ee
This generalizes the condition $C_1=0$, \eqref{eq:CFT-massless}, for two-point energy correlators. Therefore in this case the leading non-trivial three-point multipole is $C_{222}$ in $d>3$.

\subsection{The Buhmann--J\"ager unitarity bounds}

To characterize unitarity constraints on the three-point energy correlator, we follow the analysis of Buhmann and J\"ager \cite{BuhmannJager2022} (see also \cite{Musin2007}).

The basic idea is to take one of the detectors and average it over the sphere in an axisymmetric fashion. To this extent, we consider a nonnegative kernel 
\begin{equation}
\label{eq:smearker}
h(\vec n_2) = \sum_{L=0}^\infty h_L P_{L}^{(d)}( \vec e_z\cdot \vec n_2) \ge 0,
\end{equation}
where $\vec e_z$ is a unit vector which we can choose to be pointing in the direction of the north pole of the celestial sphere. The $h_L$ parametrize the freedom in choosing such a function as long as $h(\vec n_2)$ remains positive. It is clear that \eqref{eq:smearker} is invariant under rotations around the $z$-axis.

Next we define the following kernel
\begin{equation}
K_h(\vec n_1,\vec n_3)\equiv \int d\Omega_{\vec n_2} \,
\langle {\cal E}(\vec n_1)\,{\cal E}(\vec n_2)\,{\cal E}(\vec n_3)\rangle \, h(\vec n_2 ) \ , ~~~~ h(\vec n_2 ) \geq 0 . 
\end{equation}
The kernel $K_h$ is an axisymmetric (invariant under rotating $\vec n_1$ and $\vec n_3$ simultaneously) and positive-definite kernel on the celestial sphere, and
\begin{equation}
\label{eq:positivityKh}
K_h(f,f) \equiv \int d\Omega_{\vec n_1} d\Omega_{\vec n_3}\,
f^*(\vec n_1)\,K_h(\vec n_1,\vec n_3)\,f(\vec n_3)\ge 0
\end{equation}
for every test function $f$. Theorem~2 of \cite{BuhmannJager2022} gives a sharp characterization of such kernels.

For simplicity, let us restrict our attention to the most physically relevant case of $d=4$. In this case we can define
\be
K_h(Y_{J_1 m},Y_{J_3 \tilde m}) =\delta_{m \tilde m} c_m^{(h)}(J_1,J_3) ,
\ee
where axial symmetry of the kernel implies that the integral is only nonzero for $m=\tilde m$. The matrix $c_m^{(h)}(J_1,J_3)$ is Hermitian
\be
c_m^{(h)}(J_1,J_3)=\left( c_m^{(h)}(J_3,J_1) \right)^* \ .
\ee
For a given $m$, consider next a finite sequence of complex coefficients $a_J$ with $J \ge |m|$. We can apply \eqref{eq:positivityKh} to   
\begin{equation}
f(\vec n)=\sum_{J\ge |m|} a_J\,Y_{Jm}(\vec n) \ .
\end{equation}
We obtain
\be
\sum_{J_1,J_3\ge |m|} a_{J_1}^*\,c_m^{(h)}(J_1,J_3)\,a_{J_3} = K_h \left(\sum_{J\ge |m|} a_J\,Y_{Jm}(\vec n), \sum_{J\ge |m|} a_J\,Y_{Jm}(\vec n) \right) \geq 0 \ . 
\ee
This implies that for any $h(\vec n) \geq 0$ and every $m$, the expression $c_m^{(h)}(J_1,J_3)$ seen as a matrix in $J_1$ and $J_3$ is positive semidefinite
\begin{equation}
\label{eq:PSDch}
c_m^{(h)}(J_1,J_3) \succeq 0 \ .
\end{equation}
The general $d$-dimensional statement is analogous, with $J_i$ replaced by a multi-index label, see \cite{BuhmannJager2022} for details.

Let us next calculate $c_m^{(h)}(J_1,J_3)$ in terms of the spherical harmonics used in the previous section using the ansatz \eqref{eq:smearker} for $h(\vec n_2)$. For simplicity, we set $d=4$ (we also work out the case of $d=3$ in Appendix \ref{app:d3bounds}). The result is
\begin{align}
\label{eq:cFromC}
c_m^{(h)}(J_1,J_3)
&=
h_0 \left( \delta_{m,0}\delta_{J_1,0}\delta_{J_3,0}
\,+\,
\frac{C_{J_1}}{2J_1+1}\,\delta_{J_1,J_3} \right) 
\cr
&+
\delta_{m,0}
\left(
\frac{h_{J_1}\,C_{J_1}}{(2J_1+1)^{3/2}}\,\delta_{J_3,0}
+
\frac{h_{J_3}\,C_{J_3}}{(2J_3+1)^{3/2}}\,\delta_{J_1,0}
\right)
\cr
&+
\sum_{J_2=1}^{\infty}
\frac{h_{J_2}}{2J_2+1}\,
\frac{(-1)^m\,C_{J_1 J_2 J_3}}{\sqrt{(2J_1+1)(2J_3+1)}}
\begin{pmatrix}
J_1 & J_2 & J_3 \\
0 & 0 & 0
\end{pmatrix}
\begin{pmatrix}
J_1 & J_2 & J_3 \\
-m & 0 & m
\end{pmatrix} \ .
\end{align}
Due to the properties of $3j$-symbols, the sum is truncated to $|J_1 - J_3| \leq J_2 \leq J_1+J_3$.

Setting $m=0$ and $J_1=J_3=J\ge 1$ in
\eqref{eq:cFromC}, the terms proportional to $\delta_{J_i,0}$ drop out and we
find
\begin{equation}
(2J+1)\,c_0^{(h)}(J,J)
=
h_0 C_J
+
\sum_{J_2=1}^{2J}
\frac{h_{J_2}}{2J_2+1}\,
\begin{pmatrix}
J & J & J_2 \\
0 & 0 & 0
\end{pmatrix}^{\!2}\,C_{J J_2 J} \ge 0 \ .
\end{equation}

For the detector localized at the north pole, $h(\vec n_2)=\volS \delta(\vec n_2 -\hat z)$, one has
$h_{J_2}=2J_2+1$, and the bound becomes
\begin{equation}
(2J+1)\,c_0^{(\delta)}(J,J)
=
C_J+\sum_{J_2=1}^{2J} \begin{pmatrix}
J & J & J_2 \\
0 & 0 & 0
\end{pmatrix}^{\!2} \,C_{J J_2 J} \ge 0 \ .
\end{equation}
For comparison, the $m=1$ bound gives
\begin{equation}
(2J+1)\,c_1^{(\delta)}(J,J)
=
C_J
+
\sum_{J_2=1}^{2J}
\left(
1-\frac{J_2(J_2+1)}{2J(J+1)}
\right)
\begin{pmatrix}
J & J & J_2 \\
0 & 0 & 0
\end{pmatrix}^{\!2}\,C_{J J_2 J}
\ge 0 \ .
\end{equation}

\subsection{Bounds on $({C_{222} \over C_2}, {C_{224} \over C_2})$}

Let us work out fully the bounds on $({C_{222} \over C_2}, {C_{224} \over C_2})$, where we limit ourselves to the case $d=4$. We get from $c_m^{(\delta)}(2,2)$ with $0\leq m\leq 2$
\begin{align}
\nonumber
m=0:\qquad
5c_{0}^{(\delta)}(2,2)
&=
C_{2}
+\frac{2}{35}C_{222}
+\frac{2}{35}C_{224}
\geq 0,
\\[4pt]
\label{eq:twosidedC3}
m=1:\qquad
5c_{1}^{(\delta)}(2,2)
&=
C_{2}
+\frac{1}{35}C_{222}
-\frac{4}{105}C_{224}
\geq 0,
\\[4pt]
m=2:\qquad
5c_{2}^{(\delta)}(2,2)
&=
C_{2}
-\frac{2}{35}C_{222}
+\frac{1}{105}C_{224}
\geq 0.
\nonumber
\end{align}
By taking linear combinations of these bounds we get for instance the following two-sided bound on $C_{222}$: 
\be
\label{eq:simpleboundC222}
| C_{222} | \leq 25 C_2 \ . 
\ee
The bounds \eqref{eq:twosidedC3} imply an allowed region in the space of ratios $({C_{222} \over C_2}, {C_{224} \over C_2})$. This takes the form of a triangular shape, which we depict in gray in Figure \ref{fig:twosidedbispectrumbound}. However, they can be further improved by making use of the matrix condition \eqref{eq:PSDch}. Indeed, let us notice that 
\be
\label{eq:detm03pt}
m=0:\qquad
5 \begin{pmatrix}
c_{0}^{(\delta)}(0,0) & c_{0}^{(\delta)}(0,2) \\
c_{0}^{(\delta)}(2,0) & c_{0}^{(\delta)}(2,2) 
\end{pmatrix} 
= \begin{pmatrix}
5 & \sqrt{5} C_2 \\
\sqrt{5} C_2 & C_{2}
+\frac{2}{35}C_{222}
+\frac{2}{35}C_{224} 
\end{pmatrix}
\succeq 0\,,
\ee
where for the evaluation of $c_0^{(\delta)}(0,0)=1$ it is important to note that $C_0$ does not enter our parametrization \eqref{eq:threePointParam}. 
The condition \eqref{eq:detm03pt} in particular implies that the determinant is non-negative
\be
C_{2}
+\frac{2}{35}C_{222}
+\frac{2}{35}C_{224}  \geq C_2^2 \ . 
\label{eq:determinantC2sq}
\ee
In the projective limit, $ C_2  \sim C_{222}  \sim C_{224} \sim \epsilon $, this inequality reduces to the first inequality in \eqref{eq:twosidedC3}.

It is also clear that the bound cannot be improved any further if we only use constraints from positivity on two-point and three-point energy correlators. This is because the above model with
\be
C_2 = \epsilon, ~~~ C_{222} = \epsilon x,~~~C_{224} = \epsilon y, ~~~ C_4 = A , \qquad 0<A \leq {7 \over 9} 
\ee
obeys all the positivity constraints.\footnote{$A$ is taken to be a finite (w.r.t. $\epsilon$) distance from $C_4=0$ and $C_4=1$. This is needed to satisfy positivity for the full matrix $c_0^{(\delta)}(i,j)$, $i,j=0,2,4$. The upper bound $A\leq {7 \over 9}$ arises from pointwise positivity of the EEEC.} 
For any $(x,y)$ inside the gray triangle and small enough $\epsilon$, this energy correlator is unitary and pointwise positive, and the extra constraint \eqref{eq:determinantC2sq} becomes infinitesimally weak as $\epsilon\to0$.

\paragraph{Bounds from an infinite hierarchy.} Let us next show that the event distribution formula leads to strictly stronger constraints on $({C_{222} \over C_2}, {C_{224} \over C_2})$, still working in $d=4$. In other words, the fact that EEEC is part of a consistent hierarchy of energy correlators leads to strictly stronger bounds.

To show this, let us introduce the following notations
\begin{align}
M^{i j}[\varepsilon] &= \int_{S^2} \varepsilon(d \Omega_{\vec n}) n^i n^j , ~~~ Q = M - {1 \over 3} 1 \ , \\
s &= \text{Tr} Q^2, ~~~ t = \text{Tr} Q^3,~~~R=\int_{S^2} \varepsilon(d \Omega_{\vec n}) \left( n^i Q_{i j} n^j \right)^2 \ . 
\end{align}
We then have
\bea
\label{eq:momentsevents3}
C_2 &= {15 \over 2} \langle s \rangle,~~~ C_{222} = {1125 \over 2} \langle t \rangle, ~~~ C_{224}= {7875 \over 8} \langle R -{4 \over 7}t - {2 \over 15} s\rangle ,
\eea
where the averaging $\langle ... \rangle$ is over the event distribution law $\sigma(d \varepsilon)$. We derive \eqref{eq:momentsevents3} in Appendix \ref{app:moments2}.

The idea is next to derive inequalities at the level of a single event. For a given event $\varepsilon$, let us denote the eigenvalues of the matrix $Q^{i j}[\varepsilon]$ by $q_{1 \leq i \leq 3}$. The tracelessness condition and the fact that $M \succeq 0$ together imply that
\be
q_1 + q_2 + q_3 = 0 \ , ~~~ q_i \geq - {1 \over 3} \ . 
\ee
It is then easy to check that on this domain 
\be
-{s \over 6} \leq t \leq {s \over 3} . 
\ee
From $\det M \geq 0$ we also get
\be
t - {s \over 2} + {1 \over 9} \geq 0 \ . 
\ee
Averaging over the events, these constraints lead to the following bound
\be
\max\bigg( - {25 \over 2} C_2 , {75 \over 2} C_2 - {125 \over 2}\, \bigg) \leq C_{222} \leq 25 C_2 \ ,
\ee
where we recall that $0\leq C_2 \leq 5$. This inequality implies a sharp lower bound
\be
- {25 \over 2} \leq {C_{222} \over C_2} , 
\ee
which improves the lower bound \eqref{eq:simpleboundC222} by a factor of two! We thus conclude that the event-distribution formula leads to strictly stronger bounds than the ones produced by the positivity properties of the three-point energy correlator. 

To complete the analysis, in a similar fashion we get that
\be
s^2 \leq R \leq t +{s \over 3} \ . 
\ee
The lower inequality is Jensen's inequality for $n^iQ_{ij}n^j$, using that $\int \varepsilon(d\Omega_{\vec n})n^iQ_{ij}n^j=\Tr QM=\Tr(Q^2)$ (since $\Tr Q=0$), whereas the upper one follows from integrating the Cauchy--Schwarz inequality $(n^i Q_{i j} n^j)^2 \leq n^i (Q^2)_{i j} n^j $. Upon averaging the upper bound reproduces the constraint derived above, however, the lower bound leads to a stronger bound.

First, we notice that on each event $s^3\geq 6 t^2$, so we therefore get $R \geq {6 t^2 \over s}$. The key step is to notice that by the Cauchy--Schwarz inequality we have
\be
\langle t \rangle^2 \leq \langle s \rangle \langle {t^2 \over s} \rangle .
\ee
We therefore get
\be
\langle R \rangle \geq 6{\langle t \rangle^2 \over \langle s \rangle} ,
\ee
which translates into a nontrivial lower bound on ${C_{224} \over C_2}$: 
\be
{C_{224} \over C_2} \geq {7 \over 50} \left( {C_{222} \over C_2} \right)^2 - {C_{222} \over C_2} - {35 \over 2} \ . 
\ee
In this way, we get a strictly stronger bound from the event distribution compared to the one that simply follows from the positivity properties of the three-point energy correlator.

We plot the combination of these bounds in blue in Figure~\ref{fig:twosidedbispectrumbound}. 
The sharpness of the blue region can be seen constructively. The two cusps are
\begin{align}
 A&=\left(-\frac{25}{2},\frac{135}{8}\right),
 &
 B&=(25,45).
\end{align}
The point $A$ is realized, for example, by the three-atom equatorial event
\be
 \varepsilon_A
 =\frac13\sum_{i=0}^{2}
 \delta_{(\cos(2\pi i/3),\,\sin(2\pi i/3),\,0)},
\ee
for which
\be
 (C_2,C_{222},C_{224})
 =\left(\frac54,-\frac{125}{8},\frac{675}{32}\right),
\ee
whereas $B$ is realized by the antipodal two-atom event,
\be
 \varepsilon=\frac12(\delta_{\hat z}+\delta_{-\hat z}),
 \qquad
 (C_2,C_{222},C_{224})=(5,125,225).
\ee
The lowest extremum is obtained at
\be
 C=\left(\frac{25}{7},-\frac{135}{7}\right),
 \qquad
 (C_2,C_{222},C_{224})
 =\left(\frac5{49},\frac{125}{343},-\frac{675}{343}\right).
\ee
It is realized by the minimal four-atom event
\be
 \varepsilon_C
 =\frac14\sum_{\eta_1,\eta_2=\pm1}
 \delta_{\left(
 \eta_1\sqrt{2/7},\,
 \eta_2\sqrt{2/7},\,
 \eta_1\eta_2\sqrt{3/7}\right)}.
\ee
More generally, every point on the lower parabolic arc can be reached by a four-atom measure parameterized as follows: 
\be
 r=\frac{25+2(C_{222}/C_2)}{75},\qquad
 a_r=\sqrt{\frac{1-r}{2}},\qquad b_r=\sqrt r,
\ee
\be
 \varepsilon_r
 =\frac14\sum_{\eta_1,\eta_2=\pm1}
 \delta_{(\eta_1a_r,\,\eta_2a_r,\,\eta_1\eta_2b_r)}.
\ee

\subsection{General two-sided bounds}

Similarly, it is easy to see that \eqref{eq:cFromC} implies two-sided bounds on general $C_{J_1 J_2 J_3}$. To see it, we can choose
\be
h^{\pm}(\vec n_2)=1 \pm P_{J_2}(\vec n_2 \cdot \hat z) \geq 0 \ . 
\ee
For this choice we have $h_0^{\pm}=1$ and $h_{J_2}^{\pm} = \pm 1$. We can then consider the conditions (we set $d=4$ for simplicity)
\be
\begin{pmatrix}
c_m^{\pm,J_2}(J_1, J_1) & c_m^{\pm,J_2}(J_1, J_3) \\
c_m^{\pm,J_2}(J_3, J_1) & c_m^{\pm,J_2}(J_3, J_3) 
\end{pmatrix} \succeq 0 \ ,
\ee
where the off-diagonal elements are proportional to $C_{J_1 J_2 J_3}$. This leads to the bound 
\be
|C_{J_1 J_2 J_3}| \leq \min_{m}\beta_{J_1 J_2 J_3}^{(m)} \sqrt{C_{J_1} C_{J_3}} \ , ~~~ |m| \leq \min(J_1, J_3) \ ,
\ee
where
\begin{equation}
\beta_{J_1J_2J_3}^{(m)}
=
\frac{N_{J_2}^{(4)}}{
\begin{pmatrix}
J_1 & J_2 & J_3\\
0 & 0 & 0
\end{pmatrix}
\begin{pmatrix}
J_1 & J_2 & J_3\\
-m & 0 & m
\end{pmatrix}} \ .
\end{equation}

Similarly, one can show that in general $d$ for the case $J_3 = J_1 + J_2$ the following stronger bound holds,\footnote{This bound also holds for the special case $J_i =2$. As in the discussion above, the lower bound on $C_{222}$ is improved by using the event distribution formula.}
\be
\label{eq:FWthree}
|C_{J_1 J_2 J_3} |  \leq \min\left\{ N_{J_1}^{(d)} \sqrt{N_{J_2}^{(d)} N_{J_3}^{(d)} C_{J_2} C_{J_3}},  N_{J_2}^{(d)} \sqrt{N_{J_1}^{(d)} N_{J_3}^{(d)} C_{J_1} C_{J_3}}, N_{J_3}^{(d)} \sqrt{N_{J_1}^{(d)} N_{J_2}^{(d)} C_{J_1} C_{J_2}} \right\} \ .
\ee
For even $J_i$, this bound is saturated by the back-to-back two-atom event (in this sense it is an analog of the $d$-dimensional two-point Fox--Wolfram bound \eqref{eq:general_bound_cj_nj}). We prove this bound in Appendix \ref{app:stretched-triangle-bound}. 
For example, for $C_{246}$ in $d=4$, the bound above becomes
\be
|C_{246}|
\leq
\min\left\{
15 \sqrt{13} \sqrt{C_4 C_6},\,
9 \sqrt{65} \sqrt{C_2 C_6},\,
39 \sqrt{5}\sqrt{C_2 C_4}
\right\}.
\ee

Let us notice that the sharp two-sided bounds on the three-point partial-wave coefficients in terms of the two-point energy multipoles can be trivially generalized to the higher-point energy correlators using the following simple argument. Consider a set of test functions $f_i(\vec n)$ such that 
\be
\label{eq:boundff}
| f_i(\vec n) | \leq 1.
\ee
We then consider the following state
\begin{align}
C_J = \langle \psi_J | \psi_J \rangle ,
\end{align}
where $| \psi_J \rangle =\sqrt{N_J^{(d)}} \int d \Omega_{\vec n} Y_{J \mathbf{m}}^{(d)}(\vec n) {\cal E}(\vec n) | \psi \rangle$, and recall that we set $E_{\text{tot}}=1$. We then use the energy conservation Ward identity, positivity of the multi-point energy correlators and \eqref{eq:boundff} to write
\begin{align}\nonumber 
\langle \psi_J | \psi_J \rangle &\geq \int \prod_{i=1}^k d \Omega_{\vec n_i} | f_i(\vec n_i) | \langle \psi_J | {\cal E}(\vec n_1) ... {\cal E}(\vec n_k) | \psi_J \rangle \ \\
&\geq \Big| \int \prod_{i=1}^k d \Omega_{\vec n_i}  f_i(\vec n_i)  \langle \psi_J | {\cal E}(\vec n_1) ... {\cal E}(\vec n_k) | \psi_J \rangle \Big| \ . 
\end{align}
We can further consider states which are superpositions of different $J$ harmonics to probe the most general multipole coefficients.
It is clear that in this way we can parametrically bound higher-point multipoles in terms of the two-point energy correlator multipoles $C_J$ (very schematically)
\be
\label{eq:parametericbound}
| C_{J_1 ... J_k} | \lesssim \sqrt{C_{J_1} C_{J_k}} .
\ee

As we will see below, these types of bounds are naturally saturated by hard events in the event distribution formula. 

\section{Physical models of energy correlators}
\label{sec:physmodels}

In this section we consider different physical models for the energy correlators and discuss their different interpretations at the level of the underlying event distribution.

\subsection{Free gas}

First, we consider energy correlators in the state $:\phi^k:$ in the free scalar theory (massless or massive), see \cite{Chicherin:2023gxt,Firat:2023lbp}.
We will work in the center-of-momentum frame $q_{\text{tot}} = (E_{\text{tot}}, \vec{0})$.  In this section we \emph{do not} set $E_{\text{tot}} = 1$.  

In the thermodynamic limit,
\begin{align} \label{eq:thermodynamic limit}
     k \rightarrow \infty, \quad E_{\text{tot}} \rightarrow \infty, \quad \frac{E_{\text{tot}}}{k}
     \quad \text{fixed},
 \end{align}
we expect a homogeneous distribution of energy on the sphere \cite{Chicherin:2023gxt,Firat:2023lbp}.  The $l$-point energy correlator diverges in this limit as $E_{\text{tot}}^l$, so it is useful to define the normalized energy flow operators $\hat{\mathcal{E}}(\vec{n}) = \mathcal{E}(\vec{n})/\braket{\mathcal{E}(\vec{n})}$.  In the thermodynamic limit,
\begin{align} \label{eq:thermodynamic limit of normalized ec}
    \braket{\hat{\mathcal{E}}(\vec{n}_1) \dots \hat{\mathcal{E}}(\vec{n}_l)} \rightarrow \braket{\hat{\mathcal{E}}(\vec{n}_1)} \dots \braket{\hat{\mathcal{E}}(\vec{n}_l)} = 1.
\end{align}

Since correlations are suppressed in the thermodynamic limit, we expect the correlator to admit an expansion in powers of $1/k$. The structure of this expansion is very easy to state: each pair-wise correlation introduces a factor $1/k$. In this way, the corrections are polynomial terms and contact terms of the form 
\begin{align} \label{eq: z-counting rule}
     {z_{ij} \over k}& ( 1 + O(1/k) ) && \text{(polynomial terms)} ,\\
    \frac{\delta(z_{ij})}{k} &(1 + O(1/k)) && \text{(contact terms)}.
\end{align}
We call \eqref{eq: z-counting rule} \emph{the $z$-counting rule} and it straightforwardly follows from the considerations of the energy-momentum fluctuations in the thermodynamic ensemble of the free particles, and the definition of the energy correlators as we show in Appendix \ref{app:free gas details}. 

For example, by applying the $z$-counting rule, we can immediately write the leading corrections to the two-point energy correlator
\begin{align}\label{eq:free gas eec unfixed coefficients}
    \text{EEC}(z) = 1 + \frac{1}{k} \Big(a^{(1)}(\delta(z) - 1) - 3 b^{(1)} (1-2z) \Big) + O(1/k^2),
\end{align}
where we fixed the constant term corrections such that the energy-conservation Ward identity is satisfied. At the order $1/k^2$, we get $z^2$, as well as corrections to the terms that are already present in the formula above. For the three-point energy correlator, energy conservation and the $z$-counting rule fix the form of the three-point energy correlator through $O(1/k^2)$ to be
\begin{align}\label{eq:free gas eeec}
    &\text{EEEC}(z_{12}, z_{13}, z_{23}) = 1 + \frac{1}{k} \sum_{i < j}^{3} \Big[ a^{(1)}\big(\delta(z_{ij}) - 1\big) - 3 b^{(1)} (1 - 2z_{ij}) \Big] \nonumber \\
    &+ \frac{1}{k^2} \Big[ c^{(2)}_{1} + c^{(2)}_{z}\, S_z + c^{(2)}_{z^2}\, S_{z^2} + c^{(2)}_{zz}\, S_{zz} 
    + c^{(2)}_{\delta}\, S_\delta + c^{(2)}_{z\delta}\, S_{z\delta} + c^{(2)}_{\delta\delta}\, S_{\delta\delta} \Big] + O(1/k^3),
\end{align}
where the various structures are given by 
\begin{align}
\begin{split}
    S_z &= z_{12} + z_{13} + z_{23}, \\
    S_{\delta} &= \delta(z_{12}) + \delta(z_{13}) + \delta(z_{23}), \\
    S_{z^2} &= z_{12}^2 + z_{13}^2 + z_{23}^2, \\
    S_{zz} &= z_{12}\,z_{13} + z_{12}\,z_{23} + z_{13}\,z_{23} , \\
    S_{z\delta} &= \delta(z_{12})\,(z_{13} + z_{23}) + \delta(z_{13})\,(z_{12} + z_{23}) + \delta(z_{23})\,(z_{12} + z_{13}), \\
    S_{\delta\delta} &= \delta(z_{12})\,\delta(z_{13}) + \delta(z_{12})\,\delta(z_{23}) + \delta(z_{13})\,\delta(z_{23}).
\end{split}
\end{align}
The coefficients can be calculated explicitly from the energy and momentum moments of the thermal distribution of free particles. 

For the massless free gas in $d = 4$ we obtain
\begin{gather}
    \begin{aligned}
    \label{eq:twoPointLargeK}
    \text{EEC}(z)
    &= 1 + \frac{1}{k} \left[\frac{3}{2}\big(\delta(z) - 1\big) - \frac{9}{2} P_1(1-2z) \right] \\
    &+ \frac{1}{k^2} \left[-\frac{3}{2}\big(\delta(z) - 1\big) + \frac{9}{2} P_1(1-2z) + 12 P_2(1-2z) \right] + O(1/k^3) \ ,   
    \end{aligned} \\
    \begin{aligned}
    &\text{EEEC}(z_{12}, z_{13}, z_{23}) = 1 + \frac{1}{k} \sum_{i < j}^{3} \Big[ \frac{3}{2} \big(\delta(z_{ij}) - 1\big) - \frac{9}{2} (1 - 2z_{ij}) \Big] \\
    &+ \frac{1}{k^2} \Big[ 168 - 207 \, S_z + 72 \, S_{z^2} + 108 \, S_{zz} 
    -\frac{27}{2} \, S_\delta + 9 \, S_{z\delta} + S_{\delta\delta} \Big] + O(1/k^3).   
    \end{aligned}
\end{gather}
We present the detailed calculation in Appendix \ref{app:free gas details}. For the multipoles we find 
\begin{equation}
\label{eq:C_JJJ massless free gas}
\begin{split}
    C_2 &= \frac{15}{2k} + \frac{9}{2k^2} + O(1/k^3), \\
    C_{J > 2} &= \frac{3}{2}(2J + 1)
    \left(\frac{1}{k} - \frac{1}{k^2}\right) + O(1/k^3) \\
    C_{J_1 J_2 J_3}
    &= \frac{1}{k^2} \Big[
    135(\delta_{J_1,2}\delta_{J_2,1}\delta_{J_3,1}
    + \text{cyclic}) \\
    &\qquad
    -9(N^{(4)}_{J_1}N^{(4)}_{J_2}\delta_{J_3,1}+\text{cyclic})
    +3N^{(4)}_{J_1}N^{(4)}_{J_2}N^{(4)}_{J_3}
    \Big]
    + O(1/k^3).
\end{split}
\, \, \text{(massless)}
\end{equation}
$N^{(4)}_J = 2J + 1$ from \eqref{eq:NJ}.  From the last multipole we can easily verify the condition \eqref{eq:masslessEEEC} $C_{1 J J+1} = 0$ through $O(1/k^2)$.

For the two-point $\text{EEC}(z)$, we can cross-check with the expression given in equation (3.10) of \cite{Firat:2023lbp}. Specializing to $d=4$, their expression becomes
\begin{equation}
    \text{EEC}(z)=\frac3{2(k+1)}\delta(z)+\frac{(k-1)(k-2)}{(k+1)(k+2)}{_2F_1}(3,3,k+3,z)
\end{equation}
which is valid for $k=3,4,\ldots $. From it we extract the multipoles
\begin{align}
\label{eq:C2anyk}
    C_2&=\frac{15}{2(k+1)}-\frac{5 \left(6 k^3-9 k^2-11 k+7\right)}{2 (k+1)}-15k(k-1)(k-2)(S_2(k-1)-\zeta_2),
    \\
\label{eq:C3anyk}
    C_3&=\frac{21}{2(k+1)}-\frac{7 \left(60 k^4-120 k^3-65 k^2+125 k+13\right)}{2 (k+1)}
    \nonumber\\&\quad\qquad \qquad\quad -105k(k-1)(k-2)(2k-1)(S_2(k-1)-\zeta_2), 
\end{align}
etc., where $S_2(n)=\sum_{i=1}^ni^{-2}$ are the second harmonic numbers. Expanding for large $k$ we get perfect agreement with \eqref{eq:twoPointLargeK}. 

Let us also discuss the scaling properties of the energy multipoles. As in the stringy model of the next section, the distribution is almost homogeneous, with the deviation from $1$ being
\be
\label{eq:freegasmoments}
C_J \sim O(1/k), ~~~C_{J_1 J_2 J_3} \sim O(1/k^2) .
\ee
It is instructive to construct dimensionless measures of `non-Gaussianity'
\be
\label{eq:nonGaussianity}
\hat C_{J_1 J_2 J_3} = \frac{C_{J_1 J_2 J_3}}{\sqrt{C_{J_1} C_{J_2} C_{J_3}}} \ .
\ee
The behavior of the normalized non-Gaussianity \eqref{eq:nonGaussianity} in this case is
\be
\hat C_{J_1 J_2 J_3} \sim O(1/\sqrt{k}) .
\ee

Let us next discuss corrections that are nonperturbative in the
$1/k$ expansion.  In addition to the typical fluctuations around the
homogeneous energy distribution, the microcanonical free-gas ensemble
contains large-deviation tails.  Consider, for example, events in which
all particles lie inside two antipodal caps of angular radius $\delta$. To an experimentalist, this would appear as a very clean two-jet event. The probability of such an event is non-zero and is suppressed by the ratio of the corresponding phase-space volumes $\sim \left({2 \text{Area}_\delta \over \Omega_{d-2}}\right)^k$. We therefore see that while the perturbative $1/k$ corrections naturally expand around the homogeneous energy correlator, there are `rare events', see e.g. \cite{Green:2026nnw}, which are nonperturbative in $1/k$ and which exhibit jets.

\subsection{Stringy corrections}
\label{sec:stringy-corrections}

Next we discuss the computation of the energy correlators at strong coupling in the planar limit of ${\cal N}=4$ SYM \cite{Hofman:2008ar,Dempsey:2025yiv,Ren:2026zxs} in the one-particle state of the graviton supermultiplet. In this case, the leading stringy corrections to the energy multipoles $C_J$ can be calculated from the flat-space stringy scattering amplitude, see Section (4.2) in \cite{Dempsey:2025yiv}. This calculation can be trivially generalized to the case of the three detectors. The relevant worldsheet integral is
\begin{align}
{\cal I}_3(\sigma_1, \sigma_2, \sigma_3) &\equiv\left\langle 0 \left|
:\!e^{i\vec{k}_1\cdot\vec{X}(\sigma_1,0)}\!:
:\!e^{i\vec{k}_2\cdot\vec{X}(\sigma_2,0)}\!:
:\!e^{i\vec{k}_3\cdot\vec{X}(\sigma_3,0)}\!:
\right| 0 \right\rangle_{\mathrm{osc}}  \\
&=
\left|2\sin(\sigma_1-\sigma_2)\right|^{\frac{\alpha'}{2}\vec{k}_1\cdot\vec{k}_2}
\left|2\sin(\sigma_1-\sigma_3)\right|^{\frac{\alpha'}{2}\vec{k}_1\cdot\vec{k}_3}
\left|2\sin(\sigma_2-\sigma_3)\right|^{\frac{\alpha'}{2}\vec{k}_2\cdot\vec{k}_3} \ , \nonumber
\end{align}
which represents three insertions of the energy flux operator in the state that corresponds to the worldsheet oscillator vacuum. 
To calculate the stringy corrections to the energy correlators we need to integrate ${\cal I}_3(\sigma_1, \sigma_2, \sigma_3)$ over the locations of the vertex operators
\be
\mathcal{K}(u,v,w) \equiv \int_0^{2 \pi} \prod_{i=1}^3 {d \sigma_i \over 2 \pi} {\cal I}_3(\sigma_1, \sigma_2, \sigma_3),
\ee
where we introduced
\be
u={1 \over 2}\alpha' \vec k_1\!\cdot \vec k_2,
\qquad
v={1 \over 2}\alpha' \vec k_1\!\cdot \vec k_3,
\qquad
w={1 \over 2} \alpha' \vec k_2\!\cdot \vec k_3 \ . 
\ee
The integral evaluates to the simple formula
\be
\mathcal{K}(u,v,w)
=
\frac{
\Gamma(1+u)\Gamma(1+v)\Gamma(1+w)
\Gamma\!\left(1+\frac{u+v+w}{2}\right)
}{
\Gamma\!\left(1+\frac{u}{2}\right)
\Gamma\!\left(1+\frac{v}{2}\right)
\Gamma\!\left(1+\frac{w}{2}\right)
\Gamma\!\left(1+\frac{u+v}{2}\right)
\Gamma\!\left(1+\frac{u+w}{2}\right)
\Gamma\!\left(1+\frac{v+w}{2}\right)
}.
\ee
To calculate the stringy corrections to the energy correlator we expand this result at small $\alpha'$ and denote the corresponding coefficients by $d_{abc}$
\be
\mathcal{K}(u,v,w) = 1 + \sum_{a,b,c} (-1)^{a+b+c} d_{abc}{u^a v^b w^c \over a! b! c!} .
\ee
As explained in \cite{Hofman:2008ar,Dempsey:2025yiv}, these could then be uplifted to the corresponding terms in the energy correlator
\begin{align}
\mathrm{EEEC}
&=
1+
\sum_{a,b,c}
\frac{d_{abc}}{8\lambda^{(a+b+c)/2}}\,
{\Gamma(a+b+3)\Gamma(a+c+3)\Gamma(b+c+3) \over a! b! c!}\,
(\vec n_1 \cdot \vec n_2)^a (\vec n_1 \cdot \vec n_3)^b (\vec n_2 \cdot \vec n_3)^c \nonumber \\
&+\text{lower-spin terms} ,
\end{align}
where the lower-spin terms require knowledge of the AdS corrections. The information available in the formula above allows us to fix precisely the leading order behavior of the energy multipoles
\be
\label{eq:threepN4}
C_{J_1 J_2 J_3}=\frac{d_{{J_1+J_2-J_3 \over 2}{J_1+J_3-J_2 \over 2} {J_2+J_3-J_1 \over 2}}}{\lambda^{(J_1+J_2+J_3)/4}}\,
\frac{(J_1+J_2+J_3+1)!!}{8}
\prod_{i=1}^{3}
\frac{\Gamma(J_i+3)}{(2J_i-1)!!} + \text{subleading},
\ee
where $x!!=x \cdot (x-2) \cdot ... \cdot 1$ .
Similarly, the energy multipoles $C_J$ were calculated in \cite{Hofman:2008ar}
\begin{align}
C_2 &=  {4 \pi^2 \over \lambda} + \ldots   \  , ~~~ C_3 = {360 \zeta_3 \over \lambda^{3/2}}+ \ldots \ .
\end{align}
As required by consistency, the masslessness condition \eqref{eq:masslessEEEC} is satisfied and $C_{1 J J+1}=0$.

The leading three-point connected multipole was (numerically) calculated in \cite{Hofman:2008ar}
\begin{align}
C_{222} &= {1680 \zeta_3 \over \lambda^{3/2} } ,
\end{align}
which our calculation above confirms.
Using \eqref{eq:threepN4}, we trivially can get higher multipoles as well
\be
C_{224} = {360 \pi^4 \over \lambda^2} + \ldots \ , ~~~ C_{233} = {252 \pi^4 \over \lambda^2} + \ldots \ . 
\ee
These are the values used to produce Figure~\ref{fig:twosidedbispectrumbound} in the Introduction, where we used that $\lambda=(4 \pi g)^2$.

Let us notice that for the normalized non-Gaussianity \eqref{eq:nonGaussianity} in the string corrections example 
\be
\hat C_{J_1 J_2 J_3} \sim {\cal O}(1) \ ,
\ee
which is in sharp contrast to the free gas model, where it goes to zero in the $k \to \infty$ limit.

It is interesting to ask if there are nonperturbative corrections in $1/\lambda$. The exclusive amplitudes at strong coupling behave as $e^{ - \sqrt{\lambda} \text{Area}}$ \cite{Alday:2007hr}. They are infrared-divergent, however the scaling suggests that as in the free-gas model, the strong coupling probability distribution contains hard events with collimated energy flux, which are nonperturbative in $1/\lambda$. It would be interesting to understand the nonperturbative parts of the strong-coupling event distribution better.

It is also interesting to ask how the parametric bound \eqref{eq:parametericbound} constrains higher-derivative corrections to the gravitational effective actions in AdS. The calculation of energy correlators in this case is very subtle because the light-ray transform and the low-energy expansion do not commute \cite{Goncalves:2014ffa,Ren:2026zxs}. The expected behavior in the string-type tree-level gravitational EFT leads to
\be
C_{J_1 ... J_k} \sim \left( {\ell_s \over R_{AdS}} \right)^{\sum_{i=1}^k J_i} ,
\ee
which is a much stronger suppression than \eqref{eq:parametericbound}, which rather corresponds to the effective action obtained by integrating out a massive particle in AdS
\be
C_{J_1 ... J_k} \sim {1 \over c_T} \ . 
\ee
We therefore see that the bound \eqref{eq:parametericbound} is not refined enough to reproduce the expected scaling of the energy multipoles. Nevertheless it is an interesting example of a bound on the connected higher-point correlation functions in AdS/CFT.

\subsection{Loops in AdS}

\begin{figure}
    \centering
    \includegraphics[width=0.4\linewidth]{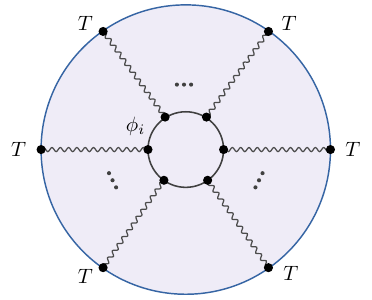}
    \caption{Matter loop contributing to energy correlators.}
    \label{fig:oneLoopHigherPtEnergyCorrelator}
\end{figure}

Let us consider next energy correlators in a gravitational theory which is given by $N_s$ minimally coupled massive scalar fields in AdS
\begin{equation}
S= \int d^{d+1}x \sqrt{g} \Big[-\frac{1}{16\pi G_N}(R-2\Lambda) +\frac{1}{2}\left( \partial_{\mu}\phi^i \partial^{\mu}\phi^i + m^2 \phi^i \phi^i \right) \Big] \,,
\end{equation}
where $i=1,\cdots N_s$. This corresponds to Einstein gravity minimally coupled to $N_s$ scalar fields dual to operators with dimension $\Delta_{\phi}$. Let us denote by
\be
\epsilon = {N_s \over c_T} \ll 1 .
\ee
The most relevant diagrams for the calculation of $n$-point energy correlators are shown in Figure \ref{fig:oneLoopHigherPtEnergyCorrelator}.

Based on the simple large $c_T$ scaling of the diagrams, we get
\be
C_J \sim \epsilon, ~~~ C_{J_1 J_2 J_3} \sim \epsilon ,
\ee
and the same scaling for the higher-point energy correlators. The normalized non-Gaussianity is
\be
\hat C_{J_1 J_2 J_3} \sim \epsilon^{-1/2} ,
\ee
which diverges in the $\epsilon \to 0$ limit in contrast to the other two models considered in this section. 
Let us also point out that perturbative AdS diagrams are known to effectively appear when studying gauge theories at weak coupling due to the instanton correction \cite{Dorey:1999pd,Maldacena:2015iua}.

\subsection{Event interpretation}

Next we would like to understand the interpretation of the models above at the level of the event distribution formula. It is convenient to organize the underlying events in two families:
\begin{itemize}
    \item \textbf{soft events:} there are events where the distribution of energy is almost homogeneous $\varepsilon(d \Omega_{\vec n}) = {d \Omega_{\vec n} \over \Omega_{d-2}} + \delta \varepsilon(d \Omega_{\vec n})$, and $\delta \varepsilon(d \Omega_{\vec n}) \ll 1$.
    \item  \textbf{hard events:} these are events which have hard constituents (e.g. atomic or multi-particle events) so that $\delta \varepsilon(d \Omega_{\vec n}) = O(1)$.
\end{itemize}
The sharp difference between the two types of events is that they lead to different predictions for the higher-point mean-subtracted energy correlators, which we can define by considering the subtracted detector operators
\be
{\cal E}_{\text{sub}}(\vec n) \equiv {\cal E}(\vec n) - \langle {\cal E}(\vec n) \rangle \ , 
\ee
or, equivalently, by studying the multipole expansion of the energy correlators. We then define
\be
\langle {\cal E}(\vec n_1) ... {\cal E}(\vec n_k) \rangle_{\text{sub}} \equiv \langle {\cal E}_{\text{sub}}(\vec n_1) ... {\cal E}_{\text{sub}}(\vec n_k) \rangle .
\ee
Applying this definition to the event distribution formula we get that
\be
\langle {\cal E}(f_1) ... {\cal E}(f_k) \rangle_{\text{sub}} = \int \sigma(d \varepsilon) \prod_{i=1}^k \int  \delta \varepsilon(d \Omega_{\vec n_i}) f_i (\vec n_i) \ . 
\ee
We now see that there are two ways in which the subtracted energy correlators can be small. Let us introduce a small parameter
\be
| \delta \varepsilon | \sim \epsilon \ll 1 \ .
\ee

First, we could have the event distribution measure to be fully localized on the soft events. In this case
\be
\label{eq:softscaling}
\langle {\cal E}(\vec n_1) ... {\cal E}(\vec n_k) \rangle_{\text{sub}} \sim O(\epsilon^k) .
\ee
Both the stringy corrections and the free-gas model are of this type. The difference between the two is the structure of $\delta \varepsilon$ itself. For the stringy corrections it is a structureless ``jelly,'' in which higher spin harmonics of the mean-subtracted energy correlators are further suppressed, see \eqref{eq:threepN4}. In the free-gas model the fluctuations are carried by soft particles (which are atomic measures on the celestial sphere), and all multipoles are of the same order, see \eqref{eq:freegasmoments}. This is also manifested in the presence of the delta-function contact terms for coincident detectors in this case.

Second, we could imagine that the event distribution admits a \emph{rare} hard fluctuation for which $\delta \varepsilon \sim O(1)$. If this is the case, we expect that 
\be
\label{eq:rareevents}
\langle {\cal E}(\vec n_1) ... {\cal E}(\vec n_k) \rangle_{\text{sub}} \sim O(\sigma_{\text{rare}}) \ . 
\ee
This is the structure we expect to get from considering matter (or graviton) loops in AdS. In this case adding detectors does not lead to the additional suppression of the energy correlator. We expect somewhat similar situations at weak coupling, where we do expect spherically symmetric energy distributions to be rarely produced and capture the instanton corrections to the underlying correlation function \cite{Maldacena:2015iua}.

Table~\ref{tab:modelsofhierarchies} in the Introduction summarizes the
event-level interpretation of the models discussed in this section.
These models are sharply distinguished by the behavior of the normalized
non-Gaussianity $\hat C_{J_1 J_2 J_3}$.

\section{Positivity in experimental energy correlators}
\label{sec:experimental}

\begin{figure}
    \centering
  \includegraphics[width=0.84\textwidth]{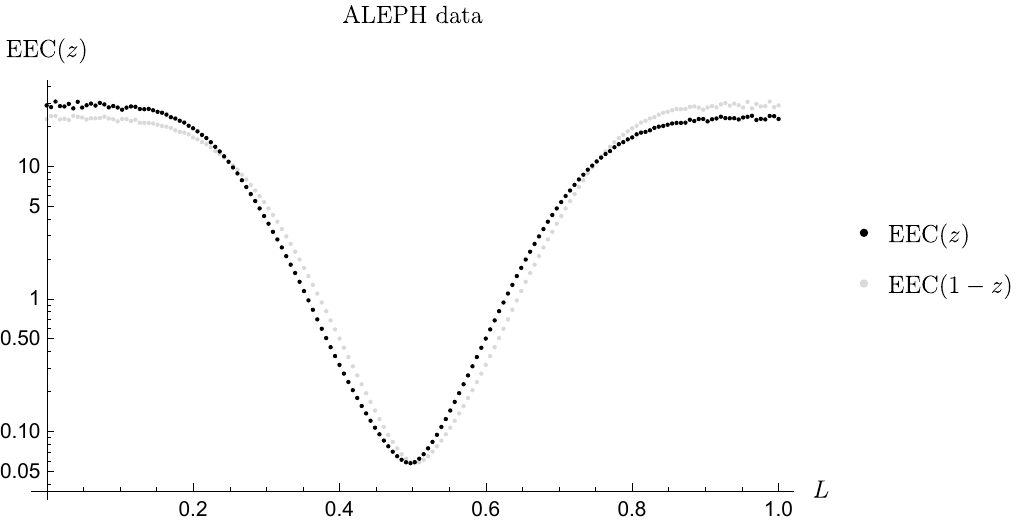}
    \caption{The ALEPH data set (showing central values only) plotted in the two-sided logarithmic scale used in \cite{Electron-PositronAlliance:2025wzh,Bossi:2025nux}. In gray we also display the mirrored plot of $\text{EEC}(1-z)$, which highlights that the continuous part of the physical energy-correlator is skewed to the right: $\text{EEC}(z)>\text{EEC}(1-z)$ for $0.5<z<0.998$ ($0.5<L<0.755$), consistent with $A_1<0$.}
    \label{fig:aleph}
\end{figure}

In this section, we consider the experimentally measured two-point energy correlator in $e^+e^-$ collisions, where we in particular will use data from the large electron-positron collider (LEP). 
In Figure~\ref{fig:aleph} we display this data in a doubly logarithmic plot.\footnote{In the plot, the horizontal axis has a two-sided logarithmic scale defined by
\begin{equation}
\label{eq:Ldefn}
    L=\frac{\log(10^5z)}{2\log(10^5/2)}, \quad z\leq\frac12, \qquad  1-L=\frac{\log(10^5(1-z))}{2\log(10^5/2)}, \quad z\geq\frac12
\end{equation}
for $0\leq z\leq 1$. }
Energy correlators in hadronic final states, introduced in
\cite{Basham:1977iq,Basham:1978bw,Basham:1978zq,Basham:1979gh}, form a rich class of observables that probe different physical regimes as the angular separations between detectors are varied. They therefore provide a powerful window into QCD dynamics, including confinement, see the recent review \cite{Moult:2025nhu}. For the two-point $\text{EEC}(z)$, in the collinear limit $z\ll1$, the effective energy scale of the process is 
\begin{equation}
    \mu^2_{\text{eff}}\sim zE_{\mathrm{tot}}^2\sim \theta^2E_{\mathrm{tot}}^2
\end{equation}
and likewise in the back–to-back limit $\mu^2_{\text{eff}}\sim (1-z)E_{\mathrm{tot}}^2$. We refer to \cite{Jaarsma:2025tck} for a detailed discussion about the physics in the two-point energy correlator and for more references. 

The experimental $\text{EEC}(z)$ has a continuous piece and contact term $A_\delta\delta(z)$. The latter is the
self-correlation contribution, in which both detector insertions register
the same particle, and is therefore localized at zero angular separation,
$z=0$. The partial wave decomposition therefore reads
\begin{equation}
\label{eq:CJwithDeltaAJ}
    \text{EEC}(z)=\sum_{J=0}^\infty C_JP_J(1-2z)=A_\delta \delta(z)+\sum_{J=0}^{\infty}A_JP_J(1-2z)
\end{equation}
where $A_J$ are the multipole coefficients of the continuous part. From \eqref{eq:CJwithDeltaAJ}, it is clear that 
\begin{equation}
    C_J=A_J+(2J+1)A_\delta
\end{equation}
and unitarity puts constraints on positivity of $C_J$ rather than $A_J$, while pointwise positivity is imposed on the continuous part.  

The key observation in this section is that the continuous part of $\text{EEC}(z)$ produces $A_J<0$ for some $J$. Notably, this happens already for $J=1$, where $A_1<0$ can be inferred from the fact that $\text{EEC}(z)$ is skewed to the right. 
Demanding that $C_J\geq0$, we can therefore derive a lower bound on $A_\delta$ by 
\begin{equation}
\label{eq:AdeltaFromSup}
A_{\delta}
\geq
\sup_{J}
\left[
-\frac{A_J}{2J+1}
\right].
\end{equation}
In this section, we will extract $A_J$ from an experimental data set \cite{Electron-PositronAlliance:2025wzh,Bossi:2025nux} from the ALEPH collaboration at LEP. This data set concerns binned data, and in order to precisely compute the $A_J$ we would need to unfold this binning. Moreover, the data for each bin is assigned a statistical uncertainty. In the rest of this section, we will therefore describe three different protocols to extract the $A_J$. We will then derive lower bounds on $A_\delta$.   

\subsection{Definitions and data sets}

Here we will use the data set of \cite{Electron-PositronAlliance:2025wzh,Bossi:2025nux}, which reports measurements of the two-point energy correlation in hadronic events in $e^+e^-$ collisions at $E_{\mathrm{tot}}=m_Z=91\,\text{GeV}$. The data we use was collected by the ALEPH collaboration \cite{ALEPH:1990ndp},\footnote{An openly available data set for the full (\emph{i.e.} not track-based) energy correlators, using 100 bins of equal angular size in $\theta$ (with $\cos\theta=1-2z$), has been produced by the OPAL collaboration \cite{OPAL:1993pnw}, see \href{https://www.hepdata.net/record/ins354188}{\texttt{hepdata.net/record/ins354188}}.} and was recently re-analyzed as a track-based observable. Measurements on tracks means that only electrically charged particles contribute to the energy correlator (corresponding to an extra indicator function $\Theta_{q\neq0}$ in \eqref{eq:crosss}). The track-based formalism \cite{Chang:2013iba,Chang:2013rca,Li:2021zcf,Chen:2022muj,Chen:2022pdu,Jaarsma:2022kdd,Jaarsma:2023ell} was employed because it allowed for an unprecedented angular resolution, in particular in the limits $z\to0$ and $z\to1$. 
The track-based energy correlator has a partial wave decomposition
\begin{equation}
    \text{EEC}_{\text{track}}(z)=\sum_{J=0}^\infty C_{J,\text{track}}\,P_J(1-2z)
\end{equation}
but compared to our usual energy correlator, we lose the overall normalization $C_0= 1$. Instead this multipole has the interpretation of the effective ratio of charged particles (see equation~(2.8) of \cite{Jaarsma:2025tck}),
\begin{equation}
\label{eq:ratioCharged}
    C_{0,\mathrm{track}}=A_{0,\mathrm{track}}+A_{\delta,\mathrm{track}}=r_{Q\neq0,\text{eff}}^2=\frac1{E_\mathrm{tot}^2}\left\langle\bigg(\sum_{i,\text{charged}}E_i\bigg)^2\right\rangle \leq 1
\end{equation}
The rest of our conditions remain:\footnote{They follow trivially from the particle representation (or the event distribution formula), see Appendix~\ref{app:general-event-shapes} for unitarity constraints on more general collider observables. } 
\begin{align}
    \text{EEC}_{\text{track}}(z)&\geq0
    \\
    C_{J,\text{track}}&\geq0
    \\
    C_{J,\text{track}}&\leq C_J^{\text{max}}C_{0,\text{track}}
\end{align}
where the upper bounds $C_J^{\text{max}}$ represent the upper bounds on the $C_J$ coefficients discussed in Section~\ref{sec:twoPoint} above. In the rest of this section we will drop the subscript ``track''. 
Note also from Section~7.6 of \cite{Jaarsma:2025tck} that the computation on tracks barely changes the shape with respect to that of the full energy correlator.

Formally, $k$-point energy correlators in $e^+e^-$ events can be described as expectation values in a state formed by the electromagnetic current
 $J_\mu$, i.e. 
\begin{equation}
    \langle \mathcal E(n_1)\cdots \mathcal E(n_k)\rangle_{\mu\nu} =\frac{1}{\sigma E_{\mathrm{tot}}^k} \int d^4xe^{iqx}\langle 0 | J_\mu(x)\mathcal E(n_1)\cdots \mathcal E(n_k)J_\nu(0)|0\rangle 
\end{equation}
where $E_{\mathrm{tot}}=91\,\text{GeV}$ is the total energy, and $\sigma_{\text{tot}} =L_{\mu\nu}\int d^4xe^{iqx}\langle 0 | J_\mu(x)J_\nu(0)|0\rangle $ is the total inclusive cross section. 
Contracting the Lorentz indices with a polarization tensor $L_{\mu\nu}$ and averaging over all angles produces an observable that is independent of the beam direction, and therefore conforms to the assumption of rotationally-invariant states used in this paper. The two-point energy correlator is then defined by
\begin{equation}
    \text{EEC}(z)=\int d^2n_1d^2n_2\delta\left(z-\frac12(1-n_1\cdot n_2)\right)L_{\mu\nu}\langle \mathcal E(n_1) \mathcal E(n_2)\rangle^{\mu\nu}
\end{equation}
We refer to \cite{Electron-PositronAlliance:2025wzh,Bossi:2025nux,Jaarsma:2025tck} for more precise definitions. 

The data set consists of a set of bin separators $b_1<b_2<\ldots<b_{N_{\mathrm{bins}}+1}$ and integrated values with uncertainties.\footnote{We thank the authors of \cite{Electron-PositronAlliance:2025wzh} and \cite{Jaarsma:2025tck} for sharing this data in raw format.} 
\begin{equation}
    \frac1{b_{i+1}-b_i}\int\limits_{b_i}^{b_{i+1}}\text{EEC}(z)dz=e_i\pm\delta_i
\end{equation}
The bins have different size, although they are symmetric in $z\to1-z$: $b_{i}=1-b_{N_{\mathrm{bins}}+2-i}$.

\subsection{Extraction of multipole coefficients and bound on delta function}

We now proceed to extracting the multipole coefficients $A_J$. 
We assume that we only have access to the continuous piece of the energy correlator, corresponding to what is presented in \cite{Electron-PositronAlliance:2025wzh,Bossi:2025nux}.\footnote{The coefficient of $\delta(z)$ is actually also measured, see \eqref{eq:measuredDelta} below, and is needed to produce Figure~21 in \cite{Jaarsma:2025tck}. We thank the authors of \cite{Electron-PositronAlliance:2025wzh} and \cite{Jaarsma:2025tck} for sharing this data in raw format.}

There is a binning procedure, which we have to unfold.

\subsubsection*{Bounding the contact term in the ALEPH data set}
\label{sec:ALEPHdataset}

We next present three protocols for extracting $A_J$. The reason that we employ more than one protocol is that there is no canonical way to deal with the following three issues:
\begin{itemize}
    \item We have only access to the bin-averaged observables $\int_{b_i}^{b_{i+1}}dz\, \text{EEC}(z)$, while to extract $A_J$ one needs also the moments $\int_{b_i}^{b_{i+1}}dz\, z^k\text{EEC}(z)$ for $k\leq J$. We will assume that the underlying function is sufficiently smooth and use interpolations. 
    \item The bins have assigned statistical error bars, so our extraction should take care of error propagation. 
    \item At the end-points, the central values between adjacent bins fluctuate due to statistical uncertainty, although we expect the underlying function to be a smooth curve. Hence we will treat the end-points with extra care.
\end{itemize}
Moreover, the first bins near the lower end are $\{b_1,b_2,b_3,b_4\}= 10^{-6}\cdot \{9.30732$, $ 10.6124$, $12.1005$, $ 13.7973\}$, which also leaves a small-angle region without any measurement.

\begin{figure}
    \centering
  \includegraphics[height=4.7cm]{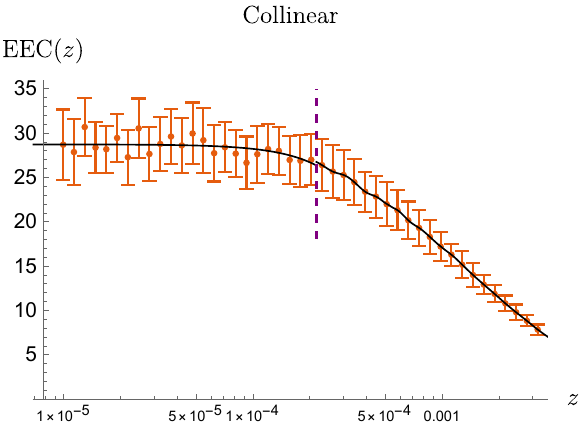}\qquad
  \includegraphics[height=4.7cm]{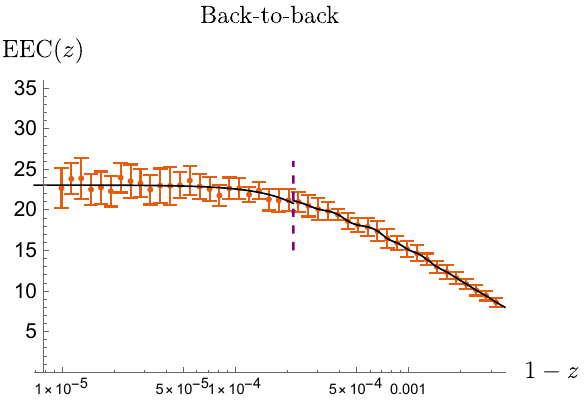}
    \caption{End-point interpolations used in protocol~1 for the collinear and back-to-back regions. Before the dashed line, a quadratic interpolation is used, see \eqref{eq:interpolP2}.}
    \label{fig:interpol}
\end{figure}

\paragraph{Protocol 1. Quadratic interpolation at end-points, and splines elsewhere}
Here we perform an interpolation of the following kind:
\begin{equation}
    g(z)_{\text{interpol}}=\begin{cases}
        A+B z^2 & \text{fitted from first 25 points}
        \\
        g(z)_{\text{splines}} & \text{cubic splines interpolating points with $25<i,n-25$} 
        \\
        C+D (1-z)^2 & \text{fitted from last 25 points}
    \end{cases}
    \label{eq:interpolP2}
\end{equation}
where we find $A=14.3509$, $ B=-2.51513\cdot 10^7$, $C=11.5147$, $D=-2.39368\cdot10^7$.\footnote{The large values for B and D are artifacts of performing the fit using very small values of $z$ and $1-z$, and the final results are largely insensitive to their values, e.g. they can even be set to zero.} In Figure~\ref{fig:interpol} we show what such an interpolation looks like, including the small discontinuities appearing at the merger of the three cases.
The integrals against the Legendre polynomials are then
\begin{equation}
    A_J=(2J+1)\int_0^1dz\,P_J(1-2z)g(z)_{\text{interpol}}\,.
\end{equation}
This protocol does not allow for a clean error propagation. 

\paragraph{Protocol 2. Fixing end-points only and integrating using midpoint rule}

Here we infer the values of the zeroth and $(n+1)$th bins to be the average of the first 10 and last 10 data points respectively, giving
\begin{equation}
    e_0\pm \delta_0=28.725\pm 3.173, \qquad e_{n+1}\pm \delta_{n+1}=23.096\pm 2.004
\end{equation}
and $b_0=0$, $b_{n+2}=1$. The error is taken to be the mean of the errors of these ten data points. 
For the rest of the points, we simply use the central values. This means that we compute the integrals against the Legendre polynomials by the sums
\begin{equation}
    A_J=(2J+1)\sum_{i=0}^{n+1}\left(b_{i+1}-b_i\right)P_J\left(1-2\frac{b_i+b_{i+1}}2\right)e_i\,,
\end{equation}
and add the errors in quadrature.

\paragraph{Protocol 3. Conservative treatment of error bars}
This is similar to protocol 2, but we use the upper and lower limit of the error bars respectively, depending on the sign of the Legendre polynomial and on whether we want to find upper and lower bound on the $A_J$. For instance, if we want to find the upper bound on $A_1$, we use $e_i+\delta_i$ when $P_1(1-2z)=1-2z>0$ (i.e. when $z<1/2$) and $e_i-\delta_i$ when $P_1(1-2z)=1-2z<0$. 
In equations, this reads
\begin{equation}
    A^\pm_J=(2J+1)\sum_{i=0}^{n+1}\left(b_{i+1}-b_i\right)P_J\left(1-2\frac{b_i+b_{i+1}}2\right)\left(e_i\pm\mathrm{sign}\left[P_J\left(1-2\frac{b_i+b_{i+1}}2\right)\right]\delta_i\right)\,.
\end{equation}

\paragraph{Results}

\begin{figure}
    \centering
  \includegraphics[width=0.45\textwidth]{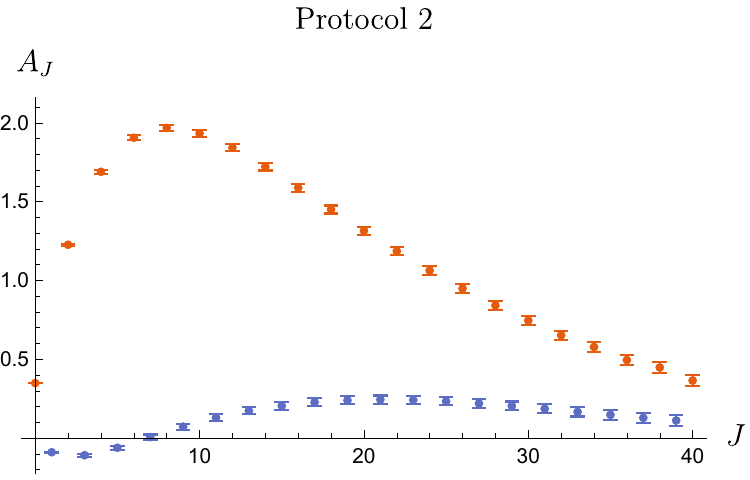}\qquad \includegraphics[width=0.45\textwidth]{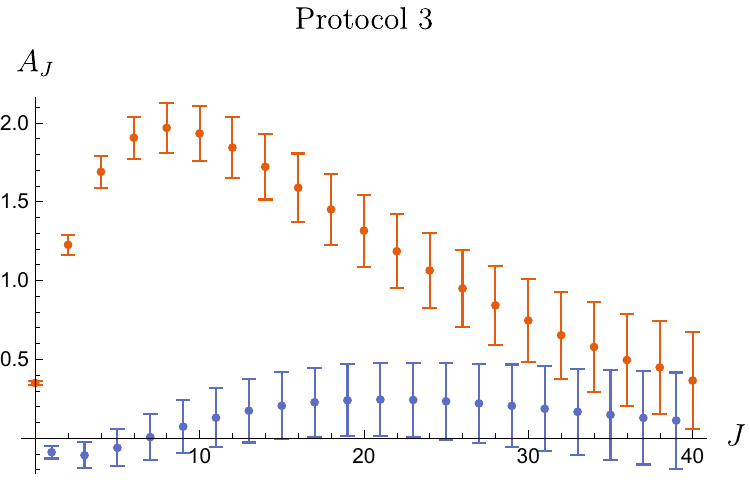}
    \caption{Coefficients $A_J$ extracted from the ALEPH data using protocols 2 and 3. The latter implements the most conservative procedure to estimate the errors.}
    \label{fig:CJcoefsProt3}
\end{figure}

\begin{figure}
    \centering
  \includegraphics[width=0.72\textwidth]{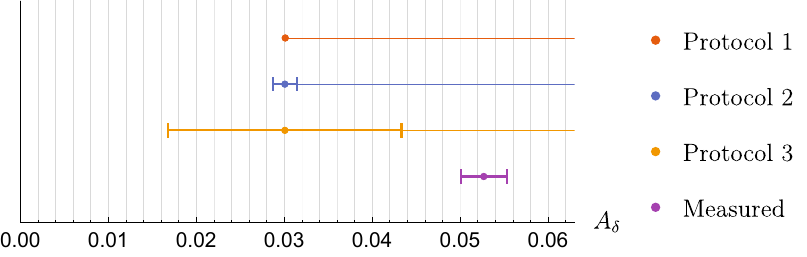}
    \caption{Final result for $A_\delta$ following three different numerical protocols. For each protocol, we have derived a lower bound on $A_\delta$ and the region to the right is allowed. For protocols 2 and 3, there is an uncertainty in the end-point of the allowed region, which we display with error bars.}
    \label{fig:aDelta-res}
\end{figure}

Using the different protocols above, we find the coefficients reported in Table~\ref{tab:AJmeasured}.
\begin{table}[]
    \centering
    \begin{tabular}{|c|r|r|r|}
    \hline
        $J$ & Protocol 1& Protocol 2& Protocol 3  \\\hline\hline
$0$ &$ 0.34916 $ &$0.3487\pm 0.0014 $ &  $ 0.349\pm 0.015 $ \\\hline 
$1$ &$-0.0903 $ & $
 -0.090\pm 0.004 $ & $  -0.09\pm 0.04 $ \\\hline $2$ &$ 1.2282 $ & $
 1.226\pm 0.007 $ & $ 1.23\pm 0.06 $ \\\hline $3$ &$-0.10945 $ & $
 -0.109\pm 0.009 $ & $ -0.11\pm 0.08 $ \\\hline $4$ &$ 1.69162 $ & $
 1.689\pm 0.011 $ & $  1.69\pm 0.10 $ \\\hline $5$ &$-0.0622 $ & $
 -0.062\pm 0.013 $ & $  -0.06\pm 0.12 $ \\\hline $6$ &$1.90908$ & $
 1.905\pm 0.015 $ & $  1.91\pm 0.13 $ \\\hline $7$ &$0.00555 $ & $ 
 0.006\pm 0.017 $ & $ 0.00\pm 0.15 $ \\\hline $8$ &$1.97101$ & $
 1.968\pm 0.018 $ & $  1.97\pm 0.16 $ \\\hline$9$ & $0.07219 $ & $ 
 0.072\pm 0.020 $ & $ 0.07\pm 0.17 $ \\\hline $10$ &$1.93656 $ & $
 1.932\pm 0.021 $ & $  1.93\pm 0.18 $ \\\hline$11$ & $0.12885 $ & $
 0.129\pm 0.022 $ & $  0.13\pm 0.19 $ \\\hline $12$ &$ 1.84566 $ & $
 1.842\pm 0.023 $ & $ 1.84\pm 0.19$
 \\\hline
    \end{tabular}
    \caption{Values of $A_J$ extracted from the measured data using three protocols.}
    \label{tab:AJmeasured}
\end{table}
In Figure~\ref{fig:CJcoefsProt3} we display the results using protocols~2 and~3. 
We note that imposing positivity on $C_1$ will determine the lower bound on $A_\delta$, with positivity on the other $C_J$ following.
We therefore need the results for $A_1$, which we summarize here:
\begin{align}
    &\text{Protocol 1}: & A_1&=-0.09038
    \\&\text{Protocol 2}: & -0.0943<A_1&< -0.0861
    \\&
    \text{Protocol 3}: & -0.12917<A_1&<-0.04987 
    \end{align}
We then use that $
     C_J=A_J+(2J+1)A_\delta
$
and find that \eqref{eq:AdeltaFromSup} is saturated at $J=1$. Thus demanding $C_1\geq 0$ we find bounds on the value of $A_\delta$ which leads to all $C_J\geq 0$. In Figure~\ref{fig:aDelta-res} we display the regions for $A_\delta$ using our three protocols. 
The measured value \cite{Electron-PositronAlliance:2025wzh,Bossi:2025nux}\footnote{We thank the authors of \cite{Electron-PositronAlliance:2025wzh,Bossi:2025nux} for providing us with this value, which can only be indirectly extracted from Figure~21 of \cite{Jaarsma:2025tck}.} is
\begin{equation}
\label{eq:measuredDelta}
    A_{\delta,\text{measured}}=0.0527
\end{equation}
with an error of roughly 5~\%. The results are visualized in Figure~\ref{fig:aDelta-res}. 

We can also study the effective ratio of charged particles, \eqref{eq:ratioCharged}. We find that our bounds imply the following 
\begin{align}
\nonumber
 &\text{Protocol 1:}
    \\ 
   & \quad A_\delta\geq 0.0301394,&r_Q&\geq  0.615867     \\\nonumber
    & \text{Protocol 2:}
   \\
&  \quad  A_\delta\geq 0.02984\pm 0.00137,& r_Q&\geq 0.615060\pm0.00162     \\\nonumber
      &\text{Protocol 3:} \\ 
    &\quad A_\delta\geq 0.02984\pm 0.01322, &r_Q&\geq 0.6151\pm 0.0161 \end{align}
Note that one expects $r_Q$ to be close to $2/3$, in agreement with the expectation that most final particles are pions of which two thirds are charged. A more precise theoretical determination of the effective ratio of charged particles could be done by a simulation, \emph{e.g.} using Pythia.\footnote{We thank G. Vita for discussions.} 
The inverse of $A_\delta$ also corresponds to an effective number of charged particles. 

Adding the measured value \eqref{eq:measuredDelta} of $A_\delta$ to the $A_J$ that we extracted, we get final values of $C_J$:
\begin{equation}
\label{eq:finalCj}
    C_0=0.40(2), \quad C_1=0.07(4), \quad C_2=1.49(6), \quad C_3=0.26(8), \quad \ldots
\end{equation}
where we used protocol 3.

\section{CFTs and AdS quantum gravity}\label{sec:Ads}

In this section, we exploit positivity of energy operators to derive results relevant for CFTs. We consider a rotation-invariant state generated by the stress tensor acting on the vacuum and derive upper bounds on the three-point functions $\lambda_{TT{\cal O}}$. By the AdS/CFT correspondence, these bounds have nontrivial implications for gravitational theories in AdS when the CFTs are holographic. Energy correlators in CFTs were first analyzed in \cite{Hofman:2008ar}, where positivity of the one-point energy correlator was analyzed. These bounds were later extended to a more general state in \cite{Cordova:2017zej,Cordova:2017dhq}, and more recently to the two-point function in \cite{Dempsey:2025yiv,Mecaj:2025ecl}.

\subsection{Positivity constraints in general CFTs}

In conformal field theories, energy correlators are calculated using integrated Wightman correlation functions of the stress-energy tensor. Using the operator product expansion \cite{Kologlu:2019bco,Dempsey:2025yiv,Mecaj:2025ecl}, they can be expressed in terms of the CFT data. A basic insight of Hofman and Maldacena \cite{Hofman:2008ar} is that the OPE data that obeys the standard unitarity bounds  \emph{does not} manifestly generate a positive hierarchy of energy correlators as discussed in the present paper. In this way, we get nontrivial constraints on the CFT data beyond the standard unitarity bounds, which we can broadly call \emph{the Hofman--Maldacena bounds}. They have been explicitly worked out in certain cases, but a complete classification of the bounds on the CFT OPE data that follows from the positivity of the energy correlators as captured by \eqref{eq:representation} is not currently known. 

In contrast to the event distribution realization of energy correlators in gapped theories, energy correlators in interacting CFTs are not generated by multi-particle (or atomic) measures on the celestial sphere. As a result, the CFT energy correlators \emph{do not} contain delta-function contact terms at coincident points, and instead exhibit a simple power-like behavior \cite{Hofman:2008ar}.\footnote{An exception to this statement are free CFTs, in which case the event distribution measure is atomic. Other exceptions are various perturbative expansions, e.g. small-coupling or large-charge/dimension expansions.}

\subsubsection{Stress-tensor unpolarized beams}
\label{sec:T-unpolarized}

To begin the discussion, let us derive simple universal bounds on CFT OPE data using the positivity of the energy-energy correlator. We consider a four-point function of stress tensors and, to simplify the analysis, work with a state ‘averaged over the beam,’ namely, the state generated by
\begin{equation}
\rho_T ={1 \over {\cal N}} \int d^{d-1} \vec n \delta (\vec n^2 -1) |T_{ij}(E_{\text{tot}}, \vec 0) \rangle \langle T_{k l}(E_{\text{tot}}, \vec 0) | n^i n^j n^k n^l .
\end{equation}

For simplicity, as we have often done in this paper, we set $E_{\text{tot}} = 1$ and we choose ${\cal N}$ such that the state is properly normalized, namely that the two-point energy correlator takes the form
\begin{equation}
\label{eq:EECnobeam}
{\rm Tr}[\rho_T {\cal E}(\vec n_1) {\cal E}(\vec n_2)] = 1 + \sum_{J=2}^\infty C_J P_J^{(d)}(\cos \theta), 
\end{equation}
where $\cos \theta = \vec n_1 \cdot \vec n_2$ is the angle between detectors and $P_J^{(d)}(x)$ are given in \eqref{eq:PJdef}.\footnote{Note that with our choice of normalization, Einstein gravity in AdS gives $1$ for the energy correlators. Equivalently, it originates from the stress-tensor contribution to the OPE, and more precisely the universal tensor structure which is completely fixed by the conformal Ward identities.}

Our task will be to calculate the contribution of various operators in the OPE of $T \times T$ to the energy-energy correlator \eqref{eq:EECnobeam}
\begin{equation}
C_J = \sum_{{\cal O}} C_{J}^{{\cal O}} ,\qquad  C_{J}^{{\cal O}} \geq 0 \ ,
\end{equation}
where $C_{J}^{{\cal O}} \geq 0$ follows from the standard CFT unitarity \cite{Kologlu:2019bco}. The problem of calculating $C_{J}^{{\cal O}}$ in terms of the OPE data has been solved for the four-point function of scalar operators in \cite{Kologlu:2019bco}, and more recently for the $\langle \phi T T \phi \rangle$ four-point function in \cite{Mecaj:2025ecl}. 

The generalization to the case of the four-point function of stress tensors follows exactly the same path. To do the calculation we have to introduce the basis for the relevant two- and three-point conformal structures $\langle T T X \rangle$. We do this in Appendix \ref{app:conformalstructures}. We will focus on the contribution of symmetric traceless currents of even spin,
and consider four cases for $X$:
\begin{itemize}
    \item scalar operator of dimension $\Delta$ (1 tensor structure, $\lambda_{TT\mathcal{O}}$)
    \item stress tensor (3 tensor structures parameterized by $(c_T,t_2,t_4)$)
        \item generic spin-2 operator (2 tensor structures $(\lambda_1, \lambda_2)$)
    \item generic spin-4 operator (3 tensor structures $(\lambda_1, \lambda_2, \lambda_3)$)
\end{itemize}
These will be sufficient for our purposes.
We will not consider the parity-odd three-point structures that appear in $d=3$. The calculation is relatively lengthy, but conceptually straightforward and follows the steps described in \cite{Kologlu:2019bco}. A general operator of spin $S$ contributes to $C_J$ with $2 \leq J \leq S+2$.

For the scalar of dimension $\Delta$, we get
\begin{align}
\label{eq:C2fromO}
C_2^{\mathcal{O}} &={\lambda_{TT\mathcal{O}}^2 \over c_T} \sin^2\left(\frac{\pi  (\Delta-2d) }{2} \right) \alpha(\Delta,d)
\end{align}
where 
\begin{equation}
 \alpha(\Delta,d)=   \frac{2 (d-1)^3 \pi ^{d-2} \Gamma \left(\frac{d}{2}+1\right)^3 \Gamma (d-2)\Gamma (\Delta ) \Gamma \left(-d+\frac{\Delta
   }{2}+1\right)^2 \Gamma \left(-\frac{d}{2}+\Delta +1\right)}{(d-2) \Gamma
   \left(\frac{\Delta }{2}+2\right)^4 \Gamma \left(\frac{d+\Delta }{2}\right)^2}.
\end{equation}
For the stress tensor, we get 
\begin{align}
C_0^{T} &= 1 \ , \cr
C_2^{T} &= \frac{(d-3)((d+3) t_2 +4 t_4)^2}{2(d-1)^2 (d+1) (d+3)} , \cr 
C_4^{T} &= \frac{2d t_4^2}{(d+1)^2 (d+3)} \ . 
\end{align}
The universal $1$ in \eqref{eq:EECnobeam} comes from $C_0^T$, which is present in any CFT by Ward identities.
The $t_2$ structure is absent in $d=3$, which is consistent with $C_2^T$ vanishing in this limit. 

We have calculated the contribution of general spin-two and spin-four operators. We present the explicit formulas in an ancillary file (the highest-spin energy multipole contribution can also be found in Appendix \ref{app:conformalstructures}), and here  focus on the simple bounds on the OPE coefficients that follow from the universal bounds on $C_J$. Indeed, given that the contribution of each operator to a given energy multipole is manifestly nonnegative we can write
\begin{equation}
\label{eq:simpleboundTTTT}
0 \leq C_J^{\mathcal{O}} \leq \sum_{{\cal O}} C_{J}^{{\cal O}} \leq C_J^{\text{max}} \ , 
\end{equation}
where $C_J^{\text{max}}$ are the upper bounds discussed in Section \ref{sec:deltafunctionmodels} (here we denote  $C_J^{\text{max}}$ the optimal upper bounds, including the modification on $N^{(d)}_J$ for odd $J$). For scalar operators these bounds were first discussed in \cite{Cordova:2017zej}.

Let us now present simple universal bounds that follow from \eqref{eq:simpleboundTTTT}. First, let us emphasize that they are not optimal, because we could have considered positivity in a more complicated state where we keep track of the polarization of the external stress tensor. We leave this to future work. Second, due to the presence of the universal factor $\sin^2\left(\frac{\pi  (\Delta-2d) }{2} \right)$, the bounds trivialize around $\Delta =2d+S+2 \mathbb{Z}_{\geq 0}$.

\begin{figure}
    \centering
  \includegraphics[width=0.45\textwidth]{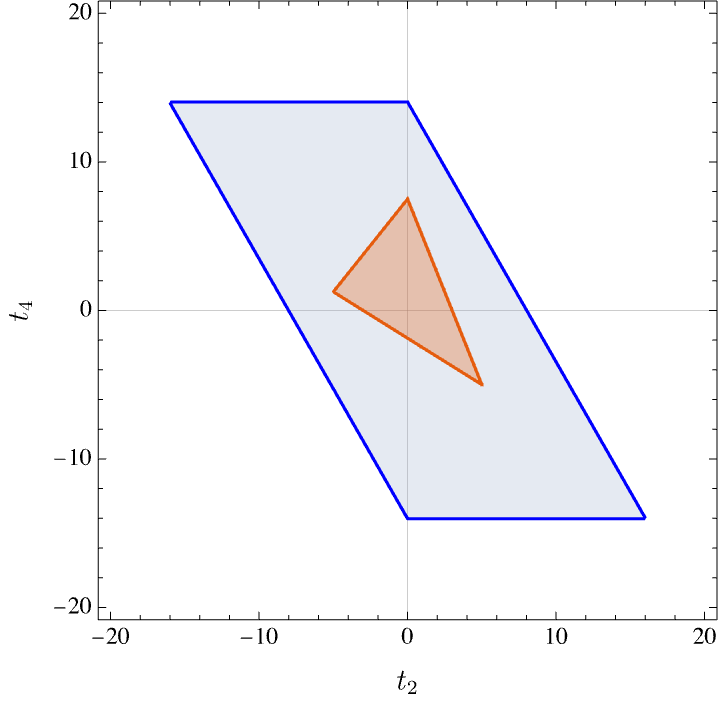}
    \caption{We plot, in blue, the bound \eqref{eq:simpleboundTTTT} applied to the OPE data in $\langle TTT\rangle$ in $d=4$. The standard Hofman--Maldacena bounds are plotted in red.}
    \label{fig:HMplot}
\end{figure}

\begin{figure}[htbp]
    \centering

    \begin{subfigure}[t]{0.48\textwidth}
        \centering
        \includegraphics[width=\linewidth]{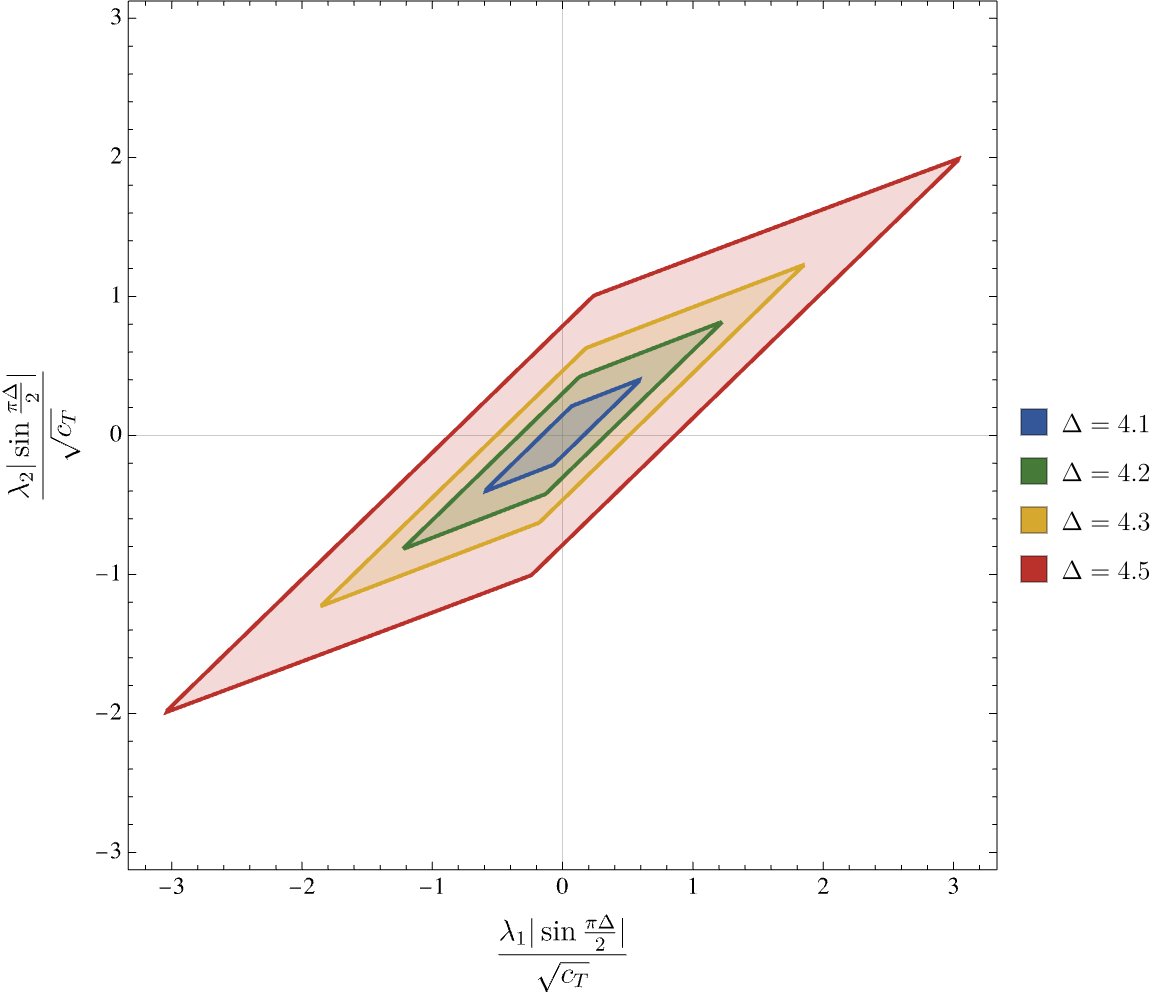}
        \caption{Light operators}
        \label{fig:left-panel_first}
    \end{subfigure}
    \hfill
    \begin{subfigure}[t]{0.48\textwidth}
        \centering
        \includegraphics[width=\linewidth]{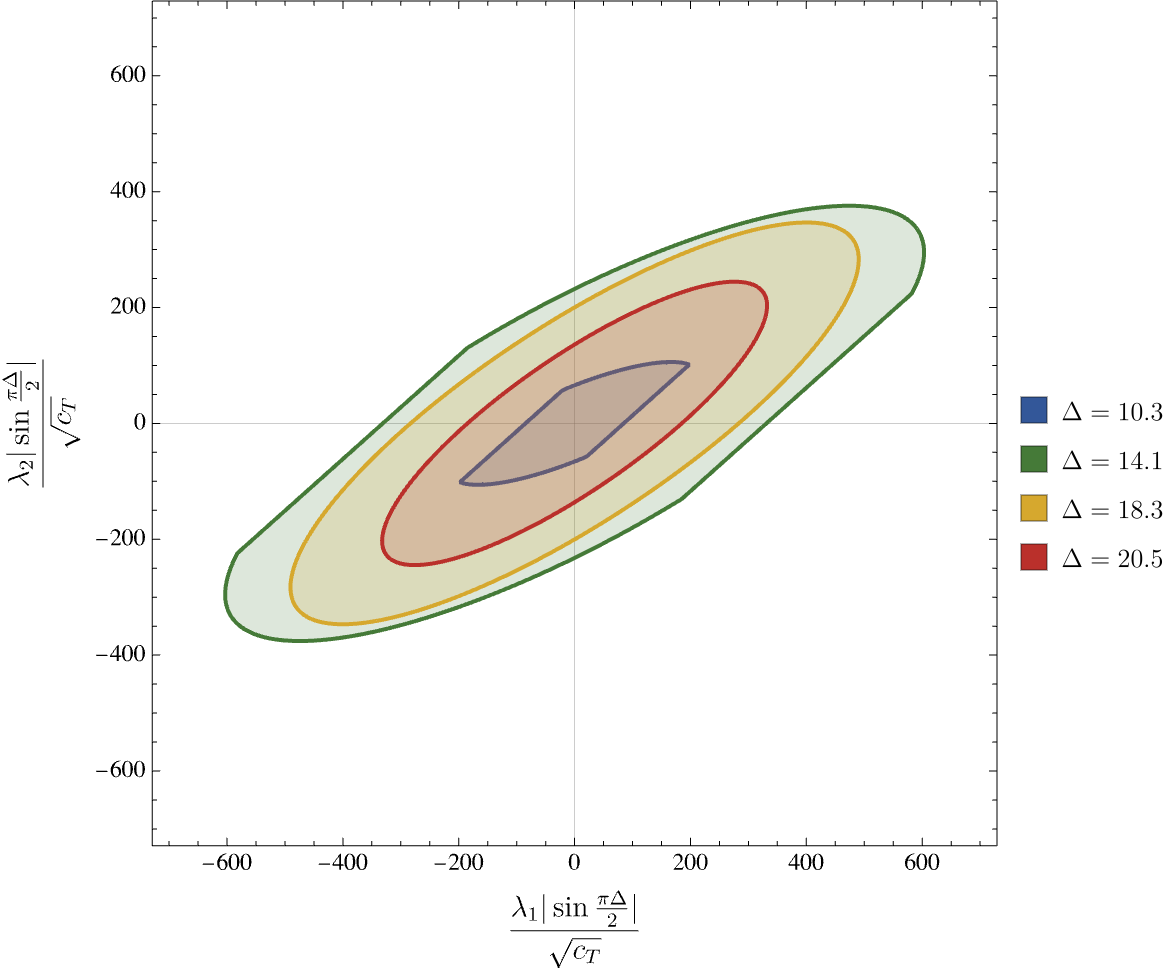}
        \caption{Heavy operators}
        \label{fig:right-panel_first}
    \end{subfigure}

    \caption{Bounds on the OPE coefficients $\lambda_{TT{\cal O}_2}$ in $d=4$, where ${\cal O}_2$ is a generic spin-two operator. For small $\Delta$ the bounds are dominated by $C_{3},C_{4}$, which are effectively linear in $\lambda_i$. For larger $\Delta$, stronger bounds instead emerge from $C_2$ which is quadratic in $\lambda_i$. Physically, it is due to the fact that in this case $C_2$ acquires contributions from states of different $SO(d-1)$ representations, see \cite{Kologlu:2019bco}.}
    \label{fig:two-panels-spin2}
\end{figure}

\begin{figure}[htbp]
    \centering

    \begin{subfigure}[t]{0.48\textwidth}
        \centering
        \includegraphics[width=\linewidth]{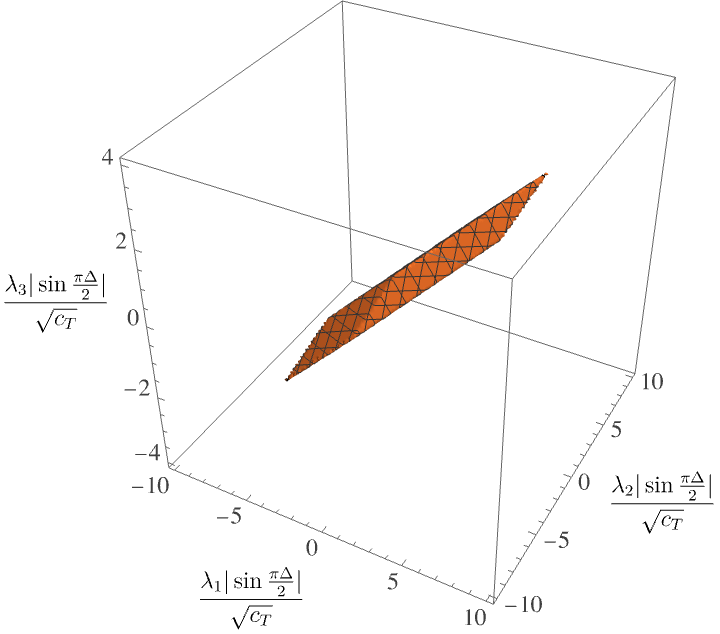}
        \caption{Light operators}
        \label{fig:left-panel_second}
    \end{subfigure}
    \hfill
    \begin{subfigure}[t]{0.48\textwidth}
        \centering
        \includegraphics[width=\linewidth]{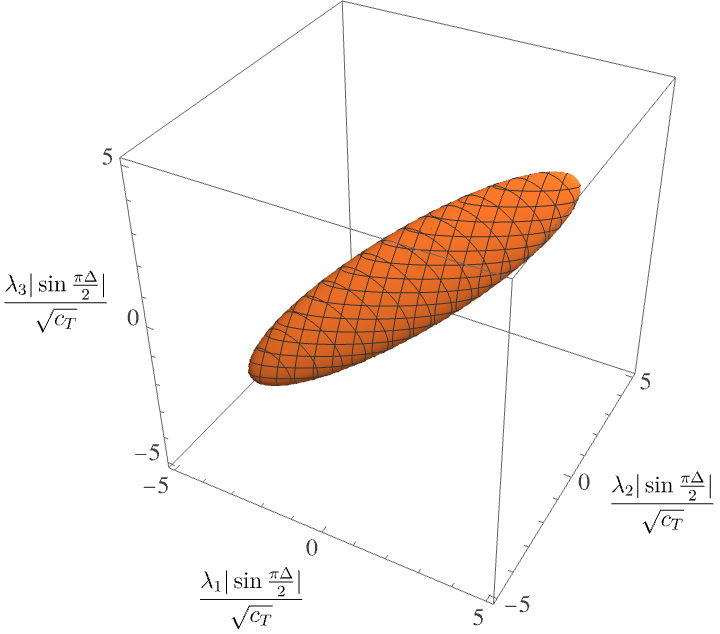}
        \caption{Heavy operators}
        \label{fig:right-panel_second}
    \end{subfigure}

    \caption{Bounds on the OPE coefficients $\lambda_{TT{\cal O}_4}$ in $d=4$, where ${\cal O}_4$ is a generic spin-four operator. For the left panel $\Delta = 6.45$, on the right panel $\Delta = 36.15$. As above linear, constraints dominate at small $\Delta$, and quadratic constraints dominate at higher $\Delta$.}
    \label{fig:two-panels-spin4}
\end{figure}

We plot the bounds on the three-point couplings: for $\langle TTT \rangle$ in Figure \ref{fig:HMplot}; for $\langle TT \mathcal{O}_2 \rangle$ in Figure \ref{fig:two-panels-spin2}; for $\langle TT \mathcal{O}_4 \rangle$ in Figure \ref{fig:two-panels-spin4}.

\paragraph{3d Ising CFT}

For the Ising CFT in three dimensions, we use \eqref{eq:C2fromO} with the values $\Delta_\epsilon=1.41262528(29),$ and $c_T=0.00899103927(40)$ \cite{Chang:2024whx}.\footnote{To determine the central charge in our conventions, one notes that $c_T=\frac1{\Omega_2^2}\frac{3}{2}\frac{C_T}{C_{B}}$, where $\frac{C_T}{C_{B}}=0.946538675(42) $ is given in \cite{Chang:2024whx}.} Solving
\begin{equation}
    \frac{\lambda_{TT\epsilon}^2}{c_T}\sin^2\left(\frac{\pi(\Delta_\epsilon-6)}2\right)\alpha(\Delta_\epsilon,d)\leq N^{(3)}_2=2\,,
\end{equation}
we find
\begin{equation}
\langle T_{\text{pol.aver.}}|\mathcal E\mathcal E|T_{\text{pol.aver.}}\rangle\big|_\epsilon: \qquad \Rightarrow \qquad \lambda_{TT\epsilon}\leq 0.01255
\end{equation}
This should be contrasted with the following bounds
\begin{align}
&    \langle \text{mix}|\mathcal E|\text{mix}\rangle: &\Rightarrow\qquad & \lambda_{TT\epsilon}\leq 0.00882
\\
&    \langle \epsilon|\mathcal E\mathcal E|\epsilon \rangle\big|_T: &\Rightarrow\qquad & \lambda_{TT\epsilon}\leq 0.00888
\end{align}
from \cite{Cordova:2017zej} and \cite{Mecaj:2025ecl} respectively. Here the notation in the first line refers to the fact that  \cite{Cordova:2017zej} considered the bounds deriving from the one-point energy correlator from interference effects in the mixed state $|\text{mix}\rangle=v_1|T\rangle+v_2|\epsilon\rangle$. 

The bootstrap value for the OPE coefficient is \cite{Chang:2024whx}\footnote{We use that $\lambda_{TT\epsilon}^\text{here}=c_T\lambda_{TT\epsilon}^\text{there}$.}
\begin{equation}
    \lambda_{TT\epsilon}=0.0085712937(38).
\end{equation}
We note that our bound is weaker than the other conformal collider bounds.

\paragraph{$\mathcal N=4$ SYM}

We can also consider $\mathcal N=4$ SYM. In this case, scalar operators in the singlet representation of the $R$ symmetry contribute, and we get a bound of the form
\begin{equation}
    \sum_{\O\in[0,0,0]}\frac{\lambda^2_{TT\O}}{c_T}\sin^2\left(\frac{\pi(\Delta_\O-8)}2\right)\alpha(\Delta_\O,4)\leq 5
\end{equation}
One might expect that the sum in this equation is dominated by the Konishi operator $\mathcal K$ with $\Delta_\mathcal K=2+3\frac{g_{YM}^2N_c}{4\pi^2}+\ldots$. This would then give an upper bound on $\lambda_{TT\mathcal K}$, which, in principle, could be compared with results for this OPE coefficient.

\begin{figure}
    \centering
  \includegraphics[width=0.44\textwidth]{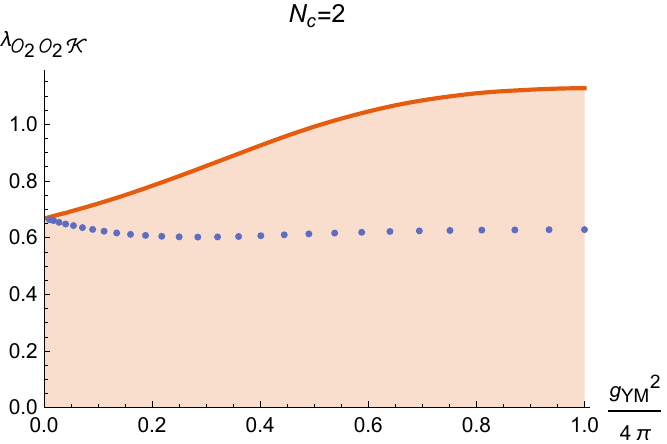}\qquad
  \includegraphics[width=0.44\textwidth]{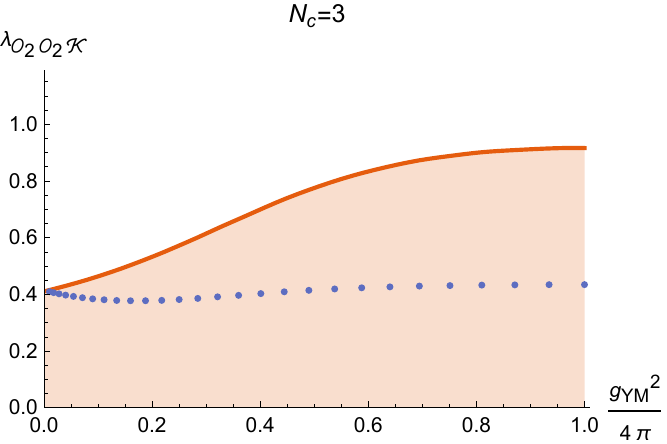}
    \caption{Upper bound on the Konishi OPE coefficient $\lambda_{\mathcal O_2\mathcal O_2\mathcal K}$ following from \eqref{eq:upperBoundKonishi}, compared with bootstrap upper bounds (dots) from \cite{Chester:2021aun,Chester:2023ehi}. Note that the self-dual value of the coupling is $g_{\text{YM}}^2/4\pi=1$.}
    \label{fig:resKonishi}
\end{figure}

It is however more convenient to consider a superconformal formalism, where $\lambda_{TT\mathcal K}$ is related to the OPE coefficient with two half-BPS operators $\mathcal O_2$ in the same supermultiplet as the energy-momentum tensor.
The relevant equation for this setup was derived in Section~3.2 of \cite{Dempsey:2025yiv}. Focusing again on the upper bound on $C_2$, the equations given there reduce to
\begin{equation}
\label{eq:upperBoundKonishi}
c\,    \lambda^2_{\mathcal O_2\mathcal O_2\mathcal K}\sin^2\left(\frac{\pi\Delta_{\mathcal K}}2\right)\frac{49152 (\Delta _{\mathcal{K}}+1){}^2 (\Delta _{\mathcal{K}}+3) \Gamma (\Delta
   _{\mathcal{K}}){}^2}{\pi^2 (\Delta _{\mathcal{K}}-2)^2 (\Delta _{\mathcal{K}}+4)^2
   (\Delta _{\mathcal{K}}+6)^2 \Gamma (\frac{\Delta _{\mathcal{K}}}{2}+2)^4}\leq N_2^{(4)}=5 \ , 
\end{equation}
where $c=\frac{N_c^2-1}4$. In the limit $\Delta_{\mathcal K}\to2$, we find $ \lambda^2_{\mathcal O_2\mathcal O_2\mathcal K}\leq\frac 1{3c}$, which is saturated by the known OPE coefficient in the limit of weak coupling. For finite coupling, we can compare with bootstrap results of \cite{Chester:2021aun,Chester:2023ehi},\footnote{We thank R. Dempsey for sharing the raw data with us.} which give precision values for $\Delta_{\mathcal K}$ and upper bound on the OPE coefficients $\lambda_{\mathcal O_2\mathcal O_2\mathcal K}$. In Figure~\ref{fig:resKonishi} we display this comparison.

\paragraph{Large-$c_T$ scaling} Consider a CFT with a large number of degrees of freedom, such that the stress-tensor two-point function
\begin{equation}
\label{eq:Ttwop}
\braket{T_{\mu\nu}(x_1)T_{\rho\sigma}(x_2)}=c_T \frac{I_{\mu\nu\rho\sigma}(x_{12})}{x_{12}^{2d}} \,, \quad \text{with}\  c_T\gg 1 \,,
\end{equation}
where $I_{\mu\nu\rho\sigma}(x_{12})$ is a function fixed by conformal invariance \cite{Osborn:1993cr}. 
In general, starting from \eqref{eq:Ttwop} we do not know what the properties of the multi-point correlation functions are, in particular, whether the simple factorization and scaling properties observed in gauge theories, see \cite{tHooft:1973alw}, must hold. Based on the identification of the central charge with the Planck constant in the bulk 
\begin{equation}
c_T \sim \left(\frac{\ell_{AdS}}{\ell_p}\right)^{d-1} \,,
\end{equation}
a natural scaling suggested by tree-level gravitational correlators in AdS is
\begin{equation}
\label{eq:univTTTT}
\braket{T(x_1) \cdots T(x_n)}_{\text{conn}} \sim c_T \, .
\end{equation}
To the best of our knowledge, the scaling \eqref{eq:univTTTT} has not been proven from first principles in CFTs.\footnote{This property follows if one assumes large $c_T$ factorization. But large $c_T$ factorization is stronger, because it implies that \textit{all} correlators should factorize, not just those of the stress-tensor. A strongly coupled QFT in AdS, weakly coupled to gravity would have a large $c_T$ CFT dual, with factorizing stress-tensor correlators, but would not obey large-$c_T$ factorization. Explicit constructions of such CFTs are known, see \cite{Aharony:2015zea,Apolo:2022pbq,Apolo:2024bmu}.}

Let us here simply point out that positivity of the energy correlators together with the energy conservation condition 
\begin{equation}
\int d\Omega_{\vec{n}_1} \cdots d\Omega_{\vec{n}_k} \braket{\psi | \mathcal{E}(\vec{n_1}) \cdots \mathcal{E}(\vec{n_k}) | \psi } \lesssim c_T \,,
\end{equation}
imply a related statement, namely that in the large $c_T \gg 1$ limit
\begin{equation}
\label{eq:eeeecuniv}
\textbf{Universality in all large $c_T$ CFTs:} \qquad \braket{\psi | \mathcal{E}(f_1) \cdots \mathcal{E}(f_k) | \psi } \lesssim c_T \,, \quad c_T\gg 1 \,,
\end{equation}
where the test functions $f_i \sim O(1)$. It would be interesting to understand whether \eqref{eq:univTTTT} can indeed be proven rigorously.

\subsection{The AdS species bound}
\label{sec:species-bound}

\begin{figure}
    \centering
    \includegraphics[width=0.3\linewidth]{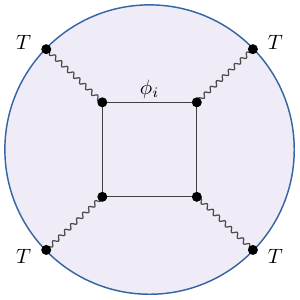} \hspace*{0.2em}
    \includegraphics[width=0.3\linewidth]{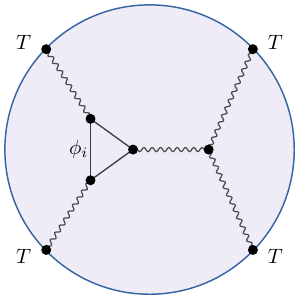} \hspace*{0.2em}
    \includegraphics[width=0.3\linewidth]{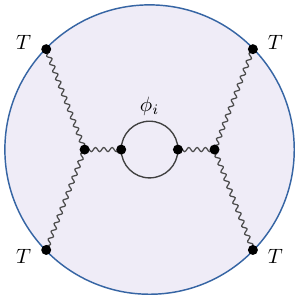}
    \caption{The basic loop diagrams that contribute to the energy-energy correlator in the AdS model. The box diagram is the only one to contribute to $C_{J>4}$.}
    \label{fig:oneLoopTTTT}
\end{figure}

Let us establish next that the bounds derived in the previous section can be interpreted as a species bound in AdS. Indeed, let us notice that if we consider the bounds \eqref{eq:simpleboundTTTT} as a function of $c_T$, they imply that
\begin{equation}
\label{eq:anecbound}
\lambda_{TT \mathcal{O}}^2 \lesssim c_T ,
\end{equation}
up to a $c_T$-independent constant. 
Now, consider the following bulk effective field theory for the bulk theory with $N_s$ scalars:
\begin{equation}
S= \int d^{d+1}x \sqrt{g} \Big[-\frac{1}{16\pi G_N}(R-2\Lambda) +\frac{1}{2}\left( \partial_{\mu}\phi^i \partial^{\mu}\phi^i + m^2 \phi^i \phi^i \right) \Big] \,,
\end{equation}
where $i=1,\cdots N_s$. This corresponds to Einstein gravity minimally coupled to $N_s$ scalar fields dual to operators with dimension $\Delta_{\phi}$. We will now show that there is a bound on the number of scalars given by\footnote{The species bound has been recently proven in flat space using dispersion relations in \cite{Caron-Huot:2024lbf}, and the same argument is expected to apply in AdS.}
\begin{equation} \label{speciesbound}
\frac{N_s}{c_T} \lesssim 1 \,.
\end{equation}
The argument is as follows. First, it is important to realize where the potential problem can come from. Consider the three one-loop diagrams represented in Fig. \ref{fig:oneLoopTTTT}. Each of them is suppressed compared to gravitational exchanges by $N_s/c_T$. But if we take $N_s/c_T\gg 1$, this can dominate over the gravitational exchange and violate \eqref{eq:anecbound}.

In fact, one must be slightly more careful. The loop expansion simply breaks down if $N_s\sim c_T$ as infinitely many bubble diagrams (shown in Fig. \ref{fig:oneLoopTTTT})  need to be resummed. Without explicitly performing this resummation, one cannot draw a strong conclusion since the resummed bubbles and triangles could in principle change the naive ${N_s \over c_T}$ scaling of the three-point functions. In flat space this resummation and the derivation of the species bound was discussed in \cite{Caron-Huot:2024lbf}. 

This problem can be sidestepped if we focus on the bound \eqref{eq:anecbound} applied to the double-trace operators of matter fields $[\mathcal{O}_i\mathcal{O}_i]_{S,n}$ ($O_i$ are the CFT single-trace operators dual to $\phi_i$). All the diagrams depicted in Fig. \ref{fig:oneLoopTTTT} contribute to this three-point function. However, the diagrams that involve infinite bubble chains only contribute to the three-point functions with spin $S \leq 2$. It can be seen by effectively performing the OPE decomposition at the level of the relevant AdS diagrams using the split representation of the spinning AdS propagators \cite{Costa:2014kfa,Carmi:2024tzj}. 
The box diagram, on the other hand, contributes to all spins. We can therefore focus on the energy multipole bounds for $C_{J>4}$ for which only operators with spin greater than $2$ contribute. The species bound \eqref{speciesbound} is therefore a direct consequence of the EEC bound \eqref{eq:anecbound} applied to double-trace operators with $S > 2$
\begin{equation}
\sum_i  \sin^2\left[\frac\pi2(2\Delta_{\phi}-2d) \right]\sum_{n, S >2} \lambda^2_{TT [\mathcal{O}_i\mathcal{O}_i]_{S,n}} \lesssim c_T \ , 
\end{equation}
where the sum over species $\sum_i$ produces the expected $N_s$ scaling, because $\lambda^2_{TT [\mathcal{O}_i\mathcal{O}_i]_{S,n}} \sim O(1)$ in the $ c_T \to \infty$ limit. 
For generic operator dimension $\Delta_{\phi}$, this gives a qualitative upper bound on the number of species.\footnote{Each term in the sum over $n$ and $S$ contributes positively, so a crude bound can be obtained by just restricting the inner sum to, for instance, $n=0$, $S=4$. }
It would be very interesting to study this bound quantitatively and apply it to other cases, for example KK towers and internal manifolds \cite{Alday:2019qrf} and confining gauge theories in AdS \cite{Kaplan:2019soo,Kaplan:2020tdz} where the extra fields have different masses.

Let us also comment that for the scalar operators in the OPE the bound \eqref{eq:anecbound} was derived and discussed in detail in \cite{Cordova:2017zej}. Their basic argument is applicable to operators of arbitrary spin, and here we essentially used the EEC to derive the (non-optimal) higher-spin versions of the bounds discussed in \cite{Cordova:2017zej}.

\section{Conclusions}

In this paper, we have studied positivity properties of energy correlators in a rotation-invariant state. The positivity naturally comes in two guises: locally (pointwise positivity of the energy flux detector), and globally (energy correlators as state norms in a unitary theory). In addition, due to energy conservation, energy correlators form a consistent hierarchy, where $n$-point energy correlators can be obtained from integrating out a detector over the celestial sphere of the $(n+1)$-point energy correlator. We argued that the most general solution to the positivity problem for the infinite hierarchy of energy correlators is captured by \emph{the event distribution formula} \eqref{eq:representationsmeared}, which 
expresses energy correlators as moments of a probability distribution
on the space of events, or equivalently on the space of
probability measures on the celestial sphere, see
also appendix~B in \cite{Hofman:2008ar}. This representation follows directly from
the basic properties of energy correlators assumed in the present paper, such as pointwise positivity, strong commutativity, and energy conservation. The event distribution formula generalizes the
familiar cross section formula of particle physics to a broader setting. It sharpens the familiar observation that energy correlators remain meaningful even in theories without asymptotic particle states: it shows, without assuming an S-matrix, that the full hierarchy of energy correlators admits an event-by-event probabilistic interpretation.

We then focused on the positivity properties of the two- and three-point energy correlators. We derived the two-sided bounds on the multipole coefficients, generalizing the previous analysis of the two-point energy correlator \cite{Fox:1978vw,Fox:1978vu,Dempsey:2025yiv,Mecaj:2025ecl,Mecaj:2026kji}, and showed that consistency of the infinite hierarchy of energy correlators leads to \emph{strictly} stronger bounds on the energy correlators than the ones that follow from the consistency of the two- and three-point energy correlators, see Figure \ref{fig:twosidedc2c3} and Figure \ref{fig:twosidedbispectrumbound}. Moreover, for a given set of moments, classical theorems, see e.g. \cite{Schmudgen2017}, guarantee existence of a finite mixture of atomic measures that fully cover the space of possible moments. In particular, the bounds explored in the present paper are saturated by simple mixtures of few-particle events.

We then discussed two main applications of the positivity properties
of energy correlators. First, we applied this analysis to ALEPH data
from $e^+e^-$ collisions at $91\,\mathrm{GeV}$. We used positivity to derive a lower bound on the contact term $\delta(z)$ of the energy correlator. Second, we considered a CFT state generated by the stress tensor acting on the vacuum. Positivity of the two-point energy correlator leads to upper bounds on the OPE coefficients of operators that appear in the OPE $T \times T \sim {\cal O}$, see e.g. \cite{Cordova:2017zej,Dempsey:2025yiv,Mecaj:2025ecl,Mecaj:2026kji}. We derived these bounds explicitly for symmetric traceless tensors of spin $0,2,4$. Positivity of energy correlators also implies a simple bound on the large $c_T$ scaling of correlation functions with stress tensors, which can be interpreted as a species bound in the dual gravitational theory.

There are many interesting open directions that we have not addressed:
\begin{itemize}
    \item The existence of the event-distribution representation implies that many standard collider observables can also be defined in CFTs and gravitational theories. It would be very interesting to study them in these settings. As an example, consider thrust \cite{Farhi:1977sg}:
    \bea
    T[\varepsilon] &=& \max_{|\vec t|=1} \int_{S^{d-2}} \varepsilon(d \Omega_{\vec n}) | \vec n \cdot \vec t | \ , \\
    T &\equiv& \int \sigma(d \varepsilon) T[\varepsilon].
    \eea
    In $d=4$, each event satisfies ${1 \over 2} \leq T[\varepsilon] \leq 1$ with $T[\varepsilon] = {1 \over 2}$ saturated by a homogeneous energy distribution, and $T[\varepsilon] = 1$ saturated by the back-to-back two-particle event.
        
    Consider next, for example, the state created by the scalar half-BPS operator $\mathcal O_{20'}$ in the planar limit of $\mathcal N=4$ SYM. In this state, at zero 't~Hooft coupling,
\be
T_{\mathcal N=4}(\lambda=0)=1,
\ee
corresponding to two infinitely narrow back-to-back jets, whereas at infinite
coupling,
\be
T_{\mathcal N=4}(\lambda=\infty)=\frac12,
\ee
corresponding to a spherical energy distribution characteristic of the emergence of a classical gravity dual. 
    \item We have solely focused on the rotation-invariant states. There are many interesting constraints that arise from studying the energy correlators in spinning states, see e.g. \cite{Hofman:2008ar,Cordova:2017zej,Riembau:2025wjc,Riembau:2025isw}. In this case, we expect that all the qualitative features discussed in the paper will still hold, but many interesting details will change. For the three-point energy correlators in rotation-invariant states, it would also be interesting to generalize the bounds obtained in the present paper to $d \neq 4$.
        \item In the context of CFTs, it would be very interesting to explore the relationship between the event-distribution formula and the conformal bootstrap. For example, deriving bootstrap bounds on the multipole coefficients, as in Figure \ref{fig:twosidedc2c3} and Figure \ref{fig:twosidedbispectrumbound}, could shed light on whether the constraints implied by the event-distribution formula go beyond those obtained by simply studying the consequences of the OPE and crossing symmetry.
       
        It would be interesting to populate these plots and other bounds with values from known theories including non-perturbative CFTs.\footnote{To check the consistency of our bounds we have produced a large number of randomly generated atomic models, which indeed all land in the blue region.} However, three-point energy correlators have not been computed in models such as the 3d Ising CFT and the 4d $\mathcal N=4$ SYM at finite $N_c$ and coupling.\footnote{For the state sourced by the half-BPS operator $\O_{\mathbf{20'}}$, the weak-coupling result \cite{Yan:2022cye} and strong-coupling results we discuss in Section~\ref{sec:stringy-corrections} would be starting points for an interpolation in the planar theory, which we expect to follow a smooth trajectory from top-right corner to the origin in Figure~\ref{fig:twosidedbispectrumbound}.}
    \item We expect that the conditions discussed in the present paper should apply to the energy correlators in nontrivial media, such as energy correlators in quark-gluon plasma \cite{Andres:2022ovj,Andres:2023xwr,Yang:2023dwc,Barata:2023bhh,Andres:2024ksi,CMS:2025ydi,Bossi:2024qho,Barata:2025fzd} or CFT heavy states \cite{Chicherin:2023gxt,Firat:2023lbp,Cuomo:2025pjp}. It would be interesting to understand whether they imply nontrivial constraints on the corresponding EFTs. 
        \item Energy correlators are the simplest examples of a broader family of detector operators \cite{Kravchuk:2018htv,Caron-Huot:2022eqs,Korchemsky:2021okt}. For example, by performing the light-ray OPE or equivalently considering a square of the smeared operator $[{\cal E}(f)]^2$, where the support of smearing on the celestial sphere $\delta \Omega \to 0$, we can construct the event distribution measure representation for other detector operators. More generally, given various algebras of the light-ray operators recently studied in the literature \cite{Cordova:2018ygx,Belin:2020lsr,Besken:2020snx,Gonzo:2020xza,
Korchemsky:2021htm,Hu:2023geb,Himwich:2025ekg,Strominger:2026yrh}, it is an interesting open question whether the existence and positivity properties of these algebras impose nontrivial constraints on the microscopic properties of the theory beyond the constraints that follow from the study of crossing symmetry and dispersion relations. It would also be interesting to understand if extended light-ray algebras lead to more refined event distribution formulas compared to the one studied in the present paper.
    \item As we briefly commented at the beginning of the paper, the condition ${\cal E}(\vec n) \geq 0$ is trivially satisfied in a theory with particles. However, the average null energy condition $\int dy^- T_{--}(y^+,y^-, \vec y) \geq 0$ is nontrivial. In CFTs, ${\cal E}(\vec n) \geq 0$ and ANEC are related by a conformal transformation and are thus equivalent. In general QFT, however, these are not equivalent (see e.g. \cite{Li:2025knf}) and it would be very interesting to explore the constraints on the space of QFTs (and properties of the RG flow)  that follow from the existence of the positive hierarchy of the ANEC operators. Recently, the ANEC operators have been used to derive familiar results on the monotonicity properties of the RG flows \cite{Hartman:2023qdn,Hartman:2023ccw}.   
       \item A very general feature of energy correlators is that they are singular at $z=0$. In finite $c_T$ CFTs it follows from the convexity property of the leading Regge trajectory and the light-ray OPE. In theories with particles it is manifest due to the presence of $\delta(z_{ij})$ terms that correspond to the same particle going through a pair of detectors. It would be very interesting to understand what are the positivity properties of the regular part of the energy correlator, since it is the one that is most naturally studied in collider experiments. 
    \item 
In this paper, we have established that CFTs with a large $c_T$ necessarily have an 't Hooft-like expansion for (smeared) energy correlators. By this, we mean that the $c_T$ scaling of energy correlators (which should be viewed as equivalent at the scaling level to connected stress-tensor correlators) cannot exceed the $c_T$ scaling of stress-tensor exchange contributions. For a CFT to be holographic, higher derivative terms in the effective action must be suppressed by a large parameter. A natural question to ask is the following: without any further assumptions, what generalizes the large gap condition on single trace operators of Heemskerk--Penedones--Polchinski--Sully (HPPS)  \cite{Heemskerk:2009pn}? 

Note that if we make no further assumptions other than large $c_T$, we cannot use the large gap condition of HPPS. The matter sector will generically be strongly coupled, and hence there will be no hierarchy in the matter sector between single and multi-trace operators, and there always exist light multi-trace operators.

We would thus like to emphasize the following two points.
\begin{enumerate}
    \item The HPPS condition/setup is sufficient but not necessary. It clearly does not accommodate for strongly coupled QFTs in AdS, weakly coupled to gravity.
    
    \item $C_J \ll 1$ is a necessary condition for all theories that are described by GR plus small corrections. It is not a gap condition on the spectrum, but should be viewed as a condition on the collective contribution of everything  coupling to the stress tensor. We conjecture that it might be a sufficient condition.

    \end{enumerate}
    
   \item In the experimental data section, we focused on the contact term $\delta(z)$ and derived a lower bound on its coefficient using the precise data for angular bins away from $z=0$. 
  One of the original motivations for this project was to explore a related but different problem: Assume that an experimental two-point energy correlator is known to good precision for a wide range of angles, but not very precisely in a certain region, for instance at very small angle due to detector resolution. This precise determination could be either theoretical or experimental. Since the unitarity condition $C_J\geq0$ is non-local in $z$, an interesting question is then whether one can derive bounds in the less precise region using the precision results for the remaining region. 

For this we have experimented with the publicly available data set by OPAL \cite{OPAL:1993pnw}, with data in angular bins of equal size. 
Within this data, we minimized/maximized the contribution to a single bin, assuming that the integrated $\text{EEC}(z)$ is confined within error bars in all other bins. However, we found that our bounds trivialize, in the sense that they correspond to ``redistributing'' the total probability onto the unknown bin (e.g. upper bound in $b_1$ is found to be $1-\sum_{i\neq1}l_i$, where $l_i$ are the lower ends of the error interval in each bin, and vice versa for the lower bound). We can now understand this by the dominance of the $\delta$ function. In fact, all residual $C_J$ are positive with some margin after adding the contribution $(2J+1)A_\delta$, see \eqref{eq:finalCj}, and it appears to us that this allows for enough wiggle room in the physical data. 

Our explorations however are not exhaustive, and in particular we did not impose consistency with our event distribution formula (the underlying cross section), or constraints from higher-point energy correlators. We hope that the interesting problem of bounding physical energy correlator data can be revisited in the future and stronger bounds be derived.  

\end{itemize}

\begin{acknowledgments}
We are grateful to Jo\~ao Barata, Dean Carmi, Ross Dempsey, Tilmann Gneiting, Diego Hofman, Gregory Korchemsky, Petr Kravchuk, Pier Monni, Ian Moult, Jo\~ao Penedones, Riccardo Rattazzi, Marc Riembau, Francesco Riva, Slava Rychkov, and Gherardo Vita for useful discussions. RP thanks the CERN Theoretical Physics Department for its hospitality during the early stages of this work. AB thanks EPFL for hospitality during the period where this work started.  NB acknowledges support from the US Department of Energy Office of Science Graduate Student Research (SCGSR) program and thanks the CERN Theoretical Physics Department for its hospitality. This project has received funding from the European Research
Council (ERC) under the European Union’s Horizon 2020 research and innovation
programme (grant agreement number 949077). We acknowledge using OpenAI's ChatGPT and Codex agents to assist with calculations, editing, literature searches, numerical checks, and figure preparation. We reviewed the AI-assisted material and take full responsibility for the
contents of the manuscript.
\end{acknowledgments}

\appendix

\section{The Hewitt--Savage theorem}
\label{app:Hewitt--Savage}

Our purpose in this appendix is not to reproduce the full proof of the Hewitt--Savage theorem. Instead, we state
the theorem as a standard result and explain the physical meaning of its
assumptions and the basic idea behind its application to energy
correlators. Throughout this appendix we normalize the energy flow by
$E_{\rm tot}$ and continue to denote the normalized operator by
$\mathcal E$, so that $\mathcal E(S^{d-2})=1$.

To discuss the theorem, it is convenient to organize the relevant mathematical objects in three levels:
\begin{enumerate}
    \item \textbf{Points on the sphere:} 
    \[
    x\in X=S^{d-2}.
    \]
    Each point $x$ is a direction on the celestial sphere. The auxiliary random variables $ (x_1,x_2,\ldots )$ used below take values in this space. These variables will not have a direct physical interpretation, but they will be mathematically useful.

    \item \textbf{One event-level energy distribution:}
    \[
    \varepsilon \in \mathcal{P}(X),
    \]
    where $\mathcal{P}(X)$ is the space of probability measures on $X$.
    $\varepsilon$ is a single normalized energy profile on the sphere. In physics language, $\varepsilon$ describes how one event distributes its energy over angles.

    \item \textbf{An ensemble of events:}
    \[
    \sigma \in \mathcal{P}(\mathcal{P}(X)).
    \]
    This is a probability measure on the space of energy profiles.  In physics language, it is the analog of a differential cross section that describes the probability to create a state with a given energy flux $\varepsilon$ in the final state.
            \end{enumerate}
Thus, $\varepsilon$ is a point in $\mathcal P(X)$, while $\sigma$ is a
probability measure on $\mathcal P(X)$. 
            
Let us next state the relevant extension of de Finetti's theorem that we reviewed in the bulk of the paper. 

\begin{theorem*}[Hewitt--Savage, \cite{Hewitt1955}]
Let $X$ be a compact metric space, and let
\[
x_1,x_2,\ldots
\]
be an exchangeable sequence of \(X\)-valued random variables on some probability space. Then there exists a unique
probability measure
\[
\sigma\in \mathcal{P}(\mathcal{P}(X))
\]
such that for every $k\ge 1$ and every collection of bounded measurable functions
$f_1,\ldots,f_k:X\to\mathbb R$, the expectation values
\begin{equation}
\mathbb E\!\left[f_1(x_1)\cdots f_k(x_k)\right]
=
\int_{\PPP(X)} \sigma(d\varepsilon)
\prod_{i=1}^k \left(\int_X f_i\,d \varepsilon \right).
\label{eq:HS-theorem-smeared}
\end{equation}
Equivalently, for all Borel sets \(A_1,\ldots,A_k\subset X\), the joint probabilities
\begin{equation}
\mathbb P(x_1\in A_1,\ldots,x_k\in A_k)
=
\int_{\PPP(X)} \sigma(d\varepsilon)\,\varepsilon(A_1)\cdots \varepsilon(A_k).
\label{eq:HS-theorem-sets}
\end{equation}
\end{theorem*}

The starting point of the Hewitt--Savage theorem is an exchangeable
sequence of random variables. To connect this statement to energy
correlators, define 
\begin{equation}
\Gamma_k(A_1,\cdots, A_k)
\equiv
\left\langle
\mathcal E(A_1)\cdots\mathcal E(A_k)
\right\rangle .
\end{equation}
We assume that the smeared energy correlators, including possible contact
terms, define nonnegative Borel probability measures $\Gamma_k$ on
$X^k$. Energy correlators are nonnegative, and satisfy the expected additivity property
\begin{equation}
{\cal E}(A_1 \cup A_2) = {\cal E}(A_1) +  {\cal E}(A_2), \qquad A_1\cap A_2=\varnothing.
\end{equation}
Finally, thanks to the fact that ${\cal E}(X)=1$ (recall that $X$ corresponds to the celestial sphere), we can consistently reduce the number of detectors, in the same way we can integrate out random variables. Finally, commutativity of the energy-flow operators leads to permutation symmetry of $\Gamma_k$.  In other words, we have
\begin{equation}
\Gamma_k(X, \cdots, X)=1,\qquad
\Gamma_{k+1}(B,X)=\Gamma_k(B),
\end{equation}
for every Borel set $B\subset X^k$, together with permutation symmetry
of $\Gamma_k$.

The Kolmogorov extension theorem, see e.g. \cite{tao2011introduction}, allows us then to go from finite-dimensional consistent probability measures $\Gamma_k$ given by energy correlators to an infinite sequence of auxiliary random variables
\begin{equation}
x_1,x_2,\ldots\in X
\label{eq:x-sequence}
\end{equation}
whose joint probability distributions satisfy
\begin{equation}\label{eq:sampled-sequence}
\mathbb{P}(x_1 \in A_1,\dots, x_k\in A_k)= \langle {\cal E}(A_1)\cdots {\cal E}(A_k)\rangle.
\end{equation}
For every permutation $\pi\in S_k$, permutation symmetry implies
\begin{equation}
\mathbb P(x_1\in A_1,\ldots,x_k\in A_k)
=
\mathbb P(x_{\pi(1)}\in A_1,\ldots,x_{\pi(k)}\in A_k).
\end{equation}
The sequence $x_1,x_2,\ldots$ is therefore exchangeable, the
Hewitt--Savage theorem applies and the desired representation follows.

Let us next briefly comment on the basic idea of the proof of the theorem above, see e.g. \cite{Kirsch2019}. The proof of \eqref{eq:representation} is constructive. It is convenient to introduce the map
\begin{equation}
S_N:X^N\to \PPP(X),
\qquad
S_N(x_1,\ldots,x_N)=\frac1N\sum_{i=1}^N\delta_{x_i}.
\label{eq:TN}
\end{equation}
The probability measure on $X^N$ given by the energy correlators \eqref{eq:sampled-sequence} then induces a measure on the space of probability measures as follows. Indeed, given a measurable set $B \subset \PPP(X)$, we define the measure $\mu_N \in \mathcal{P}(\mathcal{P}(X))$ through
\be
\mu_N(B) \equiv \Gamma_N \big( S_N^{-1}(B) \big) \ . 
\ee
Our task is then to show that the measure $\sigma(d\varepsilon)$ in \eqref{eq:representation} emerges as the $N \to \infty$ limit of $\mu_N \in \mathcal{P}(\mathcal{P}(S^{d-2}))$. The key step of the proof is to consider the moments of the measure $\mu_N$,
\begin{equation}
\int_{\PPP(X)} \mu_N(d\varepsilon) \prod_{a=1}^k\!\left(\int_X f_a\,d\varepsilon\right)
=
\int_{X^N} \Gamma_N(d x_1\cdots d x_N) \prod_{a=1}^k\!\left(\frac1N\sum_{i=1}^N f_a(x_i)\right) ,
\label{eq:moments-of-muN}
\end{equation}
where $\Gamma_N(d x_1\cdots d x_N)$ is the joint probability measure for \eqref{eq:sampled-sequence}.
Expanding the product gives a sum over the expectation values
\begin{equation}
\int_{\PPP(X)} \mu_N(d\varepsilon) \prod_{a=1}^k\!\left(\int_X f_a\,d\varepsilon\right)
=
\frac1{N^k}
\sum_{i_1,\ldots,i_k=1}^N
\mathbb E\bigl[f_1(x_{i_1})\cdots f_k(x_{i_k})\bigr].
\label{eq:moment-expansion}
\end{equation}
Among these terms, the dominant ones are those with all indices distinct. By exchangeability,
all such terms are equal, and each of them is just the original smeared correlator $\Gamma_k(f_1,\ldots,f_k) \equiv \langle {\cal E}(f_1) \cdots  {\cal E}(f_k) \rangle$. The number of distinct \(k\)-tuples is
\begin{equation}
(N)_k =N(N-1)\cdots(N-k+1).
\label{eq:pochhammer}
\end{equation}
Therefore the distinct-index contribution is
\begin{equation}
\frac{(N)_k}{N^k} \langle {\cal E}(f_1) \cdots  {\cal E}(f_k) \rangle.
\label{eq:distinct-contribution}
\end{equation}
The remaining terms involve repeated indices. Their total number grows only like $N^{k-1}$ or slower,
so after dividing by $N^k$ they disappear at large $N$.

Using the fact that $\lim_{N \to \infty} {(N)_k \over N^k} = 1$, we therefore arrive at
\begin{equation}
\lim_{N\to\infty}
\int_{\PPP(X)}\mu_N(d \varepsilon)  \prod_{a=1}^k\!\left(\int_X f_a\,d\varepsilon\right) 
=\langle {\cal E}(f_1) \cdots  {\cal E}(f_k) \rangle.
\label{eq:moment-limit}
\end{equation}
Since the celestial sphere $X=S^{d-2}$ is compact, a standard
compactness result for probability measures guarantees that the
sequence $\mu_N$ has a weakly convergent subsequence $\mu_{N_j} \rightarrow\sigma$.
Here weak convergence means convergence after integration against every continuous test function on $\mathcal P(X)$. Combining this result with \eqref{eq:moment-limit} gives the desired event
distribution formula 
\begin{equation}
\label{eq:measformula}
\langle {\cal E}(f_1)\cdots  {\cal E}(f_k) \rangle
=
\int_{\PPP(X)}\sigma(d\varepsilon) \prod_{a=1}^k\!\left(\int_X f_a\,d\varepsilon\right).
\end{equation}
The uniqueness statement in the
Hewitt--Savage theorem ensures that the result is independent of the
chosen subsequence. Agreement for products of continuous test functions identifies the two Borel measures on $X^k$. A standard extension then gives the same identity for Borel regions,
\begin{equation}
\langle {\cal E}(A_1) \cdots  {\cal E}(A_k) \rangle
=
\int_{\PPP(X)} \sigma(d\varepsilon)\,\varepsilon(A_1)\cdots \varepsilon(A_k).
\label{eq:HS-regions}
\end{equation}
We refer the reader to the dedicated mathematical literature for further details.

\section{Unitarity for general event shape observables}
\label{app:general-event-shapes}

In this appendix we discuss positivity properties of collider observables beyond the energy correlators.

Consider a class of event shapes in the center-of-mass frame which only depend on the relative angle on the celestial sphere. The simplest example would be the energy-energy correlator with an unpolarized beam. We consider the partial wave expansion of the event shape matrix
\be
\langle {\cal X}_I (z) {\cal X}_K (0)  \rangle = \sum_{J=0}^\infty (H_J)_{IK} P^{(d)}_J(1-2z) ,
\ee
where we can consider ${\cal X}_I = {{\cal E}, {\cal E}_{charged}, {\cal Q}, ...}$ etc. We first show that the partial wave event shape matrix is positive semi-definite
\be
\label{eq:goal}
H_J \succeq 0. 
\ee
We start with the expression for $H_J$:
\be
\label{eq:projectionP}
 (H_J)_{IK}  =N_J^{(d)} \int_0^{1} \frac{d\mu(z)}{dz}  \langle {\cal X}_I (z) {\cal X}_K (0)  \rangle P_J(1-2z),
\ee

If we have an underlying multi-particle event distribution, we can then write
\be
\langle {\cal X}_I (z) {\cal X}_K (0)  \rangle \equiv \sum_{n=2}^\infty \int {d \sigma_{J \to n} \over \sigma} \sum_{a,b=1}^{n} (x_{a b})_{IJ} \delta\left(z-\frac{1-\vec n_a \cdot \vec n_b}2\right) ,
\ee
where for example for the energy-energy correlator we have $(x_{a b})_{{\cal E} {\cal E}} = {E_a E_b \over E_{\mathrm{tot}}}$, for the charge-charge correlator $(x_{a b})_{{\cal Q} {\cal Q}} = {q_a q_b \over q_{tot}^2}$, etc.
The sum goes over the $n$-particle final states and we absorbed all the combinatorial factors into the differential cross section $d \sigma_{J \to n} \geq 0$. Finally, $\vec n_a$ is the direction in which the particle propagates. Therefore let us write
\be
x_{ab} = x_{a} x_{b} \,. 
\ee
Performing the projection integral \eqref{eq:projectionP} simply removes the $ \delta$ function
\be
 (H_J)_{IK}  = N_J^{(d)} \sum_{n=2}^\infty \int {d \sigma_{J \to n} \over \sigma} \sum_{a,b=1}^{n} (x_{a b})_{IJ}  P_J( \vec n_a \cdot \vec n_b ) . 
\ee
It is convenient to decompose Legendre polynomials in terms of spherical harmonics\footnote{In $d=4$, the sum is over a single $m$ with the usual formula $P_J(\vec n_a \cdot \vec n_b) = {4 \pi \over 2J+1} \sum_{m=-J}^J Y_{Jm}(\vec n_a) \overline{Y_{Jm}(\vec n_b) }$.}
\be
P_J(\vec n_a \cdot \vec n_b) = \frac{\Omega_{d-2}}{N_{J}^{(d)}} \sum_{\textbf m} Y_{J, \textbf{m}}^{(d)}(\vec n_a) \overline{Y_{J, \textbf{m}}^{(d)}(\vec n_b)} ,
\ee
to get
\be
(H_J)_{IK}  = \Omega_{d-2} \sum_{n=2}^\infty \int {d \sigma_{J \to n} \over \sigma} \sum_{\textbf m} \left(\sum_{a=1}^n (x_a)_I Y_{J, \textbf{m}}^{(d)}(\vec n_a)\right) \left( \sum_{b=1}^n (x_b)_K \overline{Y_{J, \textbf{m}}^{(d)}(\vec n_b)} \right) .
\ee
We can now define the following vector 
\be
X_{I,\textbf m} \equiv \sum_{a=1}^n (x_a)_IY_{J, \textbf{m}}^{(d)}(\vec n_a) ,
\ee
and the scalar product
\be
\langle X_I , X_K \rangle_J = \sum_{\textbf m}  X_{I,\textbf m} \overline{ X_{K,\textbf m}} \ . 
\ee
In this way we get
\be
(H_J)_{IK}  = \Omega_{d-2} \sum_{n=2}^\infty \int {d \sigma_{J \to n} \over \sigma} \langle X_I , X_K \rangle_{J,n} .
\ee
The matrix $\langle X_I , X_K \rangle_J$ is positive semi-definite
\be
\psi^I \langle X_I , X_K \rangle \bar \psi^K = \langle \psi , \psi \rangle \geq 0 . 
\ee
Given that the measure $\sum_{n=2}^\infty \int {d \sigma_{J \to n} \over \sigma} \geq 0$, we have thus proven \eqref{eq:goal}.

We can now consider the tracked energy operator 
\begin{equation}
\mathcal X=    {\cal E}_{\text{charged}}(\vec n) 
\end{equation}
and immediately conclude that
\begin{equation}
    C_{J,\text{track}}\geq0
\end{equation}

Note that the considerations here do not imply pointwise positivity of general event-shape observables. For instance, the charge-charge correlator $\langle \mathcal Q(\vec n_1)\mathcal Q(\vec n_2)\rangle$ is not pointwise positive. Positivity of the tracked energy correlators follows from the fact that
the tracked-energy operator ${\cal E}_{charged}(\vec n) $ is non-negative when acting on any asymptotic state.

\section{Bounding individual multipoles}
\label{app:multipoles}

In this Appendix, we detail how this can be used in practice to solve the extremization problem for multipoles discussed in Section~\ref{sec:twoPoint}. We follow Appendix~F of~\cite{Dempsey:2025yiv}. 

We first review the basic setup for primal/dual optimization. In constrained optimization, the standard form of a linear program is called the \textit{primal} problem,
\begin{align}
    \mathbf{Primal:}\qquad
    \max_{\mathbf{x}}\,&\mathbf{c}^{\mathsf T}\mathbf{x}
    \\
    \text{subject to}\,&
    A\mathbf{x}\leq\mathbf{b},
    \qquad \mathbf{x}\geq0,
\end{align}
where $\mathbf{c}$ is the primal objective vector and $\bm{x}$ contains the primal variables. The associated dual problem is sometimes more convenient to solve. It introduces one nonnegative variable $\mathbf{y}$ for each primal inequality. The dual problem reads
\begin{align}
    \mathbf{Dual:}\qquad
    \min_{\mathbf{y}}\,&\mathbf{b}^{\mathsf T}\mathbf{y}
    \\
    \text{subject to}\,&
    A^{\mathsf T}\mathbf{y}\geq\mathbf{c},
    \qquad \mathbf{y}\geq0.
    \label{eq:general-inequality-primal-dual}
\end{align}
Solving the dual problem gives a bound on what the primal problem can achieve. This follows from the following set of inequalities:
\begin{align}
\nonumber
    \text{dual objective}&=\mathbf b^{\mathsf T}\mathbf y
    \\\nonumber
    &\geq\mathbf x^{\mathsf T}A^{\mathsf T}\mathbf y &&\text{(primal condition)}
    \\\nonumber
    &\geq \mathbf x^T\mathbf c &&\text{(dual condition)}
    \\ &=\text{primal objective}
\end{align}

The multipole extremization problem can also be formulated as a linear program, but with equality constraints instead. In this case the primal formulation is given by 
\begin{align}
    \begin{split}
        \max_{\bm{x}} &\quad \sum_i c_i x_i \\
        \text{subject to} &\quad
        \begin{dcases}
            \sum_i A_{ki} x_i = b_k, \\
            x_i \geq 0 ,
        \end{dcases}
    \end{split}
\end{align}
and the dual problem is
\begin{align}
    \begin{split}
        \min_{\bm{y}} &\quad \sum_k b_k y_k \\
        \text{subject to} &\quad
        \begin{dcases}
            \sum_k A_{ki} y_k \geq c_i, \\
            y_k \in \mathbb{R}.
        \end{dcases}
    \end{split}
\end{align}
Assuming that one can find an optimizer $\bm{y}^*$ of this dual program, the primal optimizer is recovered according to the following steps: 
\begin{enumerate}
    \item Compute, for all $i$,
    \begin{align*}
        s_i = \sum_k A_{ki} y^*_k - c_i \geq 0
    \end{align*}
    and find the vanishing points, i.e. build the contact set $S = \{i, \, s_i = 0\}$.

    \item Set $x^*_i = 0$ for every $i \notin S$ and use the set of constraints of the primal problem to determine the remaining coordinates of $\bm{x}^*$,
    \begin{align*}
        \sum_{i \in S} A_{ki} x_i = b_k
    \end{align*}
\end{enumerate}
The objectives of both formulations are then related by
\begin{align}
    \sum_i c_i x^*_i = \sum_k b_k y^*_k
\end{align}
Hence this primal/dual correspondence allows one, when more convenient, to solve the dual program rather than the initial optimization problem, and then recover the primal optimizer.

\begin{figure}
    \centering
    \includegraphics[width=0.86\textwidth]{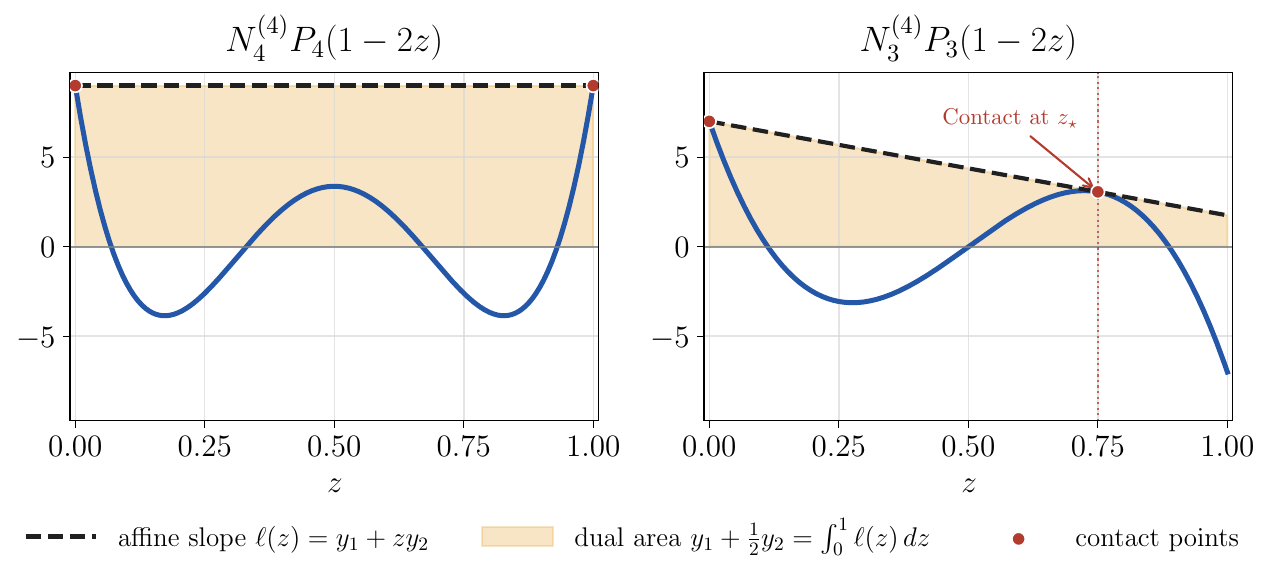}
    \caption{Geometric interpretation of the dual multipole extremization problem in $d=4$. The dashed line is the affine majorant $\ell(z)=y_1+z y_2$, and the shaded region is the dual objective $\int_0^1 \ell(z)\,dz = y_1+\frac12 y_2$. }
    \label{fig:dual_geometrical_view}
\end{figure}

We now address the problem of bounding multipoles. 
Assuming that the underlying physical theory imposes the massless condition, the extremization problem takes the form
\begin{align}
    \begin{split}
        \mathbf{Primal:} \quad \max_{\text{EEC}(z)}& \quad C_J = N_J^{(d)} \int_0^1 d\mu(z) P_J^{(d)}(1-2z)\,\text{EEC}(z) \\
    \text{subject to}& \quad
        \begin{dcases}
            \operatorname{EEC}(z) \geq 0 \quad \forall z \in [0,1],
            & \text{(positivity)}, \\
            \int_0^1 \text{EEC}(z) \, d\mu(z) = 1,
            & \text{(energy conservation)} \\
            \int_0^1 d\mu(z) z \text{EEC}(z) = \frac{1}{2} & \text{(massless condition)}
        \end{dcases}
    \end{split}
\end{align}
The integral form of the massless condition follows from $C_1^{(d)} = 0$. This form can be immediately identified with a continuous version of a primal linear program, with
\begin{align}
    c(z) \equiv N_J^{(d)} P_J^{(d)}(1-2z), \quad x(z) \equiv \frac{d\mu(z)}{dz} \operatorname{EEC}(z), \quad \bm{A}(z) = (1, z)^{\mathsf{T}}, \quad \bm{b} = \left(1, \frac{1}{2}\right)^{\mathsf{T}}
\end{align}
In particular, the variables $\operatorname{EEC}(z) (d\mu/dz)$ correspond to the primal variables. Hence, one can also formulate the dual version of the multipole extremization as
\begin{align}
    \begin{split}
        \mathbf{Dual:} \quad \min_{\bm{y} \in \mathbb{R}^2}& \quad y_1 + \frac{1}{2}y_2 \\
    \text{subject to}& \quad y_1 + z y_2 \geq N_J^{(d)} P_J^{(d)}(1-2z) \quad \forall z \in [0,1]
    \end{split}
\end{align}
This is precisely the $d$-dimensional version of the dual program given in Appendix F of~\cite{Dempsey:2025yiv}, which admits a simple geometric interpretation: the dual problem amounts to minimizing the area under the straight line $\ell(z)=y_1+z y_2$ on the interval $z\in[0,1]$, while requiring that $\ell(z)$ lies above the curve $N_J^{(d)} P_J^{(d)}(1-2z)$ for all $z\in[0,1]$. Figure~\ref{fig:dual_geometrical_view} illustrates this picture in $d=4$ for the spins $J=3$ and $J=4$.

For even spin $J$, the partial wave $P_J^{(d)}(1-2z)$ reaches its maximum value at the endpoints $z=0$ and $z=1$,
\begin{align}
\forall J \in 2\mathbb{N}, \qquad
\begin{dcases}
    P_J^{(d)}(1) = 1\,, & \\
    P_J^{(d)}(-1) = 1\,, &\\
    P_J^{(d)}(1-2z) \leq 1 & \forall z \in [0,1]\,.
\end{dcases}
\label{eq:extr_pts_partial_waves_even}
\end{align}
Hence, the dual problem is necessarily solved by the flat slope $y_1 = N_J^{(d)}$, $y_2 = 0$ (see the left panel of Figure~\ref{fig:dual_geometrical_view}), and we reproduce the Fox--Wolfram bound
\begin{equation}
    C_J\leq N_J^{(d)}, \qquad J\, \text{ even},
\end{equation}
which is optimized by the distribution
$
\text{EEC}(z) = \frac{1}{2} \frac{1}{\frac{d\mu(z)}{dz}} \Big(\delta(z) + \delta(1-z)\Big)
$. 

\paragraph{Odd spin}
For odd spin, the dual problem does not admit a simple generic closed form solution. Instead, as pointed out in \cite{Dempsey:2025yiv}, only two contact points remain between the curve $N_J^{(d)} P_J^{(d)}(1-2z)$ and the dual slope $\ell(z)$, namely at $z = 0$ and at some $z = z_* \in (0,1)$ (see right panel of Figure~\ref{fig:dual_geometrical_view} for $J=3, d=4$). Following arguments similar to the even-spin case, this gives the following extremal distribution and upper bound:
\begin{align}
\label{eq:extremalFunctionalOddSpin}
        \text{EEC}(z) &= \frac{1}{\frac{d\mu(z)}{dz}} \left(\frac{2z_*-1}{2z_*}\delta(z)+\frac1{2z_*}\delta(z-z_*) \right) \\
        C_{J \text{-odd, max}} &= N_J^{(d)} \Bigg[ 1 - \frac{1}{2z_*} \Big(1 - P_J^{(d)}(1-2z_*)\Big) \Bigg] < N_J^{(d)}
\end{align}
This shows that the bound $C_J \leq N_J^{(d)}$ can be improved for odd spins. The contact point $z_*$ with the straight line $\ell(z)$ in $(0,1)$ is fixed by requiring $\ell$ to be tangent to the curve while also passing through the endpoint $z=0$. Having a tangent at $z=z_*$ therefore requires the derivative to match the corresponding slope, which imposes the condition
\begin{align}
    \frac{d}{dz} P_J^{(d)}(1-2z) \Big|_{z = z_*} = \, \frac{P_J^{(d)}(1-2z_*) - 1}{z_*}.
    \label{eq:zstar_tangent_condition}
\end{align}
Conversely, $z_*$ is the largest real root of this equation. We can illustrate this result for $J=3$ and $J=5$. The corresponding partial waves are given by
\begin{align}
\begin{split}
    P_{J=3}^{(d)}(1-2z) &= \frac{(1-2z)\Big((d+1)(1-2z)^2 - 3\Big)}{d-2}, \\
    P_{J=5}^{(d)}(1-2z) &= \frac{(1-2z)\Big((d^2+8d+15)(1-2z)^4 - 10(d+3)(1-2z)^2 + 15\Big)}{d(d-2)}.
\end{split}
\end{align}
Using the condition~\eqref{eq:zstar_tangent_condition}, one finds that the contact points, denoted by $z_3$ and $z_5$, respectively, are the largest roots in $(0,1)$ of the equations
\begin{align}
    \begin{split}
        -\frac{4(d+1)}{d-2} \, z_3^2 (4z_3 - 3) &= 0 \qquad ( J = 3) \\
        \frac{5 (d+2)}{16 (d+5)}-\frac{5 (d+4)}{4 (d+5)} z_5 +\frac{15}8 z_5 ^2-z_5 ^3 &= 0 \qquad (J = 5)
    \end{split}
\end{align}
For $J=3$, one gets $z_3 = 3/4$, which gives a closed-form expression for the extremal distribution and maximal value
\begin{align}
    \begin{split}
    \operatorname{EEC}(z) &= \frac{1}{\frac{d\mu(z)}{dz}} \left( \frac{1}{3}\delta(z) + \frac{2}{3}\delta\left(z - \frac{3}{4}\right)\right) \\
    C_{3 ,\text{max}}^{(d)} &= N_3^{(d)} \left( 1-\frac{3(d-3)}{4(d-2)} \right)\,,
    \end{split}
\end{align}
as stated in the main text. Not only is this primally allowed, it is also compatible with the infinite positive hierarchy discussed in Section~\ref{sec:posHierarchy}, since it corresponds to an event distribution measure where all events have a triangular configuration. 

For $J=5$, the root $z_5$ is not available in a simple closed form but can be found numerically. The corresponding upper bound takes the form
\begin{equation}
\label{eq:C5Upper}
    C^{(d)}_5\leq N^{(d)}_5\frac{d+3}{8d(d-2)}\left(-10 (d-11) z_5^2+20 (d-8) z_5 -7 d+50\right)
\end{equation}
and the distribution extremizing the bound is \eqref{eq:extremalFunctionalOddSpin} with $z_*=z_5$. 
Note that $3/4<z_5<1$, which means that this distribution does not correspond to any of the simplicial models discussed in connection to Figure~\ref{fig:tetrahedra} in the main text.

\paragraph{Large odd spin.}
Following \cite{Dempsey:2025yiv}, one can also show that at asymptotically large odd $J$,
\begin{align}
    1 - \frac{1}{2z_*}\left(1-P_J^{(d)}(1-2z_*)\right) = r^{(d)} + O\left(\frac{1}{J}\right),
\end{align}
with the constant parameter 
\begin{align}
    r^{(d)} = \frac{1}{2} \left( 1
    -\Gamma\left(\frac{d-2}2\right)\left(\frac {j_{\frac{d-2}2,1}}{2}\right)^{\frac{4-d}2}
  \mathrm J_{\frac{d-4}2}\left(j_{\frac{d-2}2,1}\right)\right),
\end{align}
where $\mathrm J_\alpha$ is a Bessel function and $j_{\alpha,1}$ is its first zero (note the shifted parameter).  To obtain this asymptotic behavior, we used the Mehler--Heine formula for Jacobi polynomials $\mathsf P_J^{(\alpha,\beta)}$:
\begin{equation}
    \lim_{J\to\infty} J^{-\alpha}\mathsf P_J^{(\alpha,\beta)}\left(1-\frac{z^2}{2J^2}\right)=  \left(\frac z2\right)^{-\alpha}
   \mathrm J_\alpha(z) 
    \end{equation}
    where
    \begin{equation}
      P_J^{(d)}(x)=\frac{\Gamma(J+1)}{(\frac{d-2}2)_J}\mathsf P_J^{(\frac{d-4}2,\frac{d-4}2)}(x)  
\end{equation}
and $(a)_J$ is the Pochhammer symbol.\footnote{We also used other identities here, including
\begin{equation*}
 \frac{\Gamma(J+1)}{(\frac{d-2}2)_J}\left(\frac {1}{J}\right)^{\frac{4-d}2}=\Gamma\left(\frac{d-2}2\right)\left(1-\frac{(d-2)(d-4)}J+\ldots
\right).
\end{equation*}
}
Accordingly, one obtains the following bound at large odd spin:
\begin{equation}
    C^{(d)}_J \lesssim {N_J^{(d)}}\left( r^{(d)} +O\left( \frac{1}{J} \right) \right),
\end{equation}
and the constant parameter verifies
\begin{align}
    \begin{dcases}
        & r^{(d=4)} \approx 0.701 \\
        & r^{(d)} \underset{d \rightarrow \infty}{\longrightarrow} 0.5
    \end{dcases}
\end{align}
In particular, the behavior at large $d$ can be easily recovered using large-$d$ Legendre polynomials. Indeed, one has $P_J^{(d)}(1-2z) \approx (1-2z)^J$ in the $d\rightarrow \infty$ limit, which simplifies the multiplicative factor in the odd-$J$ bound to
\begin{align}
    1 - \frac{1}{2z_*}\left(1-P_J^{(d)}(1-2z_*)\right) &\underset{d \rightarrow \infty}{\approx} 1 - \frac{1 - (\cos\theta_*)^J}{1-\cos\theta_*}, \qquad \cos\theta_* = 1 - 2z_*
    \label{eq:mult_factor_large_d}
\end{align}
On the other hand, the self-consistent equation~\eqref{eq:zstar_tangent_condition} on $z_*$ is reduced to
\begin{align}
    (J-1)(\cos\theta_*)^J - J (\cos \theta_*)^{J-1} + 1 = 0, \qquad J \in 2\mathbb{N}+1, \quad d \rightarrow \infty
    \label{eq:zstar_large_d}
\end{align}
One can easily check that, for large odd-$J$, the unique non-trivial root of this equation verifies $-1 < \cos\theta_* < 0$ (i.e. $1/2 < z_* < 1$). Consequently,
\begin{align}
    |\cos\theta_*|^J = \frac{1}{J-1} (1-J|\cos\theta_*|^{J-1}) \leq \frac{1}{J-1},
\end{align}
which ensures that $|\cos\theta_*|^J \rightarrow 0$ at large odd $J$. Then, consistency with~\eqref{eq:zstar_large_d} implies that $\cos\theta_* \rightarrow -1$ in this limit. Equation~\eqref{eq:mult_factor_large_d} allows one to conclude that $r^{(d \rightarrow \infty)} = 1/2$.

\section{Bounding the two-delta model}
\label{app:twodeltamodel}

\begin{figure}
    \centering
   \includegraphics[width=0.55\textwidth]{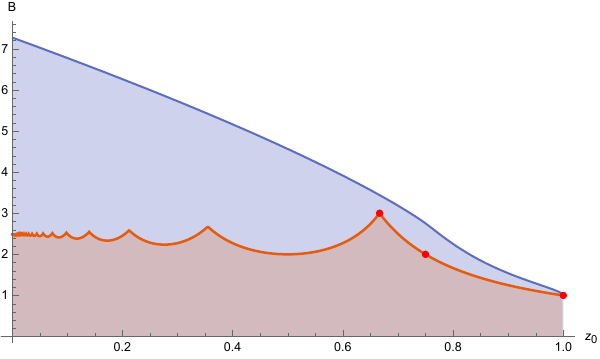}
    \caption{Bounds on the coefficient $B$ in \eqref{eq:TwoDeltas-app} in four dimensions. The red curve is the exact solution found in the main text, the blue curve is the bound following from \eqref{eq:maxAtZero-app}. The red dots are primally allowed models: back-to-back, triangle and tetrahedron configurations.}
    \label{fig:bound-delta-model}
\end{figure}

In this appendix, we derive the dual bounds on the coefficient $B$ of the model\footnote{Notice that the EEC is now normalized to $C_0 = 1$, hence the $1+B$ factor in the denominator.}
\begin{equation}
\label{eq:TwoDeltas-app}
    \text{EEC}(z)=\frac1{1+B}\big(\delta(z)+B\delta(z-z_0)\big)
\end{equation}
in $d=4$ by using the positivity constraint of equation~\eqref{eq:maxoriginsmeared}  for different angular regions $\delta\Omega_{\vec N_i}$ centered around the directions $\vec N_i$. We recall that this constraint reads
\begin{align}
\label{eq:maxAtZero-app}
    \left\langle \left(\int_{\delta \Omega_{\vec N_1}} d \vec n {\cal E}(\vec n)- \int_{\delta \Omega_{\vec N_2}} d \vec n{\cal E}(\vec n) \right)^2 \right\rangle \geq 0 \,
\end{align}
We will evaluate this in the case of infinitesimally small discs, and in the case of almost-overlapping discs. 
We will find that, away from $z=1$, this gives a weaker bound than the exact bound derived in the main text. We compare the two bounds in Figure~\ref{fig:bound-delta-model}.

\subsection{Writing the two-point energy correlator}

First of all, we need the proper expression of the two-point energy correlator as a function of spherical variables, $\langle \mathcal{E}(\vec n_1) \mathcal{E}(\vec n_2) \rangle$, to compute the integrals. 
   We use \eqref{eq:firstDefn-2pt} for the definition
\begin{equation}
      \langle \mathcal{E}(\vec n_1) \mathcal{E}(\vec n_2) \rangle=\frac{E_{\mathrm{tot}}^2}{\Omega_{d-2}^2}\text{EEC}(z) \,.
\end{equation}
We consider the two-$\delta$ model in which case $\text{EEC}(z)=\delta(z)+B\delta(z-z_0)$, where we relax the overall normalization. We work in $d=4$ and with the normalization $E_{\mathrm{tot}} = 1$. Hence, one obtains
\begin{align}
   \nonumber 
    \langle \mathcal{E}(\vec n_1) \mathcal{E}(\vec n_2) \rangle &= \frac{1}{(4\pi)^2} \text{EEC}\left(z = \frac{1 - \vec n_1 \cdot \vec n_2}{2}\right) \\ 
    & = \frac{1}{(4\pi)^2} \Bigg( \delta\left(\frac{1 - \vec n_1 \cdot \vec n_2}{2}\right) + B \, \delta\left( \frac{ 1 - \vec n_1 \cdot \vec n_2}{2}-z_0 \right) \Bigg),
\end{align}

\subsection{The bound at infinitesimal radius}
\label{app:smallRegions}

We consider first the case of infinitesimally small regions, showing, somewhat counterintuitively, that this gives an infinitely weak bound. 
We choose $\vec N_1=\mathrm N=(0,0,1)$ and $\vec N_2=X_\theta=(\sin\theta,0,\cos\theta)$ in polar coordinates, for $\cos\theta=1-2z_0$. Expanding out the left-hand side in \eqref{eq:maxoriginsmeared} we get
\begin{equation}
\begin{split}
    &\int \limits_{\delta \Omega_{\mathrm N}^2}d^2\Omega_{\vec n_1}d^2\Omega_{\vec n_2}\langle\mathcal E(\vec n_1)\mathcal E(\vec n_2)\rangle
    + \int \limits_{\delta \Omega_{X_{\theta}}^2}d^2\Omega_{\vec n_1}d^2\Omega_{\vec n_2}\langle\mathcal E(\vec n_1)\mathcal E(\vec n_2)\rangle \\
    & \hspace*{15em}-2\int \limits_{\delta \Omega_{\mathrm N}}d^2\Omega_{\vec n_1}\int \limits_{\delta \Omega_{X_\theta}}d^2\Omega_{\vec n_2}\langle\mathcal E(\vec n_1)\mathcal E(\vec n_2)\rangle
\end{split}
\end{equation}
We choose the radii of the small regions to be $\rho$. The first two terms are then only sensitive to the $\delta(z)$-term and give the same contribution, proportional to the area of the disks
\begin{align}
    \begin{split}
        \int \limits_{\delta \Omega_{\mathrm N}^2}d^2\Omega_{\vec n_1}d^2\Omega_{\vec n_2}\langle\mathcal E(\vec n_1)\mathcal E(\vec n_2)\rangle & = \frac{1}{(4\pi)^2} \int_{\delta \Omega_N}d^2\Omega_{\vec n_1} \int d\phi \sin\theta d\theta \delta\left(\frac{1 - \cos\theta}{2}\right) \\
        & = \frac{2\pi}{(4\pi)^2}\int_{d\Omega_{N}}d^2\Omega_{\vec n_1} \int 2 \delta(z) dz \\
        & = \frac{\rho^2}{4}
    \end{split}
\end{align}
The final term is more difficult to evaluate. First, one can notice that, for a cross-pair $(\vec n_1, \vec n_2) \in \delta \Omega_{\mathrm N} \times \delta \Omega_{X_{\theta}}$, one never has $\vec n_1 \cdot \vec n_2 = 1$. Therefore, the cross term is insensitive to the $\delta(z)$ contribution and only involves the part proportional to $B$ in the energy correlator. Next, we notice that in the small-$\rho$ limit, both regions $\delta \Omega_{\vec N_i}$ can be locally approximated by disks in the tangent planes at the corresponding point, namely
\begin{align}
    \begin{split}
        \delta\Omega_{\mathrm N} &\approx \Big\{ \mathrm{N} + (x_1, y_1, 0), \quad x_1^2 + y_1^2 \leq \rho^2 \ll 1\Big\} \\
        \delta \Omega_{X_{\theta}} &\approx \Big\{ X_{\theta} + (x_2 \cos \theta, y_2, - x_2 \sin \theta), \quad x_2^2 + y_2^2 \leq \rho^2 \ll 1 \Big\}
    \end{split}
\end{align}
which yields the following parameterization in the last integral
\begin{equation}
    \vec n_1=(x_1,y_1,1), \quad \vec n_2=(\sin\theta+\cos\theta x_2,y_2,\cos\theta-\sin\theta x_2).
\end{equation}
To first order in the local coordinates, one obtains
\begin{align}
    \vec n_1 \cdot \vec n_2 \approx \sin\theta (x_1 - x_2) + \cos\theta, \qquad d^2\Omega_{\vec n_i} \approx dx_i dy_i,
\end{align}
which reduces the cross term to
\begin{align}
    \begin{split}
    &\frac{B}{(4\pi)^2} \, \int_{x_1^2 + y_1^2 \leq \rho^2} dx_1 dy_1 \, \int_{x_2^2 + y_2^2 \leq \rho^2} dx_2 dy_2 \, \delta\left( \frac{\sin \theta (x_1 - x_2)}{2}\right) \\
    & = \frac{2}{\sin \theta} \frac{B}{(4\pi)^2} \, \int_{-\rho}^\rho dx_1\int_{-\sqrt{\rho^2-x_1^2}}^{\sqrt{\rho^2-x_1^2}} dy_1\int_{-\sqrt{\rho^2-x_1^2}}^{\sqrt{\rho^2-x_1^2}} dy_2 \\
    & = \frac{B}{8\pi^2\sin \theta} \times 4\int_{-\rho}^{\rho} dx_1 (\rho^2 - x_1^2) \\
    & = \frac{B}{8 \pi^2 \sin \theta} \times \frac{16\rho^3}3
    \end{split}
\end{align}
So we end up with the bound
\begin{equation}
    \frac{\rho^2}{4} - \frac{2 \rho^3}{3\pi^2 \sin\theta} B\geq 0
\end{equation}
In the limit $\rho\to0$, this bound becomes infinitely weak. Nevertheless, the computation is instructive. The inequality has the schematic form
\begin{align}
\begin{split}
    & \{\text{contribution from coincident pairs in the same region}\} \\
    & \hspace*{5em}- B\, \times \{\text{contribution from cross-pairs separated by } \theta\} \geq 0 .
\end{split}
\end{align}
To obtain a stronger bound, one should choose the smearing regions so as to minimize the possible coincident pairs in the same region, while maximizing the number of cross-pairs separated by $\theta$. At the same time, one wants to avoid pairs at separation $\theta$ inside a single region, since such terms would enter the diagonal pieces with a positive sign and weaken the bound. This is where spherical-disks configuration comes from, to which we now turn.

\subsection{Bound with a pair of spherical disks}

We now take the regions, in $d=4$,
\begin{align}
    \begin{split}
        \delta \Omega_{\mathrm N} & = \{ \vec n \in \mathbb{S}^2, \, \vec n \cdot \mathrm N > \cos\left(\theta/2\right) \}\\
        \delta \Omega_{X_{\theta}} & = \{ \vec n \in \mathbb{S}^2, \, \vec n \cdot \mathrm X_{\theta} > \cos\left(\theta/2\right) \}
    \end{split}
\end{align}
As before, there is no pair $(\vec n_1, \vec n_2)$ inside the same region such that $\vec n_1 \cdot \vec n_2 = \cos \theta$, so that the diagonal terms are still only sensitive to the $\delta(z)$ contribution of the energy correlator. Hence, the first two integrals both contribute as the area of the two regions:
\begin{align}
 I_1=   \int \limits_{\delta \Omega_{\mathrm N}^2}d\vec n_1d\vec n_2\langle\mathcal E(\vec n_1)\mathcal E(\vec n_2)\rangle 
    = 
    \int \limits_{\delta \Omega_{X_{\theta}}^2}d\vec n_1d\vec n_2\langle\mathcal E(\vec n_1)\mathcal E(\vec n_2)\rangle
    = \frac{1}{(4\pi)^2} \times 4\pi \times A_{\theta},
\end{align}
where $A_{\theta} = 2\pi(1-\cos(\theta/2))$ is the area of each spherical disk. The cross term is not sensitive to the $\delta(z)$ contribution but only to the $B$-proportional part of the energy correlator, because of similar argument to the previous one. We then want to evaluate the integral
\begin{align}
    I_{\mathrm{cross}}^{(\theta)}
    = \frac{B}{(4\pi)^2}
    \int_{\delta \Omega_{\mathrm N}} d^2\Omega_{\vec n_1}
    \int_{\delta \Omega_{X_{\theta}}} d^2\Omega_{\vec n_2}
    \,\delta\!\left(z_0-\frac{1-\vec n_1\cdot\vec n_2}{2}\right).
\end{align}
We first evaluate it in the flat approximation $\theta\ll1$. Then the geometry becomes approximately flat. We take $\theta=2\rho\ll1$ and thus $I_1=\frac{\rho^2}4$ like above. Moreover $z_0=\frac{1-\cos\theta}2\approx \rho^2$ and the integration for $ I_{\mathrm{cross}}$ is over disks of radii $\rho$ centered at $\vec 0$ and $2\rho\vec e_x$ respectively. We get
\begin{align}\nonumber
    I_{\mathrm{cross}}&=\frac B{(4\pi)^2} \int_{D_\rho(\vec 0)}d^2x \int_{D_{\rho}(2\rho\vec e_x)}d^2y\,\delta\left[\rho^2-\frac14(\vec x-\vec y)^2\right]
    \\\nonumber
    &=\frac B{(4\pi)^2} \int_{D_\rho(\vec 0)}d^2x \int_{D_{\rho}(2\rho\vec e_x)}d^2y\frac2\rho\delta\left(|\vec x-\vec y|-2\rho\right)
    \\
    &=\frac {\rho^2B}{8\pi^2} \int_{D_1(\vec 0)}d^2x \int_{D_1(2\vec e_x)}d^2y\delta\left(|\vec x-\vec y|-2\right).
\end{align}
The resulting integral can now be phrased as a geometric problem: Consider dropping a pen of length $2$ onto a plane at random. Given that one end of the pen is within one circle of radius $1$, what is the probability that the other end is within a tangent circle of radius $1$. The resulting integral can be solved exactly and gives 
\begin{equation}
    I_{\text{cross}}=\frac{\rho^2B}{4\pi^2}\left(2\mathrm{Li}_2(1/2)-2\mathrm{Li}_2(-1/2)-4+3\ln(3)\right)
\end{equation}
The resulting bound in this limit is
\begin{equation}
\label{eq:upperboundB}
    B\leq {\pi^2}\left(2\mathrm{Li}_2(1/2)-2\mathrm{Li}_2(-1/2)-4+3\ln(3)\right)^{-1}=7.27232146\,.
\end{equation}
We can now turn to the spherical version of the same geometrical problem, for arbitrary $\rho$. We keep the notation $\rho=\theta/2$ for the angular radius of the two spherical disks. In this language, the pen is simply a way of parameterizing the pairs of points selected by the delta function $\delta(z-z_0)$: given that one endpoint lies in $\delta\Omega_{\mathrm N}$, what is the probability that the other endpoint, reached by moving a geodesic distance $2\rho$ in a tangent direction, lies in $\delta\Omega_{X_{\theta}}$?

\begin{figure}
    \centering
    \includegraphics[width=0.5\linewidth]{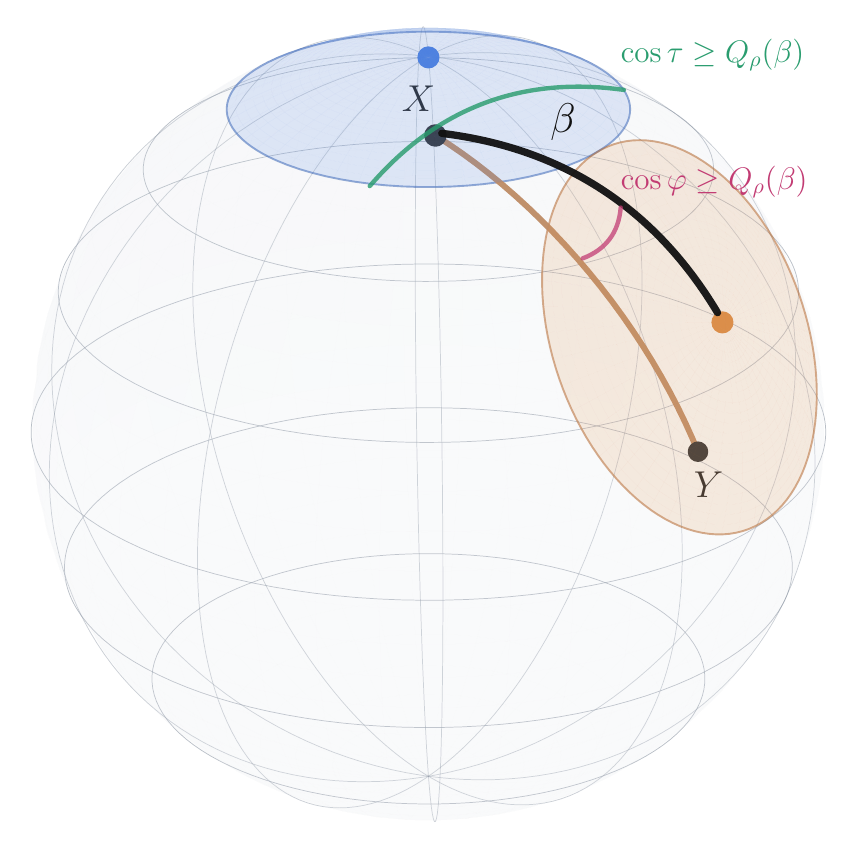}
    \caption{Parameterization for the ``spherical pen'' problem.}
    \label{fig:spherical_pen_beta_tau}
\end{figure}

More explicitly, keep the first disk centered at the north pole $\mathrm N=(0,0,1)$, and put the second center at $X_\theta=(0,\sin\theta,\cos\theta)$, with $\theta=2\rho$. We introduce polar coordinates around $X_\theta$ to parametrize the first endpoint. Let
\begin{equation}
    e_\parallel=\frac{\mathrm N-\cos\theta\,X_\theta}{\sin\theta},
    \qquad
    e_\perp=X_\theta\times e_\parallel ,
\end{equation}
so that $e_\parallel$ points from $X_\theta$ back toward $\mathrm N$ along the geodesic connecting the two centers. A point at distance $\beta$ from $X_\theta$ and azimuth $\tau$ is then
\begin{equation}
    X(\beta,\tau)= \cos\beta\,X_\theta +\sin\beta\left(\cos\tau\,e_\parallel+\sin\tau\,e_\perp\right).
\end{equation}
Since
\begin{equation}
    \mathrm N\cdot X_\theta=\cos\theta,\qquad
    \mathrm N\cdot e_\parallel=\sin\theta,\qquad
    \mathrm N\cdot e_\perp=0,
\end{equation}
we have
\begin{equation}
    \mathrm N\cdot X(\beta,\tau) = \cos\theta\cos\beta+\sin\theta\sin\beta\cos\tau.
\end{equation}
The condition that $X(\beta,\tau)$ lies in the first disk is therefore $\mathrm N\cdot X(\beta,\tau)\geq \cos\rho$, or
\begin{equation}
    \cos\tau\geq Q_\rho(\beta),\qquad
    Q_\rho(\beta)=\frac{\cos \rho-\cos(2\rho)\cos\beta}{\sin(2\rho)\sin\beta}.
\end{equation}
We now do the same thing for the other endpoint of the pen. Let $Y$ be reached from $X$ by moving a geodesic distance $2\rho$ in a tangent direction. If $\varphi$ denotes the angle of this tangent direction relative to the geodesic from $X$ to $X_\theta$, then the spherical cosine rule gives
\begin{equation}
    X_\theta\cdot Y =\cos(2\rho)\cos\beta+\sin(2\rho)\sin\beta\cos\varphi .
\end{equation}
Thus the condition $Y\in\delta\Omega_{X_\theta}$, namely $X_\theta\cdot Y\geq \cos\rho$, gives the same inequality
\begin{equation}
    \cos\varphi\geq Q_\rho(\beta).
\end{equation}
These two conditions, represented in Figure~\ref{fig:spherical_pen_beta_tau}, allow one to parameterize the counting set solely with the geodesic distance $\beta$. Indeed, if one defines
\begin{equation}
    \Theta_\rho(\beta)=
    \begin{cases}
        0, & Q_\rho(\beta)\geq1,\\
        \arccos Q_\rho(\beta), & -1\leq Q_\rho(\beta)\leq1,\\
        \pi, & Q_\rho(\beta)\leq-1,
    \end{cases}
\end{equation}
then the desired geometric probability is
\begin{equation}
    P_{\rm sph}(\rho)=
    \frac{1}{\pi^2(1-\cos \rho)}
    \int_0^\pi d\beta\,\sin\beta\,\Theta_\rho(\beta)^2.
\end{equation}
Here the normalization is the area of the first disk times the angular measure of all possible pen directions, namely $2\pi(1-\cos\rho)\times 2\pi$. The square also has a simple origin. For fixed $\beta$, the allowed starting azimuths are $-\Theta_\rho(\beta)\leq\tau\leq\Theta_\rho(\beta)$, while the allowed pen directions are $-\Theta_\rho(\beta)\leq\varphi\leq\Theta_\rho(\beta)$. Their product gives $(2\Theta_\rho(\beta))^2$, which together with the normalization above gives the prefactor in $P_{\rm sph}$. For $0<\rho<\pi/3$, this can be written without the clamps as
\begin{equation}
    P_{\rm sph}(\rho)=
    \frac{1}{\pi^2(1-\cos \rho)}
    \int_\rho^{3\rho} d\beta\,\sin\beta\,
    \arccos^2\left(
    \frac{\cos \rho-\cos(2\rho)\cos\beta}{\sin(2\rho)\sin\beta}
    \right).
\end{equation}
In terms of this probability, the cross term is
\begin{equation}
    I_{\mathrm{cross}}^{(\theta)}
    =
    B\,\frac{A_\theta}{4\pi}\,P_{\rm sph}(\rho),
\end{equation}
and the positivity inequality gives
\begin{equation}
    B\leq B_{\mathrm{dual}}(\rho), \qquad B_{\mathrm{dual}}(\rho) = \frac{1}{P_{\rm sph}(\rho)}.
\end{equation}
This is the bound depicted by the blue curve in Figure~\ref{fig:bound-delta-model}. This formula also reproduces the flat pen computation. Taking $\rho\to0$ and setting $\beta=\rho t$, one obtains
\begin{equation}
    P_{\rm flat}
    =
    \lim_{\rho\to0}P_{\rm sph}(\rho)
    =
    \frac{2}{\pi^2}\int_1^3 dt\,t\,
    \arccos^2\left(\frac{t^2+3}{4t}\right).
\end{equation}
Equivalently, exchanging the order of integration gives the planar overlap-area computation for two tangent disks. This gives
\begin{equation}
    P_{\rm flat}
    =
    \frac{2\operatorname{Li}_2(1/2)-2\operatorname{Li}_2(-1/2)-4+3\log3}{\pi^2}
    =0.1375076728\ldots,
\end{equation}
which recovers the bound on $B$ in \eqref{eq:upperboundB} in the $z_0 \rightarrow 0$ limit.

\section{Multipole coefficients and event distributions}

In this appendix we present technical details on using the event distribution formula to constrain various moments that are relevant for the evaluation of the multipole coefficients in the main text.

\subsection{Derivation of \eqref{eq:tgtr} and \eqref{eq:uless}}
\label{app:moments1}

In this section, we prove the event-level inequalities
Eqs.~\eqref{eq:tgtr} and~\eqref{eq:uless}.  We work in $d=4$, so that
the event-level energy distribution $\varepsilon$ is a probability
measure on $S^2$.

It is convenient to write
\begin{equation}
 \mathbb E_\varepsilon[F(\vec n)]
 \equiv
 \int_{S^2}\varepsilon(d\Omega_{\vec n})\,F(\vec n).
\end{equation}
The masslessness condition $C_1=0$, together with the event
distribution formula, implies
\begin{equation}
 C_1
 =
 3\int\sigma(d\varepsilon)
 \left|
   \int_{S^2}\varepsilon(d\Omega_{\vec n})\,\vec n
 \right|^2
 =0.
\end{equation}
Since the integrand is nonnegative, we conclude that
\begin{equation}
 \mathbb E_\varepsilon[\vec n]=0.
 \label{eq:appendix-event-centering}
\end{equation}

For one such event, we have defined in the main text
\begin{equation}
 M_{ij} \equiv \mathbb E_\varepsilon[n_i n_j],
 \qquad
 Q_{i j} \equiv M_{i j}- \frac13 \delta_{ij},
 \qquad
 T_{ijk} \equiv\mathbb E_\varepsilon[n_i n_j n_k],
 \label{eq:appendix-MQT-definitions}
\end{equation}
and
\begin{equation}
 s \equiv \Tr Q^2,
 \qquad
 t \equiv \Tr Q^3,
 \qquad
 u \equiv \sum_{i,j,k=1}^3T_{ijk}^2.
 \label{eq:appendix-stu-definitions}
\end{equation}
Because $|\vec n|=1$, the matrix $M$ satisfies
\begin{equation}
 M\succeq0,
 \qquad
 \Tr M=1.
 \label{eq:appendix-M-properties}
\end{equation}

Let $\lambda_1,\lambda_2,\lambda_3\geq0$ be the eigenvalues of $M$,
and introduce the elementary symmetric polynomials
\begin{equation}
 p \equiv \lambda_1+\lambda_2+\lambda_3=\Tr M,
 \qquad
 X \equiv \sum_{i<j}\lambda_i\lambda_j,
 \qquad
 Y \equiv \lambda_1\lambda_2\lambda_3=\det M.
\end{equation}
Since the expression is
symmetric, we can order the eigenvalues as
$\lambda_1\geq\lambda_2\geq\lambda_3$.  We then have
\begin{align}
 p^3+9Y-4pX
 &=
 \sum_{\mathrm{cyc}}
 \lambda_1(\lambda_1-\lambda_2)(\lambda_1-\lambda_3)
 \nonumber\\
 &=
 (\lambda_1-\lambda_2)^2
 (\lambda_1+\lambda_2-\lambda_3)
 +
 \lambda_3(\lambda_1-\lambda_3)(\lambda_2-\lambda_3)
 \geq0.
 \label{eq:appendix-schur-proof}
\end{align}
Using $\Tr M=1$, Eq.~\eqref{eq:appendix-schur-proof} becomes
\begin{equation}
 1-4X+9Y\geq0.
 \label{eq:appendix-schur-normalized}
\end{equation}

We next prove that
\begin{equation}
 \displaystyle
 u\leq\Tr M^2-\Tr M^3.
 \label{eq:appendix-u-bound}
\end{equation}
First suppose that $M$ is positive definite.  Introduce the centered
quadratic random tensor
\begin{equation}
 Z_{jk} \equiv n_j n_k-M_{jk}.
\end{equation}
Using the masslessness condition \eqref{eq:appendix-event-centering}, its two-point function with $n_i$
is
\begin{equation}
 \mathbb E_\varepsilon[n_iZ_{jk}]
 =
 \mathbb E_\varepsilon[n_i n_j n_k]
 =T_{ijk}.
 \label{eq:appendix-cross-covariance}
\end{equation}
We subtract from \(Z_{jk}\) its linear projection onto the
components of $\vec n$:
\begin{equation}
 \widehat Z_{jk}
 \equiv
 Z_{jk}
 -
 T_{i'jk}(M^{-1})_{i'i}\,n_i,
 \label{eq:appendix-residual}
\end{equation}
such that $\mathbb E_\varepsilon[n_i \widehat Z_{jk}] = 0$.

Now define the $3\times3$ matrix
\begin{equation}
 {\cal P}_{km}
 \equiv
 \mathbb E_\varepsilon
 \left[
   \sum_{j=1}^3
   \widehat Z_{jk}\widehat Z_{jm}
 \right].
 \label{eq:appendix-P-definition}
\end{equation}
This matrix is positive semidefinite.  Indeed, for every real vector
\(v_k\),
\begin{equation}
 v_k{\cal P}_{km}v_m
 =
 \mathbb E_\varepsilon
 \left[
   \sum_j
   \left(v_k\widehat Z_{jk}\right)^2
 \right]
 \geq0.
 \label{eq:appendix-P-positive}
\end{equation}

The two-point function of the unprojected quadratic tensor is
\begin{align}
 \sum_j\mathbb E_\varepsilon[Z_{jk}Z_{jm}]
 &=
 \mathbb E_\varepsilon
 \left[
   (\sum_jn_j^2)n_kn_m
 \right]
 -(M^2)_{km}
 \nonumber\\
 &=M_{km}-(M^2)_{km},
 \label{eq:appendix-Z-covariance}
\end{align}
where we used \(\sum_jn_j^2=1\).  Expanding
\eqref{eq:appendix-P-definition} and using
\eqref{eq:appendix-cross-covariance}, we obtain
\begin{equation}
 {\cal P}=M-M^2-K\succeq0,
 \label{eq:appendix-schur-complement}
\end{equation}
where
\begin{equation}
 K_{km}
 \equiv
 \sum_{j,i,i'}
 T_{ijk}(M^{-1})_{ii'}T_{i'jm}.
 \label{eq:appendix-K-definition}
\end{equation}

Since $M\succeq0$ and ${\cal P}\succeq0$,
\begin{equation}
 \Tr(M{\cal P})\geq0.
\end{equation}
Consequently,
\begin{equation}
 \Tr M^2-\Tr M^3
 \geq\Tr(MK).
 \label{eq:appendix-trace-step}
\end{equation}
Choose an eigenbasis of \(M\),
\begin{equation}
 M=\operatorname{diag}(\lambda_1,\lambda_2,\lambda_3).
\end{equation}
Then
\begin{equation}
 \Tr(MK)
 =
 \sum_{i,j,k}
 \frac{\lambda_k}{\lambda_i}\,T_{ijk}^2.
 \label{eq:appendix-MK-eigenbasis}
\end{equation}
The tensor $T_{ijk}$ is completely symmetric.  Relabeling
$i\leftrightarrow k$ in the sum therefore gives
\begin{align}
 \Tr(MK)
 &=
 \frac12
 \sum_{i,j,k}
 \left(
   \frac{\lambda_k}{\lambda_i}
   +\frac{\lambda_i}{\lambda_k}
 \right)T_{ijk}^2\geq
 \sum_{i,j,k}T_{ijk}^2
 =u,
 \label{eq:appendix-ratio-bound}
\end{align}
where we used
\begin{equation}
 \frac12\left(\frac ab+\frac ba\right)\geq1,
 \qquad a,b>0.
\end{equation}
Combining \eqref{eq:appendix-trace-step} and
\eqref{eq:appendix-ratio-bound} proves
\eqref{eq:appendix-u-bound} when \(M\) is invertible.

It remains to remove the assumption that $M$ is invertible.  This can
be done by a simple continuity argument.  Let
\[
\varepsilon_{\rm iso}(d\Omega_{\vec n})
=
\frac{d\Omega_{\vec n}}{4\pi}
\]
be the uniform event, and, for $0<\eta<1$, define
\begin{equation}
\varepsilon_\eta
=
(1-\eta)\varepsilon+\eta\varepsilon_{\rm iso}.
\end{equation}
The uniform event has vanishing first and third moments and second
moment equal to $\delta_{ij}/3$.  Therefore,
\begin{equation}
M_\eta
=
(1-\eta)M+\frac{\eta}{3} 1,
\qquad
T_\eta=(1-\eta)T.
\end{equation}
In particular,
\be
M_\eta\succeq\frac{\eta}{3} 1,
\ee
so $M_\eta$ is positive definite for every $\eta>0$.  The result proved
above can therefore be applied to $\varepsilon_\eta$:
\begin{equation}
(1-\eta)^2u
\leq
\Tr M_\eta^2-\Tr M_\eta^3.
\end{equation}
Taking the continuous limit $\eta\to0$ gives
\begin{equation}
u\leq\Tr M^2-\Tr M^3.
\end{equation}
Thus the inequality holds whether or not $M$ is invertible.

\subsection{Derivation of \eqref{eq:momentsevents3}}
\label{app:moments2}

In this section, we derive the event-representation expressions
\begin{equation}
 C_2=\frac{15}{2}\langle s\rangle,
 \qquad
 C_{222}=\frac{1125}{2}\langle t\rangle,
 \qquad
 C_{224}=\frac{7875}{8}
 \left\langle R-\frac47t-\frac{2}{15}s\right\rangle .
 \label{eq:appC222C224-target}
\end{equation}
Recall that we have $E_{\rm tot}=1$.  Thus, for every event,
$\varepsilon(d\Omega_{\vec n})$ is a normalized positive measure,
\begin{equation}
 \int_{S^2}\varepsilon(d\Omega_{\vec n})=1,
 \qquad
 \langle X\rangle
 \equiv
 \int\sigma(d\varepsilon)\,X[\varepsilon] .
 \label{eq:appC222C224-event-average}
\end{equation}

\paragraph{Event-level projection formulas.}

The event distribution formula for the normalized three-point energy
correlator in $d=4$ can be written as
\begin{equation}
 \frac{\operatorname{EEEC}(z_{12},z_{13},z_{23})}{(4\pi)^3}
 \prod_{a=1}^3 d\Omega_{\vec n_a}
 =
 \int\sigma(d\varepsilon)
 \prod_{a=1}^3\varepsilon(d\Omega_{\vec n_a}) .
 \label{eq:appC222C224-event-formula}
\end{equation}
For the parity-even triples considered here, the three-point partial wave
of the main text is
\begin{equation}
 \Phi^{(4)}_{J_1J_2J_3}(z_{12},z_{13},z_{23})
 =
 \int_{S^2}\frac{d\Omega_{\vec r}}{4\pi}
 \prod_{a=1}^3
 P_{J_a}^{(4)}(\vec n_a\!\cdot\!\vec n),
 \qquad
 z_{ab}=\frac{1-\vec n_a\!\cdot\!\vec n_b}{2} .
 \label{eq:appC222C224-Phi-definition}
\end{equation}
Combining Eq.~\eqref{eq:appC222C224-event-formula} with the three-point
projection formula in the main text gives, for a fixed reference direction
$\vec n$,
\begin{equation}
 C_{J_1J_2J_3}
 =\frac{N_{J_1}^{(4)}N_{J_2}^{(4)}N_{J_3}^{(4)}}
 {\Phi^{(4)}_{J_1J_2J_3}(0,0,0)}
 \left\langle
 \int\prod_{a=1}^3\varepsilon(d\Omega_{\vec n_a})
 \prod_{a=1}^3P_{J_a}^{(4)}(\vec n_a\!\cdot\!\vec n)
 \right\rangle .
 \label{eq:appC222C224-fixed-axis-projector}
\end{equation}
The event law is rotation invariant, so the right-hand side is independent
of the chosen direction $\vec n$.  Averaging $\vec n$ over $S^2$ and
using Eq.~\eqref{eq:appC222C224-Phi-definition}, we obtain the manifestly
rotation-invariant event-level projector
\begin{equation}
 C_{J_1J_2J_3}
 =
 \frac{N_{J_1}^{(4)}N_{J_2}^{(4)}N_{J_3}^{(4)}}
 {\Phi^{(4)}_{J_1J_2J_3}(0,0,0)}
 \left\langle
 \int\prod_{a=1}^3\varepsilon(d\Omega_{\vec n_a})
 \Phi^{(4)}_{J_1J_2J_3}(z_{12},z_{13},z_{23})
 \right\rangle .
 \label{eq:appC222C224-event-projector}
\end{equation}
The corresponding two-point identity is
\begin{equation}
 C_J
 =N_J^{(4)}
 \left\langle
 \int_{S^2}\varepsilon(d\Omega_{\vec n_1})
 \int_{S^2}\varepsilon(d\Omega_{\vec n_2})
 P_J^{(4)}(\vec n_1\!\cdot\!\vec n_2)
 \right\rangle .
 \label{eq:appC222C224-two-point-projector}
\end{equation}

\paragraph{Event tensors and contractions.}

For a fixed event, as in the previous section, we use the following definitions
\begin{equation}
 \begin{aligned}
 M_{ij}[\varepsilon]
 &=\int_{S^2}\varepsilon(d\Omega_{\vec n})\,n_in_j,
 &\qquad
 Q&=M-\frac13 \mathbf 1,
 \\
 s&=\Tr Q^2,
 &\qquad
 t&=\Tr Q^3,
 &\qquad
 R&=\int_{S^2}\varepsilon(d\Omega_{\vec n})
       \left(n^iQ_{ij}n^j\right)^2 .
 \end{aligned}
 \label{eq:appC222C224-tensor-definitions}
\end{equation}
Here $\mathbf 1$ is the $3\times3$ identity matrix.  Since
$\vec n^{\,2}=1$ and the event measure is normalized,
\begin{equation}
 \Tr M=1,
 \qquad
 \Tr Q=0 .
 \label{eq:appC222C224-traces}
\end{equation}
Expanding $M=Q+\mathbf 1/3$ then gives
\begin{equation}
 \Tr M^2=s+\frac13,
 \qquad
 \Tr M^3=t+s+\frac19,
 \qquad
 \Tr(MQ)=s .
 \label{eq:appC222C224-trace-identities}
\end{equation}

For three directions drawn independently from the same event, introduce
\begin{equation}
 x=\vec n_1\!\cdot\!\vec n_2,
 \qquad
 y=\vec n_1\!\cdot\!\vec n_3,
 \qquad
 z=\vec n_2\!\cdot\!\vec n_3 .
 \label{eq:appC222C224-xyz}
\end{equation}
We abbreviate
$\prod_{a=1}^3\varepsilon(d\Omega_{\vec n_a})$ by
$\varepsilon_1\varepsilon_2\varepsilon_3$.  Direct contraction of
indices gives
\begin{equation}
 \int\varepsilon_1\varepsilon_2\varepsilon_3\,x^2
 =M^{ij}M^{ij}=\Tr M^2 .
 \label{eq:appC222C224-x2-contraction}
\end{equation}
\begin{equation}
 \int\varepsilon_1\varepsilon_2\varepsilon_3\,xyz
 =M^{ij}M^{jk}M^{ki}=\Tr M^3 .
 \label{eq:appC222C224-xyz-contraction}
\end{equation}
The first identity is unchanged if $x^2$ is replaced by $y^2$ or
$z^2$.  The remaining contraction needed below is
\begin{equation}
 \begin{aligned}
 \int\varepsilon_1\varepsilon_2\varepsilon_3\,y^2z^2
 &=\int_{S^2}\varepsilon(d\Omega_{\vec n})
       \left(\vec n^{\,T}M\vec n\right)^2=R+\frac23s+\frac19 .
 \end{aligned}
 \label{eq:appC222C224-y2z2-contraction}
\end{equation}
Indeed,
\begin{equation}
 \vec n^{\,T}M\vec n
 =\vec n^{\,T}Q\vec n+\frac13,
 \qquad
 \int_{S^2}\varepsilon(d\Omega_{\vec n})
 \vec n^{\,T}Q\vec n
 =\Tr(MQ)=s,
 \label{eq:appC222C224-linear-Q-contraction}
\end{equation}
which proves Eq.~\eqref{eq:appC222C224-y2z2-contraction}.

\paragraph{Derivation of $C_2$.}

In $d=4$,
\begin{equation}
 P_2^{(4)}(x)=\frac12(3x^2-1),
 \qquad
 N_2^{(4)}=5 .
\end{equation}
Using Eq.~\eqref{eq:appC222C224-two-point-projector} and
$\int\varepsilon_1\varepsilon_2\,x^2=\Tr M^2$, we find
\begin{equation}
 \begin{aligned}
 C_2
 &=5\left\langle\frac32\Tr M^2-\frac12\right\rangle=\frac{15}{2}\langle s\rangle .
 \end{aligned}
 \label{eq:appC222C224-C2}
\end{equation}

\paragraph{The low-spin three-point partial waves.}

The normalized even moments of a unit vector on $S^2$ are
\begin{equation}
 \int_{S^2}\frac{d\Omega_{\vec r}}{4\pi}
 r_{i_1}\cdots r_{i_{2m}}
 =\frac{1}{(2m+1)!!}
 \sum_{\text{pairings}}
 \delta_{i_{a_1}i_{b_1}}\cdots
 \delta_{i_{a_m}i_{b_m}} .
 \label{eq:appC222C224-isotropic-moments}
\end{equation}
Expanding
\begin{equation}
 P_2^{(4)}(x)=\frac12(3x^2-1),
 \qquad
 P_4^{(4)}(x)=\frac18(35x^4-30x^2+3),
\end{equation}
and using Eq.~\eqref{eq:appC222C224-isotropic-moments}, one obtains
\begin{equation}
 \Phi^{(4)}_{222}(x,y,z)
 =\frac{2-3(x^2+y^2+z^2)+9xyz}{35} .
 \label{eq:appC222C224-Phi222}
\end{equation}
\begin{equation}
 \Phi^{(4)}_{224}(x,y,z)
 =\frac{1+2x^2-5y^2-5z^2-20xyz+35y^2z^2}{140} .
 \label{eq:appC222C224-Phi224}
\end{equation}
In Eq.~\eqref{eq:appC222C224-Phi224}, the spin-four leg is attached to
the third detector $\vec n_3$.  At coincident detector directions,
$z_{12}=z_{13}=z_{23}=0$, or equivalently $x=y=z=1$, so
\begin{equation}
 \Phi^{(4)}_{222}(0,0,0)
 =\Phi^{(4)}_{224}(0,0,0)
 =\frac{2}{35} .
 \label{eq:appC222C224-Phi-normalizations}
\end{equation}

\paragraph{Derivation of $C_{222}$.}

Equations~\eqref{eq:appC222C224-Phi222},
\eqref{eq:appC222C224-x2-contraction}, and
\eqref{eq:appC222C224-xyz-contraction} give, on a single event,
\begin{equation}
 \begin{aligned}
 \int\varepsilon_1\varepsilon_2\varepsilon_3\,
 \Phi^{(4)}_{222}
 &=\frac{2-9\Tr M^2+9\Tr M^3}{35}=\frac{9}{35}\,t .
 \end{aligned}
 \label{eq:appC222C224-Phi222-average}
\end{equation}
Using
$N_2^{(4)}=5$ and
$\Phi^{(4)}_{222}(0,0,0)=2/35$ in
Eq.~\eqref{eq:appC222C224-event-projector}, we obtain
\begin{equation}
 \begin{aligned}
 C_{222}
 &=\frac{5^3}{2/35}
 \left\langle\frac{9}{35}t\right\rangle=\frac{1125}{2}\langle t\rangle .
 \end{aligned}
 \label{eq:appC222C224-C222}
\end{equation}

\paragraph{Derivation of $C_{224}$.}

For $C_{224}$, Eqs.~\eqref{eq:appC222C224-Phi224} and
\eqref{eq:appC222C224-x2-contraction}--
\eqref{eq:appC222C224-y2z2-contraction} give
\begin{equation}
 \begin{aligned}
 &\int\varepsilon_1\varepsilon_2\varepsilon_3\,
 \Phi^{(4)}_{224}
 \\
 &\quad=\frac1{140}\left[
 2\Tr M^2-20\Tr M^3
 +35\left(R+\frac23s+\frac19\right)
 -10\Tr M^2+1\right]
 \\
 &\quad=\frac1{140}
 \left(35R-20t-\frac{14}{3}s\right)=\frac14
 \left(R-\frac47t-\frac{2}{15}s\right) .
 \end{aligned}
 \label{eq:appC222C224-Phi224-average}
\end{equation}
All constant terms cancel in the second line.  Finally,
$N_2^{(4)}=5$, $N_4^{(4)}=9$, and
$\Phi^{(4)}_{224}(0,0,0)=2/35$, so
\begin{equation}
 \begin{aligned}
 C_{224}
 &=\frac{5^2\cdot9}{2/35}
 \left\langle
 \frac14\left(R-\frac47t-\frac{2}{15}s\right)
 \right\rangle
 \\
 &=\frac{7875}{8}
 \left\langle R-\frac47t-\frac{2}{15}s\right\rangle .
 \end{aligned}
 \label{eq:appC222C224-C224}
\end{equation}
Combining Eqs.~\eqref{eq:appC222C224-C2},
\eqref{eq:appC222C224-C222}, and
\eqref{eq:appC222C224-C224} proves
Eq.~\eqref{eq:appC222C224-target}.  Notice that no masslessness condition
was used in this derivation.

\section{Derivation of the bound \eqref{eq:FWthree}}
\label{app:stretched-triangle-bound}

In this appendix, we derive the bound \eqref{eq:FWthree} presented in Section~\ref{sec:threePoint}.  We first treat  $d\geq 4$; the
$d=3$ case is summarized at the end.  Let $(a,b,c)$ be an admissible
parity-even triple, and regard $b$ as the spin of the detector localized at
the north pole.  We use
\begin{equation}
 \alpha_{abc}^{(d)}
 \equiv \Phi_{abc}^{(d)}(0,0,0)
 =\int_0^1d\mu(z)\,
 P_a^{(d)}(1-2z)P_b^{(d)}(1-2z)P_c^{(d)}(1-2z)>0.
 \label{eq:stretched-alpha}
\end{equation}
Below we suppress the superscript $(d)$ on $P_J$ and $N_J$.

\paragraph{The Buhmann--J\"ager blocks.}
Fixing the north pole leaves an $SO(d-2)$ symmetry.  Accordingly, the
spin-$J$ harmonics on $S^{d-2}$ decompose as
\begin{equation}
 \mathcal H_J(S^{d-2})
 =\bigoplus_{k=0}^{J}\mathcal H_k(S^{d-3}),
  \qquad D_k=N_k^{(d-1)},
 \label{eq:stretched-branching}
\end{equation}
where $D_k$ is the multiplicity of branch $k$, such that $N_J^{(d)}=\sum_{k=0}^{J}N_k^{(d-1)}$. Here $k$ labels the transverse $SO(d-2)$ angular momentum. To run the argument, we need the polar-angle dependence of the
corresponding $k$th harmonic, which takes the form
\begin{equation}
R_{Jk}^{(d)}(t)
=
\left[
\frac{N_J^{(d)}(J-k)!}
     {J!\,(J+d-3)_k}
\right]^{1/2}
(1-t^2)^{k/2}
\frac{d^k}{dt^k}P_J^{(d)}(t),
\qquad 0\leq k\leq J .
\label{eq:stretched-radial-wave}
\end{equation}
These are normalized as follows
\be
 \int_0^1d\mu(z)\,
 R_{Jk}^{(d)}(1-2z)^2=1 \ . 
\ee

For later use, let us also define the radial matrix element
\begin{equation}
 M_b^{ac}(k)
 \equiv\int_0^1d\mu(z)\,
 R_{ak}^{(d)}(1-2z)P_b^{(d)}(1-2z)R_{ck}^{(d)}(1-2z).
 \label{eq:stretched-M}
\end{equation}
In particular,
\begin{equation}
 M_b^{ac}(0)=\sqrt{N_aN_c}\,\alpha_{abc}^{(d)}.
 \label{eq:stretched-M0}
\end{equation}

Let us next consider the detector to be normalized on the north pole  $h(\vec n_2)=\volS \delta(\vec n_2 -\hat z)$.  After a trivial rescaling, the
corresponding Buhmann--J\"ager block in branch $k$ takes the form
\begin{equation}
\widehat c_k(J,J')
=
C_J\delta_{JJ'}
+
\sum_{\substack{
1\leq L\leq J+J'\\
L\geq |J-J'|\\
J+J'+L\in2\mathbb Z
}}
C_{JLJ'}\sqrt{\alpha_{JLJ'}^{(d)}}\,
g_L^{JJ'}(k),
\qquad
\widehat c_k\succeq0.
\label{eq:stretched-positive-block}
\end{equation}
where
\begin{equation}
 g_L^{ac}(k)
 =\frac{M_L^{ac}(k)}{
 \sqrt{N_aN_c\alpha_{aLc}^{(d)}}}.
 \label{eq:stretched-g}
\end{equation}
The two harmonic identities needed below are
\begin{align}
 \sum_{k=0}^{\min(a,c)}D_k\,
 g_L^{ac}(k)g_{L'}^{ac}(k)
 &=\frac{\delta_{LL'}}{N_L},
 \label{eq:stretched-gaunt-orthogonality}\\
 \sum_{k=0}^{J}D_k\,\widehat c_k(J,J)
 &=N_JC_J.
 \label{eq:stretched-trace-identity}
\end{align}

\paragraph{Positivity and projection.}
Suppose first that $a\neq c$, and set
\begin{equation}
 A_k=\widehat c_k(a,a),\qquad
 B_k=\widehat c_k(a,c),\qquad
 E_k=\widehat c_k(c,c).
\end{equation}
Every principal minor is positive semidefinite, and therefore
\begin{equation}
 A_k,E_k\geq0,
 \qquad |B_k|\leq\sqrt{A_kE_k}.
 \label{eq:stretched-minor}
\end{equation}
Equations~\eqref{eq:stretched-positive-block} and
\eqref{eq:stretched-gaunt-orthogonality} give
\begin{equation}
 \sum_k^{\min(a,c)} D_k\,g_b^{ac}(k)B_k
 =\frac{\sqrt{\alpha_{abc}^{(d)}}}{N_b}\,C_{abc}.
 \label{eq:stretched-projector}
\end{equation}

Let $K=\min(a,c)$; all sums in the following chain run from $k=0$ to
$K$.  Starting from \eqref{eq:stretched-projector}, we obtain the bound one
step at a time:
\begin{align}
 |C_{abc}|
 &=\frac{N_b}{\sqrt{\alpha_{abc}^{(d)}}}
 \left|\sum_kD_k\,g_b^{ac}(k)B_k\right|
 \nonumber\\
 &\leq\frac{N_b}{\sqrt{\alpha_{abc}^{(d)}}}
 \sum_kD_k\,|g_b^{ac}(k)|\,|B_k|
 &&\text{(triangle inequality)}
 \nonumber\\
 &\leq\frac{N_b}{\sqrt{\alpha_{abc}^{(d)}}}
 \max_k|g_b^{ac}(k)|\sum_kD_k\sqrt{A_kE_k}
 &&\text{\eqref{eq:stretched-minor}}
 \nonumber\\
 &\leq\frac{N_b}{\sqrt{\alpha_{abc}^{(d)}}}
 \max_k|g_b^{ac}(k)|
 \sqrt{\left(\sum_kD_kA_k\right)
       \left(\sum_kD_kE_k\right)}
 &&\text{(Cauchy--Schwarz)}
 \nonumber\\
 &\leq\frac{N_b}{\sqrt{\alpha_{abc}^{(d)}}}
 \max_k|g_b^{ac}(k)|\sqrt{N_aN_cC_aC_c}
 &&\text{\eqref{eq:stretched-trace-identity}}
 \nonumber\\
 &=
 \frac{N_b}{\alpha_{abc}^{(d)}}
 \max_k|M_b^{ac}(k)|\sqrt{C_aC_c}
 &&\text{\eqref{eq:stretched-g}.}
 \label{eq:stretched-general-branch-bound}
\end{align}
In the trace step we used positivity of every diagonal branch to restrict
the complete traces in \eqref{eq:stretched-trace-identity} to their common
range:
\begin{equation}
 \sum_{k=0}^{K}D_kA_k\leq N_aC_a,
 \qquad
 \sum_{k=0}^{K}D_kE_k\leq N_cC_c.
 \label{eq:stretched-partial-traces}
\end{equation}
When $a=c$, the same result follows directly from the diagonal conditions
$\widehat c_k(a,a)\geq0$: the constant term cancels in
\eqref{eq:stretched-projector}, and
\eqref{eq:stretched-trace-identity} replaces the $2\times2$ minor argument.

\paragraph{The special case $J_3 = J_1 + J_2$.}
Suppose now that one spin is the sum of the other two.  If the triggered spin is the largest one, $b=a+c$, the
Gegenbauer integral in \eqref{eq:stretched-M} gives
\begin{equation}
 \left|\frac{M_b^{ac}(k)}{M_b^{ac}(0)}\right|^2
 =\prod_{s=0}^{k-1}
 \frac{(a-s)(c-s)}{(a+2\lambda+s)(c+2\lambda+s)}\leq1, ~~~ \lambda = {d-3 \over 2} \ . 
 \label{eq:stretched-trigger-largest}
\end{equation}
If instead an external spin is the largest, say $c=a+b$, one finds
\begin{equation}
 \left|\frac{M_b^{ac}(k)}{M_b^{ac}(0)}\right|^2
 =\prod_{s=0}^{k-1}
 \frac{(a-s)(c+2\lambda+s)}{(c-s)(a+2\lambda+s)}\leq1.
 \label{eq:stretched-external-largest}
\end{equation}
The second inequality follows factor by factor from
\begin{equation}
 (c-s)(a+2\lambda+s)-(a-s)(c+2\lambda+s)
 =2(c-a)(\lambda+s)>0.
\end{equation}
The empty product at $k=0$ equals one.  Hence in either case
\begin{equation}
 \max_k|M_b^{ac}(k)|
 =|M_b^{ac}(0)|
 =\sqrt{N_aN_c}\,\alpha_{abc}^{(d)}.
\end{equation}
Substitution into \eqref{eq:stretched-general-branch-bound} leads to the
desired inequality
\begin{equation}
 |C_{abc}|
 \leq N_b\sqrt{N_aN_cC_aC_c}.
 \label{eq:stretched-one-trigger}
\end{equation}

\section{Unitarity bounds on EEEC in $d=3$}
\label{app:d3bounds}

In $d=3$ the celestial sphere is a circle $S^1$, and the discussion of the unitarity constraints on the three-point energy correlator simplifies.  We parameterize the celestial circle by
\begin{equation}
\vec n(\phi)=(\cos\phi,\sin\phi),
\qquad 0\leq \phi<2\pi ,
\end{equation}
and the Gegenbauer partial waves reduce to Chebyshev polynomials
\begin{equation}
 P_J^{(3)}(\cos\theta)=T_J(\cos\theta)=\cos(J\theta),\qquad N_J^{(3)}=2,\quad J\geq1.
\end{equation}
The two-point energy multipoles can be obtained, using $z= \sin^2\frac{\theta}{2}$, by
\begin{equation}
C_J
=
\frac{2}{\pi}
\int_0^\pi d\theta\,
\cos(J\theta)\,
\text{EEC}\!\left(\sin^2\frac{\theta}{2}\right),
\qquad J\geq1 .
\end{equation}
At the level of the three-point function, if we order spins as $J_1 \leq J_2 \leq J_3$ the only nonzero partial waves are the ones that satisfy
\be
J_3 = J_1 + J_2 \ . 
\ee
The corresponding parity-even partial wave functions take the form\footnote{In $d=3$, parity acts as $\phi_i \to - \phi_i$.}
\begin{equation}
 \Phi^{(3)}_{J_1,J_2,J_1+J_2}
 =\frac14\cos\!\left[J_1(\phi_1-\phi_3)+J_2(\phi_2-\phi_3)\right].
 \label{eq:PhiSimple}
\end{equation}
Introducing $\alpha \equiv \phi_1-\phi_3$ and $\beta = \phi_2-\phi_3$ as independent variables, we have
\begin{equation}
 z_{13}=\frac{1-\cos\alpha}{2},\qquad
 z_{23}=\frac{1-\cos\beta}{2},\qquad
 z_{12}=\frac{1-\cos(\alpha-\beta)}{2}.
\end{equation}
We can define
\be
\text{EEEC}(\alpha,\beta) = \text{EEEC}\left( \frac{1-\cos(\alpha-\beta)}{2} , \frac{1-\cos\alpha}{2} ,  \frac{1-\cos\beta}{2}  \right) \ . 
\ee
The three-point multipoles then take the form 
\begin{equation}
 C_{J_1,J_2,J_1+J_2}
 =8\int_0^{2\pi}\frac{d\alpha}{2\pi}
   \int_0^{2\pi}\frac{d\beta}{2\pi}\,
   \text{EEEC}(\alpha,\beta)\,
   \cos(J_1\alpha+J_2\beta) .
 \label{eq:projection3d}
\end{equation}
After imposing the masslessness condition $C_{1,J,J+1}=0$, the simplest nontrivial three-point multipoles are $C_{224}$ and $C_{235}$.

\subsection{Unitarity}

Next we would like to derive unitarity constraints on the three-point energy correlator in $d=3$. The measure and the volume simply become
\begin{equation}
d\Omega_\phi=d\phi,
\qquad
\Omega_1=\int_{S^1}d\Omega_\phi=2\pi .
\label{eq:d3AngularMeasure}
\end{equation}
It is convenient to introduce the orthonormal Fourier modes for this measure as follows
\begin{equation}
u_m(\phi)=\frac{e^{i m\phi}}{\sqrt{2\pi}},
\qquad m\in\ZZ,
\qquad
\int_{S^1}d\Omega_\phi\,
\overline{u_m(\phi)}u_n(\phi)=\delta_{mn} \ . 
\label{eq:d3FourierModes}
\end{equation}

For a real nonnegative function $h:S^1\to\mathbb R_{\geq0}$, we use the
Fourier convention
\begin{equation}
\widehat h_r
\equiv
\int_{S^1}d\Omega_\phi\,h(\phi)e^{-i r\phi},
\qquad
h(\phi)=\frac{1}{2\pi}\sum_{r\in\ZZ}\widehat h_r e^{i r\phi},
\qquad
\widehat h_{-r}=\widehat h_r^* .
\label{eq:d3hFourier}
\end{equation}
With the definition
${\cal E}(h)=\int_{S^1}d\Omega_\phi\,h(\phi){\cal E}(\phi)$, rotation
invariance and energy conservation fix
\begin{equation}
\langle {\cal E}(h)\rangle
=\frac{E_{\rm tot}}{2\pi}\,\widehat h_0
=\frac{\widehat h_0}{2\pi}
\qquad (E_{\rm tot}=1).
\label{eq:d3SmearedOnePoint}
\end{equation}

Specializing the main-text formulas for $K_h$ to $d=3$ gives
\begin{align}
K_h(\phi_1,\phi_3)
&\equiv
\int_{S^1}d\Omega_{\phi_2}\,h(\phi_2)
\langle {\cal E}(\phi_1){\cal E}(\phi_2){\cal E}(\phi_3)\rangle
\notag\\
&=
\frac{E_{\rm tot}^3}{(2\pi)^3}
\int_0^{2\pi}d\phi_2\,h(\phi_2)
\operatorname{EEEC}(z_{12},z_{13},z_{23}),
\label{eq:d3PhysicalKernel}
\end{align}
where $z_{ij}=(1-\cos(\phi_i-\phi_j))/2$.  The corresponding quadratic-form
formula \eqref{eq:positivityKh} becomes
\begin{equation}
K_h(f,f)
\equiv
\int_{S^1}d\Omega_{\phi_1}d\Omega_{\phi_3}\,
\overline{f(\phi_1)}K_h(\phi_1,\phi_3)f(\phi_3)
\geq0
\label{eq:d3KernelQuadraticForm}
\end{equation}
for every test function $f$.  Hence $K_h$ is a positive-semidefinite
Hermitian kernel.  Its Fourier expansion and its coefficients are
\begin{align}
K_h(\phi_1,\phi_3)
&=
\sum_{m,n\in\ZZ}\kappa_{mn}^{(h)}
u_m(\phi_1)\overline{u_n(\phi_3)},
\notag\\
\kappa_{mn}^{(h)}
&=
\int_{S^1}d\Omega_{\phi_1}d\Omega_{\phi_3}\,
\overline{u_m(\phi_1)}K_h(\phi_1,\phi_3)u_n(\phi_3).
\label{eq:d3KernelFourier}
\end{align}
For a finite Fourier series $f(\phi)=\sum_m a_m u_m(\phi)$,
\begin{equation}
K_h(f,f)
=\sum_{m,n\in\ZZ}\overline{a_m}\kappa_{mn}^{(h)}a_n\geq0 .
\end{equation}
Consequently,
\begin{equation}
K_h\text{ is a positive-semidefinite kernel}
\qquad\Longleftrightarrow\qquad
\left(\kappa_{mn}^{(h)}\right)_{m,n\in\ZZ}\succeq0,
\label{eq:d3KernelPSD}
\end{equation}
where the matrix condition means that every finite principal submatrix is
positive semidefinite.

Substituting the two- and three-point multipole expansions into
\eqref{eq:d3PhysicalKernel} gives the following explicit matrix entries:
\begin{equation}
\label{eq:d3FullFourierMatrix}
\frac{(2\pi)^2}{E_{\rm tot}^3}\,\kappa_{mn}^{(h)}
=
\begin{cases}
\widehat h_0,
&m=n=0,\\[2mm]
\dfrac{\widehat h_m}{2}\,C_{|m|},
&m\neq0,\ n=0,\\[2mm]
\dfrac{\widehat h_{-n}}{2}\,C_{|n|},
&m=0,\ n\neq0,\\[2mm]
\dfrac{\widehat h_0}{2}\,C_{|m|},
&m=n\neq0,\\[2mm]
\dfrac{\widehat h_{m-n}}{8}\,
C_{|m|\,|m-n|\,|n|},
&m\neq0,\ n\neq0,\ m\neq n.
\end{cases}
\end{equation}
Here $C_J$ denotes the $d=3$ two-point multipole, and the ordering of the
indices of $C_{J_1J_2J_3}$ follows the detector ordering
$(\phi_1,\phi_2,\phi_3)$.  
Reality of $h$ and permutation symmetry of $C_{J_1J_2J_3}$ make
\eqref{eq:d3FullFourierMatrix} manifestly Hermitian.  The unitarity constraints
are
\begin{equation}
\left(\kappa_{mn}^{(h)}\right)_{m,n\in\ZZ}\succeq0
\qquad\text{for every real }h\geq0.
\label{eq:d3AllPSDConstraints}
\end{equation}

\subsection{Two-sided unitarity bounds}

The axisymmetric discussion in the main text is recovered by taking $h$ to be
reflection-even about $\phi=0$,
\begin{equation}
h(\phi)=h_0+\sum_{J=1}^{\infty}h_J\cos(J\phi).
\end{equation}
With the convention \eqref{eq:d3hFourier}, its complex Fourier
coefficients are
\begin{equation}
\widehat h_0=2\pi h_0,
\qquad
\widehat h_{\pm J}=\pi h_J
\qquad (J\geq1).
\label{eq:d3EvenFunction}
\end{equation}

Finally, let us consider $h_\star(\phi)=2 \pi \delta(\phi)=1+2\sum_{J\geq1}\cos(J\phi)$ in the
distributional sense, so $h_0=1$ and $h_J=2=N_J^{(3)}$.  For $J\geq1$, the
principal submatrix of \eqref{eq:d3FullFourierMatrix} on the modes
$\{J,-J\}$ is
\begin{equation}
\left.
\left(\kappa_{mn}^{(h_\star)}\right)
\right|_{\{J,-J\}}
=
\frac{E_{\rm tot}^3}{2\pi}
\left(
\begin{array}{cc}
\dfrac12 C_J & \dfrac18 C_{J\,J\,2J} \\
\dfrac18 C_{J\,J\,2J} & \dfrac12 C_J
\end{array}
\right).
\label{eq:d3PointMeasureBlock}
\end{equation}
The overall factor is positive and therefore does not affect the eigenvalue
conditions,
\begin{equation}
C_J\pm\frac14 C_{J\,J\,2J}\geq0,
\end{equation}
or, equivalently,
\begin{equation}
\label{eq:bound3dJJ2J}
|C_{J\,J\,2J}|\leq4C_J.
\end{equation}

Stronger two-sided bounds follow by retaining the zero mode.  For
$m,n\neq0$, $m\neq n$, the principal submatrix on
$\{0,m,n\}$  is
\begin{equation}
 \left.\mathsf M\right|_{\{0,m,n\}}
 =
 \left(
 \begin{array}{ccc}
 1 & \frac12 C_a & \frac12 C_c\\[1mm]
 \frac12 C_a & \frac12 C_a & \frac18 C_{abc}\\[1mm]
 \frac12 C_c & \frac18 C_{abc} & \frac12 C_c
 \end{array}
 \right) \succeq0\,,
\end{equation}
where $a=|m|$, $b=|m-n|$, and $c=|n|$. 
Its determinant can be written as
\begin{equation}
 64\det\left.\mathsf M\right|_{\{0,m,n\}}
 =
 4C_aC_c(2-C_a)(2-C_c)
 -
 \left(C_{abc}-2C_aC_c\right)^2 .
\end{equation}
Consequently,
\begin{equation}
 \left(C_{abc}-2C_aC_c\right)^2
 \leq
 4C_aC_c(2-C_a)(2-C_c).
\label{eq:d3ShiftedThreeModeBound}
\end{equation}
For example, the modes $ \{ 0,J,-J \}$ give
\begin{equation}
 4C_J(C_J-1)
 \leq
 C_{JJ\,2J}
 \leq
 4C_J,
\label{eq:d3JJ2JTwoSidedBound}
\end{equation}
while the modes $ \{0,J,2J\} $  give the complementary condition
\begin{equation}
 \left(C_{JJ\,2J}-2C_JC_{2J}\right)^2
 \leq
 4C_JC_{2J}(2-C_J)(2-C_{2J}).
\label{eq:d3JJ2JMixedBound}
\end{equation}
Similarly, one can explore the constraints that arise from the event distribution formula and compare them to the bounds derived above. 

\subsection{$C_2 + C_3 \leq 2$ from event distribution}

Let us next derive an improved upper bound on $(C_2,C_3)$, displayed in Figure \ref{fig:twosidedc2c3}, that follows from the event distribution formula. 

To connect to the event distribution formula we write
\begin{equation}
\frac{d\mu(z)}{dz}\,\text{EEC}(z)
=
\int \sigma(d\varepsilon)
\int \varepsilon(d\phi_1)\,\varepsilon(d\phi_2)\,
\delta\!\left(z-z_{12}\right),
\qquad
z_{12}=\frac{1-\cos(\phi_1-\phi_2)}{2}.
\end{equation}
This implies that on a single event, 
\be
C_J[\varepsilon] = 2 \int \varepsilon(d\phi_1) \varepsilon(d \phi_2) \cos J(\phi_1 - \phi_2) = 2 \Big| \int \varepsilon(d\phi) e^{i J \phi} \Big|^2 \equiv 2 | m_J(\varepsilon) |^2 , ~~~ J \geq 1.  
\ee
Energy conservation implies that $| m_0(\varepsilon) |^2=1$ and the masslessness condition is $| m_1(\varepsilon) |^2=0$. 

Define the scalar product on the Hilbert space of square-integrable functions on $S^1$
\be
\langle f , g \rangle_{\varepsilon} = \int \varepsilon(d \phi) f^*(\phi) g(\phi) \ . 
\ee
With this definition
\be
\langle 1 , e^{3 i \phi} \rangle_{\varepsilon} = m_3[\varepsilon], ~~~\langle e^{i \phi} , e^{3 i \phi} \rangle_{\varepsilon} = m_2[\varepsilon] .
\ee
We also have $\langle 1 , e^{i \phi} \rangle_{\varepsilon} =0$ due to the masslessness condition.

We can then apply Bessel's inequality to get the desired bound
\be
\label{eq:C3bound2d}
{C_2[\varepsilon] + C_3[\varepsilon] \over 2} = | m_2(\varepsilon) |^2 + | m_3(\varepsilon) |^2=| \langle 1 , e^{3 i \phi} \rangle_{\varepsilon} |^2 + | \langle e^{i \phi} , e^{3 i \phi} \rangle_{\varepsilon} |^2 \leq | \langle e^{i 3\phi} , e^{3 i \phi} \rangle_{\varepsilon} |^2 = 1 \ . 
\ee
It is very easy to populate the allowed triangle by actual events. Let us consider
\begin{align}
 \mu_O &=\frac14\sum_{r=0}^{3}\delta_{\phi_0+r\pi/2},
 &(C_2,C_3)&=(0,0),
 \label{eq:d3-square}\\
 \mu_\triangle
 &=\frac13\sum_{r=0}^{2}\delta_{\phi_0+2\pi r/3},
 &(C_2,C_3)&=(0,2),
 \label{eq:d3-triangle}\\
 \mu_B
 &=\frac12\left(\delta_{\phi_0}+\delta_{\phi_0+\pi}\right),
 &(C_2,C_3)&=(2,0).
 \label{eq:d3-back-to-back}
\end{align}
The uniform circle may replace the square at the origin. Given any
$(x,y)$ satisfying $x,y\geq 0$ and $x+y\leq 2$, the event distribution
\begin{equation}
 \sigma_{x,y}
 =\left(1-\frac{x+y}{2}\right)\delta_{\mu_O}
  +\frac{y}{2}\delta_{\mu_\triangle}
  +\frac{x}{2}\delta_{\mu_B}
 \label{eq:d3-explicit-law}
\end{equation}
realizes $(C_2,C_3)=(x,y)$ and defines a positive, consistent hierarchy.  Averaging each shape over $\phi_0$ makes the event
law manifestly rotation invariant without changing any $C_J$.

\subsection{Fake EEC}

Let us comment on the kink in the gray region in Figure~\ref{fig:twosidedc2c3}. It is possible to check that it corresponds to the following delta-function model:
\begin{equation}
\label{eq:fakeAtom}
    \langle \mathcal E(\theta)\mathcal E(0)\rangle \sim \frac1{\phi+2}\left(\phi\,\delta(\theta)+2\delta(\theta-\tfrac{4\pi}5)
    \right)
\end{equation}
where $\phi=\frac12(1+\sqrt5)$ is the golden ratio. The corresponding coefficients $C_J$ read
\begin{equation}
    C_J=\begin{cases}
2-\delta_{J0}& J=0\ \text{ (mod 5)},
        \\
        0&J=1 \text{ or } 4\ \text{ (mod 5)},
\\
2\phi-2 &J=2 \text{ or } 3\ \text{ (mod 5)},
    \end{cases}
\end{equation}
This gives the point $C_2=C_3=\sqrt5-1=1.23607$ shown as a gray dot in Figure~\ref{fig:twosidedc2c3}. However, it does not satisfy the bound $C_2+C_3\leq2$ derived above, showing that \eqref{eq:fakeAtom} does not derive from a consistent hierarchy. Also from the point of view of atomic events, it is clear why \eqref{eq:fakeAtom} is not good. The structure of the $\delta$ functions indicates an opening angle of particles of $4\pi/5$, but such an angle would also imply an opening angle of $2\pi/5$, as in a pentagonal event. It can be checked that the pentagonal event gives $C_J\geq0$ and lies inside our bound (in fact it has $C_2=C_3=0$).

\subsection{Free theory}

For completeness, we also reproduce here the $C_J$ extracted from the energy correlator in the state sourced by $\phi^k$ in the theory of a free scalar, $k>2$. Specializing to $d=3$, the expression in \cite{Firat:2023lbp} reads
\begin{equation}
\left(\pi z(1-z)\right)^{-1/2}    \text{EEC}(z)=\frac2{2+k}\delta(z)+\frac1{\pi\sqrt{z(1-z)}}\frac{k(k-2)}{(k+1)(k+3)}\,{_2F_1}\left(2,2,\tfrac{k+5}2,z\right)
\end{equation}
where we divided by $d\mu(z)/dz$. From it, we extract the relevant partial waves for the plot in Figure~\ref{fig:twosidedc2c3}: $C_0=1$, $C_1=0$, and
\begin{align}
    \label{eq:C2in3d}
    C_2&=\frac{4k^2-2k+6}{k-1}-8k(k-2)\Psi(k),
    \\
    \label{eq:C3in3d}
    C_3&=\frac{16k^3-8k^2+8k+16}{k-1}-32k^2(k-2)\Psi(k)
\end{align}
where $\Psi(k)=\frac12(\psi(\frac{k-1}2)-\psi(\frac{k-2}2))$ and $\psi(x)=\frac d{dx}\log\Gamma(x)$ is the digamma function.

\section{Details of free gas computation}
\label{app:free gas details}

In this section, we show how the free gas energy correlator is computed.  We begin by deriving the $z$-counting rule \eqref{eq: z-counting rule}, setting up an expansion in powers of $1/k$, with coefficients given in terms of moments in the canonical ensemble of the particles.  Next we compute the coefficients of this expansion to determine the EEC and EEEC.  Finally, we check consistency with the bounds previously outlined.  

\subsection{Derivation of the $z$-counting rule}

Recall that we work in the center-of-mass frame and $E_{\text{tot}} = ku$.  The thermodynamic limit is
\begin{align} \label{app:thermodynamic limit}
     k \rightarrow \infty, \quad E_{\text{tot}} \rightarrow \infty, \quad \frac{E_{\text{tot}}}{k} = u \, \, \, \text{fixed}.
 \end{align}

Starting from the event distribution representation of the energy correlator \eqref{eq:crosss}, we write the normalized energy correlator as
\begin{align} \label{eq: free gas with explicit sum}
    &\braket{\hat{\mathcal{E}}(\vec{n}_1) \dots \hat{\mathcal{E}}(\vec{n}_l)} \nonumber \\
    & \quad = \left( \frac{\Omega_{d-2}}{ku} \right)^l \sum_{j_1 \dots j_l = 1}^k \braket{ E_{j_1} \dots E_{j_l} \, \delta^{(d-2)}(\hat{q}_{j_1} - \vec{n}_1) \dots \delta^{(d-2)}(\hat{q}_{j_l} - \vec{n}_l)}_{\text{MCE}_k}.
\end{align}
$\braket{ \, }_{\text{MCE}_k}$ denotes an average with respect to the probability distribution 
\begin{align}\label{eq:mce}
    \rho_{\text{MCE}_k}(q_1 \dots q_k) = \frac{1}{\Omega(k, \sqrt{q^2_{\text{tot}}})} \prod_{i = 1}^k \left( \delta_+(q_i^2 - m^2) \right)\delta^d(q_{\text{tot}} - \sum_{i = 1}^k q_i).
\end{align}
The normalization $\Omega(k, \sqrt{q_{\text{tot}}^2})$ is the total phase space of $k$ particles with total energy-momentum vector $q_{\text{tot}}$. \eqref{eq:mce} is the probability distribution of the microcanonical ensemble (MCE) for $k$ particles, since all configurations of $q_1 \dots q_k$ with total energy-momentum vector $q_{\text{tot}}$ are equally likely.

The key point about \eqref{eq: free gas with explicit sum} is that in each term, the observable averaged in $\braket{ \dots }_{\text{MCE}_k}$ is in a \emph{subsystem} of the total $k$-particle system.  For each term, let $l' = |\{j_1 ,\dots ,j_l \}|$
denote the number of distinct values among the indices $j_1 \dots j_l$.  $l'$ counts the number of particles in the subsystem.  The rest of the particles act as a large heat bath, and we may treat the $l'$-particle subsystem as sitting in the canonical ensemble instead of the microcanonical ensemble.  The ensemble equivalence will hold up to corrections that can be organized as an expansion in powers of $1/k$. 

The $l'$-particle subsystem is governed by a probability distribution obtained by integrating out $k-l'$ momenta from \eqref{eq:mce}, 
\begin{align} \label{eq:marginal dist}
    \rho_{kl'}(q_1 \dots q_{l'}) = \frac{\Omega(k_{\text{bath}}, \sqrt{q_{\text{bath}}^2})}{\Omega(k, \sqrt{q_{\text{tot}}^2})} \prod_{i = 1}^{l'} \delta_+(q_i^2 - m^2), \quad k_{\text{bath}} = k - l', \, q_{\text{bath}} = q_{\text{tot}} - Q_{l'},
\end{align}
where $Q_{l'} = q_1 + q_2 + \dots + q_{l'}$ is the energy-momentum vector of the $l'$-particle subsystem. The log of the numerator in \eqref{eq:marginal dist} is the entropy of the bath  $S(k_{\text{bath}},\sqrt{q^2_{\text{bath}}})$. When $Q_{l'} = 0$, $\sqrt{q_{\text{bath}}^2} = E_{\text{tot}}$, so we define the energy shift $\Delta E$ through
\begin{align} \label{eq:defining Delta E}
    \sqrt{q_{\text{bath}}^2} = E_{\text{tot}} - \Delta E.  
\end{align}
Expanding $S(k_{\text{bath}}, E_{\text{tot}} - \Delta E)$ about $E_{\text{tot}}$, we may write the probability distribution for the subsystem \eqref{eq:marginal dist} as
\begin{align} \label{eq:entropy expansion of marginal dist}
    \rho_{kl'}(q_1 \dots q_{l'}) \propto \left( \prod_{i = 1}^{l'} \delta_+(q_i^2 - m^2) \right) \exp \left [\sum_{n = 1}^\infty \frac{1}{n!} (-\Delta E)^n \frac{\partial^n }{\partial E_{\text{tot}}^n} S(k_{\text{bath}}, E_{\text{tot}}) \right],
\end{align}
with the proportionality constant fixed by normalization. 

In Appendix~\ref{app:saddle point approx} we will show the entropy of the free gas takes the form
\begin{align} \label{eq:form of entropy}
    S(N, \sqrt{p^2}) = N s(u) + \sum_{n \geq 0} \frac{1}{N^n} s_n(u) + \text{$u$-independent}, \quad u \equiv \frac{\sqrt{p^2}}{N}.
\end{align}
Recalling $E_{\text{tot}} = ku$, \eqref{eq:form of entropy} gives the scaling of the derivatives in the entropy expansion \eqref{eq:entropy expansion of marginal dist},
\begin{align} \label{eq:derivatives of entropy}
    \frac{\partial}{\partial E_{\text{tot}}} S(k_{\text{bath}}, E_{\text{tot}}) = \beta(u) + O(1/k), \quad \frac{\partial^n}{\partial E_{\text{tot}}^n} S(k_{\text{bath}}, E_{\text{tot}}) = O(1/k^{n-1}), \, \, n > 1.
\end{align}
The first equation defines the inverse temperature.  To obtain the overall scaling of the terms we also need the scaling of $\Delta E$.  From \eqref{eq:defining Delta E},
\begin{align} \label{eq:leading behavior of delta E}
    \Delta E = Q_{l'}^0 + O(1/k).
\end{align}
From \eqref{eq:derivatives of entropy} and \eqref{eq:leading behavior of delta E}, the terms in the entropy expansion \eqref{eq:entropy expansion of marginal dist} are $O(1/k^{n-1})$, so all terms $n > 1$ vanish in the thermodynamic limit \eqref{app:thermodynamic limit}.  The $n = 1$ term is
\begin{align} \label{eq:n=1 term}
    -\Delta E  \frac{\partial}{\partial E_{\text{tot}}} S(k_{\text{bath}}, E_{\text{tot}}) = -\beta(u) Q_{l'}^0 + O(1/k).
\end{align}
Keeping just the $O(1)$ part of the $n=1$ term, we see that $\rho_{kl'} \rightarrow \rho_{\text{CE}_{l'}}$ in the thermodynamic limit \eqref{app:thermodynamic limit}, where
\begin{align} \label{eq:canonical ensemble}
    \rho_{\text{CE}_{l'}}(q_1 \dots q_{l'}) = \frac{1}{Z^{l'}(\beta(u))} \prod_{i = 1}^{l'} \delta_+(q_i^2 - m^2) e^{-\beta(u) p^0_i}, \quad Z(\beta(u)) = \int d^dq \, \delta_+(q^2 - m^2) e^{-\beta(u) q^0}.
\end{align}
$Z(\beta(u))$ is the partition function of a single particle. 

If we do \emph{not} take the thermodynamic limit, then all the terms in the sum in \eqref{eq:entropy expansion of marginal dist} contribute.   However, we may still factor out $e^{-\beta(u) Q_{l'}^0}$ from the right-hand side of \eqref{eq:entropy expansion of marginal dist}, and then the full distribution is fixed from normalization,
\begin{equation}
\label{eq:expand ratio}
\begin{gathered}
    \rho_{kl'}(q_1 \dots q_{l'})
    =
    \rho_{\text{CE}_{l'}}(q_1 \dots q_{l'}) \times 
   \frac{e^A}{\langle e^A\rangle_{\mathrlap{\mathrm{CE}_{l'}}}}
    \\
    A
    =
    \beta(u) Q_{l'}^0
    + \sum_{n=1}^{\infty}
    \frac{1}{n!}
    \left(-\Delta E\right)^n
    \frac{\partial^n}{\partial E_{\text{tot}}^n}
    S(k_{\text{bath}},E_{\text{tot}})
\end{gathered}
\end{equation}
where $\braket{ \, }_{\text{CE}_{l'}}$ denotes the average with respect to $\rho_{\text{CE}_{l'}}$. Note that although $\beta(u) Q_{l'}^0$ is $O(1)$, it cancels against $-\beta(u) Q_{l'}^0$ in the $n = 1$ term \eqref{eq:n=1 term}.  \emph{All} the terms in $A$ vanish in the thermodynamic limit, and we may write
\begin{align} \label{eq:one plus corrections}
    \frac{e^A}{\langle e^A\rangle_{\mathrlap{\mathrm{CE}_{l'}}}} = 1 + \text{corrections}
\end{align}
where the corrections vanish as powers of $1/k$. We will show an example in the next section.  

We can now write the correlator as 
\begin{align} \label{eq: free gas with corrections}
    &\braket{\hat{\mathcal{E}}(\vec{n}_1) \dots \hat{\mathcal{E}}(\vec{n}_l)} \nonumber \\
    &\quad = \left( \frac{\Omega_{d-2}}{u} \right)^l \frac{1}{k^l} \sum_{j_1 \dots j_l = 1}^k \braket{ E_{j_1} \dots E_{j_l} \, \delta^{(d-2)}(\hat{q}_{j_1} - \vec{n}_1) \dots \delta^{(d-2)}(\hat{q}_{j_l} - \vec{n}_l) \times (1 + \text{corrections})}_{\text{CE}_{l'}}
\end{align}
and again $l' = |\{j_1 ,\ldots, j_l \}|$ \footnote{For each term in \eqref{eq: free gas with corrections} we can re-label the distinct momenta appearing as $q_1 \dots q_{l'}$.}.  The correlator is a sum of a smooth part when all the indices $j_1 \dots j_l$ take distinct values, plus contact terms proportional to delta functions when some of the index values coincide.  When computing the correlator, it is useful to first compute the smooth part which consists of all terms in the sum with $l' = l$, and then fix the contact terms from the energy conservation Ward identity. 

Next we check that the correlator as written in \eqref{eq: free gas with corrections} reproduces the correct thermodynamic limit.    The number of non-contact terms is 
\begin{align} 
    \frac{k!}{(k-l)!} = k^l + O(k^{l-1}).
\end{align}
The sum of all the non-contact terms produces the smooth part of the correlator.  Factoring in the $1/k^l$ prefactor from \eqref{eq: free gas with corrections}, the smooth part of the correlator is $O(1)$.  The number of contact terms with $D$ index coincidences of a given type \footnote{For example, $j_1 = j_2$ with all remaining indices distinct has $D = 1$.} is
\begin{align} \label{eq:contact term prefactor}
    \frac{k!}{(k-(l-D))!} = k^{l-D} + O(k^{l-D-1}).
\end{align}
Factoring in the $1/k^l$ prefactor from \eqref{eq: free gas with corrections}, the contact contribution is $O(1/k)$. Isolating the smooth part of the correlator, we recover the correct thermodynamic limit
\begin{align}
    \braket{\hat{\mathcal{E}}(\vec{n}_1) \dots \hat{\mathcal{E}}(\vec{n}_l)} = \prod_{i = 1}^l \frac{\Omega_{d-2}}{u} \braket{E_{i} \delta^{(d-2)}(\hat{q}_i - \vec{n}_i)}_{\text{CE}_1} + O(1/k) = 1 + O(1/k). 
\end{align}

Now we will derive the $z$-counting rule.  A contact term with $D$ index coincidences is proportional to $D$ delta functions.  Combined with the prefactor \eqref{eq:contact term prefactor} divided by $k^l$ we recover the $z$-counting rule for the contact terms.  The $z$-counting rule for the polynomial terms comes from corrections in \eqref{eq: free gas with corrections}, which in turn come from expanding the ratio \eqref{eq:one plus corrections}.  First consider the contribution to the corrections coming from the $n = 1$ term in $A$.  From \eqref{eq:defining Delta E} we have
\begin{align} \label{eq:gives polynomial structure in z}
    \begin{split}
    \Delta E &= Q_{l'}^0 + \frac{|\vec{Q}_{l'}|^2}{2(ku - Q_{l'}^0)} + \frac{|\vec{Q}_{l'}|^4}{8(ku - Q_{l'}^0)^3} + \dots \\
    &= Q_{l'}^0 + \frac{|\vec{Q}_{l'}|^2}{2ku}\left(1 + O(1/k) \right) + \frac{|\vec{Q}_{l'}|^4}{8(ku)^3}\left(1 + O(1/k) \right) + \dots
    \end{split} 
\end{align}
The appearance of $|\vec{Q}_{l'}|^2$ determines the $z$-dependence of the corrections. This is because
\begin{align} \label{eq:Q-vec-sq expansion}
    |\vec{Q}_{l'}|^2 = \sum_{i = 1}^{l'} |\vec{q}_i|^2 + \sum^{l'}_{i \neq j} \vec{q}_i \cdot \vec{q}_j.
\end{align}
Upon insertion into \eqref{eq: free gas with corrections}, the detector delta functions identify $\hat{q}_i = \vec{n}_a$ and $\hat{q}_j = \vec{n}_b$, so each power of $|\vec{Q}_{l'}|^2$ generates terms in the correlator proportional to $z$.  From \eqref{eq:gives polynomial structure in z}, we see that the $O(1/k^r)$ term in the correlator contains all powers $z^{p}$ with $0 \leq p\leq r$, which is exactly the $z$-counting rule for the polynomial terms.  Next consider contributions to the corrections coming from terms with $n > 1$ in $A$.  The same argument applies, except the entropy derivatives are $O(1/k^{n-1})$, but these are \emph{additional} factors of $1/k$ so they don't break the rule.

\subsection{Computation of the free gas EEC and EEEC}
\label{app:free gas eec and eeec}

Here we compute the massless EEC and EEEC through $O(1/k^2)$ and the massive EEC through $O(1/k)$ in $d = 4$. Owing to the $z$-counting rule, this is all we will need to verify both the two-point and three-point bounds previously outlined.  

From the $z$-counting rule we can immediately write down the form of the two-point energy correlator,
\begin{align} \label{eq:free gas eec}
    \text{EEC}(z)
    &= 1 + \frac{1}{k} \left[a^{(1)}\big(\delta(z) - 1\big) - 3b^{(1)} P_1(1-2z) \right] \nonumber \\
    &+ \frac{1}{k^2} \left[a^{(2)}\big(\delta(z) - 1\big) - 3 b^{(2)} P_1(1-2z) + c^{(2)} P_2(1-2z) \right] + O(1/k^3),
\end{align}
where we fixed the constant terms ahead of time such that the energy-conservation Ward identity is automatically satisfied.  Next we will compute the coefficients.

The smooth part of the EEC is the sum of all the terms in \eqref{eq: free gas with corrections} with $l' = l = 2$. We need to expand the ratio \eqref{eq:one plus corrections} to $O(1/k^2)$, which requires the entropy of the free gas.  The massless case is particularly simple because the phase space is fixed by dimensional analysis, 
\begin{align}
    \Omega(N, \sqrt{q^2}) = C_N (\sqrt{q^2})^{2N - 4} \implies S(N, \sqrt{q^2}) = (2N - 4) \log \frac{\sqrt{q^2}}{N} + \dots \, \text{(massless)}
\end{align}
where the terms in $\dots$ only depend on $N$.  For the smooth part of the EEC we have $N = k_{\text{bath}} = k - 2$, $\sqrt{q^2} = E_{\text{tot}} = ku$.  From $\partial S/\partial E_{\text{tot}} = \beta(u) + O(1/k)$ we obtain the inverse temperature
\begin{align}
    \beta(u) = \frac{2}{u} \quad \text{(massless)}.
\end{align}
Computing further derivatives of the entropy and expanding $\Delta E$ \eqref{eq:defining Delta E}, we obtain the numerator of \eqref{eq:one plus corrections}
\begin{align}
    e^A = 1 + \frac{A^{(1)}}{k} + \frac{A^{(2)}}{k^2} + O(1/k^3)
\end{align}
where
\begin{align}
\begin{split}
A^{(1)}
&=
\frac{8Q_2^0}{u}
-\frac{(Q_2^0)^2+|\vec Q_2|^2}{u^2},
\\
A^{(2)}
&=
\frac{1}{2}(A^{(1)})^2
+\frac{4\left[(Q_2^0)^2+|\vec Q_2|^2\right]}{u^2}
-\frac{2Q_2^0\left[(Q_2^0)^2+3|\vec Q_2|^2\right]}{3u^3}
\end{split}
\qquad
\text{(massless)}
\end{align}
and $Q_2 = q_1 + q_2$.  We may now compute the denominator of \eqref{eq:one plus corrections} by averaging in the canonical ensemble $\rho_{\text{CE}_2}$ \eqref{eq:canonical ensemble} with $m = 0$.  Expanding the ratio \eqref{eq:one plus corrections} to $O(1/k^2)$ and substituting into \eqref{eq: free gas with corrections}, it is straightforward to compute 
\begin{equation} \label{eq:massless free gas eec coefficients}
\begin{gathered}
    a^{(1)} = \frac{3}{2}, \quad b^{(1)} = \frac{3}{2}, \quad a^{(2)} = -\frac{3}{2}, \quad b^{(2)} = -\frac{3}{2}, \quad c^{(2)} = 12 \quad \text{(massless)}.
\end{gathered}
\end{equation}

For the massive free gas we cannot fix the phase space from dimensional analysis.  One can compute the entropy by saddle point approximation (see \ref{app:saddle point approx}).  We obtain
\begin{align} \label{app a1, b1}
    a^{(1)} = \frac{\braket{p_0^2}_{u}}{\braket{p^0}_{u}^2}, \quad b^{(1)} = \frac{\braket{p^0 |\vec{p}|}^2_{u}}{\braket{p^0}^2_{u} \braket{|\vec{p}|^2}_{u}} \quad \text{(massive, massless)}.
\end{align}
$\braket{ \, }_u$ is an average in the \emph{single-particle} canonical ensemble with partition function $Z(\beta(u))$ \eqref{eq:canonical ensemble}.  $\braket{p_0}_u = u$.  We write the coefficient functions with the more explicit $\braket{p_0}_u$ to be suggestive.  Note that in the massless limit we recover the result \eqref{eq:massless free gas eec coefficients} $a^{(1)} = b^{(1)} = 3/2$.

\begin{figure}
\centering
\includegraphics[width=0.55\textwidth]{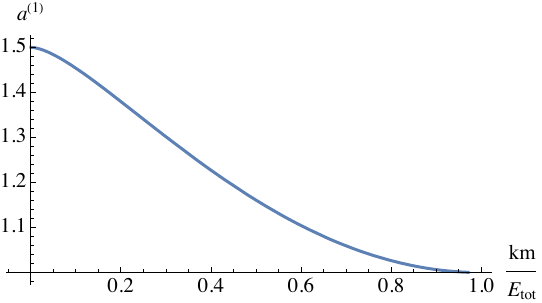}
    \caption{The coefficient $a^{(1)}$ in the EEC \eqref{eq:free gas eec} and EEEC \eqref{app:free gas eeec}.}
    \label{fig:a1plot}
\end{figure}

Energy conservation and the $z$-counting rule fix the form of the three-point energy correlator through $O(1/k^2)$ to be
\begin{align}\label{app:free gas eeec}
    &\text{EEEC}(z_{12}, z_{13}, z_{23}) = 1 + \frac{1}{k} \sum_{i < j}^{3} \Big[ a^{(1)}\big(\delta(z_{ij}) - 1\big) - 3 b^{(1)} (1 - 2z_{ij}) \Big] \nonumber \\
    &+ \frac{1}{k^2} \Big[ c^{(2)}_{1} + c^{(2)}_{z}\, S_z + c^{(2)}_{z^2}\, S_{z^2} + c^{(2)}_{zz}\, S_{zz} 
    + c^{(2)}_{\delta}\, S_\delta + c^{(2)}_{z\delta}\, S_{z\delta} + c^{(2)}_{\delta\delta}\, S_{\delta\delta} \Big] + O(1/k^3),
\end{align}
where the various structures are given by 
\begin{align}
\begin{split}
    S_z &= z_{12} + z_{13} + z_{23} \\
    S_{\delta} &= \delta(z_{12}) + \delta(z_{13}) + \delta(z_{23}) \\
    S_{z^2} &= z_{12}^2 + z_{13}^2 + z_{23}^2  \\
    S_{zz} &= z_{12}\,z_{13} + z_{12}\,z_{23} + z_{13}\,z_{23}  \\
    S_{z\delta} &= \delta(z_{12})\,(z_{13} + z_{23}) + \delta(z_{13})\,(z_{12} + z_{23}) + \delta(z_{23})\,(z_{12} + z_{13}) \\
    S_{\delta\delta} &= \delta(z_{12})\,\delta(z_{13}) + \delta(z_{12})\,\delta(z_{23}) + \delta(z_{13})\,\delta(z_{23}) 
\end{split}
\end{align}
Here we have not imposed the energy conservation Ward identity ahead of time.  We can again determine the smooth part of the correlator by focusing only on those terms in \eqref{eq: free gas with corrections} with $l' = l = 3$.  Then we fix the coefficients of the contact terms by imposing the energy conservation Ward identity. We find
\begin{equation} \label{eq:massless free gas eeec coefficients}
\begin{gathered}
    c_1^{(2)} = 168, \quad c_z^{(2)} = -207, \quad
    c_{z^2}^{(2)} = 72, \quad c_{zz}^{(2)} = 108, \\
    c_{\delta}^{(2)} = -\frac{27}{2}, \quad
    c_{z\delta}^{(2)} = 9, \quad c_{\delta\delta}^{(2)} = 1.
\end{gathered}
\qquad \text{(massless)}
\end{equation}

\subsection{Consistency with bounds}

In this section we compute the multipoles $C_J$ and $C_{J_1 J_2 J_3}$ from the results of the previous section and check their consistency with the bounds previously outlined.  

We can extract the coefficients $C_J$ from \eqref{eq:free gas eec} 
\begin{align}
\label{eq:C_J free gas}
    \begin{split}
    C_1
    &= \frac{3}{k} \left[a^{(1)} - b^{(1)}\right]  + \frac{3}{k^2} \left[a^{(2)} - b^{(2)}\right] + O(1/k^3) \\
    C_{J > 1}
    &= \frac{(2J + 1)}{k} a^{(1)} + O(1/k^2).
    \end{split}
\end{align}
For the massless free gas $a^{(1)} = b^{(1)} = 3/2$ and $a^{(2)} = b^{(2)} = -3/2$, so we recover $C_1 = 0$ through $O(1/k^2)$. 

Next we check $C_J > 0$ for $0 < m_{\text{gas}} < u$ \footnote{In this section we rename the mass $m_{\text{gas}}$ to avoid confusion with the $m$ in the Buhmann--J\"ager Bounds.}.  $C_1$ satisfies this constraint in the large-$k$ limit from Cauchy--Schwarz,
\begin{align}
    b^{(1)} = \frac{\braket{p^0 |\vec{p}|}^2_{u}}{\braket{p^0}^2_{u} \braket{|\vec{p}|^2}_{u}} < \frac{\braket{p_0^2}_{u} \braket{|\vec{p}|^2}_{u}}{\braket{p^0}^2_{u} \braket{|\vec{p}|^2}_{u}} = a^{(1)}.
\end{align}
$C_{J > 1}> 0$ in the large $k$ limit for all allowed values of $m_{\text{gas}}$ simply because $a^{(1)} > 0$.    

Next we check the multipoles $C_{J_1 J_2 J_3}$.  From \eqref{eq:massless free gas eeec coefficients}, 
\begin{align} \label{eq:C_JJJ massless free gas_appendix}
    C_{J_1 J_2 J_3} 
    &= \frac{1}{k^2} \left[135 (\delta_{J_1, 2} \, \delta_{J_2, 1} \, \delta_{J_3, 1} + \text{cyclic}) - 9(N^{(4)}_{J_1} N^{(4)}_{J_2} \delta_{J_3, 1} + \text{cyclic}) + 3N^{(4)}_{J_1} N^{(4)}_{J_2} N^{(4)}_{J_3}\right] \nonumber \\
    &+ O(1/k^3) \quad \text{(massless)},
\end{align}
where $N^{(4)}_J = 2J+1$ \eqref{eq:NJ}.  From \eqref{eq:C_JJJ massless free gas_appendix} it is easy to verify $C_{1 J J+1} = 0$ through $O(1/k^2)$.  

Finally we check the Buhmann--J\"ager Bounds \eqref{eq:PSDch}, which we will express as 
\begin{align} \label{eq: quadratic form}
    &\sum_{J_1 J_3} a_{J_1} c_m^{(\delta)}(J_1, J_3) a_{J_3} \nonumber \\
    &= a_0^2 \, \delta_{m, 0}
    + 2 a_0 \sum_{J \geq 1} a_J \, c_m^{(\delta)}(0, J) 
    + \sum_{J \geq 1} a_J^2 \, c_m^{(\delta)}(J, J) 
    + \sum_{\substack{J_1 \neq J_3 \\ J_1, J_3 \geq 1}} a_{J_1} \, c_m^{(\delta)}(J_1, J_3) \, a_{J_3} \geq 0 \quad \forall a.
\end{align}
The second and third terms are $O(1/k)$ since they involve $C_J$'s, and the last term is $O(1/k^2)$ since it only involves $C_{J_1 J_2 J_3}$'s, which are $O(1/k^2)$ by the $z$-counting rule.

First we check \eqref{eq: quadratic form} for the massive free gas.  In the $m = 0$ block, if $a_0$ is non-zero, then the leading term is the first term, and it's positive.  If instead $a_0 = 0$, then the third term is leading, and it's also positive.  For the $m \geq 1$ blocks, the first two terms vanish, and the third term is leading and positive. This verifies that the Buhmann--J\"ager bounds hold for the massive free gas in the large $k$ limit.  For the massless free gas, $C_1 = 0$ and $C_{1, J, J+1} = 0$ through $O(1/k^2)$ implies all terms containing $a_1$ in \eqref{eq: quadratic form} vanish to this order, and then the verification of the Buhmann--J\"ager bounds from the massive case carries over.

\subsection{Saddle-point approximation of the phase space}
\label{app:saddle point approx}

A key point in the derivation of the $z$-counting rule was that the entropy takes the form \eqref{eq:form of entropy}.  For the massless case this was immediate from dimensional analysis. In this section we compute the phase space of $N$ free scalar particles by saddle-point approximation and show it is consistent with the form of the entropy given in \eqref{eq:form of entropy} for general mass.  Similar techniques were used in \cite{Cuomo:2025pjp} to compute energy correlators sourced by heavy operators in CFTs.    

The phase space of $N$ scalar particles with total energy-momentum $p$ is
\begin{align} \label{number of states}
    \Omega(N,\sqrt{p^2})
    =
    \int d^dq_1 \dots d^dq_N\,
    \prod_{i=1}^N \delta_+(q_i^2-m^2)\,
    \delta^d\left(p-\sum_{i=1}^N q_i\right).
\end{align}
Using the Fourier representation of the delta function gives
\begin{align} 
    \Omega(N,\sqrt{p^2})
    =
    \frac{1}{(2\pi)^d}
    \int d^dx\,e^{ix\cdot p} Z^N(ix),
\end{align}
where the single-particle partition function is
\begin{align} \label{general partition function}
    Z(x)
    =
    \int d^dq\,\delta_+(q^2-m^2)e^{-x\cdot q},
    \qquad
    \braket{q_\mu}_x
    =
    -\partial_\mu\log Z(x).
\end{align}

The integral defining $Z(x)$ converges when $\operatorname{Re}(x)$ lies
inside the future lightcone. Since $Z(ix)$ lies on the boundary of this
domain, we shift the contour $x\rightarrow x-i\beta$, with $\beta$ inside
the future lightcone. Equivalently,
\begin{align} \label{correct delta function identity}
    \delta^d\left(p-\sum_{i=1}^N q_i\right)
    =
    \frac{1}{(2\pi)^d}
    \int d^dx\,
    e^{i(x-i\beta)\cdot\left(p-\sum_{i=1}^N q_i\right)} .
\end{align}
This gives
\begin{align} \label{well-defined number of states}
    \Omega(N,\sqrt{p^2})
    =
    e^{\beta\cdot p}
    \frac{1}{(2\pi)^d}
    \int d^dx\,e^{ix\cdot p} Z^N(\beta+ix).
\end{align}

To apply the saddle-point approximation, write
\begin{align}
    p=N\bar p.
\end{align}
Then
\begin{align} \label{number of states set up for saddle point approximation}
    \Omega(N,\sqrt{p^2})
    =
    \frac{e^{N\beta\cdot\bar p}}{(2\pi)^d}
    \int d^dx\,e^{Nf(x,\beta)},
    \qquad
    f(x,\beta)
    =
    ix\cdot\bar p+\log Z(\beta+ix).
\end{align}
The saddle condition is
\begin{align}
    \partial_\mu f(x,\beta)
    =
    i\left(
        \bar p_\mu-\braket{q_\mu}_{\beta+ix}
    \right)=0.
\end{align}
The saddle lies at $x=0$ and we must choose $\beta$ such that
\begin{align} \label{defines beta}
    \braket{q_\mu}_\beta=\bar p_\mu.
\end{align}
For instance, if we wish to compute the bath entropy $S(k_{bath}, E_{\text{tot}})$ discussed in the previous section, we take $N = k_{\text{bath}} = k - l'$, $\vec{\beta} = \vec{0}$, and we choose $\beta^0$ such that $\braket{q_0}_{\beta} = \frac{ku}{k-l'}$.  

To check that the integral \eqref{number of states set up for saddle point approximation} passes through the saddle in the correct direction, note that 
\begin{align}
    |Z(\beta+ix)|
    &=
    \left|
    \int d^dq\,\delta_+(q^2-m^2)e^{-(\beta+ix)\cdot q}
    \right| \leq
    \int d^dq\,\delta_+(q^2-m^2)e^{-\beta\cdot q}
    =
    Z(\beta),
\end{align}
so $\operatorname{Re}(f)$ is maximized at $x=0$.  Expanding around the saddle,
\begin{align}
    f(x,\beta)
    =
    \log Z(\beta)
    -\frac{1}{2}
    x^\mu x^\nu
    \partial_\mu\partial_\nu\log Z(x)\big|_{x=\beta}
    +
    \sum_{n\geq3}
    \frac{i^n}{n!}
    x^{\mu_1}\dots x^{\mu_n}
    \partial_{\mu_1}\dots\partial_{\mu_n}
    \log Z(x)\big|_{x=\beta}.
\end{align}
The derivatives of $\log Z$ are the single-particle cumulants,
\begin{align}
    \braket{q_{\mu_1}\dots q_{\mu_n}}_{\beta,c}
    =
    (-1)^n
    \partial_{\mu_1}\dots\partial_{\mu_n}
    \log Z(x)\big|_{x=\beta}.
\end{align}

Defining
\begin{align}
    A_{\mu\nu}(\beta)
    =
    \braket{q_\mu q_\nu}_{\beta,c},
\end{align}
we obtain
\begin{align} \label{app: saddle point approx of phase space} \Omega(N,\sqrt{p^2}) &= \frac{\left(e^{\beta\cdot\bar p}Z(\beta)\right)^N}{(2\pi)^{d/2} N^{d/2} \sqrt{|A(\beta)|}} \left\langle \exp\left[ \sum_{n\geq3} \frac{(-i)^n}{n!N^{(n-2)/2}} \braket{q_{\mu_1}\dots q_{\mu_n}}_{\beta,c} x^{\mu_1}\dots x^{\mu_n} \right] \right\rangle_{A(\beta)} \end{align}
where $\braket{\,}_{A(\beta)}$ denotes the Gaussian average with weight
$e^{-\frac12 x\cdot A(\beta)\cdot x}$.  Taking the logarithm reproduces an entropy of the form \eqref{eq:form of entropy}.

\section{Conformal structures}
\label{app:conformalstructures}

Here we collect various conformal structures, which are needed to calculate the contribution of a given operator to the beam-averaged EEC in Section \ref{sec:Ads}.

We normalize the general two-point function as follows 
\begin{equation}
\langle {\cal O}(x_1, z_1)  {\cal O}(x_2, z_2) \rangle = { \left( z_1 \cdot I(x_{12}) \cdot z_2 \right)^s \over x_{12}^{2 \Delta}} ,
\end{equation}
where $I_{\mu \nu}(x) = \eta_{\mu \nu} - 2{x_{\mu} x_{\nu} \over x^2}$. For the stress tensor two-point function we have
\begin{equation}
\langle T(x_1, z_1)  T(x_2, z_2) \rangle = c_T { \left( z_1 \cdot I(x_{12}) \cdot z_2 \right)^2 \over x_{12}^{2d}} .
\end{equation}
We will then need to parameterize several different three-point structures. 

\paragraph{Scalar: $\langle T(x_1, z_1)  T(x_2, z_2) {\cal O}(x_3) \rangle$.}
There is a unique conformally invariant structure that we parameterize as follows
\begin{equation}
\langle T T {\cal O} \rangle = \lambda_{TT{\cal O}} \frac{
V_{1,23}^{2}V_{2,31}^{2}
-\frac{2\bigl((d-1)(d-\Delta)+2\bigr)}{(d-2)(\Delta+2)}\,H_{12}V_{1,23}V_{2,31}
+\frac{(d-1)\Delta(\Delta-2d)+d(d-2)(d+1)}{(d-2)\Delta(\Delta+2)}\,H_{12}^{2}
}{
X_{12}^{\,d+2-\frac{\Delta}{2}}\,X_{13}^{\,\frac{\Delta}{2}}\,X_{23}^{\,\frac{\Delta}{2}}
} \,,
\end{equation}
where we use the structures $H_{ij}$ and $V_{i,jk}$ defined in \cite{Costa:2011mg}.  
The contribution of the scalar operator to the averaged EEC is given in \eqref{eq:C2fromO}.

\paragraph{Stress tensor: $\langle T(x_1, z_1)  T(x_2, z_2) T(x_3, z_3) \rangle$.}
In the three-point function of stress tensors there are in general three parity-even structures, which reduce to two in $d=3$. A convenient basis for the three-point structures in the context of the collider experiment was introduced in \cite{Hofman:2008ar}: we choose the three-point structure fixed by the conformal Ward identities to be the one in Einstein gravity and we parameterize the most general corrections by $t_2$ and $t_4$, see Appendix C.3 in \cite{Hofman:2016awc}.

\paragraph{Spin-2: $\langle T(x_1, z_1)  T(x_2, z_2) \mathcal{O}_2(x_3, z_3) \rangle$.}
The three-point function with a general spin-2 operator has two tensor structures, see e.g. Table I in \cite{Costa:2011mg}. We choose the basis of the three-point structures as follows 
\begin{equation}
\langle T T {\cal O}_2 \rangle = \frac{\lambda_1 V_{1,23}^2V_{2,31}^2 V_{3,12}^2 + \lambda_2 V_{1,23} V_{2,31}  V_{3,12} (V_{1,23}H_{23}+V_{2,31}H_{13}) + \dots }{
X_{12}^{\,d+1-\frac{\Delta}{2}}\,
X_{13}^{\,1+\frac{\Delta}{2}}\,
X_{23}^{\,1+\frac{\Delta}{2}}
} \ .
\end{equation}
where the terms in $\ldots$ are fixed by conservation of the stress tensor. 
The full contribution of this operator to the EEC is presented in the ancillary file. Here we only quote its contribution to the highest spin-four energy multipole
\begin{equation}
\begin{aligned}
C_4^{\mathcal O_2}
&={\sin^2\!\left(\frac{\pi(\Delta-2d)}{2}\right) \over c_T}
\frac{ 
 (d^2-1)\pi^{d-2}(\Delta+4)^2 
}{2(d+3)(\Delta-1)\Delta\, D(d,\Delta)^2
}
\Bigl(2\lambda_1 A(d,\Delta)+\lambda_2 B(d,\Delta)\Bigr)^2
\\
&\frac{\Gamma \left(\frac{d}{2}+1\right) \Gamma \left(\frac{d}{2}+2\right)^2 \Gamma (d+1)
   \Gamma (\Delta +2) \Gamma \left(\frac{\Delta }{2}-d\right)^2 \Gamma
   \left(-\frac{d}{2}+\Delta +1\right)}{\Gamma \left(\frac{\Delta }{2}+2\right)^2 \Gamma
   \left(\frac{\Delta }{2}+3\right)^2 \Gamma \left(\frac{1}{2} (d+\Delta +2)\right)^2} \ , \\
A(d,\Delta)
&=
(d^2-1)\Delta^2 +(3d^2-6d+7)\Delta -2(d-1)(d+3),
\\
B(d,\Delta)
&=
8d(d-1)-2(3d^2-5d+4)\Delta-(3d^2-3d-2)\Delta^2,
\\
D(d,\Delta)
&=
32(d-1)-2(3d^2-d+6)\Delta-(d-2)(d+1)\Delta^2.
\end{aligned}
\end{equation}

\paragraph{Spin-4: $\langle T(x_1, z_1)  T(x_2, z_2) \mathcal{O}_4(x_3, z_3) \rangle$.}

The three-point function with a general spin-4 operator has three independent conformal structures. We choose them as follows

\begin{equation}
\langle T T {\cal O}_4 \rangle = \frac{\lambda_1 V_{1,23}^2V_{2,31}^2 V_{3,12}^4 + \lambda_2 V_{1,23} V_{2,31} V_{3,12}^3 (V_{1,23}H_{23}+V_{2,31}H_{13}) + \lambda_3 V_{1,23} V_{2,31} V_{3,12}^2 H_{13} H_{23} +  \dots }{
X_{12}^{\,d-\frac{\Delta}{2}}\,
X_{13}^{\,2+\frac{\Delta}{2}}\,
X_{23}^{\,2+\frac{\Delta}{2}}
} \ .
\end{equation}
The full contribution of this operator to the EEC is presented in the ancillary file. Here we only quote its contribution to the highest spin-six energy multipole

\begingroup
\small

\begin{equation}
\begin{aligned}
C_6^{\mathcal{O}_4}
&={\sin^2\!\left(\frac{\pi(\Delta-2d)}{2}\right) \over c_T}
\frac{
8 \pi^{d-2}(d^2-1)(d+3)(d+4)^2(\Delta+6)^2
}{ (d+5)(d+7)(\Delta-1)\Delta(\Delta+1)(\Delta+2)^3
} \frac{1}{
\Gamma\!\left(\frac{\Delta+2}{2}\right)^2
\Gamma\!\left(\frac{\Delta+8}{2}\right)^2
\Gamma\!\left(\frac{d+\Delta-4}{2}\right)^2
}
\\[2pt]
&\quad\times
\frac{
\Gamma\!\left(1+\frac d2\right)\Gamma\!\left(2+\frac d2\right)^2
\Gamma(d+4)\Gamma(\Delta+4)\Gamma\!\left(1-\frac d2+\Delta\right)
\Gamma\!\left(\frac{\Delta}{2}-d-1\right)^2 \,
P(d,\Delta)^2
}{
(\Delta-2d)^2(d+\Delta-4)^2(d+\Delta-2)^2(d+\Delta)^2(d+\Delta+2)^2
\,Q(d,\Delta)^2 \, .
}
\end{aligned}
\end{equation}

\begin{align}
P(d,\Delta)
&= A_1(d,\Delta)\,\lambda_1 + A_2(d,\Delta)\,\lambda_2 + A_3(d,\Delta)\,\lambda_3,
\\[3pt]
A_1(d,\Delta)
&= 8\Bigl[
(2\Delta^3+17\Delta^2+18\Delta-8)d^3
-(\Delta^4+7\Delta^3+6\Delta^2-12\Delta+56)d^2
\notag\\
&\qquad\quad
-(3\Delta^4-\Delta^3-19\Delta^2-54\Delta+80)d
+4\Delta^4-16\Delta^3-22\Delta^2+52\Delta
\Bigr],
\\[3pt]
A_2(d,\Delta)
&= -(33\Delta^3+218\Delta^2+184\Delta-128)d^3
+(21\Delta^4+137\Delta^3+150\Delta^2-136\Delta+512)d^2
\notag\\
&\qquad\quad
-(11\Delta^4+104\Delta^3-28\Delta^2+448\Delta-512)d
-30\Delta^4+148\Delta^3+192\Delta^2-544\Delta,
\\[3pt]
A_3(d,\Delta)
&= (\Delta+4)(2d+2-\Delta)
\Bigl[
(9\Delta^2+10\Delta-8)d^2
-(17\Delta^2-2\Delta+16)d
-2\Delta(\Delta-10)
\Bigr],
\\[3pt]
Q(d,\Delta)
&= d^2\Delta(\Delta+6)+2d(\Delta^2+14\Delta-16)-8(\Delta^2-6\Delta+8).
\end{align}
\endgroup
All other spins can be found in the ancillary file that accompanies the submission.

\clearpage 
\let\oldbibliography\thebibliography
\renewcommand{\thebibliography}[1]{  \oldbibliography{#1}  \setlength{\itemsep}{0pt}}
\bibliographystyle{JHEP}
\bibliography{biblio}

@article{Kaplan:2020tdz,
    author = "Kaplan, Jared and Kundu, Sandipan",
    title = "{Causality constraints in large N QCD coupled to gravity}",
    eprint = "2009.08460",
    archivePrefix = "arXiv",
    primaryClass = "hep-th",
    doi = "10.1103/PhysRevD.104.L061901",
    journal = "Phys. Rev. D",
    volume = "104",
    number = "6",
    pages = "L061901",
    year = "2021"
}

@article{Kaplan:2019soo,
    author = "Kaplan, Jared and Kundu, Sandipan",
    title = "{A Species or Weak-Gravity Bound for Large $N$ Gauge Theories Coupled to Gravity}",
    eprint = "1904.09294",
    archivePrefix = "arXiv",
    primaryClass = "hep-th",
    doi = "10.1007/JHEP11(2019)142",
    journal = "JHEP",
    volume = "11",
    pages = "142",
    year = "2019"
}

@article{Aharony:2015zea,
    author = "Aharony, Ofer and Berkooz, Micha and Rey, Soo-Jong",
    title = "{Rigid holography and six-dimensional $ \mathcal{N}=\left(2,0\right) $ theories on AdS$_{5} \times  \mathbb{S}^{1}$}",
    eprint = "1501.02904",
    archivePrefix = "arXiv",
    primaryClass = "hep-th",
    reportNumber = "SNUST-15-01, WIS-01-15-JAN-DPPA",
    doi = "10.1007/JHEP03(2015)121",
    journal = "JHEP",
    volume = "03",
    pages = "121",
    year = "2015"
}

@article{Apolo:2022pbq,
    author = "Apolo, Luis and Belin, Alexandre and Bintanja, Suzanne and Castro, Alejandra and Keller, Christoph A.",
    title = "{Conformal field theories dual to quantum gravity with strongly coupled matter}",
    eprint = "2212.07436",
    archivePrefix = "arXiv",
    primaryClass = "hep-th",
    doi = "10.1103/PhysRevD.108.L061901",
    journal = "Phys. Rev. D",
    volume = "108",
    number = "6",
    pages = "L061901",
    year = "2023"
}

@article{Apolo:2024bmu,
    author = "Apolo, Luis and Belin, Alexandre and Bintanja, Suzanne",
    title = "{Searching for strongly coupled AdS matter with multi-trace deformations}",
    eprint = "2401.15141",
    archivePrefix = "arXiv",
    primaryClass = "hep-th",
    doi = "10.21468/SciPostPhys.19.5.139",
    journal = "SciPost Phys.",
    volume = "19",
    number = "5",
    pages = "139",
    year = "2025"
}

@article{Caron-Huot:2024lbf,
    author = "Caron-Huot, Simon and Li, Yue-Zhou",
    title = "{Gravity and a universal cutoff for field theory}",
    eprint = "2408.06440",
    archivePrefix = "arXiv",
    primaryClass = "hep-th",
    doi = "10.1007/JHEP02(2025)115",
    journal = "JHEP",
    volume = "02",
    pages = "115",
    year = "2025"
}

@article{Kologlu:2019bco,
    author = "Kologlu, Murat and Kravchuk, Petr and Simmons-Duffin, David and Zhiboedov, Alexander",
    title = "{Shocks, Superconvergence, and a Stringy Equivalence Principle}",
    eprint = "1904.05905",
    archivePrefix = "arXiv",
    primaryClass = "hep-th",
    reportNumber = "CALT-TH 2019-012, CERN-TH-2019-040",
    doi = "10.1007/JHEP11(2020)096",
    journal = "JHEP",
    volume = "11",
    pages = "096",
    year = "2020"
}

@Article{Yan:2022cye,
  author        = {Yan, Kai and Zhang, Xiaoyuan},
  journal       = {Phys. Rev. Lett.},
  title         = {{Three-point energy correlator in $\mathcal N=4$ supersymmetric Yang-Mills theory}},
  year          = {2022},
  number        = {2},
  pages         = {021602},
  volume        = {129},
  archiveprefix = {arXiv},
  doi           = {10.1103/PhysRevLett.129.021602},
  eprint        = {2203.04349},
  primaryclass  = {hep-th},
}

@article{Faulkner:2016mzt,
    author = "Faulkner, Thomas and Leigh, Robert G. and Parrikar, Onkar and Wang, Huajia",
    title = "{Modular Hamiltonians for deformed half-spaces and the Averaged Null Energy Condition}",
    eprint = "1605.08072",
    archivePrefix = "arXiv",
    primaryClass = "hep-th",
    doi = "10.1007/JHEP09(2016)038",
    journal = "JHEP",
    volume = "09",
    pages = "038",
    year = "2016"
}

@Article{Basham:1978zq,
  author       = {Basham, C. L. and Brown, L. S. and Ellis, S. D. and Love, S. T.},
  journal      = {Phys. Rev. D},
  title        = {{Energy correlations in electron-positron annihilation in quantum chromodynamics: Asymptotically free perturbation theory}},
  year         = {1979},
  pages        = {2018--2045},
  volume       = {19},
  doi          = {10.1103/PhysRevD.19.2018},
  reportnumber = {RLO-1388-761},
}

@article{ALEPH:1990ndp,
    author = "Decamp, D. and others",
    collaboration = "ALEPH",
    title = "{ALEPH: A detector for electron-positron annihilations at LEP}",
    reportNumber = "CERN-EP-90-25",
    doi = "10.1016/0168-9002(90)91831-U",
    journal = "Nucl. Instrum. Meth. A",
    volume = "294",
    pages = "121--178",
    year = "1990",
    note = "[Erratum: {\em Nucl. Instrum. Meth. A} {\bf 303} (1991) 393]"
}

@Article{Basham:1979gh,
  author       = {Basham, C. Louis and Brown, Lowell S. and Ellis, Stephen D. and Love, Sherwin T.},
  journal      = {Phys. Lett. B},
  title        = {{Energy correlations in perturbative quantum chromodynamics: A conjecture for all orders}},
  year         = {1979},
  pages        = {297--299},
  volume       = {85},
  doi          = {10.1016/0370-2693(79)90601-4},
  reportnumber = {RLO-1388-786},
}

@Article{Basham:1977iq,
  author       = {Basham, C. Louis and Brown, Lowell S. and Ellis, S. D. and Love, S. T.},
  journal      = {Phys. Rev. D},
  title        = {{Electron-positron annihilation energy pattern in quantum chromodynamics: Asymptotically free perturbation theory}},
  year         = {1978},
  pages        = {2298},
  volume       = {17},
  doi          = {10.1103/PhysRevD.17.2298},
  reportnumber = {RLO-1388-746},
}

@Article{Firat:2023lbp,
  author        = {Firat, Eren and Monin, Alexander and Rattazzi, Riccardo and Walters, Matthew T.},
  journal       = {JHEP},
  title         = {{Flux correlators and semiclassics}},
  year          = {2024},
  pages         = {067},
  volume        = {03},
  archiveprefix = {arXiv},
  doi           = {10.1007/JHEP03(2024)067},
  eprint        = {2309.14428},
  primaryclass  = {hep-th},
}

@article{Bellazzini:2020cot,
    author = "Bellazzini, Brando and Elias Mir{\'o}, Joan and Rattazzi, Riccardo and Riembau, Marc and Riva, Francesco",
    title = "{Positive moments for scattering amplitudes}",
    eprint = "2011.00037",
    archivePrefix = "arXiv",
    primaryClass = "hep-th",
    doi = "10.1103/PhysRevD.104.036006",
    journal = "Phys. Rev. D",
    volume = "104",
    number = "3",
    pages = "036006",
    year = "2021"
}

@article{Chen:2024iuv,
    author = "Chen, Hao and Karlsson, Robin and Zhiboedov, Alexander",
    title = "{Energy correlations and Planckian collisions}",
    eprint = "2404.15056",
    archivePrefix = "arXiv",
    primaryClass = "hep-th",
    reportNumber = "CERN-TH-2024-050",
    doi = "10.1007/JHEP03(2026)078",
    journal = "JHEP",
    volume = "03",
    pages = "078",
    year = "2026"
}

@Article{Chen:2022swd,
  author        = {Chen, Hao and Moult, Ian and Thaler, Jesse and Zhu, Hua Xing},
  journal       = {JHEP},
  title         = {{Non-Gaussianities in collider energy flux}},
  year          = {2022},
  pages         = {146},
  volume        = {07},
  archiveprefix = {arXiv},
  doi           = {10.1007/JHEP07(2022)146},
  eprint        = {2205.02857},
  primaryclass  = {hep-ph},
  reportnumber  = {MIT-CTP 5430},
}

@Article{Budhraja:2026pyi,
  author        = {Budhraja, Ankita and Pels, Isabelle and Waalewijn, Wouter J.},
  title         = {{Higher-point energy correlators: Factorization in the back-to-back limit and non-perturbative effects}},
  year          = {2026},
  archiveprefix = {arXiv},
  eprint        = {2603.16996},
  primaryclass  = {hep-ph},
}

@Article{Alday:2019qrf,
  author        = {Alday, Luis F. and Perlmutter, Eric},
  journal       = {JHEP},
  title         = {{Growing extra dimensions in AdS/CFT}},
  year          = {2019},
  pages         = {084},
  volume        = {08},
  archiveprefix = {arXiv},
  doi           = {10.1007/JHEP08(2019)084},
  eprint        = {1906.01477},
  primaryclass  = {hep-th},
  reportnumber  = {CALT-TH 2019-018},
}

@Article{Chicherin:2023gxt,
  author        = {Chicherin, Dmitry and Korchemsky, Gregory P. and Sokatchev, Emery and Zhiboedov, Alexander},
  journal       = {JHEP},
  title         = {{Energy correlations in heavy states}},
  year          = {2023},
  pages         = {134},
  volume        = {11},
  archiveprefix = {arXiv},
  doi           = {10.1007/JHEP11(2023)134},
  eprint        = {2306.14330},
  primaryclass  = {hep-th},
  reportnumber  = {CERN-TH-2023-109, IPhT-T23/051, LAPTH-034/23},
}

@article{Osborn:1993cr,
    author = "Osborn, H. and Petkou, A. C.",
    title = "{Implications of conformal invariance in field theories for general dimensions}",
    eprint = "hep-th/9307010",
    archivePrefix = "arXiv",
    reportNumber = "DAMTP-93-31",
    doi = "10.1006/aphy.1994.1045",
    journal = "Annals Phys.",
    volume = "231",
    pages = "311--362",
    year = "1994"
}

@article{Heemskerk:2009pn,
    author = "Heemskerk, Idse and Penedones, Joao and Polchinski, Joseph and Sully, James",
    title = "{Holography from Conformal Field Theory}",
    eprint = "0907.0151",
    archivePrefix = "arXiv",
    primaryClass = "hep-th",
    reportNumber = "NSF-KITP-09-110",
    doi = "10.1088/1126-6708/2009/10/079",
    journal = "JHEP",
    volume = "10",
    pages = "079",
    year = "2009"
}

@Article{Andres:2022ovj,
  author        = {Andres, Carlota and Dominguez, Fabio and Kunnawalkam Elayavalli, Raghav and Holguin, Jack and Marquet, Cyrille and Moult, Ian},
  journal       = {Phys. Rev. Lett.},
  title         = {{Resolving the scales of the quark-gluon plasma with energy correlators}},
  year          = {2023},
  number        = {26},
  pages         = {262301},
  volume        = {130},
  archiveprefix = {arXiv},
  doi           = {10.1103/PhysRevLett.130.262301},
  eprint        = {2209.11236},
  primaryclass  = {hep-ph},
}

@Article{Gneiting2011,
  author        = {Tilmann Gneiting},
  journal       = {Bernoulli},
  title         = {{Strictly and non-strictly positive definite functions on spheres}},
  year          = {2013},
  number        = {4},
  pages         = {1327--1349},
  volume        = {19},
  archiveprefix = {arXiv},
  doi           = {10.3150/12-BEJSP06},
  eprint        = {1111.7077},
  primaryclass  = {math.PR},
}

@Article{Cordova:2017zej,
  author        = {Cordova, Clay and Maldacena, Juan and Turiaci, Gustavo J.},
  journal       = {JHEP},
  title         = {{Bounds on OPE coefficients from interference effects in the conformal collider}},
  year          = {2017},
  pages         = {032},
  volume        = {11},
  archiveprefix = {arXiv},
  doi           = {10.1007/JHEP11(2017)032},
  eprint        = {1710.03199},
  primaryclass  = {hep-th},
}

@Article{Moult:2025nhu,
  author        = {Moult, Ian and Zhu, Hua Xing},
  title         = {{Energy correlators: A journey from theory to experiment}},
  year          = {2025},
  archiveprefix = {arXiv},
  eprint        = {2506.09119},
  primaryclass  = {hep-ph},
}

@Article{Mecaj:2025ecl,
  author        = {Me{\c{c}}aj, Bianka and Moult, Ian and Walters, Matthew T. and Xin, Yuan},
  title         = {{Energy correlator conformal blocks and positivity}},
  year          = {2025},
  archiveprefix = {arXiv},
  eprint        = {2512.09986},
  primaryclass  = {hep-th},
  reportnumber  = {LA-UR-25-31137},
}

@Article{Dempsey:2025yiv,
  author        = {Dempsey, Ross and Karlsson, Robin and Pufu, Silviu S. and Zahraee, Zahra and Zhiboedov, Alexander},
  title         = {{Conformal collider bootstrap in $\mathcal N=4$ SYM}},
  year          = {2025},
  archiveprefix = {arXiv},
  eprint        = {2512.10796},
  primaryclass  = {hep-th},
  reportnumber  = {CERN-TH-2025-212, MIT-CTP/5979, PUPT-2659},
}

@article{Jaarsma:2025tck,
    author = "Jaarsma, Max and Li, Yibei and Moult, Ian and Waalewijn, Wouter J. and Zhu, Hua Xing",
    title = "{From DGLAP to Sudakov: Precision predictions for energy-energy correlators}",
    eprint = "2512.11950",
    archivePrefix = "arXiv",
    primaryClass = "hep-ph",
    reportNumber = "MITP-24-091",
    year = "2025"
}

@Article{Caron-Huot:2022eqs,
  author        = {Caron-Huot, Simon and Kologlu, Murat and Kravchuk, Petr and Meltzer, David and Simmons-Duffin, David},
  journal       = {JHEP},
  title         = {{Detectors in weakly-coupled field theories}},
  year          = {2023},
  pages         = {014},
  volume        = {04},
  archiveprefix = {arXiv},
  doi           = {10.1007/JHEP04(2023)014},
  eprint        = {2209.00008},
  primaryclass  = {hep-th},
  reportnumber  = {CALT-TH 2022-31},
}

@article{Riembau:2025isw,
    author = "Riembau, Marc and Son, Minho",
    title = "{Energy correlators of spinning sources}",
    eprint = "2512.16985",
    archivePrefix = "arXiv",
    primaryClass = "hep-ph",
    reportNumber = "CERN-TH-2025-260",
    doi = "10.1007/JHEP08(2026)047",
    journal = "JHEP",
    volume = "08",
    pages = "047",
    year = "2026"
}

@Article{Riembau:2025wjc,
  author        = {Riembau, Marc and Son, Minho},
  journal       = {Phys. Rev. Lett.},
  title         = {{Flow between extremal one-point energy correlators in QCD}},
  year          = {2026},
  number        = {10},
  pages         = {101901},
  volume        = {136},
  archiveprefix = {arXiv},
  doi           = {10.1103/lc5m-kwgv},
  eprint        = {2509.02669},
  primaryclass  = {hep-ph},
  reportnumber  = {CERN-TH-2025-175},
}

@Article{Basham:1978bw,
  author       = {Basham, C. Louis and Brown, Lowell S. and Ellis, Stephen D. and Love, Sherwin T.},
  journal      = {Phys. Rev. Lett.},
  title        = {{Energy correlations in electron--positron annihilation: Testing QCD}},
  year         = {1978},
  pages        = {1585--1588},
  volume       = {41},
  doi          = {10.1103/PhysRevLett.41.1585},
  reportnumber = {RLO-1388-759},
}

@Article{Kravchuk:2018htv,
  author        = {Kravchuk, Petr and Simmons-Duffin, David},
  journal       = {JHEP},
  title         = {{Light-ray operators in conformal field theory}},
  year          = {2018},
  pages         = {102},
  volume        = {11},
  archiveprefix = {arXiv},
  doi           = {10.1007/JHEP11(2018)102},
  eprint        = {1805.00098},
  primaryclass  = {hep-th},
  reportnumber  = {CALT-TH 2018-018},
}

@Article{Hofman:2008ar,
  author        = {Hofman, Diego M. and Maldacena, Juan},
  journal       = {JHEP},
  title         = {{Conformal collider physics: energy and charge correlations}},
  year          = {2008},
  pages         = {012},
  volume        = {05},
  archiveprefix = {arXiv},
  doi           = {10.1088/1126-6708/2008/05/012},
  eprint        = {0803.1467},
  primaryclass  = {hep-th},
}

@article{Electron-PositronAlliance:2025wzh,
    author = "Bossi, Hannah and Chen, Yu-Chen and Chen, Yi and Zhang, Jingyu and Innocenti, Gian Michele and Badea, Anthony and Baty, Austin and Maggi, Marcello and McGinn, Chris and Lee, Yen-Jie",
    collaboration = "Electron-Positron Alliance",
  title         = {{Analysis note: measurement of energy-energy correlator in $e^{+}e^{-}$ collisions at $91$ GeV with archived ALEPH data}},
  year          = {2025},
  archiveprefix = {arXiv},
  eprint        = {2505.11828},
  primaryclass  = {hep-ex},
}

@Article{Bossi:2025nux,
  author        = {Hannah Bossi and Yi Chen and Yu-Chen Chen and Max Jaarsma and Yibei Li and Jingyu Zhang and Ian Moult and Wouter Waalewijn and HuaXing Zhu and Anthony Badea and Austin Baty and Christopher McGinn and Gian Michele Innocenti and Marcello Maggi and Yen-Jie Lee},
  title         = {{Energy correlators from partons to hadrons: Unveiling the dynamics of the strong interactions with archival ALEPH data}},
  year          = {2025},
  archiveprefix = {arXiv},
  eprint        = {2511.00149},
  primaryclass  = {hep-ph},
  reportnumber  = {MITP-25-057, MITHIG-MOD-24-001},
}

@Article{Chang:2013iba,
  author        = {Chang, Hsi-Ming and Procura, Massimiliano and Thaler, Jesse and Waalewijn, Wouter J.},
  journal       = {Phys. Rev. D},
  title         = {{Calculating track thrust with track functions}},
  year          = {2013},
  pages         = {034030},
  volume        = {88},
  archiveprefix = {arXiv},
  doi           = {10.1103/PhysRevD.88.034030},
  eprint        = {1306.6630},
  primaryclass  = {hep-ph},
  reportnumber  = {MIT--CTP-4476},
}

@Article{Li:2021zcf,
  author        = {Li, Yibei and Moult, Ian and van Velzen, Solange Schrijnder and Waalewijn, Wouter J. and Zhu, Hua Xing},
  journal       = {Phys. Rev. Lett.},
  title         = {{Extending precision perturbative QCD with track functions}},
  year          = {2022},
  number        = {18},
  pages         = {182001},
  volume        = {128},
  archiveprefix = {arXiv},
  doi           = {10.1103/PhysRevLett.128.182001},
  eprint        = {2108.01674},
  primaryclass  = {hep-ph},
}

@Article{Chen:2022muj,
  author        = {Chen, Hao and Jaarsma, Max and Li, Yibei and Moult, Ian and Waalewijn, Wouter J. and Zhu, Hua Xing},
  journal       = {Phys. Rev. D},
  title         = {{Collinear parton dynamics beyond Dokshitzer-Gribov-Lipatov-Altarelli-Parisi framework}},
  year          = {2025},
  number        = {7},
  pages         = {076021},
  volume        = {111},
  archiveprefix = {arXiv},
  doi           = {10.1103/PhysRevD.111.076021},
  eprint        = {2210.10061},
  primaryclass  = {hep-ph},
}

@Article{Caron-Huot:2020cmc,
  author        = {Caron-Huot, Simon and Van Duong, Vincent},
  journal       = {JHEP},
  title         = {{Extremal effective field theories}},
  year          = {2021},
  pages         = {280},
  volume        = {05},
  archiveprefix = {arXiv},
  doi           = {10.1007/JHEP05(2021)280},
  eprint        = {2011.02957},
  primaryclass  = {hep-th},
}

@Article{Tolley:2020gtv,
  author        = {Tolley, Andrew J. and Wang, Zi-Yue and Zhou, Shuang-Yong},
  journal       = {JHEP},
  title         = {{New positivity bounds from full crossing symmetry}},
  year          = {2021},
  pages         = {255},
  volume        = {05},
  archiveprefix = {arXiv},
  doi           = {10.1007/JHEP05(2021)255},
  eprint        = {2011.02400},
  primaryclass  = {hep-th},
}

@Article{Arkani-Hamed:2020blm,
  author        = {Arkani-Hamed, Nima and Huang, Tzu-Chen and Huang, Yu-Tin},
  journal       = {JHEP},
  title         = {{The EFT-hedron}},
  year          = {2021},
  pages         = {259},
  volume        = {05},
  archiveprefix = {arXiv},
  doi           = {10.1007/JHEP05(2021)259},
  eprint        = {2012.15849},
  primaryclass  = {hep-th},
  reportnumber  = {NCTS-TH/2014, CALT-TH 2020-061},
}

@Article{Chen:2022pdu,
  author        = {Chen, Hao and Jaarsma, Max and Li, Yibei and Moult, Ian and Waalewijn, Wouter J. and Zhu, Hua Xing},
  journal       = {JHEP},
  title         = {{Multi-collinear splitting kernels for track function evolution}},
  year          = {2023},
  pages         = {185},
  volume        = {07},
  archiveprefix = {arXiv},
  doi           = {10.1007/JHEP07(2023)185},
  eprint        = {2210.10058},
  primaryclass  = {hep-ph},
}

@Article{Jaarsma:2022kdd,
  author        = {Jaarsma, Max and Li, Yibei and Moult, Ian and Waalewijn, Wouter and Zhu, Hua Xing},
  journal       = {JHEP},
  title         = {{Renormalization group flows for track function moments}},
  year          = {2022},
  pages         = {139},
  volume        = {06},
  archiveprefix = {arXiv},
  doi           = {10.1007/JHEP06(2022)139},
  eprint        = {2201.05166},
  primaryclass  = {hep-ph},
}

@Article{Jaarsma:2023ell,
  author        = {Jaarsma, Max and Li, Yibei and Moult, Ian and Waalewijn, Wouter J. and Zhu, Hua Xing},
  journal       = {JHEP},
  title         = {{Energy correlators on tracks: resummation and non-perturbative effects}},
  year          = {2023},
  pages         = {087},
  volume        = {12},
  archiveprefix = {arXiv},
  doi           = {10.1007/JHEP12(2023)087},
  eprint        = {2307.15739},
  primaryclass  = {hep-ph},
}

@Article{Fox:1978vw,
  author       = {Fox, Geoffrey C. and Wolfram, Stephen},
  journal      = {Nucl. Phys. B},
  title        = {{Event shapes in $e^+ e^-$ annihilation}},
  year         = {1979},
  note         = {[Erratum: {\em Nucl. Phys. B} {\bf 157} (1979) 543]},
  pages        = {413},
  volume       = {149},
  doi          = {10.1016/0550-3213(79)90120-2},
  reportnumber = {CALT-68-678},
}

@article{Fox:1978vu,
    author = "Fox, Geoffrey C. and Wolfram, Stephen",
    title = "{Observables for the analysis of event shapes in $e^+ e^-$ annihilation and other processes}",
    reportNumber = "CALT-68-680",
    doi = "10.1103/PhysRevLett.41.1581",
    journal = "Phys. Rev. Lett.",
    volume = "41",
    pages = "1581",
    year = "1978"
}

@Article{Hartman:2016lgu,
  author        = {Hartman, Thomas and Kundu, Sandipan and Tajdini, Amirhossein},
  journal       = {JHEP},
  title         = {{Averaged null energy condition from causality}},
  year          = {2017},
  pages         = {066},
  volume        = {07},
  archiveprefix = {arXiv},
  doi           = {10.1007/JHEP07(2017)066},
  eprint        = {1610.05308},
  primaryclass  = {hep-th},
}

@Article{OPAL:1993pnw,
  author        = {Acton, P. D. and others},
  journal       = {Z. Phys. C},
  title         = {{A determination of $\alpha_s (M_{Z_0})$ at LEP using resummed QCD calculations}},
  year          = {1993},
  pages         = {1--20},
  volume        = {59},
  collaboration = {OPAL},
  doi           = {10.1007/BF01555834},
  reportnumber  = {CERN-PPE-93-38},
}

@Article{Chen:2019bpb,
  author        = {Chen, Hao and Luo, Ming-Xing and Moult, Ian and Yang, Tong-Zhi and Zhang, Xiaoyuan and Zhu, Hua Xing},
  journal       = {JHEP},
  title         = {{Three point energy correlators in the collinear limit: symmetries, dualities and analytic results}},
  year          = {2020},
  number        = {08},
  pages         = {028},
  volume        = {08},
  archiveprefix = {arXiv},
  doi           = {10.1007/JHEP08(2020)028},
  eprint        = {1912.11050},
  primaryclass  = {hep-ph},
}

@Article{Gong:2025jqi,
  author        = {Gong, Jianyu and Pokraka, Andrzej and Yan, Kai and Zhang, Xiaoyuan},
  title         = {{Toward the analytic bootstrap of energy correlators}},
  year          = {2025},
  archiveprefix = {arXiv},
  eprint        = {2509.22782},
  primaryclass  = {hep-ph},
  reportnumber  = {MIT-CTP 5912},
}

@Article{Bossi:2024qho,
  author        = {Bossi, Hannah and Kudinoor, Arjun Srinivasan and Moult, Ian and Pablos, Daniel and Rai, Ananya and Rajagopal, Krishna},
  journal       = {JHEP},
  title         = {{Imaging the wakes of jets with energy-energy-energy correlators}},
  year          = {2024},
  pages         = {073},
  volume        = {12},
  archiveprefix = {arXiv},
  doi           = {10.1007/JHEP12(2024)073},
  eprint        = {2407.13818},
  primaryclass  = {hep-ph},
  reportnumber  = {MIT-CTP-5739},
}

@article{BuhmannJager2022,
    author = "Buhmann, Martin and J{\"a}ger, Janin",
    title = "{Strict Positive Definiteness of Convolutional and Axially Symmetric Kernels on d-Dimensional Spheres}",
    journal = "J. Fourier Anal. Appl.",
    volume = "28",
    pages = "40",
    year = "2022",
    doi = "10.1007/s00041-022-09913-x",
      archiveprefix = "arXiv",
  eprint        = "2105.02586",
  primaryclass  = "math.NA",
}

@article{Tchakaloff1957,
  author  = {Tchakaloff, V.},
  title   = {Formules de cubatures m{\'e}caniques {\`a} coefficients non n{\'e}gatifs},
  journal = {Bull. Sci. Math.},
  volume  = {81},
  year    = {1957},
  pages   = {123--134}
}

@article{Winkler1988,
  author  = {Winkler, G.},
  title   = {Extreme Points of Moment Sets},
  journal = {Math. Oper. Res.},
  volume  = {13},
  number  = {4},
  year    = {1988},
  pages   = {581--587},
  doi     = {10.1287/moor.13.4.581}
}

@Article{Kirsch2019,
  author        = {Kirsch, Werner},
  journal       = {Statist. Probab. Lett.},
  title         = {{An elementary proof of de Finetti's theorem}},
  year          = {2019},
  pages         = {84--88},
  volume        = {151},
  archiveprefix = {arXiv},
  doi           = {10.1016/j.spl.2019.03.014},
  eprint        = {1809.00882},
  primaryclass  = {math.PR},
}

@Article{Franken:2025gwr,
  author        = {Franken, Victor and Kaya, Sami and Rondeau, Fran{\c{c}}ois and Shahbazi-Moghaddam, Arvin and Tran, Patrick},
  journal       = {JHEP},
  title         = {{Tests of restricted quantum focusing and a new CFT bound}},
  year          = {2026},
  pages         = {111},
  volume        = {05},
  archiveprefix = {arXiv},
  doi           = {10.1007/JHEP05(2026)111},
  eprint        = {2510.13961},
  primaryclass  = {hep-th},
}

@book{tao2011introduction,
  title={An introduction to measure theory},
  author={Tao, Terence},
  volume={126},
  year={2011},
  publisher={American Mathematical Soc.}
}

@article{Hartman:2023qdn,
    author = "Hartman, Thomas and Mathys, Gr{\'e}goire",
    title = "{Averaged null energy and the renormalization group}",
    eprint = "2309.14409",
    archivePrefix = "arXiv",
    primaryClass = "hep-th",
    doi = "10.1007/JHEP12(2023)139",
    journal = "JHEP",
    volume = "12",
    pages = "139",
    year = "2023"
}

@Article{Schoenberg1942,
  author  = {Schoenberg, I.J.},
  journal = {Duke Math. J.},
  title   = {{Positive definite functions on spheres}},
  year    = {1942},
  number  = {1},
  pages   = {96--108},
  volume  = {9},
  doi     = {10.1215/S0012-7094-42-00908-6},
}

@article{Hartman:2023ccw,
    author = "Hartman, Thomas and Mathys, Gr{\'e}goire",
    title = "{Null energy constraints on two-dimensional RG flows}",
    eprint = "2310.15217",
    archivePrefix = "arXiv",
    primaryClass = "hep-th",
    doi = "10.1007/JHEP01(2024)102",
    journal = "JHEP",
    volume = "01",
    pages = "102",
    year = "2024"
}

@article{Hofman:2016awc,
    author = "Hofman, Diego M. and Li, Daliang and Meltzer, David and Poland, David and Rejon-Barrera, Fernando",
    title = "{A Proof of the Conformal Collider Bounds}",
    eprint = "1603.03771",
    archivePrefix = "arXiv",
    primaryClass = "hep-th",
    doi = "10.1007/JHEP06(2016)111",
    journal = "JHEP",
    volume = "06",
    pages = "111",
    year = "2016"
}

@article{Costa:2011mg,
    author = "Costa, Miguel S. and Penedones, Joao and Poland, David and Rychkov, Slava",
    title = "{Spinning Conformal Correlators}",
    eprint = "1107.3554",
    archivePrefix = "arXiv",
    primaryClass = "hep-th",
    reportNumber = "LPTENS-11-22, NSF-KITP-11-128",
    doi = "10.1007/JHEP11(2011)071",
    journal = "JHEP",
    volume = "11",
    pages = "071",
    year = "2011"
}

@article{Costa:2014kfa,
    author = "Costa, Miguel S. and Gon{\c{c}}alves, Vasco and Penedones, Jo{\~a}o",
    title = "{Spinning AdS Propagators}",
    eprint = "1404.5625",
    archivePrefix = "arXiv",
    primaryClass = "hep-th",
    doi = "10.1007/JHEP09(2014)064",
    journal = "JHEP",
    volume = "09",
    pages = "064",
    year = "2014"
}

@article{Komargodski:2016gci,
    author = "Komargodski, Zohar and Kulaxizi, Manuela and Parnachev, Andrei and Zhiboedov, Alexander",
    title = "{Conformal Field Theories and Deep Inelastic Scattering}",
    eprint = "1601.05453",
    archivePrefix = "arXiv",
    primaryClass = "hep-th",
    doi = "10.1103/PhysRevD.95.065011",
    journal = "Phys. Rev. D",
    volume = "95",
    number = "6",
    pages = "065011",
    year = "2017"
}

@article{Sveshnikov:1995vi,
    author = "Sveshnikov, N. A. and Tkachov, F. V.",
    editor = "Levchenko, B. B. and Savrin, V. I.",
    title = "{Jets and quantum field theory}",
    eprint = "hep-ph/9512370",
    archivePrefix = "arXiv",
    doi = "10.1016/0370-2693(96)00558-8",
    journal = "Phys. Lett. B",
    volume = "382",
    pages = "403--408",
    year = "1996"
}

@article{Korchemsky:1997sy,
    author = "Korchemsky, Gregory P. and Oderda, Gianluca and Sterman, George F.",
    editor = "Repond, Jos{\'e} and Krakauer, Daniel",
    title = "{Power corrections and nonlocal operators}",
    eprint = "hep-ph/9708346",
    archivePrefix = "arXiv",
    reportNumber = "ITP-SB-97-41, LPTHE-ORSAY-97-40",
    doi = "10.1063/1.53732",
    journal = "AIP Conf. Proc.",
    volume = "407",
    number = "1",
    pages = "988",
    year = "1997"
}

@article{Korchemsky:1999kt,
    author = "Korchemsky, Gregory P. and Sterman, George F.",
    title = "{Power corrections to event shapes and factorization}",
    eprint = "hep-ph/9902341",
    archivePrefix = "arXiv",
    reportNumber = "ITP-SB-98-73, LPT-ORSAY-98-80",
    doi = "10.1016/S0550-3213(99)00308-9",
    journal = "Nucl. Phys. B",
    volume = "555",
    pages = "335--351",
    year = "1999"
}

@article{Cordova:2018ygx,
    author = "C{\'o}rdova, Clay and Shao, Shu-Heng",
    title = "{Light-ray Operators and the BMS Algebra}",
    eprint = "1810.05706",
    archivePrefix = "arXiv",
    primaryClass = "hep-th",
    doi = "10.1103/PhysRevD.98.125015",
    journal = "Phys. Rev. D",
    volume = "98",
    number = "12",
    pages = "125015",
    year = "2018"
}

@article{Hewitt1955,
  title={Symmetric measures on Cartesian products},
  author={Hewitt, Edwin and Savage, Leonard J},
  journal={Trans. Amer. Math. Soc.},
  volume={80},
  number={2},
  pages={470--501},
  year={1955},
  doi={10.1090/S0002-9947-1955-0076206-8}
}

@Article{Musin2007,
  author        = {Oleg R. Musin},
  journal       = {Contemp. Math.},
  title         = {{Multivariate positive definite functions on spheres}},
  year          = {2014},
  pages         = {177--190},
  volume        = {625},
  archiveprefix = {arXiv},
  doi           = {10.1090/conm/625},
  eprint        = {math/0701083},
  primaryclass  = {math.CO},
}

@Article{Adams:2006sv,
  author        = {Adams, Allan and Arkani-Hamed, Nima and Dubovsky, Sergei and Nicolis, Alberto and Rattazzi, Riccardo},
  journal       = {JHEP},
  title         = {{Causality, analyticity and an IR obstruction to UV completion}},
  year          = {2006},
  pages         = {014},
  volume        = {10},
  archiveprefix = {arXiv},
  doi           = {10.1088/1126-6708/2006/10/014},
  eprint        = {hep-th/0602178},
  reportnumber  = {CERN-PH-TH-2006-033, HUTP-06-A0005},
}

@Article{Chang:2024whx,
  author        = {Chang, Cyuan-Han and Dommes, Vasiliy and Erramilli, Rajeev S. and Homrich, Alexandre and Kravchuk, Petr and Liu, Aike and Mitchell, Matthew S. and Poland, David and Simmons-Duffin, David},
  journal       = {JHEP},
  title         = {{Bootstrapping the $3$d Ising stress tensor}},
  year          = {2025},
  pages         = {136},
  volume        = {03},
  archiveprefix = {arXiv},
  doi           = {10.1007/JHEP03(2025)136},
  eprint        = {2411.15300},
  primaryclass  = {hep-th},
  reportnumber  = {CALT-TH 2024-047},
}

@Article{Chester:2021aun,
  author        = {Chester, Shai M. and Dempsey, Ross and Pufu, Silviu S.},
  journal       = {JHEP},
  title         = {{Bootstrapping $ \mathcal{N} = 4$ super-Yang-Mills on the conformal manifold}},
  year          = {2023},
  pages         = {038},
  volume        = {01},
  archiveprefix = {arXiv},
  doi           = {10.1007/JHEP01(2023)038},
  eprint        = {2111.07989},
  primaryclass  = {hep-th},
  reportnumber  = {PUPT-2627},
}

@Article{Chester:2023ehi,
  author        = {Chester, Shai M. and Dempsey, Ross and Pufu, Silviu S.},
  journal       = {JHEP},
  title         = {{Level repulsion in $ \mathcal{N} = 4 $ super-Yang-Mills via integrability, holography, and the bootstrap}},
  year          = {2024},
  pages         = {059},
  volume        = {07},
  archiveprefix = {arXiv},
  doi           = {10.1007/JHEP07(2024)059},
  eprint        = {2312.12576},
  primaryclass  = {hep-th},
  reportnumber  = {PUPT-2650},
}

@Article{Chang:2013rca,
  author        = {Chang, Hsi-Ming and Procura, Massimiliano and Thaler, Jesse and Waalewijn, Wouter J.},
  journal       = {Phys. Rev. Lett.},
  title         = {{Calculating track-based observables for the LHC}},
  year          = {2013},
  pages         = {102002},
  volume        = {111},
  archiveprefix = {arXiv},
  doi           = {10.1103/PhysRevLett.111.102002},
  eprint        = {1303.6637},
  primaryclass  = {hep-ph},
  reportnumber  = {MIT-CTP-4449, MIT--CTP-4449},
}

@Article{Devinatz1959,
  author  = {Devinatz, Allen},
  journal = {Acta Math.},
  title   = {{On the extensions of positive definite functions}},
  year    = {1959},
  pages   = {109--134},
  volume  = {102},
  doi     = {10.1007/BF02559570},
}

@book{Schmudgen2012,
  author    = {Schm{\"u}dgen, Konrad},
  title     = {{Unbounded Self-adjoint Operators on Hilbert Space}},
  series    = {Graduate Texts in Mathematics},
  volume    = {265},
  publisher = {Springer},
  year      = {2012},
  doi       = {10.1007/978-94-007-4753-1}
}

@article{Herrmann:2024yai,
    author = "Herrmann, Enrico and Kologlu, Murat and Moult, Ian",
    title = "{Energy correlators in perturbative quantum gravity}",
    eprint = "2412.05384",
    archivePrefix = "arXiv",
    primaryClass = "hep-th",
    doi = "10.1007/JHEP07(2026)126",
    journal = "JHEP",
    volume = "07",
    pages = "126",
    year = "2026"
}

@article{Li:2025knf,
    author = "Li, Yue-Zhou and Simmons-Duffin, David",
    title = "{Regge trajectories, detectors, and distributions in the critical O(N) model}",
    eprint = "2506.06419",
    archivePrefix = "arXiv",
    primaryClass = "hep-th",
    doi = "10.1007/JHEP02(2026)149",
    journal = "JHEP",
    volume = "02",
    pages = "149",
    year = "2026"
}

@article{Ren:2026zxs,
    author = "Ren, Lecheng and Wang, Bo and Wen, Congkao",
    title = "{Energy-energy correlator from the AdS Virasoro-Shapiro amplitude}",
    eprint = "2601.05312",
    archivePrefix = "arXiv",
    primaryClass = "hep-th",
    doi = "10.1103/m9q6-47z2",
    journal = "Phys. Rev. D",
    volume = "113",
    number = "10",
    pages = "106013",
    year = "2026"
}

@article{Maldacena:2015iua,
    author = "Maldacena, Juan and Simmons-Duffin, David and Zhiboedov, Alexander",
    title = "{Looking for a bulk point}",
    eprint = "1509.03612",
    archivePrefix = "arXiv",
    primaryClass = "hep-th",
    doi = "10.1007/JHEP01(2017)013",
    journal = "JHEP",
    volume = "01",
    pages = "013",
    year = "2017"
}

@article{Goncalves:2014ffa,
    author = "Gon{\c{c}}alves, Vasco",
    title = "{Four point function of $\mathcal{N}=4$ stress-tensor multiplet at strong coupling}",
    eprint = "1411.1675",
    archivePrefix = "arXiv",
    primaryClass = "hep-th",
    doi = "10.1007/JHEP04(2015)150",
    journal = "JHEP",
    volume = "04",
    pages = "150",
    year = "2015"
}

@article{Costa:2017twz,
    author = "Costa, Miguel S. and Hansen, Tobias and Penedones, Jo{\~a}o",
    title = "{Bounds for OPE coefficients on the Regge trajectory}",
    eprint = "1707.07689",
    archivePrefix = "arXiv",
    primaryClass = "hep-th",
    doi = "10.1007/JHEP10(2017)197",
    journal = "JHEP",
    volume = "10",
    pages = "197",
    year = "2017"
}

@article{Cuomo:2025pjp,
    author = "Cuomo, Gabriel and Firat, Eren and Nardi, Filippo and Ricci, Lorenzo",
    title = "{Conformal collider physics at large charge}",
    eprint = "2503.21867",
    archivePrefix = "arXiv",
    primaryClass = "hep-th",
    doi = "10.1007/JHEP02(2026)223",
    journal = "JHEP",
    volume = "02",
    pages = "223",
    year = "2026"
}

@article{Cordova:2017dhq,
    author = "Cordova, Clay and Diab, Kenan",
    title = "{Universal Bounds on Operator Dimensions from the Average Null Energy Condition}",
    eprint = "1712.01089",
    archivePrefix = "arXiv",
    primaryClass = "hep-th",
    doi = "10.1007/JHEP02(2018)131",
    journal = "JHEP",
    volume = "02",
    pages = "131",
    year = "2018"
}

@article{Carmi:2024tzj,
    author = "Carmi, Dean",
    title = "{Loops in AdS: from the spectral representation to position space. Part III}",
    eprint = "2402.02481",
    archivePrefix = "arXiv",
    primaryClass = "hep-th",
    doi = "10.1007/JHEP08(2024)193",
    journal = "JHEP",
    volume = "08",
    pages = "193",
    year = "2024"
}

@article{Tkachov:1995kk,
    author = "Tkachov, Fyodor V.",
    title = "{Measuring multijet structure of hadronic energy flow or What is a jet?}",
    eprint = "hep-ph/9601308",
    archivePrefix = "arXiv",
    reportNumber = "FERMILAB-PUB-95-191-T-REV, FERMILAB-PUB-95-191-T",
    doi = "10.1142/S0217751X97002899",
    journal = "Int. J. Mod. Phys. A",
    volume = "12",
    pages = "5411--5529",
    year = "1997"
}

@article{Komiske:2017aww,
    author = "Komiske, Patrick T. and Metodiev, Eric M. and Thaler, Jesse",
    title = "{Energy flow polynomials: A complete linear basis for jet substructure}",
    eprint = "1712.07124",
    archivePrefix = "arXiv",
    primaryClass = "hep-ph",
    reportNumber = "MIT-CTP-4965",
    doi = "10.1007/JHEP04(2018)013",
    journal = "JHEP",
    volume = "04",
    pages = "013",
    year = "2018"
}

@article{Green:2026nnw,
    author = "Green, Daniel and Gupta, Kshitij and Premkumar, Akhil",
    title = "{The Quantum Mechanics of Rare Events: From Quantum Walks to Stochastic Inflation}",
    eprint = "2608.06319",
    archivePrefix = "arXiv",
    primaryClass = "hep-th",
    month = "8",
    year = "2026"
}

@article{Alday:2007hr,
    author = "Alday, Luis F. and Maldacena, Juan Martin",
    title = "{Gluon scattering amplitudes at strong coupling}",
    eprint = "0705.0303",
    archivePrefix = "arXiv",
    primaryClass = "hep-th",
    reportNumber = "SPIN-07-16, ITP-UU-07-24",
    doi = "10.1088/1126-6708/2007/06/064",
    journal = "JHEP",
    volume = "06",
    pages = "064",
    year = "2007"
}

@article{Dorey:1999pd,
    author = "Dorey, Nicholas and Hollowood, Timothy J. and Khoze, Valentin V. and Mattis, Michael P. and Vandoren, Stefan",
    title = "{Multi-instanton calculus and the AdS / CFT correspondence in N=4 superconformal field theory}",
    eprint = "hep-th/9901128",
    archivePrefix = "arXiv",
    doi = "10.1016/S0550-3213(99)00193-5",
    journal = "Nucl. Phys. B",
    volume = "552",
    pages = "88--168",
    year = "1999"
}

@article{tHooft:1973alw,
    author = "'t Hooft, Gerard",
    editor = "Taylor, J. C.",
    title = "{A Planar Diagram Theory for Strong Interactions}",
    reportNumber = "CERN-TH-1786",
    doi = "10.1016/0550-3213(74)90154-0",
    journal = "Nucl. Phys. B",
    volume = "72",
    pages = "461",
    year = "1974"
}

@article{Farhi:1977sg,
    author  = {Farhi, Edward},
    title   = {A QCD Test for Jets},
    journal = {Phys. Rev. Lett.},
    volume  = {39},
    pages   = {1587--1588},
    year    = {1977},
    doi     = {10.1103/PhysRevLett.39.1587}
}

@book{Schmudgen2017,
  title={{The Moment Problem}},
  author={Schm{\"u}dgen, Konrad},
  volume={9},
  year={2017},
  publisher={Springer},
  doi={10.1007/978-3-319-64546-9}
}

@article{Korchemsky:2021okt,
    author = "Korchemsky, Gregory P. and Sokatchev, Emery and Zhiboedov, Alexander",
    title = "{Generalizing event shapes: in search of lost collider time}",
    eprint = "2106.14899",
    archivePrefix = "arXiv",
    primaryClass = "hep-th",
    reportNumber = "CERN-TH-2021-090, IPhT--T21/031, IPhT{\textendash}T21/031, LAPTH-022/21",
    doi = "10.1007/JHEP08(2022)188",
    journal = "JHEP",
    volume = "08",
    pages = "188",
    year = "2022"
}

@article{Belin:2020lsr,
    author = "Belin, Alexandre and Hofman, Diego M. and Mathys, Gr{\'e}goire and Walters, Matthew T.",
    title = "{On the stress tensor light-ray operator algebra}",
    eprint = "2011.13862",
    archivePrefix = "arXiv",
    primaryClass = "hep-th",
    reportNumber = "CERN-TH-2020-200",
    doi = "10.1007/JHEP05(2021)033",
    journal = "JHEP",
    volume = "05",
    pages = "033",
    year = "2021"
}

@article{Besken:2020snx,
    author = "Be{\c{s}}ken, Mert and De Boer, Jan and Mathys, Gr{\'e}goire",
    title = "{On local and integrated stress-tensor commutators}",
    eprint = "2012.15724",
    archivePrefix = "arXiv",
    primaryClass = "hep-th",
    doi = "10.1007/JHEP07(2021)148",
    journal = "JHEP",
    volume = "21",
    pages = "148",
    year = "2021"
}

@article{Gonzo:2020xza,
    author = "Gonzo, Riccardo and Pokraka, Andrzej",
    title = "{Light-ray operators, detectors and gravitational event shapes}",
    eprint = "2012.01406",
    archivePrefix = "arXiv",
    primaryClass = "hep-th",
    reportNumber = "SAGEX-20-24-E",
    doi = "10.1007/JHEP05(2021)015",
    journal = "JHEP",
    volume = "05",
    pages = "015",
    year = "2021"
}

@article{Korchemsky:2021htm,
    author = "Korchemsky, Gregory P. and Zhiboedov, Alexander",
    title = "{On the light-ray algebra in conformal field theories}",
    eprint = "2109.13269",
    archivePrefix = "arXiv",
    primaryClass = "hep-th",
    reportNumber = "CERN-TH-2021-135, IPhT-T21/061",
    doi = "10.1007/JHEP02(2022)140",
    journal = "JHEP",
    volume = "02",
    pages = "140",
    year = "2022"
}

@article{Hu:2023geb,
    author = "Hu, Yangrui and Pasterski, Sabrina",
    title = "{Detector operators for celestial symmetries}",
    eprint = "2307.16801",
    archivePrefix = "arXiv",
    primaryClass = "hep-th",
    doi = "10.1007/JHEP12(2023)035",
    journal = "JHEP",
    volume = "12",
    pages = "035",
    year = "2023"
}

@article{Himwich:2025ekg,
    author = "Himwich, Elizabeth and Pate, Monica",
    title = "{Light-ray Operators and the ${\rm w}_{1+\infty}$ Algebra}",
    eprint = "2512.18973",
    archivePrefix = "arXiv",
    primaryClass = "hep-th",
    month = "12",
    year = "2025"
}

@article{Strominger:2026yrh,
    author = "Strominger, Andrew and Wei, Hongji",
    title = "{Every CFT$_3$ has an $\mathcal{L}_{\Lambda} w_{1+\infty}$ symmetry}",
    eprint = "2603.26459",
    archivePrefix = "arXiv",
    primaryClass = "hep-th",
    year = "2026"
}

@article{Mecaj:2026kji,
    author = "Me{\c{c}}aj, Bianka and Moult, Ian and Stoffels, Hidde and Walters, Matthew T. and Xin, Yuan",
    title = "{The EEC-Hedron: Positivity Bounds on Energy Correlators}",
    eprint = "2608.19322",
    archivePrefix = "arXiv",
    primaryClass = "hep-th",
    year = "2026"
}

@article{Andres:2023xwr,
    author = "Andres, Carlota and Dominguez, Fabio and Holguin, Jack and Marquet, Cyrille and Moult, Ian",
    title = "{A coherent view of the quark-gluon plasma from energy correlators}",
    eprint = "2303.03413",
    archivePrefix = "arXiv",
    primaryClass = "hep-ph",
    doi = "10.1007/JHEP09(2023)088",
    journal = "JHEP",
    volume = "09",
    pages = "088",
    year = "2023"
}

@article{Yang:2023dwc,
    author = "Yang, Zhong and He, Yayun and Moult, Ian and Wang, Xin-Nian",
    title = "{Probing the Short-Distance Structure of the Quark-Gluon Plasma with Energy Correlators}",
    eprint = "2310.01500",
    archivePrefix = "arXiv",
    primaryClass = "hep-ph",
    doi = "10.1103/PhysRevLett.132.011901",
    journal = "Phys. Rev. Lett.",
    volume = "132",
    number = "1",
    pages = "011901",
    year = "2024"
}

@article{Barata:2023bhh,
    author = "Barata, Jo{\~a}o and Caucal, Paul and Soto-Ontoso, Alba and Szafron, Robert",
    title = "{Advancing the understanding of energy-energy correlators in heavy-ion collisions}",
    eprint = "2312.12527",
    archivePrefix = "arXiv",
    primaryClass = "hep-ph",
    reportNumber = "CERN-TH-2023-243",
    doi = "10.1007/JHEP11(2024)060",
    journal = "JHEP",
    volume = "11",
    pages = "060",
    year = "2024"
}

@article{Andres:2024ksi,
    author = "Andres, Carlota and Dominguez, Fabio and Holguin, Jack and Marquet, Cyrille and Moult, Ian",
    title = "{Towards an interpretation of the first measurements of energy correlators in the quark-gluon plasma}",
    eprint = "2407.07936",
    archivePrefix = "arXiv",
    primaryClass = "hep-ph",
    doi = "10.1007/JHEP03(2025)166",
    journal = "JHEP",
    volume = "03",
    pages = "166",
    year = "2025"
}

@article{CMS:2025ydi,
    author = "Chekhovsky, Vladimir and others",
    collaboration = "CMS",
    title = "{Observation of nuclear modification of energy-energy correlators inside jets in heavy ion collisions}",
    eprint = "2503.19993",
    archivePrefix = "arXiv",
    primaryClass = "nucl-ex",
    reportNumber = "CMS-HIN-23-004, CERN-EP-2025-014",
    doi = "10.1016/j.physletb.2025.139556",
    journal = "Phys. Lett. B",
    volume = "866",
    pages = "139556",
    year = "2025"
}

@article{Barata:2025fzd,
    author = "Barata, Jo{\~a}o and Moult, Ian and Sadofyev, V., Andrey and Silva, Jo{\~a}o M.",
    title = "{Dissecting Jet Modification in the Quark-Gluon Plasma with Multipoint Energy Correlators}",
    eprint = "2503.13603",
    archivePrefix = "arXiv",
    primaryClass = "hep-ph",
    reportNumber = "CERN-TH-2025-029",
    doi = "10.1103/9jzc-7jcv",
    journal = "Phys. Rev. Lett.",
    volume = "137",
    number = "5",
    pages = "052302",
    year = "2026"
}

\end{document}